\documentclass[a4paper,12pt]{report}
\usepackage{amsmath,amssymb}
\usepackage{graphicx}
\usepackage{color}
\usepackage{mathrsfs}
\usepackage{bm}
\usepackage{multirow}
\usepackage{booktabs}
\usepackage{cite}

\newcommand{\Part}[3]{ \frac{ \partial^{#3} #1 }{ \partial #2^{#3} } }
\newcommand{\V}[1]{\bm{#1} } 
\newcommand{\Tr}[1]{ \mathop{\rm Tr}_{ #1 } }
\newcommand{\bra}[1]{\langle {#1}| } 
\newcommand{\ket}[1]{|{#1}\rangle } 
\newcommand{\braket}[2]{\langle {#1}|{#2}\rangle }
\newcommand{\Ave}[1]{\left\langle {#1} \right\rangle} 
\newcommand{\Wt}[1]{ {\widetilde {#1}} } 
\newcommand{\Wh}[1]{\widehat {#1}} 
\newcommand{\GI}[1]{\int\hspace{-1.mm}D{#1}} 
\newcommand{\sgn}[1]{{\rm sgn}\left({#1} \right)} 
\newcommand{\llsim}{\ $\raisebox{-.7ex}{$\stackrel{\textstyle <}{\sim}$}$\,\ }
\newcommand{\gsim}{\ $\raisebox{-.7ex}{$\stackrel{\textstyle >}{\sim}$}$\,\ }
\newcommand{\Extr}[1]{ \mathop{\rm Extr}_{ #1 } }
\newcommand{\Abs}[1]{ \left| #1 \right| }
\newcommand{\mC}{\mathbb{C}}
\newcommand{\mR}{\mathbb{R}}
\newcommand{\mN}{\mathbb{N}}

\title{{\Large Thesis} \vspace*{2cm} \\
{\Large \bf 
Role of the Finite Replica Analysis in 
the Mean-Field Theory of Spin Glasses 
}}
\author{\vspace*{5cm} \\
{\Large \bf Tomoyuki Obuchi} \vspace*{2cm} \\
{\Large Department of Physics, Tokyo Institute of Technology}}
\date{\vspace*{1.5cm} \Large February, 2010}

\begin{document}
\setlength{\baselineskip}{16pt}
\maketitle
\begin{abstract}
In this thesis, we review and examine the replica method from several viewpoints. The replica method is a mathematical technique to calculate general moments of stochastic variables. 
This method provides a systematic way to evaluate physical quantities and 
becomes one of the most important tools in the theory of spin glasses and 
in the related discipline including information processing tasks.

In spite of the effectiveness of the replica method, it is known that several 
problems exist in the procedures of the method itself.
The replica symmetry breaking is the central topic of those problems and is the main issue of this thesis. 
To elucidate this point, we first review the 
recent progress about the replica symmetry breaking including its physical 
and mathematical descriptions in detail.

Second, 
we accept the recent descriptions of the replica symmetry breaking 
and investigate the Ising perceptron, which 
is a model of a neuron, 
by intensively utilizing those descriptions, 
to examine the consequence of the replica symmetry breaking. 
Our result is consistent to the replica 
symmetry breaking description, but shows that the replica method cannot 
correctly describe the phase-space structure of this system. 
This is because the phase space of this system constitutes an exceptional 
structure which violates the usual assumption of the replica method, but
it is also explained that this problem is not directly related to the replica symmetry breaking itself.

Next, without employing the replica symmetry breaking, 
we investigate the zeros of the averaged $n$th moment of the partition function with respect to the complex replica number $n$ for the purpose of 
investigating the analyticity 
 breaking of the generating function, which is related to the replica symmetry 
breaking, appearing in the replica procedures. 
For this purpose, we employ the $\pm J$ model on 
some tree systems and ladders. Our result implies that 
the zeros actually signal a certain kind of analyticity breaking, 
but is irrelevant to any replica symmetry breaking. 

To investigate the origin of the irrelevance of zeros of the 
$n$th moment to the replica symmetry breaking, 
we finally study the partition function zeros with respect to 
the external field and temperature for the tree systems. 
The result shows that the two dimensional part of the zeros density 
continuously touches the real axis in a certain range of the objective 
parameters. 
This result implies that 
the invisibility of the replica symmetry breaking by the previous formulation 
is not due to the peculiarity of the tree systems but due to 
the problem of the formulation itself. 
Another interesting implication of the result  
 is that it leads to the instability of the system 
against deviations with respect to the external field and temperature 
in the spin-glass phase.
This is the first direct evidence that the system shows singular behavior 
in the spin-glass phase.

The above results reveal several aspects and problems of the replica method 
and contribute to underpinning this method and the mean-field theory of 
spin glasses, which consequently inspire 
future study and understanding of spin glasses and the related topics.

\end{abstract}

\setcounter{page}{3}
\chapter*{Acknowledgment}
First of all, I would like to express my sincerest gratitude to 
my supervisor, Professor Hidetoshi Nishimori for
his guidance, continuous encouragements and spiritual support.
I also greatly acknowledge Professor Yoshiyuki Kabashima for his guidance, fruitful discussions and collaborations on my studies. 

I express my gratitude to the members of the Nishimori group and the condensed matter theory group of Tokyo Institute of Technology 
for stimulating discussions and warm atmosphere by them.
Especially, Yoshiki Matsuda kindly taught me some technical details of numerical calculations and I am particularly appreciative of him.

I also thank Doctor Antonello Scardicchio, 
Doctor Markus M\"uller, and Professor Federico Ricci-Tersenghi 
for fruitful discussions and kindness when I visited Italy for research 
collaborations. 
Professor Eric Vincent kindly taught me some
experimental properties of spin glasses and discussed my studies. 
I am also grateful to him. 

This work was 
supported by the Japan Society for the Promotion of Science (JSPS) Research Fellowship for Young Scientists program, 
CREST(JST), the 21th Century COE Program `Nanometer-Scale Quantum Physics' 
and the Global COE Program `Nanoscience and Quantum Physics'
at Tokyo Institute of Technology, and by
the Grant-in-Aid for Scientific Research on
the Priority Area ``Deepening and Expansion of Statistical Mechanical
Informatics'' by  the Ministry of Education, Culture, Sports, Science
and Technology.
A portion of the numerical
calculations have been performed on the TSUBAME Grid Cluster at the Global Scientific Information and Computing Center (GSIC), Tokyo Institute of Technology.

Finally, I thank my family for spiritual support.

\tableofcontents
\chapter{Introduction}
Random spin systems are one of the major research subjects 
in statistical physics. 
The spatial randomness of conflicting interactions 
introduces random frustration into the system. 
The frustration can produce many metastable states 
and the system becomes affected by such states.
If the structure of the metastable states is highly complex, 
the system shows strange behaviors like the extraordinary 
susceptibility against external fields, the extremely slow dynamics,
 and so on. 
Spin systems with such behaviors are called spin glasses.  

At present time, 
one of the most important progresses of the theory of spin glasses is its 
mean-field description. 
The mean-field theory is constructed by several mathematical tools. 
The replica method is one of the most powerful tools and 
the main purpose of this thesis is to review and examine this method.
Because of the intricacy of the replica theory, there still remain
 many problems in the theory itself. 
We review recent progress about this topic and 
provide some new approaches to reveal the mysteries related to 
the replica method. 
Before going into the details, we start with reviewing histories and 
topics on spin glasses and related subjects.

\section{Spin glasses}\label{sec1:spinglass}
A simple description of spin glasses is by 
its low temperature state.
For ferro and anti-ferro magnets, the low temperature states are uniform or periodic, but those of spin glasses are spatially disordered and frozen ones. 
The existence of such states usually needs two requirements;
one is the existence of competing interactions among spins which 
leads to no single configuration favored by all the interactions 
(this is called the frustration). 
Second is the spatial randomness of the interactions, 
which is necessary for disordered configurations stable at 
low temperatures. 

Experimentally, materials satisfying the above two requirements are 
found
in several alloys consisting of noble metals (Au, Ag, Cu, etc.) weakly 
diluted with transition metals (Fe, Mn, etc.).
Interactions between the moments of the impurity atoms (spins) 
effectively appear through the interactions with 
the conducting electrons.  
The effective interaction is called the RKKY interaction 
\cite{Rude,Yosh}
\begin{equation}
J_{ij} \propto \frac{1}{r_{ij}^3}\cos(2k_{F}r_{ij}),
\end{equation}
where $r_{ij}$ is the distance between two 
impurity atoms and $k_{F}$ is the Fermi wave number.
The RKKY interaction takes both positive and negative values
of $J_{ij}$ due to randomness in the position of impurities 
and random frustration is introduced between the localized spins.

One of the important problems is 
how to characterize these disordered frozen states.
To characterize a ferromagnetic state, the averaged 
magnetization $m=(\sum_{i}\Ave{S_{i}})/N$ is sufficient, 
where the brackets $\Ave{\cdots}$ 
denote 
the thermal average over the Boltzmann factor 
$ \exp(-\beta \mathscr{H})$ and $i$ denotes the site index, and 
$\beta$ is the inverse temperature $\beta=1/T$, where we put 
the Boltzmann constant to unity $k_{B}=1$.
Meanwhile, to precisely describe the disordered frozen states,  
more detailed information 
about the local spontaneous magnetization 
$m_{i}=\Ave{S_{i}}$ is required. 
If the spin-glass phase exists, 
the characteristic property that spins 
randomly freeze should be reflected in that phase. 
The local magnetization
$m_{i}=\Ave{S_{i}}$ is expected to take a 
non-zero value in such a situation 
but the sign is randomly
distributed from site to site, which implies that 
the global magnetization vanishes
$m=(\sum_{i}\Ave{S_{i}})/N=0$. 
Hence, to detect the spin-glass ordering, 
the square value of the local magnetization $m_{i}^{2}$ is useful.
This speculation leads to an order parameter $q$ defined as
\begin{equation}
q=\frac{1}{N}\sum_{i}m_{i}^2.
\end{equation}
This quantity $q$ was
first introduced by Edwards and Anderson \cite{Edwa}
 and is called the spin-glass order parameter. 
The disordered frozen phase at low temperatures 
is called the spin-glass phase and 
is characterized by $m=0, q>0$, while the paramagnetic phase has $m=q=0$ and 
the ferromagnetic phase is characterized by $m,q>0$.

\begin{figure}[htbp]
\hspace{-2mm}
\begin{minipage}{0.48\hsize}
\begin{center}
\includegraphics[height=80mm,width=80mm]{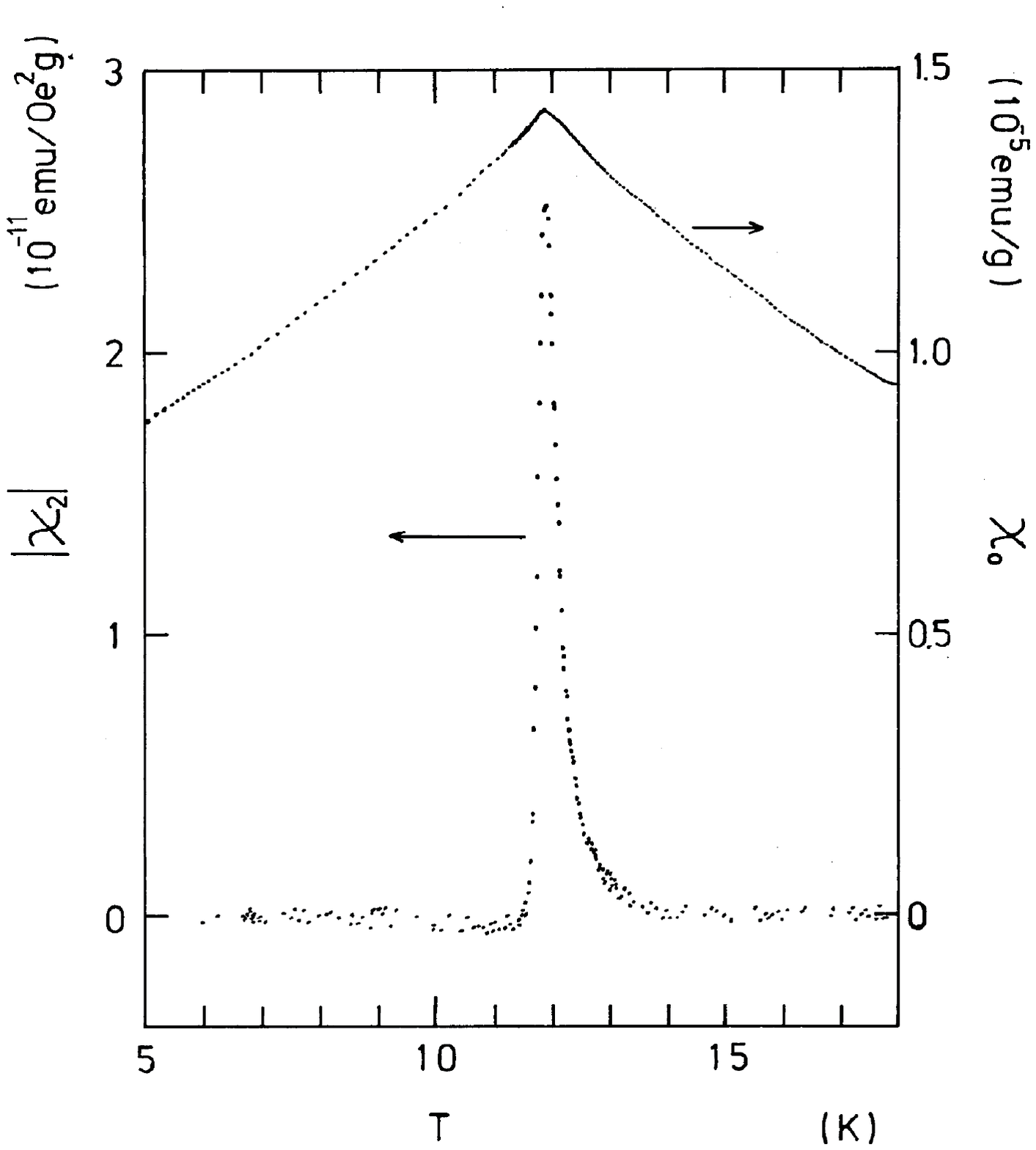}
 \caption{Temperature dependence of the linear 
$\chi_{0}=\left. \left(\partial m/\partial H \right) \right|_{H=0}$ and nonlinear 
$\chi_{2}=\chi_{\rm nl}$ 
susceptibilities for AuFe (1.5-at. \% Fe). A cusp and divergence are observed 
in $\chi_{0}$ and $\chi_{2}$, respectively, at a critical temperature $T_{c}$. From \cite{Tani}.}
 \label{fig:chinl}
\end{center}
\end{minipage}
\hspace{4mm}
 \begin{minipage}{0.48\hsize}
\begin{center}
   \includegraphics[height=60mm,width=70mm]{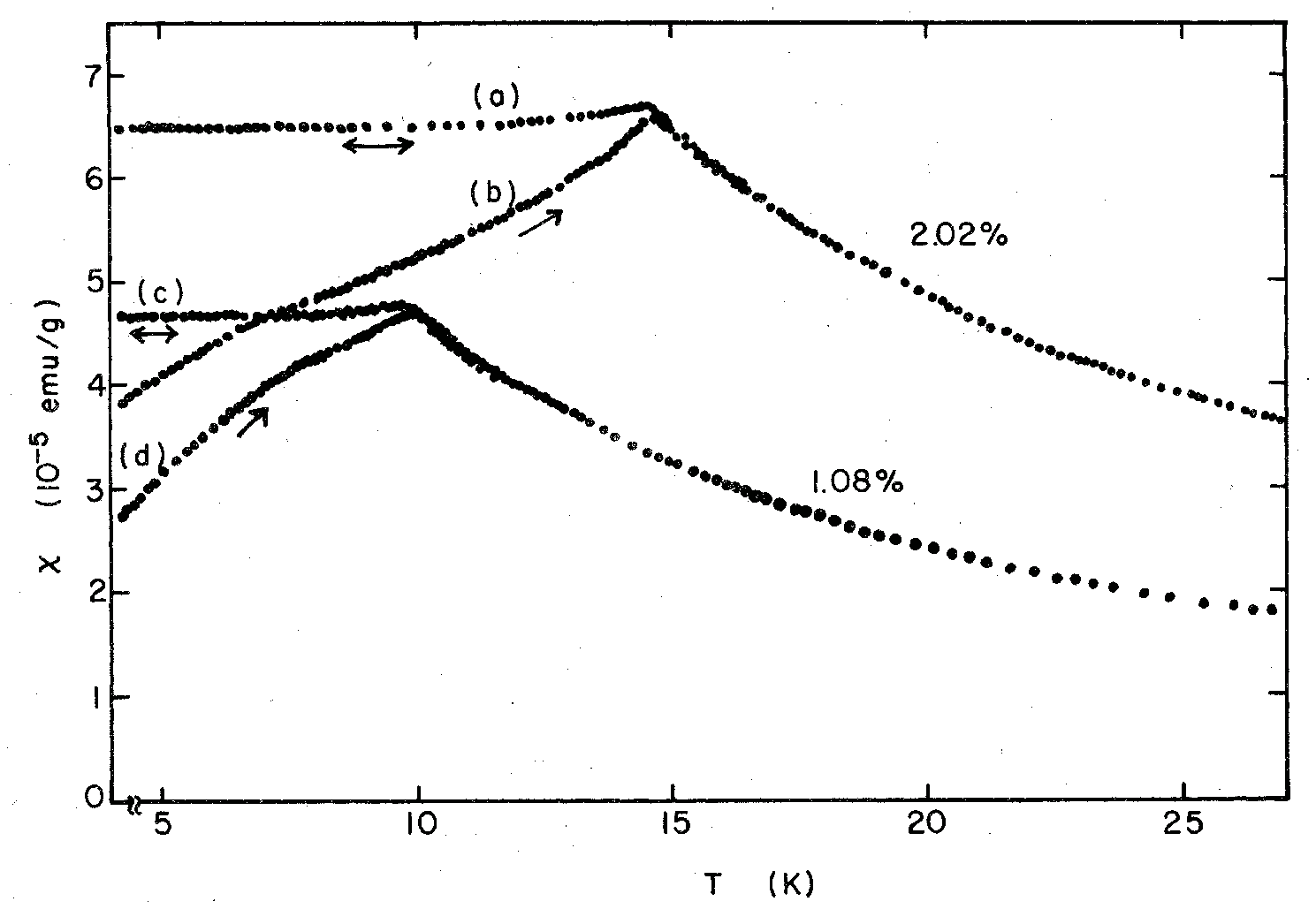}
 \caption{Linear susceptibility $\chi$ of CuMn vs temperature 
for 1.08-, 2.02-at. \% Mn.
After zero-field cooling ($H<0.05$G), 
the susceptibilities in a field $H=5.90$G
(b) and (d) increase 
as the temperature grows,
while 
the susceptibilities (a) and 
(c) in the field $H=5.90$G, which 
applied above the critical temperature $T_{c}$ before cooling, 
are almost constants. From \cite{Naga}.}
 \label{fig:chiaging}
\end{center}
 \end{minipage}
\end{figure}

In a ferromagnet, the ferromagnetic phase transition is accompanied by 
a rapid increase of the spin-correlation length which diverges at 
the critical temperature. 
This can be observed in the divergence of the linear susceptibility $\chi$
\begin{equation}
\frac{\chi}{\beta}=\Part{m(H)}{(\beta H)}{}=
\frac{1}{N}\sum_{i,j}\Part{m_{i}}{(\beta h_{j})}{}=
\frac{1}{N}\sum_{i,j}\left(
\Ave{S_{i}S_{j}}-\Ave{S_{i}}\Ave{S_{j}}
\right)
=
\frac{1}{N}\sum_{i,j}\chi_{ij},
\end{equation}
where $h_{j}$ denotes the local field 
applied to the site $j$.
For the spin-glass problem we can also expect 
a similar phenomenon coming 
from the spin-glass ordering, but the simple spin-correlation function 
$\chi_{ij}$
does not show the long range correlation 
because of the frustration. 
Instead, the square of $\chi_{ij}$ can diverge at the critical temperature 
of the spin-glass ordering, which leads to the   
spin-glass susceptibility $\chi_{SG}$ the definition 
of which is given by
\begin{equation}
\chi_{SG}=
\frac{1}{N}\sum_{i,j}
\left(
\chi_{ij}
\right)^2
=
\frac{1}{N}\sum_{i,j}
\left(
\Part{m_{i}}{(\beta h_{j})}{}
\right)^2
=
\frac{1}{N}
\sum_{i,j}
\left(
\Ave{S_{i}S_{j}}-\Ave{S_{i}}\Ave{S_{j}}
\right)^2.
\end{equation}
Although the spin-glass susceptibility $\chi_{SG}$ 
is a suitable quantity to detect 
the spin-glass ordering,
it cannot be directly observed in experiments.
A more tractable quantity in experiments is the nonlinear susceptibility 
whose definition and the relation with $\chi_{SG}$ is given as 
\begin{equation}
\chi_{\rm nl}\equiv
-\left.
\Part{m(H)}{H}{3}
\right|_{H=0}
=
\beta^3
\left(\chi_{SG}-\frac{2}{3}
\right),\label{eq1:chinl}
\end{equation}
where $H$ is the uniform external field. 
This equation shows that the divergence of $\chi_{SG}$ 
also leads to the divergence of $\chi_{\rm nl}$.
For example, a plot of $\chi_{\rm nl}$ for AuFe near the transition point 
is given in fig.\ \ref{fig:chinl}.

Another characteristic property of spin glasses is the strong 
hysteresis phenomena. 
We depict an example of the linear susceptibility of CuMn alloys 
in fig.\ \ref{fig:chiaging}.
To obtain this figure, the linear susceptibility was observed 
in two different situations. 
One is the so-called zero-field cooling. In this situation,  
the system is first cooled down below the critical temperature $T_c$
in the absence of the external field. 
After that, a constant field ($H=5.90$G in this experiment) 
is applied and the linear susceptibility is observed at 
several temperatures with increasing temperature.
The other is the field cooling, in which 
a constant field ($H=5.90$G) is applied to the system above $T_{c}$ and 
the linear susceptibility is observed at several temperatures 
with decreasing temperature. 
Figure \ref{fig:chiaging} shows clear differences 
in those two situations, which implies the ergodicity breaking of 
the spin-glass system at low temperatures.
The ergodicity breaking means that the system cannot 
explore all the states and the state of the system 
even after a long time depends on 
the initial condition. 
Once the spin-glass transition occurs, 
the ergodicity breaks and the system shows the strong dependence 
on the initial condition, which can be observed in the hysteresis of the 
linear susceptibility as fig.\ \ref{fig:chiaging}.
This interesting behavior of spin glasses 
strongly motivated the researchers in the 1970-80's,
and theoretical interpretations were intensively explored.
In the next section,
we will briefly see the mean-field description of spin glasses.
\section{Mean-field theory}
In the previous section, 
we have seen some concepts and quantities to characterize the spin-glass
behaviors, and some
experimental observations were provided. 
The next step for understanding 
spin glasses is to define a model of spin glasses to deal with 
the problem in theoretical ways.
The so-called EA model introduced by Edwards and Anderson 
\cite{Edwa} is the first model of spin glasses.
In this model,
the interactions are introduced between the nearest neighbor spins and 
are assumed to be the random variables drawn from an 
identical independent distribution. 
The EA Hamiltonian takes the following form 
\begin{equation}
\mathscr{H}=-\sum_{\Ave{i,j}}J_{ij}S_{i}S_{j},
\end{equation}
where $S$ denotes an Ising variable and $J_{ij}$ is the 
random interaction. The symbol $\Ave{i,j}$ means that the summation 
runs over all the pairs of neighboring spins. 
Edwards and Anderson solved this model within a 
mean-field scheme, but the detailed properties were unclear.
Then, to investigate the EA model in more detail, Sherrington
and Kirkpatrick introduced the infinite-range version of the EA model, which 
is called the SK model \cite{Sher}.  
The SK Hamiltonian is
\begin{equation}
\mathscr{H}=-\sum_{i<j} J_{ij}S_{i}S_{j},
\end{equation}
where the summation runs over all pairs.
The SK model is more tractable than the EA model 
because all the spins are fully connected and are symmetric.
By analyses of the SK model, it was shown that a second order phase
transition occurs and the spin-glass phase
appears at a finite temperature $T_{c}$.   
In the SK model, the cusp of the linear susceptibility and 
the divergence of the spin-glass susceptibility are observed, 
and some aging phenomena are also confirmed by some analytical and 
numerical results 
\cite{Kirk,Somp,Petr}. 
Because of this good agreement with experimental results, 
it is considered that 
the mean-field model extracts some essence of spin-glass behaviors 
and nowadays the SK model is regarded as a standard model of spin glasses. 
 
An especially remarkable property of the SK model is the coexistence of 
a huge number of thermodynamical states.
These states are called ``pure states" and 
the properties are considered as the essence of the 
mean-field description of spin glasses. 
In the next section, the notion of 
pure states and some related properties are explained.
\section{Pure state and its implications}\label{sec1:pure}
\subsection{A simple introduction to the concept of pure states}
A pure state is a somewhat abstract concept and represents 
a certain unit of the state of the system in the thermodynamic limit.
To explain this concept, let us first 
review a pure ferromagnetic system as an example. In a ferromagnetic system, there are two possible thermodynamic states at low temperatures. The first one is the state for which the magnetization is positive, 
and for the other one the magnetization is negative. 
If there is no element favoring one of signs of the magnetization 
(i.e. boundary condition, magnetic field etc.), 
the partition function of the system includes both contributions equally and we can express this statement symbolically
\begin{equation}
Z=Z_{+}+Z{-}, \,\, (Z_{+}=Z_{-}),\label{eq1:unZ}
\end{equation}
where $Z_{+}$ is the summation of $e^{-\beta H}$ over the spin configurations for which the magnetization is positive, and similarly for $Z_{-}$.
However, when a perturbation favoring a direction of the magnetization is 
applied to the system, e.g. the positive magnetic field, 
the value of $Z_{+}$ exceeds $Z_{-}$. 
This difference becomes larger and larger 
in the thermodynamic limit even when 
the magnetic field is a positive infinitesimal,
and the partition function converges to 
\begin{equation}
Z \to Z_{+}.
\end{equation}
If the magnetic field is negative, 
the partition function $Z$ converges 
to $Z_{-}$.
These facts indicate that there are two distinguishable states 
in the thermodynamic limit.
These distinguishable states are the `pure states', 
and the number of pure states for this system is two.

Meanwhile, 
the present mean-field theory of spin glasses, 
which was mainly constructed by Sherrington and Kirkpatrick \cite{Sher} and 
solved by Parisi \cite{Pari1,Pari2},
tells that the number of pure states of spin glasses
increases exponentially as the system size grows. 
Let us assume that an index $\gamma$ denotes a pure state and 
the corresponding partition function is expressed by $Z_{\gamma}$. 
Each pure state has its own free energy $f_{\gamma}=-T\log Z_{\gamma}$. 
In the large system limit $N\to \infty$,
the mean-field description tells us that the number of pure states with the free energy value $f$, $\mathscr{N}(f)$, is scaled as 
\begin{equation}
\mathscr{N}(f)=\sum_{\gamma} \delta(f-f_{\gamma})\sim e^{N\Sigma (f)},\label{eq1:complexity}
\end{equation}
where the characteristic exponent $\Sigma (f)$ 
is called the `complexity' or `configurational entropy'. 
For the ferromagnetic case, the unbiased partition function (\ref{eq1:unZ}) includes two pure states, $Z_{+}$ and $Z_{-}$, 
and both of them give the same value of the free energy. For spin-glass systems, however, each pure state can yield a different value of the free energy. From a naive speculation of thermodynamics, it seems that the actual system prefers the lowest free energy state than the other states with higher values of free energy, but it is not the case.
A reasonable explanation to this fact is to remember that 
each pure state cannot access each other because of infinitely high free energy barrier. 
This is the same situation as in the ferromagnetic case. Thus, 
when the system is once stuck in a pure state, 
the system never reaches other pure states without external forces. In the statistical mechanical model, all of such states are included in the partition function, and both the free energy value and the number of pure states are important. For general spin-glass systems, using eq.\ (\ref{eq1:complexity}), 
we can write the partition function as
\begin{equation}
Z=\sum_{\gamma}Z_{\gamma}
=\sum_{\gamma}e^{-N\beta f_{\gamma}}
=\int df \mathscr{N}(f)e^{-N\beta f}
\sim \int df e^{N(-\beta f +\Sigma (f))}.\label{eq:Z}
\end{equation}
This equation indicates that the equilibrium value of free energy 
$-\beta f_{\rm eq}=\lim_{N\to \infty} \log Z/N$ 
 is equal to the saddle point value $\max_{f}\{-\beta f+\Sigma(f) \}$. 
This relation is quite similar to the conventional microcanonical one
$-\beta f=\max_{u}\{-\beta u+s(u)\}$, where $u$ is the internal energy and 
$s(u)$ is the corresponding entropy, and 
implies that 
the complexity $\Sigma(f)$ plays the role as 
the effective `entropy' for the `internal' 
free energy $f$.

Before ending this subsection, 
we note that the notion of the pure state is different from that of the 
metastable state. 
Intuitively, metastable states are simply local minima of the free energy 
in the phase space.
The system with many metastable states shows the 
slow dynamics at low temperatures but it ultimately relaxes to 
its equilibrium state because the metastable states 
are not completely separated by free energy barriers and 
can be accessed each other in a sufficiently long term. 
On the other hand, pure states are also minima but they are 
separated by infinitely high free energy barriers. 
The system with many pure states also shows slow relaxation but 
never reaches its equilibrium state due to those 
infinitely high free energy barriers.
Hence, pure states are very stable in comparison with simple metastable states,
and the discrimination of pure and metastable states is the key concept of 
the mean-field theory of spin glasses.

However, when we perform real and numerical experiments, we can 
only treat finite size systems in a finite time scale.  
This leads to a difficulty in identifying the origin of the slow dynamics of 
spin glasses, because the phase spaces are never completely separated in finite size systems and 
the observation in a finite time scale cannot reject the possibility 
that the observed slow dynamics is due to the shortness of the observing time.
Hence, we cannot easily distinguish pure and metastable states.
This difficulty is an essential problem for studying spin glasses. To support 
the mean-field description, we should examine more detailed properties 
predicted from the mean-field theory. 
In the following subsections, we present some numerical and real experiments 
supporting the mean-field description. 

\subsection{Relations between pure states and dynamical behaviors}
Existence of many pure states directly explains the ergodicity breaking of 
spin glasses. 
At high temperatures, the system has only one pure state and the system 
relaxes to the pure state without depending on the initial state.
Below the critical temperature, however, 
the system goes into the spin-glass phase and many pure states appear. 
Since each pure state has different properties (order parameters) 
and is separated 
by infinitely high free energy barriers, 
once the system is stuck in one of the pure states, 
properties of the system is determined only by that state 
and the system never reaches other states without external perturbations. 

\begin{figure}[htbp]
\hspace{-2mm}
\begin{minipage}{0.48\hsize}
\begin{center}
\includegraphics[height=50mm,width=60mm]{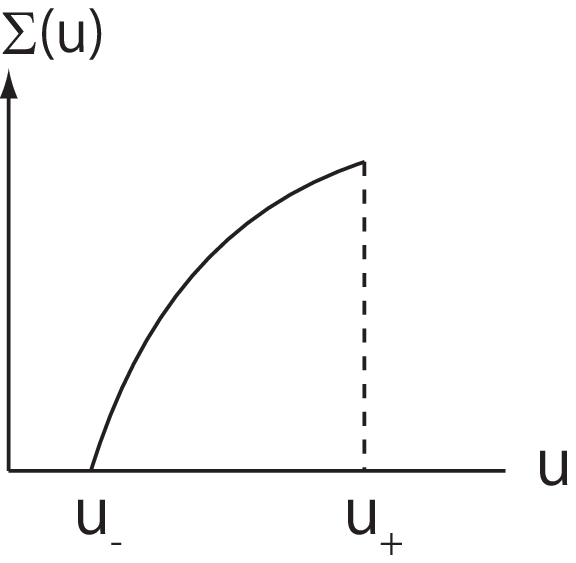}
 \caption{The complexity at zero temperature for the typical spin-glass system. 
The complexity $\Sigma(u)$ takes the maximum at the highest energy $u_{+}$, 
which implies that the pure states at $u_{+}$ dominate the dynamical behavior 
of the system.}
 \label{fig:typcomp}
\end{center}
\end{minipage}
\hspace{4mm}
 \begin{minipage}{0.48\hsize}
\begin{center}
   \includegraphics[height=50mm,width=60mm]{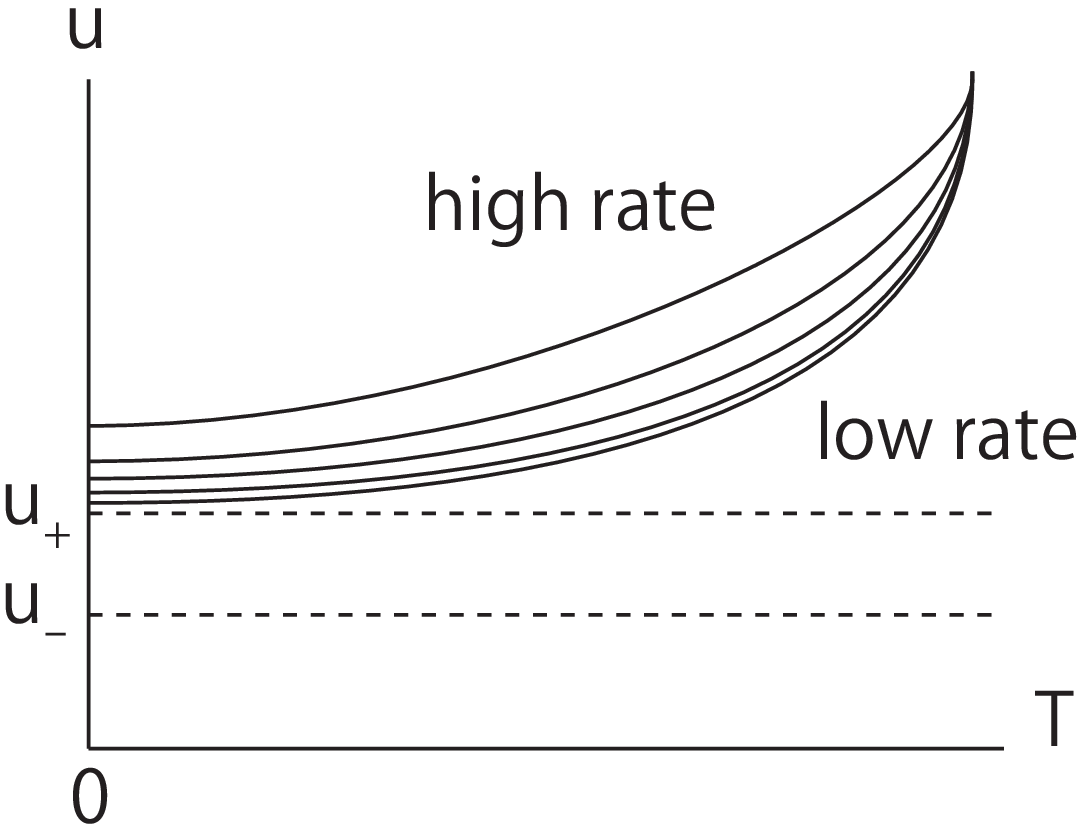}
 \caption{Schematic diagram of the internal energy 
for a temperature range in the simulated annealing. 
The actual ground-state energy 
is given by $u_{-}$ but the system cannot reach the state, since many pure states with higher energies exist. Consequently, 
in decreasing the cooling rate $v_{c}$,  
the energy seems to converge the highest energy $u_{+}$ 
where pure states exist.
}
 \label{fig:SA2}
\end{center}
 \end{minipage}
\end{figure}

This description can be examined by some numerical experiments 
for the theoretical models of spin glasses. 
We here elucidate a simple example of such experiments. 
To this end, let us consider an annealing experiment of glassy and 
non-glassy spin systems. 
Such numerical experiments are called simulated annealing.
The actual procedures are as follows:
\begin{enumerate}
\item{Set the temperature so high that the system is in the equilibrium state.}
\item{Decrease the temperature to zero gradually.}
\item{Observe the energy of the system at zero temperature.}
\end{enumerate}
When the decreasing rate $v_{c}$ is too large, 
the system cannot follow the equilibrium state and 
cannot reach the ground state at zero temperature. 
On the other hand, if the decreasing rate $v_{c}$ 
is sufficiently small, we can find that 
a typical non-glassy spin system is always in the equilibrium state 
and finally reaches the ground state at zero temperature.
For a typical spin-glass system, however, this procedure does not work 
and the system cannot reach the ground state. 
In an idealized situation (the system is infinitely large), 
this is the case even in the slow cooling limit $v_{c}\to 0$.
An important point is that 
we can qualitatively predict the value of the energy which
the spin-glass system takes in the slow cooling limit $v_{c}\to 0$, 
by accepting the description of pure states.
For the typical spin-glass system,
the complexity $\Sigma(f)$ takes finite values in a range 
$f_{-}\leq f \leq f_{+}$, and can be assessed in the statistical mechanical 
framework without any considerations about the dynamics. 
In the zero temperature limit, contributions from entropy vanish 
and the complexity becomes a function of the 
energy $u$, $\Sigma(u)$, defined in a range $u_{-}\leq u \leq u_{+}$.
The typical functional form of $\Sigma (u)$ is given in 
fig.\ \ref{fig:typcomp}.  
As in fig.\ \ref{fig:typcomp}, 
the complexity usually takes its maximum at the maximum of the energy range,
which means that it is highly possible that the system is trapped by the 
pure states with the energy $u_{+}$. Actually, 
for the annealing experiments conducted in some infinite range models,
it is observed that the energy which the system finally takes 
is equal to $u_{+}$ as in fig.\ \ref{fig:SA2} \cite{MontEPJ2003,MontPRB2004}.
This accordance in the qualitative level 
between the analysis of pure states and the dynamical behavior in annealing experiments strongly supports the validation of 
the present mean-field theory of spin glasses in a certain kind of spin-glass 
models.

\subsection{Hierarchical structure of pure states in the mean-field model}
\begin{figure}[htbp]
\begin{center}
   \includegraphics[height=46mm,width=100mm]{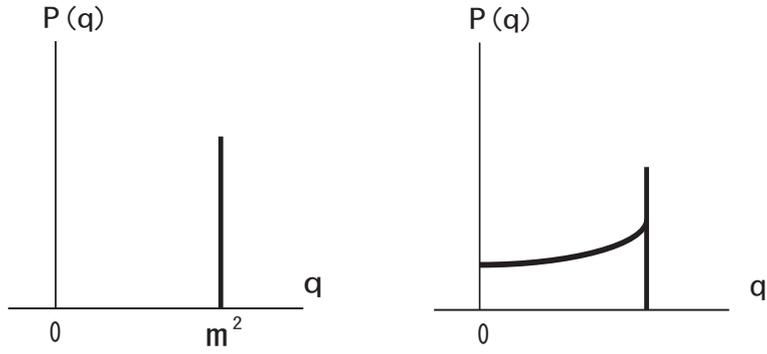}
 \caption{Schematic diagram of $P(q)$ for normal ferromagnetic 
systems (left)
 and the SK model (right).}
 \label{fig:Pofq}
\end{center}
\end{figure}

\begin{figure}[htbp]
\begin{center}
   \includegraphics[height=70mm,width=80mm]{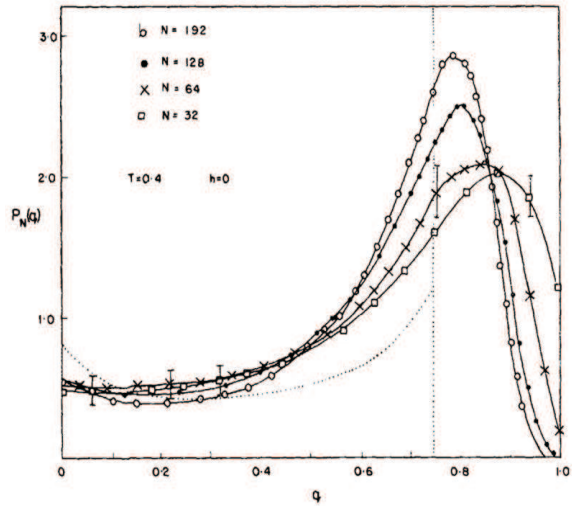}
 \caption{The distribution of spin-glass order parameter $P_{N}(q)$ for finite 
$N$ of the SK model with $T=0.4$ and zero external field $h=0$. 
The dotted line is the analytical prediction given in \cite{Pari1,Pari2}. 
From \cite{Youn1}.}
 \label{fig:Pofqnum}
\end{center}
\end{figure}

A fact that many pure states coexist in a system 
also implies that 
the values of a certain kind of order parameters can vary 
from sample to sample of materials, or from trial to trial of 
experiments. 
To precisely explain this matter, 
let us consider a situation that 
we have two replicated spin-glass systems, 
where these systems share the same interactions $\{J_{ij}\}$ but the 
states of spins independently vary, 
and 
compare the behaviors of these systems. 
This situation is unfeasible in real experiments 
but is easily realizable in numerical experiments.
In this situation, the following quantity, 
which observes the mutual overlap between the two 
replicated systems labeled by $(1)$ and $(2)$, 
becomes important for revealing the emergence of many pure states
\begin{equation}
q^{12}=\frac{1}{N}\sum_{i}^{N}S_{i}^{(1)}S_{i}^{(2)},
\end{equation}
where $S_{i}^{(k)}$ denotes the $i$th spin of the system $(k)$.
If there is only one pure state, the systems $(1)$ and $(2)$ 
are in the same pure state at equilibrium and 
$q^{12}$ always takes the value independent of 
the initial conditions and trials of experiments.
Hence, 
the typical distribution of $q^{12}$, 
$P(q)=[\Ave{\delta(q-q^{12})} ]_{\V{J}}$ where the brackets 
$[\cdots]_{\V{J}}$ denote the average over the random interactions $\V{J}$ and 
are called the configurational average, 
shows a delta function 
$P(q)=\delta(q-q_{0})$ with a certain average value $q_{0}$.
However, if there are many pure states, 
each system will be trapped in a different state since the dynamics 
of the systems $(1)$ and $(2)$ are independent each other. 
Hence, $q^{12}$ takes a different value at every  
experiment and 
the distribution $P(q)$ becomes a broad function. 
Interestingly, this $P(q)$ can be analytically assessed 
in the mean-field theory \cite{Pari1,Pari2} 
and the schematic diagram of $P(q)$ is presented 
in fig.\ \ref{fig:Pofq}.
Of course, the distribution $P(q)$ can also be assessed in 
numerical experiments and 
we refer to a numerical result in fig.\ \ref{fig:Pofqnum}.
The correspondence between the analytical and numerical results is good 
and this fact justifies the present 
mean-field solution for the SK model.

\begin{figure}[htbp]
\begin{center}
   \includegraphics[height=50mm,width=60mm]{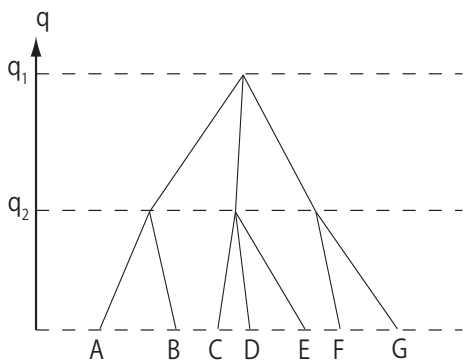}
 \caption{A schematic diagram of the
hierarchical structure in the space of pure states labeled by A-G. 
The distance between C and D equals $q_2$ being 
equal to that between C and E and to that D and E, which is smaller 
than $q_1$ being equal to that between A and C and to that C and F.
This hierarchical structure satisfies the ultrametricity.}
 \label{fig:RSBtree}
\end{center}
\end{figure}

The broad $P(q)$ implies 
that the overlap between two pure states $\alpha$ and $\beta$, 
i.e. 
$q^{\alpha \beta}=(1/N)\sum_{i}\Ave{S_{i}}_{\alpha}\Ave{S_{i}}_{\beta}$ 
where 
$\Ave{(\cdots)}_{\alpha}=(1/Z_{\alpha})\Tr{\alpha}(\cdots)e^{-\beta H}$ 
denotes the thermal average over the pure state $\alpha$,
takes different values depending on the selection of two pure states. 
The mean-field theory can provide not only the behavior of $P(q)$ 
but also some useful information of the structure of the pure states.
Let us consider a three-point probability distribution 
$P(q_{1},q_{2},q_{3})$, which represents a joint probability of the 
overlaps between three arbitrary pure states; 
for three arbitrary pure states $\alpha,\beta$ and 
$\gamma$, $P(q_{1},q_{2},q_{3})$ represents 
the probability that 
their mutual overlaps $q^{\alpha \beta},q^{\alpha \gamma}$ and 
$q^{\beta \gamma}$ take the values $q_1,q_{2}$ and $q_{3}$:
\begin{equation}
P(q_{1},q_{2},q_{3})=\left[
\sum_{\alpha,\beta,\gamma}
\omega_{\alpha}\omega_{\beta}\omega_{\gamma}
\delta(q_{1}-q^{\alpha \beta})
\delta(q_{2}-q^{\alpha \gamma})
\delta(q_{3}-q^{\beta \gamma})
\right]_{\V{J}}
\end{equation}
where $\omega_{\alpha}$ denotes the probability weight of the pure state 
$\alpha$.
In the mean-field solution of the SK model, 
it is known that this distribution shows a characteristic behavior
\begin{eqnarray}
&&P(q_{1},q_{2},q_{3})=
\frac{1}{2}P(q_{1}) x(q_1)
\delta(q_{1}-q_{2})
\delta(q_{1}-q_{3})\nonumber \\
&&
\hspace{-5mm}+\frac{1}{2}
\left\{
P(q_{1})P(q_{2})
\Theta(q_1-q_2)
\delta(q_2-q_3)
+({\rm two \hspace{1.5mm} terms \hspace{1.5mm} with \hspace{1.5mm} 1,2,3 \hspace{1.5mm} permuted)}
\right\} \label{eq1:ultra}
\end{eqnarray}
where $x(q)=\int_{0}^{q} dq' P(q')$ and $\Theta(x)$ is the step function which takes $\Theta(x)=1$ for $x>0$ and $\Theta(x)=0$ otherwise.
The first term on the right hand side is non-vanishing only if the three 
overlaps are equal to each other, and the second term requires that 
the overlaps be the edges of an isosceles triangle 
$(q_{1}>q_{2}, q_{2}=q_{3})$. This means that the distances between three pure states should form either equilateral or isosceles triangles. 
This property is called the ultrametricity, and the ultrametricity
implies that 
the pure states consist a hierarchical structure as in fig.\ \ref{fig:RSBtree}.

The ultrametricity seems to be highly mathematical and to have 
no useful information about real experiments. 
However, there are some experiments indicating 
that real spin glasses have hidden 
hierarchical structures in the phase space. 
Next, we present some examples of such experiments.
\subsection{Experimental supports of the hierarchical structure}
Two characteristic effects are observed in spin-glass materials.
One is the `rejuvenation' effect and the other is the `memory' effect. 
These are considered to be the evidence that 
the states of real spin glasses constitute the hierarchical structures 
in their phase spaces. 
We here give a simple explanation about these two effects. 

\begin{figure}[htbp]
\hspace{-2mm}
\begin{center}
\includegraphics[height=70mm,width=100mm]{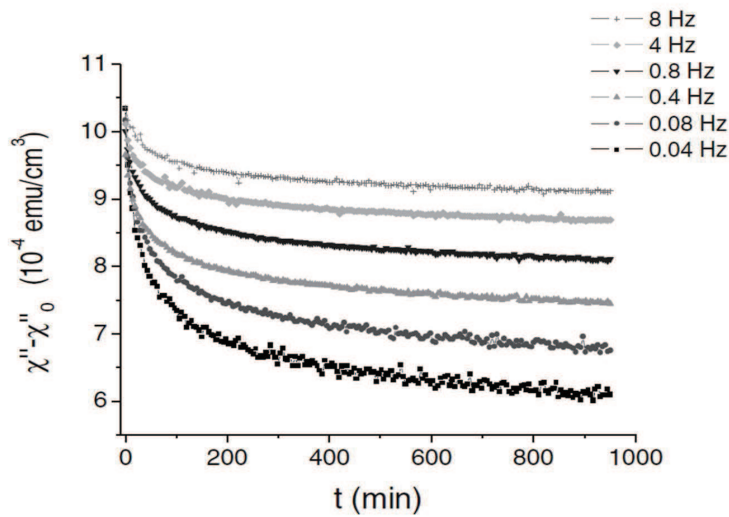}
 \caption{Relaxation of the out-of phase susceptibility $\chi''$ of 
CdC${\rm r}_{1.7}$I${\rm n}_{0.3}$${\rm S}_{4}$ for 
different frequencies. The curves are shifted vertically by an amount 
$\chi_{0}''$ for the sake of clarity. From \cite{Vinc1}.}
 \label{fig:chipp}
\end{center}
\end{figure}

Consider an experimental situation that 
a small {\it ac} field is applied to a spin-glass material.
The magnetic {\it ac} response is observed. 
If the {\it ac} response is delayed, the susceptibility has two components; 
an in-phase one $\chi'$ and an out-of-phase one $\chi''$.
This out-of-phase $\chi''$ is zero in the paramagnetic phase  
but becomes finite in the spin-glass phase and relaxes slowly as 
fig.\ \ref{fig:chipp}, which signals the aging of the system.

\begin{figure}
\begin{center}
   \includegraphics[height=70mm,width=100mm]{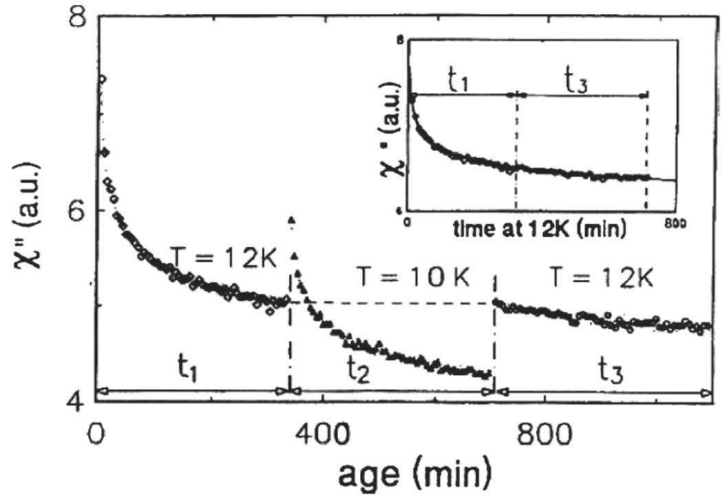}
 \caption{Relaxation of the out-of phase susceptibility $\chi''$ (frequency $0.01$ Hz) of 
CdC${\rm r}_{1.7}$I${\rm n}_{0.3}$${\rm S}_{4}$  during 
the temperature cycle given in the text. 
The rejuvenation effect occurs at the point where the temperature is 
suddenly changed from $T=12$K to $10$K, 
and the memory effect is observed at $T=12$K, 
despite of the rejuvenation at $10$K. The inset is the jointed 
figure of both parts of $12$K, which clearly shows the memory effect.
From \cite{Vinc2}.}
 \label{fig:rej-memory}
\end{center}
\end{figure}

In the above experiment, 
we can change the temperature and observe the response of the system.
Figure \ref{fig:rej-memory} shows the behavior of $\chi''$ in a cycle of 
temperature;
the temperature is kept 
a constant $T=12$K$<T_{c}$ 
for $t_{1}=350$min.. During in the aging, 
the temperature is suddenly decreased from $12$ to $10$K and kept at $T=10$K 
for $t_{2}=350$min., and the temperature is increased to the initial temperature $T=12$K 
and $\chi''$ is observed for $t_{3}=350$min..
This figure \ref{fig:rej-memory} clearly shows two extraordinary behaviors. 
One is that 
the relaxation of $\chi''$ shows a sudden jump when the temperature is 
decreased, which implies that the aging restarts from 
a `younger state'. This is the rejuvenation effect. 
The other is the continuity of the two separated curves at $T=12$K, 
although the system has experienced the rejuvenation along the way. 
This phenomenon indicates that the spin glass has held the perfect 
memory about the past state at $T=12$K. This is the memory effect.

\begin{figure}
\begin{center}
   \includegraphics[height=50mm,width=90mm]{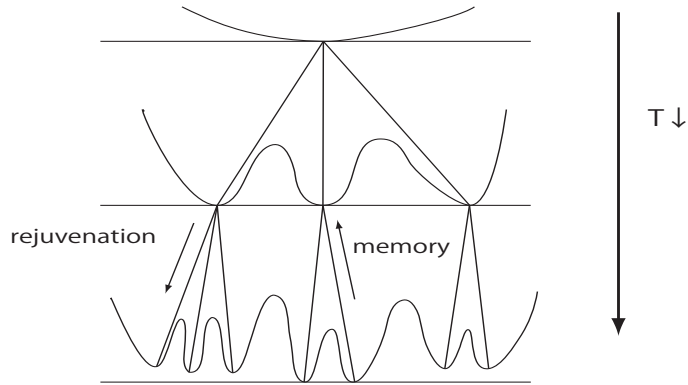}
 \caption{Schematic picture of the hierarchical structure of the metastable 
states of a spin glass}
 \label{fig:hierarchy}
\end{center}
\end{figure}

These phenomena are characteristic behaviors of spin glasses, and 
can be interpreted by assuming that 
the metastable states of a spin glass 
constitute a hierarchical structure.
This structure splits into many states as the temperature decreases 
and becomes more and more complicated as fig.\ \ref{fig:hierarchy}. 
According to this description, 
the rejuvenation effect occurs as the result of splitting 
of the metastable states when the temperature decreases. 
Decreasing temperature produces new more stable states 
by splitting the space of the states 
and the system restarts relaxing to those stable states. 
On the other hand, 
the memory effect is due to the vanishing of the metastable states 
by increasing temperature.
The states which the system explores at a low temperature 
are integrated into the same state when the temperature increases.
This means that the process of the system
 at the low temperature does not influence 
the state at the high temperature, which causes the memory effect 
of the temperature.
This 
hierarchical structure of the metastable states is considered to be 
related to the hierarchical structure of pure states in 
the mean-field theory.

\subsection{The mean-field theory and more}
In the previous subsections, we have seen some evidence supporting the 
mean-field theory of spin glasses.
In addition,
a recent mathematical study \cite{Tala} shows that the mean-field solution 
constructed by Parisi \cite{Pari1,Pari2} is rigorous for the SK model.
These facts constitute the firm foundation of  
the mean-field description of spin glasses.

At this point, there are some ways to proceed with the spin glass theory 
further.

One way is to go beyond the mean-field theory. 
Treating more realistic models, 
e.g. finite dimensional systems or
systems with correlated interactions $\{J_{ij} \}$ etc., 
is a naive one to better 
understand real spin-glass systems. 
However, in spite of the success of the SK model, 
direct knowledge on finite-dimensional spin glasses or correlated systems 
is yet limited. 
Analytical approaches in finite dimensions are 
rather difficult and many researches 
rely on numerical calculations. 
However, 
even numerical approaches often do not work well for spin glasses 
because of the slow dynamics 
at and below the critical temperature.
Frustrated systems like spin glasses often have many metastable states, 
and these metastable states prevent the system from relaxing 
to the equilibrium states, which means that 
the correct sampling of equilibrium states for 
frustrated systems at low temperatures often fails.
Although 
the spin-glass transition of the SK model is definitely  
the equilibrium one and the phase space splits into many pure states 
at the transition point, 
the numerical simulations cannot identify the appearance of the pure states as 
the origin of the slow dynamics.
Consequently, 
we do not have sufficient theoretical descriptions of finite dimensional 
spin glasses and there are some different arguments about 
their low-temperature behaviors \cite{Mari,Fish,Pala}. 
These are very important and interesting problems,  but
we do not treat them in this thesis.

Another way is to refine the mean-field theory of spin glasses and to 
apply it to other subjects. 
Actually, in the last two decades,
the spin-glass theory has been applied to many subjects  
and compared with other approaches in different disciplines \cite{SPIN}. 
The notion of pure states can be generalized to other problems 
and many research areas are included in the scope of the mean-field 
 theory of spin glasses. For example, structural 
glasses \cite{SLOW}, super-cooled liquids \cite{Scio}, information processing 
tasks \cite{STAT,INFO} and so on.
Especially, 
 a statistical mechanical approach based on the spin-glass theory
to the information tasks has provided many fruitful results 
which influence not only the informatics 
but also the spin-glass theory itself.

As the basis of such interdisciplinary researches, 
the mean-field theory of spin glasses 
provides some useful concepts and mathematical tools. 
The replica method is the most powerful one of such tools.
Expressing in a simple term, we can identify 
the replica method as a mathematical technique to 
calculate the averaged quantities which are hard to assess by other 
usual methods. 
Many problems can be treated in a unified way in the framework 
of the replica method \cite{STAT}.
On the other hand, in spite of many successes of the replica method, 
the mathematical foundation and the physical interpretation 
of this method still have some mysterious parts.
Such mysteries of the replica method are the main subject of this 
thesis. 
Next, we briefly explain the basic concept of the replica method and 
how it is used in actual situations, and eventually state the objective of 
this thesis.

\section{The replica method}\label{sec1:replica}
For the study of spin-glass models, the exchange interactions $\V{J}$ 
between spins should be treated as the random variables.
There are some ways to treat the randomness 
based on the time scales of the spins and the exchange interactions. 
The time scale of spins are determined by how often the spin flipping 
occurs in the system, while the 
time scale of the exchange interactions is dominated by the 
spatial motion of the atoms (impurities in alloys).
If time scales of spins and interactions are almost the same,
we should take both the averaging simultaneously. 
This situation is called the annealed case. 
The annealed partition function becomes
\begin{equation}
Z_{\rm anne}=\left[ 
\Tr{\V{S}}e^{-\beta \mathscr{H}(\V{S},\V{J})}
\right]_{\V{J}},
\end{equation} 
where $\Tr{\V{S}}$ means the summation over the spins. The angular 
brackets $[\cdots]_{\V{J}}$ denote the average over the random interactions 
$\V{J}$ and are called the configurational average. 
Some frustrated systems like spin glasses are, however, 
not correctly described by the annealed average, 
because the frustration is relaxed by motions of the atoms.
To correctly describe spin glass behaviors, we must consider a situation that 
the time scale of spins is drastically shorter than that of interactions.
This situation is called the quenched case, 
in which the interactions are fixed as constants, 
and the partition function becomes 
\begin{equation}
Z( \V{J} )=\Tr{S} e^{ -\beta \mathscr{H}( \V{S} | \V{J} ) }.
\end{equation} 
In quenched systems, 
we should treat the partition function leaving the random variable 
dependence $\V{J}$, but this treatment is quite difficult.
For a class of quantities, however, 
the values of the quantities depending on random variables
converge to the averaged values with the probability $1$ in the 
large system size limit $N\to \infty$.
These quantities are called the self-averaging quantities, and 
an important point is 
that the free energy is also one of such quantities. 
Accordingly, we just calculate the averaged free energy $f$ as
\begin{equation}
-\beta f=\lim_{N\to \infty}
\left[ \frac{1}{N}\log Z(\V{J}) \right]_{\V{J}}. \label{eq1:f} 
\end{equation}
Assessing eq.\ (\ref{eq1:f}) is, however, seriously difficult. 
The replica method is employed to overcome this difficulty. 
To precisely state the details, 
we here define the generating 
function $\phi(n)$ as 
\begin{eqnarray}
&&\phi_{N}(n)=\frac{1}{N}\log[( Z(\V{J}) )^n]_{\V{J}},\\
&&\phi(n)=\lim_{N\to \infty}\phi_{N}(n).
\end{eqnarray}
The basic idea of the replica method is the following identity 
expressed in terms of $\phi_{N}(n)$ as 
\begin{equation}
-\beta f=\lim_{N\to\infty} \frac{1}{N}[\log Z(\V{J})]_{\V{J}}=
\lim_{N\to \infty}\lim_{n \to 0} 
\Part{  }{ n }{}\phi_{N}(n)
. \label{eq1:replica}
\end{equation}
Thanks to this identity, 
the average of the logarithm $[\log Z(\V{J})]_{\V{J}}$ 
is replaced by the average of the $n$th moment $[Z^n]_{\V{J}}$.
Unfortunately, assessing the average $[Z^n]_{\V{J}}$ 
for general $n \in \mR$ is still difficult.  
To avoid this difficulty, the replica method additionally 
performs the following procedures;
\begin{itemize}
\item{The asymptotic behavior of 
$[(Z(\V{J}))^n]_{\V{J}}$ in the limit $N\to \infty$ is calculated 
for natural numbers $n=1,2,\cdots \in \mN$ by using some analytical techniques 
like the saddle-point method.}
\item{The obtained solution of $[(Z(\V{J}))^n]_{\V{J}}$ for 
$n\in \mN$ is extended to $n \in \mR$ by using 
their analytical continuation.}
\end{itemize}
The first step is necessary because
the direct assessment of $[(Z(\V{J}))^n]_{\V{J}}$ 
is unfeasible even for $n \in {\mN }$, and 
the second one is inevitable to take the $n\to 0$ limit.

The procedures of the replica method mentioned above are necessary 
but appear rather mysterious.  
Actually, these procedures cause two possible problems:
\begin{enumerate}
\item{The uniqueness of 
the analytical continuation from natural to real
numbers. Even if all the moments of $[(Z(\V{J}))^n]_{\V{J}}$ are given 
for $n \in \mN$, it is impossible to uniquely 
continue the analytical expressions for $n\in \mN$ to 
$n \in {\mR}$ (or $\mC$).}
\item{The analyticity breaking
of $\phi(n)=\lim_{N\to \infty}\phi_{N}(n)$. In general, even if 
$\phi_N(n)$ is analytic
with respect to $n$ for finite $N$, the analyticity 
of $\phi(n)=\lim_{N \to \infty} \phi_N(n)$
can be broken.}
\end{enumerate} 
Especially, the second problem is essentially related to the complex 
behavior of spin glasses. 
It is known that the analyticity breaking with respect to 
$n$ actually occurs in some spin-glass models and such analyticity 
breaking relates to the replica symmetry breaking (RSB).
Nowadays, we have some prescriptions to handle the RSB and 
we can obtain a correct solution, which is proposed by 
Parisi in 1979 \cite{Pari1,Pari2}, 
even though the RSB occurs. 
Unfortunately, however, the Parisi solution requires some mysterious 
procedures in the evaluation of the free energy, 
and its full comprehension is not still obtained.

The subject of this thesis is to investigate the problems related to the 
RSB from 
some perspectives. 
We have mainly two results; 
one is about the interpretation of the RSB. 
The RSB is considered to correspond to the emergence of many pure states. 
We reexamine this description by treating a specific model and resolve 
some new aspects of the replica method.
Second is a proposition of a new method to detect the RSB as 
singularities with respect to $n$. This method is based on the 
Lee-Yang description of phase transitions \cite{LeeYang} and is applicable to any 
singularities with respect to $n$. 
Our results will be useful for supporting 
the correctness of the replica method 
and making the meaning of the RSB clearer.

\section{Overview of thesis}
In this thesis, we treat the problems related to 
the replica method and the RSB. Some considerations and new methods 
are proposed to reveal the mysteries of the RSB.

To state the problem precisely, 
we introduce the replica method with a detailed explanation
in the next chapter. 
Its practical applications are also presented by treating 
some typical models. 
To reconsider the replica result from a microscopic viewpoint,
we also present the so-called TAP equation \cite{TAP}. 
This method reproduces the replica result and 
is useful to understand 
the microscopic properties of pure states. 

In chapter 3, 
we reexamine recent progresses about the replica method.
Particular ingredients of the progresses are summarized as 
follows: 
\begin{enumerate}
\item{Relationship between the generating function $\phi(n)$ and 
the probability $P(f)$ 
that the free energy $-\log Z/N \beta$ takes a value $f$.}
\item{Connection with complex structures of phase spaces (pure states) 
and a formalism of the RSB.}
\end{enumerate}
To extensively investigate these two statements, 
we concentrate on the investigation 
of the Ising perceptron \cite{STAT} 
by using the replica method with the RSB.
The Ising perceptron is a model of a neuron which generates 
a map from $\mR^N$ to $\{+1,-1\}$, and is characterized 
by the value of ratio $\alpha=M/N$, 
where $M$ is the number of random patterns stored in the perceptron. 
The reason why we treat this model is that 
the meaning of complexity 
for the perceptrons of finite size is clearer 
than that for other systems. 
For the fully connected models, including the Ising perceptrons, 
a pure state at zero temperature can be identified with a stable cluster, 
the detailed definition of which will be given in the same chapter, 
with respect to single spin flips \cite{Cocc,Biro1}. 
For samples of small systems, the size of the clusters can be numerically 
evaluated by exhaustive enumeration without any ambiguity. 
This property is extremely useful for justifying theoretical predictions 
through numerical experiments.  
Investigating the Ising perceptron analytically and numerically,
we have found that the replica method cannot correctly 
detect the phase-space structure of the Ising perceptron, 
which is due to the strange distribution of the cluster size. This 
 also affects the probability distribution of 
the free energy $P(f)$ and the replica method 
cannot correctly calculate $P(f)$ 
in a certain region. The origin of this failure is also explained 
in the same chapter.

In chapter 4, apart from the notion of pure states, we concentrate on the 
analyticity of $\phi(n)$ with respect to the replica number $n$. 
For this purpose, we propose a new scheme based on the Lee-Yang theory 
about phase transitions \cite{LeeYang}. In particular, we observe zeros 
with respect to $n$ of the averaged $n$th moment $[Z^n]=0$. 
In terms of the analyticity, the RSB can be regarded as 
an analyticity breaking of $\phi(n)$. 
This motivates us to observe the zeros of $[Z^n]$ to obtain useful information 
about the RSB. 
To investigate these zeros, we treat some tractable systems, 
i.e. the $\pm J$ models with 
a symmetric distribution on two types of lattices, 
ladder systems and Cayley trees with random fields on the boundary. 
There are two reasons for using these models: 
Firstly, these models can be investigated 
in a feasible computational time by the cavity method 
\cite{MezaRevisit,Bowm}. 
Especially, at zero temperature this approach gives 
a simple iterative formula to yield the
partition function. Employing the replica method and the cavity method, 
we can perform symbolic calculations of  
the $n$th moment of the partition function $[Z^n]$, 
which enables us 
to directly solve the
equation of the objective zeros $[Z^n]=0$. 
The second reason is the
existence of the spin-glass phase. It is known that the spin-glass phase is 
present for Cayley trees \cite{Chay,Mott,Carl,Lai} 
and is absent for ladder systems. Therefore, we can 
compare the behavior of the zeros, which are considered to be 
dependent on the spin-glass ordering.  
Our results indicate that the zeros approach the real axis of $n$ for 
some Cayley trees but not for 
ladder systems, which seems to be consistent 
with the presence and absence of spin glass ordering. 
However, further investigations reveal that the singularities speculated from the zeros of Cayley trees are not related to the RSB. 
This topic is quite complicated and 
the details are given in the main texts of chapter 4.
We also discuss a general possibility that the zeros of $[Z^n]$ with respect 
to $n$ cannot examine the full-step RSB, but 
this argument has some uncertainties 
and we cannot give a reliable conclusion.
To investigate this point futher,
we also examine the zeros of $[Z^n]$ with respect to $n$ 
for the so-called regular random graph. 
The regular random
 graph is locally similar to the Cayley tree, which enables us 
to treat this system in a manner similar 
to the Cayley tree in the large system limit,
but has some global loops, 
which introduce nontrivial correlations into the system and 
is considered to lead to the RSB at low temperatures. 
However, any appealing result cannot be obtained by this investigation because 
of the computational difficulty for assessing $[Z^n]$ of the regular random 
graph.  
The RSB is still 
a great mystery both from mathematical and physical points of view.

To remove some ambiguities of the discussions given in chapter 4, 
we tackle the RSB by investigating some tree-like systems in another way 
in chapter 5.
In this chapter, we observe the zeros of the partition function 
with respect to the external field $H$ and temperature $T$ for 
the $\pm J$ model on Bethe lattices. 
A Bethe lattice is 
an interior part of the 
infinitely-large Cayley tree and shares similar properties to 
those of the Cayley tree. 
The result shows that an extraordinary response to deviations of 
the external field and temperature
exists for spin glasses on Bethe lattices, 
which is often identified with the
RSB and implies the existence of the RSB in these models. 
This observation supports our previous discussion that the zeros of $[Z^n]$ 
with respect to $n$ cannot detect the full-step RSB, which means that 
we need some modifications to obtain information about the RSB from the 
zeros with respect to $n$. 
Investigation along this line is an important future work. 


The last chapter is devoted to a conclusion of this thesis.

\chapter{Review of the replica method and its interpretations}
In this chapter, we review the replica method in detail. 
This method requires some complicated and mysterious prescriptions 
in the calculations. 
Hence, 
the meaning of the mathematical manipulations and 
physical interpretations have been examined from various perspectives 
for a long time. 
Our review in this chapter includes the recent progress in this point.
In particular, 
a relation between the replica symmetry breaking 
and pure state statistics,
and 
a relation with the large deviation theory, 
are quite important for the interpretations of the replica method. 
We will explain these topics by 
demonstrating the actual calculations for some typical models.

\section{Replica calculations for the fully-connected $p$-spin interacting model}\label{sec2:RM}
The SK model presented in the previous 
chapter is the basic model of spin glasses. 
This model has long-range interactions and 
each spin is connected to all the other spins.
This property is useful for the exact treatment of the model
because all the spins are symmetric 
and can be equally treated. 
The detailed analyses of the SK model 
developed many useful concepts and analytical tools, 
and eventually provided a comprehensive view of spin glasses.

When investigating spin systems, we only treat the two-spin interactions usually. This is because the interactions between 
the nearest neighboring spins 
are considered to be 
sufficient to capture the behavior of many physical systems. 
Actually, the SK model also has only the two-spin interactions.
However, to demonstrate the replica method, we here treat  
the fully-connected $p$-spin interacting model as an example.
The reason is the following:
\begin{enumerate}
\item{The SK model $(p=2)$ is naturally included.}
\item{It is known that there are qualitatively different phase transitions 
for $p\geq 3$.}
\item{A particular limit $(p\to \infty)$ makes the analysis easier, which 
is suitable for the demonstration of the replica method.}
\end{enumerate}
Thus, we hereafter concentrate on the analysis of 
the fully-connected $p$-spin interacting 
model.
The Hamiltonian of this system can be written as
\begin{equation}
\mathscr{H}(\V{S}|\V{J})=-\sum_{i_{1}<\ldots <i_{p}} J_{i_{1}\ldots i_{p}} S_{i_{1}} \ldots S_{i_{p}}, \label{eq2:pHam}
\end{equation}
where $i$ is the site index and $S$ is the Ising
spin variable. The interaction $ J_{i_{1}\ldots i_{p}}$ is a
quenched random variable, the distribution function of which is given by
\begin{equation}
P(J_{i_{1}\ldots i_{p}}) = \left( \frac{N^{p-1}}{J^2 \pi p!} \right)^{\frac{1}{2}}  \exp \left\{ -\frac{N^{p-1}}{J^2 p!} \left(J_{i_{1}\ldots i_{p}} \right)^2 \right\}.\label{eq2:distJ}
\end{equation} 
We adopt appropriate normalizations $N^{p-1}$ and $p!$.
This is because the physical quantities should be appropriately scaled, 
e.g. the average of the Hamiltonian is extensive, 
in the limits 
$N\to \infty$ and $p \rightarrow \infty$ which we will take afterward. 
If we set $p=2$, this model is reduced to the SK model.
Following the prescription of the replica method, we 
calculate the generating function 
$\phi(n)=\lim_{N\to \infty}( \log[Z^n]_{\V{J}})/N$. 
Assuming $n\in \mN$,
the $n$th moment of the partition function can be assessed as
\begin{eqnarray}
 &&[Z^n]_{\V{J}}=
\int \prod_{ i_{1}< \ldots <i_{p} } dJ_{ i_{1} \ldots i_{p} } 
  P( J_{ i_{1} \ldots i_{p} } ) Z^n = \Tr{} \exp 
  \left\{ \frac{\beta^2 J^2 p!}{4N^{p-1}} \sum_{i_{1}<\ldots <i_{p}}
   \left(
\sum_{\mu=1}^{n}S_{i_{1}}^{\mu} \ldots S_{i_{p}}^{\mu}
\right)^2 
  \right\} \notag \\
&&=\Tr{} \exp 
  \left\{ \frac{\beta^2 J^2 N}{2}\sum_{\mu<\nu} 
   \left(\frac{1}{N}\sum_{i}S_{i}^{\mu} S_{i}^{\nu} \right)^p   
+\frac{1}{4}\beta^2J^2 N n  
 \right\}, \label{eq2:2-2-3}
   \end{eqnarray}
 where the symbol $\Tr{}$ denotes the trace over all the spins 
 and $\mu$ and $\nu$ represent the replica indices.
In deriving the final expression in eq.\ (\ref{eq2:2-2-3}), 
we have used the following relation 
\begin{equation}
\frac{1}{N^{p-1}}\sum_{i_{1}<\ldots <i_{p}} 
 S_{i_{1}} \ldots S_{i_{p}}
 =\frac{N}{p!}\left(\frac{1}{N}\sum_{i}S_{i} \right)^{p}+O(N^0),
\label{eq2:2-2-4}
\end{equation}
and left only the leading term.
It is convenient to introduce the variables 
\begin{equation}
q^{\mu \nu}=\frac{1}{N}\sum_{i}S_{i}^{\mu}S_{i}^{\nu}=
\frac{\V{S}^{\mu} \cdot \V{S}^{\nu}}{N}
\label{eq2:qori}.
\end{equation} 
to replace the spin product terms 
$\sum_{i}S_{i}^{\mu} S_{i}^{\nu}/N $ in
eq.\ (\ref{eq2:2-2-3}).
We treat $q^{\mu \nu} $ as 
a dummy integrating variable and 
use delta functions 
$
\delta\left(
\V{S}^{\mu} \cdot \V{S}^{\nu}-Nq^{\mu \nu} \right )$
to satisfy the constraint (\ref{eq2:qori}).
We also use the Fourier-transformed expression of 
the delta function 
\begin{eqnarray}
\delta\left(
\V{S}^{\mu}  \cdot \V{S}^{\nu}-Nq^{\mu \nu} \right )
=\int_{-{\rm i}\infty}^{+{\rm i}\infty}
\frac{d \Wh{q}^{\mu \nu }}{2 \pi}
\exp \left (
\Wh{q}^{\mu \nu} (\V{S}^{\mu} \cdot \V{S}^{\nu}-Nq^{\mu \nu}) 
\right ). 
\end{eqnarray}
Substituting these expressions, we rewrite 
eq.\ (\ref{eq2:2-2-3}) as
\begin{eqnarray}
 &&[Z^n]_{\V{J}}=\Tr{} \int \prod_{ \mu < \nu } dq^{\mu\nu}d\Wh{q}^{\mu\nu}
  \exp\Bigg\{
  \frac{\beta^2 J^2 N}{2}\sum_{\mu < \nu}( q^{\mu\nu} )^p
  -N\sum_{\mu<\nu}q^{\mu\nu}\Wh{q}^{\mu\nu} \notag \\
 && \quad\quad\quad\quad\quad\quad\quad\quad\quad\quad\quad\quad 
  +\sum_{\mu<\nu}\Wh{q}^{\mu\nu}
 \left(
  \sum_{i}S_{i}^{\mu}S_{i}^{\nu}
 \right)
 +\frac{1}{4}\beta^2 J^2 N n
 \Bigg\}. \label{eq2:2-2-6}
\end{eqnarray}
The spin trace can now be independently taken at each $i$ 
\begin{equation}
\Tr{}e^{\sum_{\mu<\nu}\Wh{q}^{\mu\nu}
 \left(
  \sum_{i}S_{i}^{\mu}S_{i}^{\nu}
 \right)}
=\left(
\Tr{}e^{
\sum_{\mu<\nu}\Wh{q}^{\mu\nu}
 S^{\mu}S^{\nu}
}
\right)^N
=
\exp
\left\{
N \log \Tr{} 
e^{\sum_{\mu<\nu}
\Wh{q}^{\mu\nu}
 S^{\mu}S^{\nu}
}
\right\}.
\end{equation}
Using the saddle point method, we obtain the generating function 
$\phi(n)=\lim_{N \to \infty}\log [Z^n]_{\V{J}}/N$ for $n\in \mN$ as 
\begin{eqnarray}
\phi(n)=\Extr{
q^{\mu\nu},
\Wh{q}^{\mu\nu}
}
\left\{ \frac{\beta^2 J^2 }{2}\sum_{\mu < \nu}( q^{\mu\nu} )^p
      -\sum_{\mu<\nu}q^{\mu\nu}\Wh{q}^{\mu\nu} 
  +\frac{1}{4}\beta^2 J^2  n
  +\log \Tr{} 
  e^{\sum_{\mu<\nu}\Wh{q}^{\mu\nu}
 S^{\mu}S^{\nu}}
\right\},\label{eq2:phi}
\end{eqnarray}
where the symbol $\Extr{x}$ represents to take the 
extremization with respect to $x$.
The extremization condition yields
\begin{equation}
q^{\mu\nu}=\frac{\Tr{}S^{\mu}S^{\nu} e^{\sum_{\mu<\nu}\Wh{q}^{\mu\nu}
 S^{\mu}S^{\nu}} }{\Tr{} e^{\sum_{\mu<\nu}\Wh{q}^{\mu\nu}
 S^{\mu}S^{\nu}} } \, ,
 \, \Wh{q}^{\mu\nu}=\frac{1}{2}p \beta^2 J^2 (q^{\mu\nu})^{p-1}.
 \label{eq2:qmat}
\end{equation}
Taking this extremization condition is, 
however, quite difficult in the general form.  
Hence, we need some ansatz to reduce this extremization problem to 
a tractable one. 
\subsection{Replica symmetric ansatz}
To proceed further, we are required to 
determine the explicit dependence of $q^{\mu\nu}$
on the replica indices $\mu$ and $\nu$ at the saddle point. 
For $n \in \mN$, an equality $[\Ave{S_{i}^{\mu}S_{i}^{\nu}}_{n}]_{\V{J}}= 
[\Ave{S_{i}^{\gamma}S_{i}^{\omega}}_{n} ]_{\V{J}}$, where $\Ave{\cdots}_{n}$ denotes 
the average over the replicated Boltzmann factor
 $e^{-\beta \sum_{\mu}\mathscr{H}(\V{S}^{\mu}|\V{J})}$, 
 holds for any different combinations of replicas ($\mu\neq \nu $ and 
$\gamma\neq  \omega$), which is due to the permutation symmetry of 
$\sum_{\mu}\mathscr{H}(\V{S}^{\mu}|\V{J})$ with respect to the replica 
indices. 
This observation naturally leads to an ansatz that the saddle point of 
eq.\ (\ref{eq2:phi})
has also the same symmetry, which is the so-called replica symmetry (RS).
Under the RS, the order parameter matrix $q^{\mu\nu}$ becomes 
\begin{equation}
q^{\mu\nu}=q, \,\,\Wh{q}^{\mu\nu}=\Wh{q},
\end{equation} 
and the physical
meaning of $q^{\mu\nu}$ is easily 
understood. 
To see this, we note that eq.\ (\ref{eq2:qmat}) can be written as 
\begin{equation}
q^{\mu\nu}=[\Ave{S_{i}^{\mu}S_{i}^{\nu}}_{n}]_{\V{J}},
\end{equation}
which is almost clear from the definition of $q^{\mu\nu}$ (\ref{eq2:qori}).
In the RS ansatz, each replica is independent and equivalent,
which means that the contributions from replicas $\gamma\neq \mu,\nu$
are canceled out and the averaged quantity belonging to different replicas 
gives the same value 
\begin{eqnarray} 
q^{\mu\nu}=
[\Ave{S_{i}^{\mu}S_{i}^{\nu}}_{n}]_{\V{J}}
=\left[
\frac{
\Tr{}S^{\mu}_i 
e^{
- \beta \mathscr{H}(\V{S}^{\mu}|\V{J})
}
}{
\Tr{} e^{- \beta \mathscr{H}(\V{S}^{\mu}|\V{J}) }
}
\frac{
\Tr{}S^{\nu}_i e^{-\beta \mathscr{H}(\V{S}^{\nu}|\V{J}) }
}{
\Tr{} e^{-\beta \mathscr{H}(\V{S}^{\nu}|\V{J})}
} 
\right]_{\V{J}}
=[\Ave{S_{i}}^2]
,\label{eq2:qmunu}
\end{eqnarray}
which is the spin-glass order parameter explained 
in chapter 1.

Using the RS, we can easily calculate the terms in eq.\ (\ref{eq2:phi}) 
to derive 
\begin{eqnarray}
&&\sum_{\mu<\nu} q^{\mu\nu} \Wh{q}^{\mu\nu}=\frac{1}{2}n(n-1)\Wh{q}q,\\ 
&&\Tr{}  e^{ \sum_{ \mu<\nu }\Wh{q}^{\mu\nu}S^{\mu}S^{\nu} }
=\Tr{}  e^{ \Wh{q}/2\left\{
\left(\sum_{\mu}S^{\mu} \right)^2 -n 
\right\}
}
=e^{-\frac{1}{2}n\Wh{q}}\GI{z} \left(2\cosh\sqrt{\Wh{q}}z \right)^n,
\end{eqnarray}
where the last equation is derived by using the 
 Hubbard-Stratonovich transformation 
\begin{equation}
\exp\left(\frac{1}{2}x^2\right)=\int_{-\infty}^{\infty} \frac{dz}{\sqrt{2 \pi }}\exp\left(-\frac{z^2}{2}+zx\right)=\GI{z} \exp\left(zx\right), 
\end{equation}
where $Dz$ is the Gaussian measure 
$dz e^{-\frac{z^2}{2}}/\sqrt{2 \pi } $ and 
we hereafter assume that the domain of integration of $\GI{z}$ 
is $]-\infty, \infty[$ if there is no explicit indication.
Substituting the above expressions, 
we can reduce $\phi(n)$ in eq.\ (\ref{eq2:phi}) to $\phi_{\rm RS}(n)$ 
as
\begin{eqnarray}
&&\hspace{-1cm}\phi_{{\rm RS}}(n)=\Extr{q,\Wh{q}}
\Biggl\{
\frac{n(n-1)}{4}\beta^2  q^p-\frac{n(n-1) }{2}q\Wh{q}+\frac{1}{4}\beta^2 n \nonumber \\
&&\hspace{3cm}-\frac{1}{2}n\Wh{q}+\log \GI{z} \left(2\cosh \sqrt{\Wh{q}}z \right)^n
\Biggr\}\label{eq2:phiRS},
\end{eqnarray} 
where we hereinafter put $J=1$ for simplicity of the notation. 
Taking the extremization condition 
with respect to $q$ and $\Wh{q}$, 
we get 
\begin{eqnarray}
&&q=\frac{
\GI{z} \left( \cosh \sqrt{\Wh{q}}z \right)^n  
\left( \tanh \sqrt{\Wh{q}}z \right)^2 
}{
\GI{z} \left( \cosh \sqrt{\Wh{q}}z \right)^n 
}
,\,\, \Wh{q}=\frac{1}{2}p \beta^2 q^{p-1}
\label{eq2:q}.
\end{eqnarray}
Fortunately, under 
the RS the expression of 
$\phi_{\rm RS}(n)$ can be extended to $n\in \mR$, which enables us to  
take the $n\to 0$ limit. 
The resultant free energy $f_{\rm RS}$ can be derived as 
\begin{eqnarray}
&&-\beta f_{\rm RS} 
=\lim_{n\to 0} \Part{\phi_{\rm RS}(n)}{n}{} \nonumber \\
&&=\Extr{q,\Wh{q}}
\left\{
-\frac{1}{4}\beta^2  q^{p}+\frac{1}{2}q\Wh{q}+\frac{1}{4}\beta^2
-\frac{1}{2}\Wh{q}+\GI{z} \log2\cosh(\sqrt{\Wt{q}}z)
\right\}
. \label{eq2:fRS}
\end{eqnarray}
The extremization condition gives the state equations 
(\ref{eq2:q}) in the limit $n\to 0$, and now 
the problem is solved in the RS level.
Unfortunately, this RS solution gives unphysical behaviors, e.g. 
negative entropy, 
at low temperatures. Next, we study this point 
in more detail by treating the particular limit $p\to \infty$.

\section{The solution in the limit $p\to \infty$}\label{sec2:REM}
\subsection{Failure of the RS solution}
To demonstrate the failure of the RS ansatz, we 
here analyze the $p\to \infty$ limit. 
In this limit, the system becomes rather simple and we can derive 
the exact solution without employing the replica method.
First, let us study the RS solutions. In the limit $p\to \infty$, 
there are two possible solutions to eqs.\ (\ref{eq2:q}), 
namely the paramagnetic solution $(q,\Wh{q})=(0,0)$ and 
the spin-glass solution $(q,\Wh{q})=(1,+\infty)$.   
Substituting these solutions into eq.\ (\ref{eq2:fRS}), 
we obtain the paramagnetic and spin-glass free energies, 
$f_{P}$ and $f_{{\rm RS}SG}$ respectively, as 
\begin{eqnarray}
&&-\beta f_{P}=\frac{1}{4}\beta^2 +\log2 , \label{eq2:fp} \\
&&-\beta f_{{\rm RS}SG}\rightarrow \sqrt{\frac{ 2\Wh{q} }{\pi} }
 \rightarrow +\infty.
\label{eq2:fsg}
\end{eqnarray}
The divergence of $f_{{\rm RS}SG}$ clearly 
shows an inconsistency of this spin-glass solution. 
On the other hand, the paramagnetic solution also 
shows the unphysical behavior at low temperatures. 
The entropy of the paramagnetic solution  $s_{P}$ is
\begin{equation}
s_{P}=\log2-\frac{1}{4}\beta^2.
\end{equation}
This clearly becomes negative below $T_{c}=1/(2\sqrt{\log2})$, 
which implies that there should be a phase transition 
above $T_{c}$. 
At least in the RS level,
 we do not have any candidate to correctly express the 
low temperature behavior of this system. 
Before seeing the correct solution in the framework of 
the replica theory,  
we employ the microcanonical approach in order to obtain the 
correct low temperature behavior. 

\subsection{Microcanonical approach}\label{sec2:micro}
The microcanonical approach starts from calculations of the energy 
distribution of 
the system (\ref{eq2:pHam}) in the limit
$p \rightarrow \infty$ as the microcanonical ensemble
\begin{equation}
P(E(\V{S}))=[\delta(E-H( \V{S}|\V{J} ))]_{\V{J}}.\label{eq2:def-distE}
\end{equation}
The configurational average can be carried out by using the Fourier 
expression of the delta function. The result becomes a Gaussian distribution 
and does not depend on the spin configuration 
\begin{equation}
P(E(\V{S}))=P(E)=\frac{1}{\sqrt{N\pi }}\exp\left(-\frac{E^2}{N}\right).\label{eq2:distE}
\end{equation}
The two-point probability distribution
$P(E_{1}(\V{S}_{1}),E_{2}(\V{S}_{1}))$ 
for two independent configurations of 
spins $\V{S}^1 $ and ${\V{S}}^2$ is similarly
calculated 
\begin{eqnarray}
&&P(E_{1}(\V{S}^1),E_{2}(\V{S}^2))=
 [\delta(E_1-\mathscr{H}( \V{S}^1|\V{J} ))\delta(E_2-\mathscr{H}( \V{S}^2|\V{J} ))]_{\V{J}}
 \notag\\
 &&=\frac{1}{N\pi \sqrt{(1+q^p)(1-q^p)}}\exp\left(-\frac{E_{1}^2+E_{2}^2}{2N(1+q^p)}-\frac{E_{1}^2-E_{2}^2}{2N(1-q^p)}\right),\label{eq2:2-3-3}
\end{eqnarray}
where $q=\V{S}^{1} \cdot \V{S}^{2}/N$. 
Under the assumption that the spin configurations $\V{S}^1$ and $\V{S}^2$ 
are thermodynamically distinguishable,  
the value of $|q|$ is smaller than $1$, which means that 
$q^p$ goes to $0$ in the limit $p \to \infty$ and 
\begin{equation}
P(E_{1}(\V{S}^1),E_{2}(\V{S}^2)) \to P(E_{1})P(E_{2}).\label{eq2:2-3-4}
\end{equation} 
Hence, the energy distribution of two different spin configurations 
is completely independent for each configuration. 
We can prove that the same property holds 
for any multi-point distributions by following similar discussions. 
Consequently, each energy level for 
each spin configuration can be treated as an 
identically independently distributed (i.i.d.) Gaussian random variable,
 the distribution of which is given by eq.\ (\ref{eq2:distE}).
Due to this property, 
this model is called the random energy model (REM) \cite{DerrREM} and  
the free energy of the REM can be assessed by the microcanonical 
approach. 

\begin{figure}[htbp]
\begin{center}
   \includegraphics[height=40mm,width=60mm]{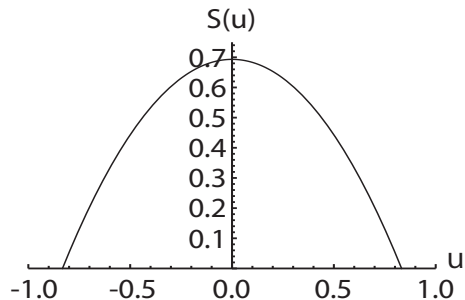}
 \caption{The shape of entropy $s(u)$ of the REM. }
 \label{fig2:REMent}
\end{center}
\end{figure}

The microcanonical approach utilizes the entropy to yield the 
thermodynamical behavior of the system.
Because each energy level is i.i.d. for the REM, 
the number of states with energy $E$ can be calculated as 
\begin{equation}
n(u)=2^N P(Nu)=\frac{1}{\sqrt{N\pi }}e^{N\left(\log2-u^2\right)}.\label{eq2:2-3-5}
\end{equation}
where we put $u=E/N$, which represents the energy per spin. 
This relation implies that 
in the limit $N \rightarrow \infty$, there are very many
states for the range 
$|u|<\sqrt{\log2} \equiv u_{0}$ but none in the
other range $|u| \geq u_{0}$, and 
the entropy for $|u|< u_{0}$ is given by
\begin{equation} 
s(u)=\frac{1}{N}\log n(E)=
\log2-u^2.\label{eq2:2-3-6}
\end{equation}
An illustration of the entropy $s(u)$ of the REM is given in
fig.\ \ref{fig2:REMent}.
Using this entropy function $s(u)$, we can construct the free energy 
$f(T)$ from the definition 
\begin{equation}
f(T)=\max_{|u|<u_{0}}
\left\{
u-Ts(u)
\right\}
=\max_{|u|<u_{0}}
\left\{
u-T(\log2-u^2)
\right\}
.\label{eq2:fmicREM}
\end{equation}
where $T$ is the temperature. 
In the usual case, the temperature $T$ is equal to 
$(\partial s(u)/\partial u)^{-1}$ as the definition of temperature in 
the microcanonical ensemble. This is actually the case for 
$T>T_{c}=1/(2\sqrt{\log2})$. 
In this region, eq.\ (\ref{eq2:fmicREM}) gives 
\begin{equation}
f_{P}=-\frac{1}{4}\beta-T \log 2 \,\,\,\,\, (T>T_{c}).\label{eq2:f-highT}
\end{equation}
On the other hand, for $T\leq T_{c}$, the maximum of eq.\ 
(\ref{eq2:fmicREM}) is given by $u=-u_{0}$ and the free energy is 
\begin{equation}
f_{SG}=-\sqrt{\log2}=-u_{0} \,\,\,\,\, (T<T_{c}).\label{eq2:f-lowT}
\end{equation}
These free energies (\ref{eq2:f-highT}) and (\ref{eq2:f-lowT}) 
are the exact solution of the REM. 

\begin{figure}[t]
\begin{tabular}{cc}
\hspace{-5mm}
\begin{minipage}[t]{0.50\hsize}
\begin{center}
 \includegraphics[height=50mm,width=60mm]{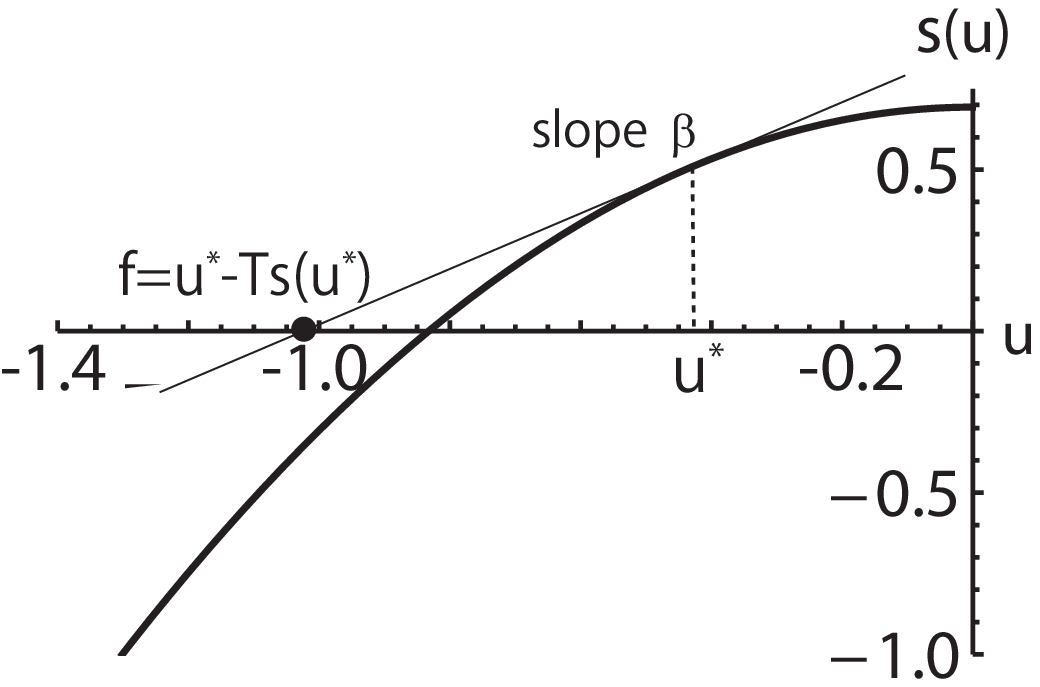}
 \caption{A pictorial expression of the derivation of the free energy 
for $T>T_{c}$. The free energy is given by the $u$-intercept of 
the tangent to the curve $s(u)$ with slope $\beta$. }
\label{fig2:REMenthighT}
\end{center}
\end{minipage}
\hspace{2mm}
 \begin{minipage}[t]{0.50\hsize}
\begin{center}
\includegraphics[height=50mm,width=60mm]{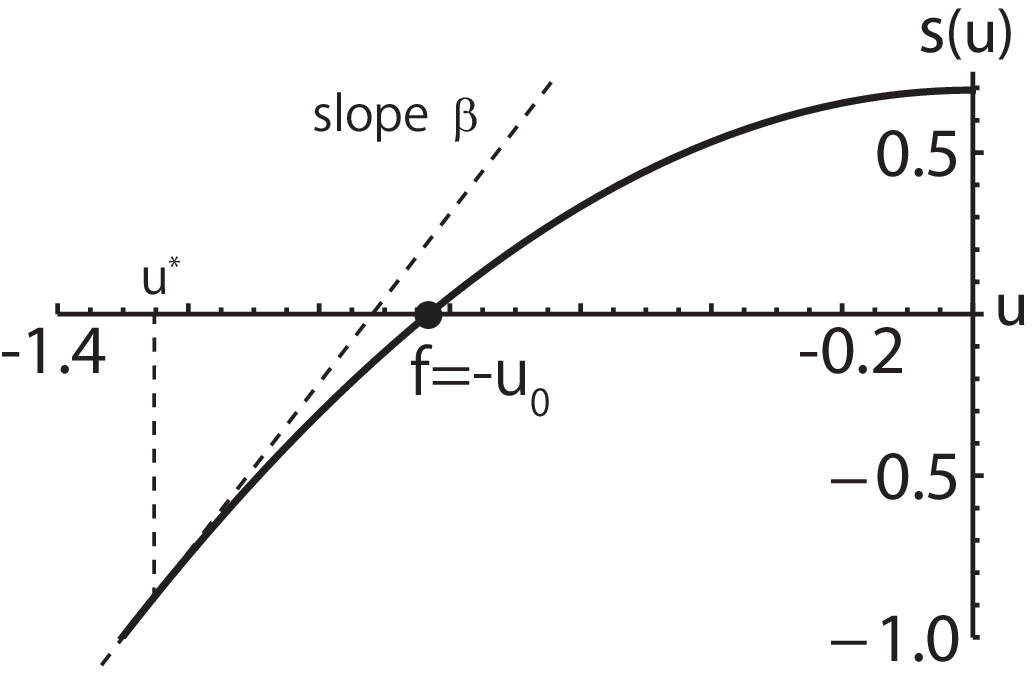}
\caption{A pictorial expression of the derivation of the free energy 
for $T \leq T_{c}$. The free energy is given by the nearest value of $u$ 
to the tangent of $s(u)$ with slope $\beta$.}
\label{fig2:REMentlowT}
\end{center}
\end{minipage}
\end{tabular}
\end{figure}

The above discussion is 
easily understood by using pictorial expressions given 
in figs.\ \ref{fig2:REMenthighT} and \ref{fig2:REMentlowT}.
What we should do is to find a tangent to the curve $s(u)$ with slope 
$\beta$. We denote the abscissa of the tangent point as $u^*$. 
For $T>T_{c}$, this abscissa $u^*=-\beta/2$ 
is in the range $|u|<u_{0}$, and then
the maximum in eq.\ (\ref{eq2:fmicREM}) is given by $u^{*}$. 
The figure corresponding to this situation is fig.\ \ref{fig2:REMenthighT}. 
On the other hand, for $T\leq T_{c}$, this value $u^{*}$ is out of the range 
$|u|<u_{0}$, which means that there is no physical state for the energy level
$u^{*}$. In such a situation, the maximum of 
eq.\ (\ref{eq2:fmicREM}) is the nearest value of $u$ to $u^{*}$ under the 
condition $|u|<u_{0}$, which leads to $f=-u_{0}-Ts(-u_{0})=-u_{0}$.  

Before ending of this subsection, 
let us consider relations between the 
microcanonical solution (\ref{eq2:fmicREM})
 and the RS result.
As we can see easily, for $T>T_{c}$
the paramagnetic solution of the RS (\ref{eq2:fp}) 
is identical to the microcanonical solution. 
As the temperature decreases, the entropy of the paramagnetic solution decreases and vanishes at $T=T_{c}$, and  
below $T_{c}$ 
the free energy is kept as a constant at the value of $T=T_{c}$. 
These observations imply that the spin glass transition of the REM
is due to the vanishing entropy of the paramagnetic phase. 
This is a characteristic feature of the REM and 
this transition is sometimes called the frozen transition. 

To obtain the correct solution in the replica theory, it is known that 
the so-called replica symmetry breaking (RSB), 
which allows $q^{\mu\nu}$ to depend on the replica indices, 
is required. Next we introduce the RSB in a general framework.
\section{Replica symmetry breaking}\label{sec2:RSB}
In order to obtain the low temperature spin-glass phase, 
we must treat $q^{\mu\nu}$ in more general forms. 
Although the correct form of
$q^{\mu\nu}$ is unknown, Parisi has proposed a particular ansatz which
describes a hierarchical breaking of replica symmetry \cite{Pari1,Pari2}. 
The Parisi scheme is constructed from the following recursive algorithm:

(i) First step: The $n$ replicas are grouped in $n/m_{1}$ clusters
of $m_{1}$ replicas. Within the same cluster, each 
$q^{\mu\nu} \, (\mu \neq \nu)$ has a value $q_{1}$. Two
replicas in different clusters have an overlap $q^{\mu\nu}=q_{0}\leq
q_{1}$.

(ii) Second step: Each cluster of size $m_{1}$ is broken up into
$m_{1}/m_{2}$ sub-clusters of $m_{2}$ replicas. Any two replicas in the same
sub-cluster have overlap $q^{\mu\nu}=q_{2}\geq q_{1}$. The other
overlaps remain unchanged.

Continuing this procedure, we obtain the general $k$-step RSB ($k$RSB) 
situation
\begin{equation}
n \geq m_{1} \geq m_{2} \cdots \geq m_{k} \geq 1, \,\, q_{k} \geq q_{k-1} \geq \cdots \geq q_{1} \geq q_{0}. \label{eq2:RSB}
\end{equation}
Each $q_k$ is interpreted as a hierarchically constructed 
spin-glass order parameter and $m_{k}$ is called the breaking 
parameter.
The following example is for the case of the $2$-step RSB such that
$n=12,m_{1}=6,m_{2}=3$,
\begin{equation}
\{q^{\mu\nu}\}=
 \left(\begin{array}{c|c}
  \begin{array}{c|c}	      
   \begin{array}{ccc}
    0 & q_{2} & q_{2} \\
	  q_{2} & 0 & q_{2} \\
	  q_{2} & q_{2} & 0
	 \end{array}
    & q_{1}  \\ \hline
	 q_{1} & \begin{array}{ccc}
	  0 & q_{2} & q_{2} \\
		  q_{2} & 0 & q_{2} \\
		q_{2} & q_{2} & 0
		 \end{array}
  \end{array} & q_{0} \\ \hline
	q_{0} & \begin{array}{c|c}	
	 \begin{array}{ccc}
	  0 & q_{2} & q_{2} \\
		q_{2} & 0 & q_{2} \\
		q_{2} & q_{2} & 0
	       \end{array} 
	  & q_{1} \\ \hline
	       q_{1} & \begin{array}{ccc}
		0 & q_{2} & q_{2} \\
			q_{2} & 0 & q_{2} \\
			q_{2} & q_{2} & 0
		       \end{array}
	\end{array} 
       \end{array}
 \right). \label{eq2:2-2-17}
\end{equation}
Taking the limit $n \rightarrow 0$, we assume that 
the parameters  $\{m_{k}\}$ become
continuous and the following inequalities hold;
\begin{equation}
n\leq m_{1} \leq m_{2} \leq \cdots \leq 1, 
\end{equation}
which are the inverse relations of eq.\ (\ref{eq2:RSB}).
All the parameters $\{m_{k}\},\{q_{k}\}$ are determined by taking the 
extremization condition of $\phi(n)$.
The above procedures are all the ingredients of the Parisi scheme. 
The full-step ($k=\infty$-step) RSB (FRSB) 
is also constructed in a similar manner 
but the concrete procedure is quite complicated and we here only refer to
\cite{STAT}. 

Physically, the RSB is considered to be related to the emergence of enormous 
number of pure states. 
According to this speculation, the RS ansatz describes situations 
that only a small number (subexponential number of the system size $N$) of 
pure states exist, and the failure of the RS solution at low temperatures 
is due to the neglect of many pure states in the spin-glass phase.

Curiously, it is known that 
 most models only show two types of RSB: The 1RSB and FRSB.   
In either case, it is considered that 
there are many pure states but the intricacy of the 
phase space structure is different depending on the step of RSB. 
Generally speaking, 
as the step of RSB becomes higher and higher, 
the phase space structure is considered to become 
more and more complicated.

Although our knowledge about the FRSB is still rather poor 
because of intricacy of the FRSB construction, 
the 1RSB solution has been intensively studied and 
its relation to pure states is well understood nowadays \cite{Mona1995}. 
In the following, we mainly treat the 1RSB solution and 
 explain the relation with many pure states in detail.

\subsection{The 1RSB solution for the $p$-spin interacting model}\label{sec2:1RSBp-glass}
Under the 1RSB ansatz, the replica indices are divided into
$n/m$ groups of identical size $m$, and 
$q^{\mu\nu}$ and $\Wh{q}^{\mu\nu}$ are parameterized as 
\begin{equation}
(q^{\mu\nu},\Wh{q}^{\mu \nu})=
\left\{
\begin{array}{ll}
(q_{1},\Wh{q}_{1}) & (\,\, {\rm {\it \mu}\,\, and\,\, {\it \nu}\,\, belong\,\, to\,\,  the\,\, same\,\, group} \,\,)\\
(q_{0},\Wh{q}_{0}) & (\,\, {\rm otherwise}\,\,)
\end{array}
\right.
\end{equation}
Under this assumption, we can 
transform the term $\sum_{\mu<\nu}\Wh{q}^{\mu\nu}S^{\mu}S^{\nu}$ as
\begin{equation}
\sum_{\mu<\nu}\Wh{q}^{\mu\nu}S^{\mu}S^{\nu}
=\frac{1}{2}\left\{
	     \Wh{q}_{0}\left(\sum_{\mu}^{n}S^{\mu} \right)^2
	     +(\Wh{ q}_{1}-\Wh{ q}_{0})\sum_{{\rm block}}^{n/m}
	     \left(\sum_{\mu \in {\rm block}}^{m} S^{\mu}\right)^2 
	     -n\Wh{ q}_{1}
	    \right\}, \label{eq2:Tr1RSB}
\end{equation}
where the first term on the right-hand side fills up all
matrix elements of $\{q^{\mu\nu}\}$ with
$q_{0}$. The second term replaces the elements in the block part with
$q_{1}$ and the last term sets the diagonal elements to zero.
Similarly, we derive
\begin{eqnarray}
&&\sum_{\mu < \nu} (q^{\mu\nu})^{p} 
=\frac{1}{2}n(m-1)q_{1}^{p}+
\frac{1}{2}n(n-m)q_{0}^{p},
\\
&&\sum_{\mu<\nu} q^{\mu \nu} \Wh{q}^{\mu\nu}
=
\frac{1}{2}n(m-1)\Wh{q}_{1}q_{1}+
\frac{1}{2}n(n-m)\Wh{q}_{0}q_{0}
.
\end{eqnarray}
Combining the Hubbard-Stratonobich transformation and eq.\ (\ref{eq2:Tr1RSB}),
we obtain 
\begin{eqnarray}
e^{\sum_{\mu<\nu}\Wh{q}^{\mu\nu}S^{\mu}S^{\nu}}=
e^{-\Wh{ q}_{1}n/2}\int Dz_{0} \, \prod_{{\rm block}}^{n/m}\left\{ \int Dz_{1} \, 
\prod_{\mu \in {\rm block }}^{m}
e^{\left(
\sqrt{\Wh{ q}_{0}}z_{0}+\sqrt{ \Wh{ q}_{1}-\Wh{ q}_{0} }z_{1}
\right)S^{\mu}
}
\right\}.
\end{eqnarray}
In this expression, we can perform the spin trace 
for each replica spin independently. 
Substituting all the results into eq.\ (\ref{eq2:phi}), we get
\begin{eqnarray}
&&\hspace{-5mm}\phi_{{\rm 1RSB}}(n,m)
=
\Extr{q_{1},q_{0},\Wh{q}_{1},\Wh{q}_{0}}
\Biggl\{
\frac{\beta^2}{4}(n (m-1)q_{1}^{p}+n(n-m)q_{0}^{p})
+\frac{\beta^2}{4}n
-\frac{1}{2}n\Wh{q}_{1}
\notag
\\
&&\hspace{-5mm}-\frac{1}{2}n
(m-1)\Wh{q}_{1}q_{1}-\frac{1}{2}n(n-m)\Wh{q}_{0}q_{0}
+
\log 
\int Dz_{0} 
\left(
\int Dz_{1} (2\cosh h)^m
\right)^{n/m} 
\Biggr\}
.\label{eq2:phi1RSB}
\end{eqnarray}
where we put $h=\sqrt{\Wh{q}_{0}}z_{0}+\sqrt{\Wh{q}_{1}-\Wh{q}_{0}}z_{1}$.
Taking the extremization condition, we get the following equations of 
states:
\begin{eqnarray}
 &&
q_{1}=\frac{
\GI{z_{0}} \left(\GI{z_{1}}(\cosh h)^m \right)^{n/m}
\frac{\GI{z_{1}}(\tanh h)^2(\cosh h)^m}{\GI{z_{1}}(\cosh h)^m}
}{
\GI{z_{0}} \left(\GI{z_{1}}(\cosh h)^m \right)^{n/m}
}\label{eq2:q1}\\
&&
q_{0}=\frac{
\GI{z_{0}} \left(\GI{z_{1}}(\cosh h)^m \right)^{n/m}
\left(
\frac{\GI{z_{1}}\tanh h (\cosh h)^m}{\GI{z_{1}}(\cosh h)^m}
\right)^2
}{
\GI{z_{0}} \left(\GI{z_{1}}(\cosh h)^m \right)^{n/m}
}\label{eq2:q0}\\
&& 
\Wh{q}_1=\frac{1}{2}p \beta^2   q_{1}^{p-1}, \,\, 
\Wh{q}_0=\frac{1}{2}p\beta^2  p q_{0}^{p-1}.\label{eq2:qhat}
\end{eqnarray}
If we set $q_{0}=q_{1}=q$, the RS solution (\ref{eq2:phiRS}) and
(\ref{eq2:q}) is reproduced.
In the limit $n\to 0$, we can derive the free energy parameterized by 
$m$
\begin{eqnarray}
&&-\beta f_{{\rm 1RSB}}(m)
=\lim_{n\to 0}\Part{}{n}{}\phi_{\rm 1RSB}(n,m)
=
\Extr{q_{1},q_{0},\Wh{q}_{1},\Wh{q}_{0}}
\Biggl\{
\frac{\beta^2}{4}((m-1)q_{1}^{p}-mq_{0}^{p})
+\frac{\beta^2}{4}
\notag
\\
&&
-\frac{1}{2}\Wh{q}_{1}
-\frac{1}{2}
(m-1)\Wh{q}_{1}q_{1}+\frac{1}{2}m\Wh{q}_{0}q_{0}
+\frac{1}{m}
\int Dz_{0} \log 
\left(
\int Dz_{1} (2\cosh h)^m
\right)
\Biggr\}
.\label{eq2:f1RSB}
\end{eqnarray}
The extremization condition re-derives 
eqs.\ (\ref{eq2:q1})-(\ref{eq2:qhat}) with the condition $n=0$. 
Although the physical meaning of $m$ is unclear at this moment, we 
here treat $m$ as an artificial parameter and take the extremization condition
to erase the $m$-dependence of the free energy at the final step 
of calculations. 
We will come back to this problem later 
after deriving the correct solution of the 
REM in the 1RSB level.
\subsection{The 1RSB solution in the $p\to \infty$ limit}\label{sec2:1RSBREM}
It is time to solve the REM in the framework of the replica theory. 
What we should do at first is to find the solution of 
eqs.\ (\ref{eq2:q1})-(\ref{eq2:qhat}). 
There are three solutions for these equations in the limit 
$p \to \infty$, namely 
$(q_{1},q_{0})=(0,0),(1,1)$ and $(1,0)$.   
For the 1RSB solution to be
non-trivial, an inequality $q_{0}<q_{1}, \,(\Wh{ q}_{0}<\Wh{ q}_{1})$ 
must hold, which means that $(q_{1},q_{0})=(1,0)$ is the appropriate 
solution. 
Substituting this solution, we get 
\begin{equation}
-\beta f_{\rm 1RSB}(m) =\frac{\beta^2 }{4}m+\frac{1}{m}\log2.
\end{equation}
Variation with respect to $m$ gives
\begin{equation}
(m\beta )^2=4\log2. \label{eq2:ext-m}
\end{equation}
Substituting this value, we obtain
\begin{equation}
f_{SG}=-\sqrt{\log2}, \label{eq2:2-2-23}
\end{equation}
which is identical to the microcanonical 
solution (\ref{eq2:f-lowT}) below the 
the transition temperature $T_{c}$. 
In the replica theory, 
 the transition temperature is determined by the 
direct comparison of the free energy value 
$f_{SG}=f_{P}$, which is surely identical to that of the 
microcanonical approach. Hence, for the REM, the 1RSB solution 
gives the exact result.

The above 
is the conventional construction of the 1RSB solution without 
recent perspectives about the physical meaning of $m$.
Actually, the RSB solution has more information about the phase space structure
of the system. We elucidate this point in the next section.

\section{Pure state statistics of the 1RSB level}\label{sec2:purestat}
\begin{figure}[htbp]
\begin{center}
   \includegraphics[height=50mm,width=70mm]{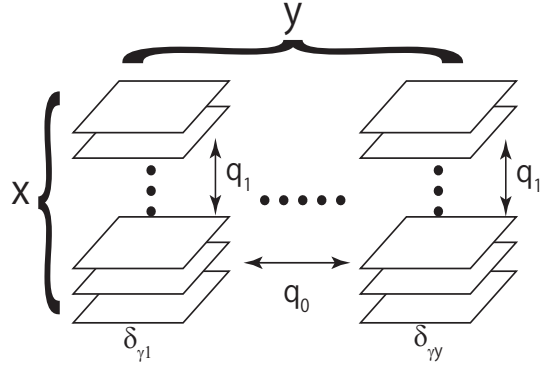}
 \caption{Schematic diagram of the 1RSB description of the factor $\prod_{\nu=1}^{y}
\left(
\prod_{\mu=1}^{x}\delta_{\gamma^{\nu}}(\V{S}^{\nu \mu})
\right)$  }
 \label{fig2:1RSB}
\end{center}
\end{figure}

As already mentioned in chapter 1, 
in the mean-field description of spin glasses, 
there are many pure states in a spin-glass system and 
the partition function becomes the summation over contributions from 
all the pure states. 
Each pure state $\gamma$ has its own free energy 
$-\beta f_{\gamma}=(\log Z_{\gamma})/N$
and the number of pure states having the free energy value $f$, 
$\mathscr{N}(f)$, is scaled as
\begin{equation}
\mathscr{N}(f)\sim e^{N\Sigma(f)},
\end{equation}
where $\Sigma(f)\sim O(1)$ is called the complexity.
Note that an inequality $\Sigma(f)\geq 0$ must hold, because 
the complexity is the logarithm of the number of pure states, which 
cannot become negative. This is the same reason as the 
entropy cannot become negative. 
Under these assumptions, we can write the partition function as
\begin{equation}
Z=\sum_{\gamma}Z_{\gamma}\sim \int df e^{N(-\beta f +\Sigma (f))},\label{eq2:Z}
\end{equation}
where $\gamma$ is the index of pure states and 
\begin{equation}
Z_{\gamma}=\Tr{\V{S} \in \gamma}e^{-\beta H(\V{S}|\V{J})}=\Tr{\V{S}}e^{-\beta H(\V{S}|\V{J})}\delta_{\gamma}(\V{S})\label{pure_state_volume},
\end{equation}
where $\delta_{\gamma}(\V{S})$ is the indicator function where $\delta_{\gamma}(\V{S})=1$ when $\V{S}\in \gamma$ and $\delta_{\gamma}(\V{S})=0$ otherwise.
Equation (\ref{eq2:Z}) indicates that the partition function 
is dominated by the saddle point of $(-\beta f +\Sigma (f))$. 
For convenience, we define another generating function $g(x|J)$ as
\begin{eqnarray}
e^{Ng(x|\V{J})}=\sum_{\gamma}Z_{\gamma}^{x}\sim \int df e^{N(-\beta x f +\Sigma(f))}\Rightarrow
g(x|J)=\max_{f_{-}\leq f\leq f_{+}}\{ -\beta x f +\Sigma(f)\},
\label{eq2:gdef}
\end{eqnarray}
where the bounds $f_{-}$ and $f_{+}$ 
for the range of $f$ come from the constraint $\Sigma(f)\geq 0$.
Equation (\ref{eq2:gdef}) implies that, 
when the complexity $\Sigma(f)$ is a convex upward function, 
the complexity can be calculated from the generating function $g(x|\V{J})$
in the parameterized form as follows:
\begin{equation}
-\beta f(x)=\Part{g(x|J)}{x}{},\,\, \Sigma (f(x))=g(x|J)-x\Part{g(x|J)}{x}{}.\label{eq2:f(x)}
\end{equation}
So far, we are discussing about one sample of quenched disorder. We here assume that $g(x|J)$ is a self-averaging quantity and we can replace $g(x|J)$ by $g(x)=[g(x|J)]_{\V{J}}$. This averaged function $g(x)$ can be expressed by the replica method as
\begin{eqnarray}
&&Ng(x)=[Ng(x|J)]_{\V{J}}=\left[
\log \left(
\sum_{\gamma}
Z_{\gamma}^x
\right)
\right]_{\V{J}}
=\lim_{y\to 0}\Part{}{y}{}\log
\left[
\left(
\sum_{\gamma}
Z_{\gamma}
^{x}
\right)^{y}
\right]_{\V{J}} \label{eq2:replica_trick_g}
\end{eqnarray}
Although exact evaluation of the right-hand side of eq.\ (\ref{eq2:replica_trick_g})
is difficult, for $x,y \in \mN$ we can derive the following expression:
\begin{eqnarray}
\left [ \left (\sum_\gamma Z_\gamma^x \right )^y \right ]_{\V{J}}  =
\sum_{\gamma^{1}\cdots \gamma^{y}}
\Tr{}
\left[
\exp\left(
-\beta \sum_{\nu=1}^{y}\sum_{\mu=1}^{x} H(\V{S}^{\nu \mu}|\V{J})
\right)
\prod_{\nu=1}^{y}
\prod_{\mu=1}^{x}\delta_{\gamma^{\nu}}(\V{S}^{\nu \mu})
\right]_{\V{J}}.
\label{g_1RSB}
\end{eqnarray}
For the evaluation of this equation,
the following observations are important. 
\begin{itemize}
\item The summation is taken over all possible configurations 
of $xy$ replica spins. 
\item However, the factor $\prod_{\nu=1}^y \prod_{\mu=1}^x 
\delta_{\gamma^\nu}(\V{S}^{\nu \mu})$
allows only contributions from configurations in which $xy$ replicas 
are equally assigned to $y$ pure states by $x$. 
\end{itemize}
These observations are nothing more than the physical situation 
of the 1RSB ansatz, 
in assessing $\left [Z^n(\V{J}) \right ]_{\V{J}}$
with substitution of $n=xy$ and $m=x$
(fig.\ \ref{fig2:1RSB}).
Accepting this interpretation, we obtain the following expression:
\begin{eqnarray}
\frac{1}{N} 
\log \left [ \left (\sum_\gamma Z_\gamma^x \right )^y \right ]_{\V{J}} 
=\phi_{\rm 1RSB}(xy,x), 
\label{eq2:phig}
\end{eqnarray}
where $\phi_{\rm 1RSB}(n,m)$ is the 1RSB solution considered in the previous 
section.
Then, combining eqs.\ (\ref{eq2:replica_trick_g}) and (\ref{eq2:phig}), 
we obtain the generating function $g(x)$ in the form
\begin{eqnarray}
g(x)=\lim_{y\to 0}\Part{}{y}{} \phi_{{\rm 1RSB}}(n=xy,m=x) 
=-x \beta f_{\rm 1RSB}(x) ,
\end{eqnarray}
where $f_{\rm 1RSB}(x)$ is the 1RSB free energy shown in the previous section.
Hence, the RSB breaking parameter $m$ 
corresponds to the parameter $x$ which controls the generating function 
$g(x)$ of complexity, which leads to the following relations 
from eq.\ (\ref{eq2:f(x)}); 
\begin{equation}
-\beta f(x)=-\beta \Part{(xf_{\rm 1RSB}(x))}{x}{},\,\, 
\Sigma (f(x))=-\beta x f_{\rm 1RSB}(x)+\beta x\Part{(x f_{\rm 1RSB}(x))}{x}{}.\label{eq2:f(x)_2}
\end{equation}

\subsection{Pure states of the limit $p\to \infty$}\label{sec2:REMpure}
Here, let us see the complexity of the REM as an example.
According to eqs.\ (\ref{eq2:f1RSB}) and (\ref{eq2:f(x)_2}), 
the generating function of the $p$-spin model 
is given by 
\begin{eqnarray}
&&g(x)=\frac{1}{4}x(x-1)\beta^2  q_{1}^p-\frac{1}{2}x(x-1)q_{1} 
\Wh{q}_{1}+\frac{1}{4}\beta^2  x 
\nonumber \\
&& \hspace{1.5cm}
 -\frac{1}{2}x\Wh{q}_{1}+\log \GI{z} \left(2\cosh \sqrt{\Wh{q}_{1}}z \right)^x,\label{eq2:gx_p}
\end{eqnarray}
where we put $q_{0}=0$ 
because the nontrivial 1RSB solution exists only under this condition.
The value of $q_{1}$ is determined by eq.\ (\ref{eq2:q1}). 
Taking the limits $p\to \infty$ and $q_{1}\to 1$, we obtain
\begin{equation}
g(x)=\frac{1}{4}x^2\beta^2+\log 2.\label{eq2:REMg(x)}
\end{equation}
The parameterized free energy and the complexity are derived from 
eq.\ (\ref{eq2:f(x)}) as
\begin{eqnarray}
-\beta f(x)=\frac{1}{2}x\beta^2  ,
\hspace{2mm}\Sigma(f(x))=\log 2-\frac{1}{4}x^2 \beta^2 =\log 2
 -f^2.\label{eq2:REMcomplexity}
\end{eqnarray}
Clearly, this complexity $\Sigma(f)$ is identical to the 
microcanonical entropy $s(u)$ if we identify $u=f$.
This can be understood if we remember the construction of the REM. 
For this model, each spin configuration is completely independent of each other as we see the independence of each energy level in section \ref{sec2:micro}.
This means that each pure state of the REM is constructed 
from only one configuration, which indicates that the entropy of 
a pure state defined as the logarithm of the number of configurations
belonging to the state, is $0$. Hence, the free energy of a pure state 
corresponds to the energy of the state, which leads to the agreement between 
the complexity $\Sigma(f)$ and the entropy $s(u)$. As might be expected, 
this is a peculiar property of the REM, and in general
the complexity $\Sigma(f)$ is different from the entropy $s(u)$.  

\begin{figure}[t]
\begin{tabular}{cc}
\hspace{-5mm}
\begin{minipage}[t]{0.50\hsize}
\begin{center}
 \includegraphics[height=50mm,width=60mm]{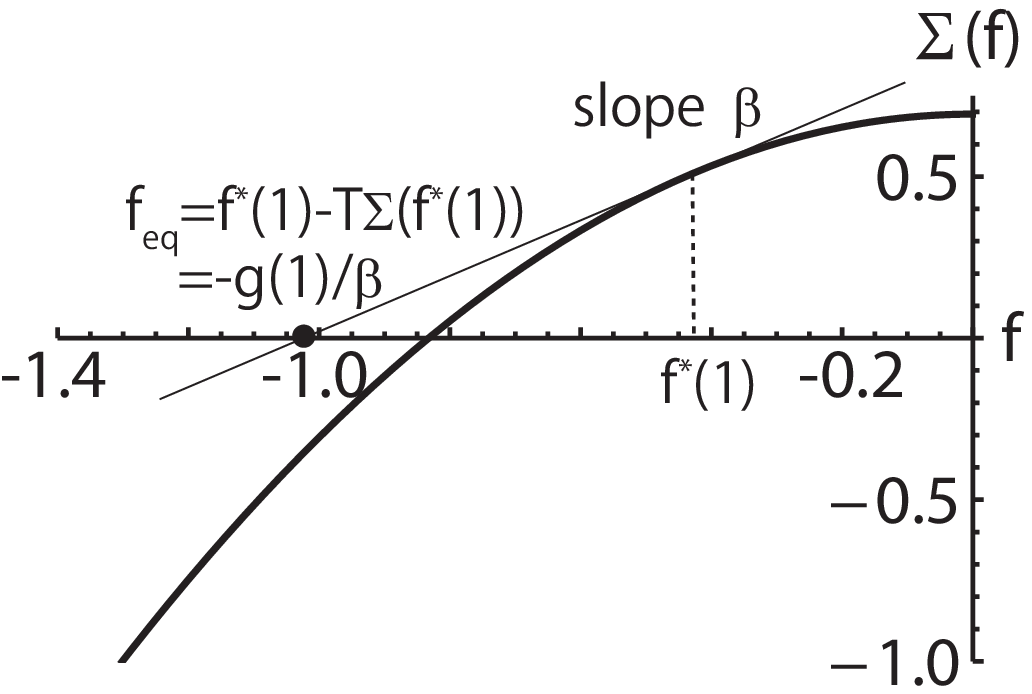}
 \caption{An expression of the derivation of $f_{\rm eq}$ 
for $T>T_{c}$. The free energy is given by the $f$-intercept of 
the tangent to the curve $\Sigma(f)$ with slope $\beta$.  }
\label{fig2:REMcomphighT}
\end{center}
\end{minipage}
\hspace{2mm}
 \begin{minipage}[t]{0.50\hsize}
\begin{center}
\includegraphics[height=50mm,width=60mm]{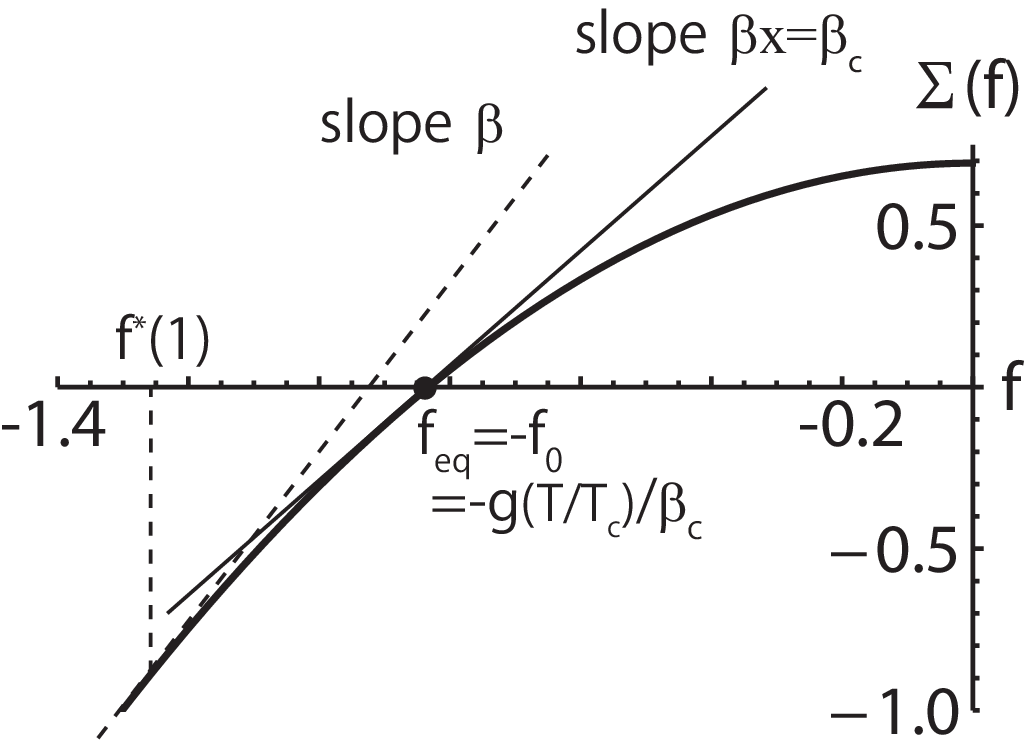}
\caption{An expression of the derivation of $f_{\rm eq}$ 
for $T\leq T_{c}$. The point $-f_{0}$ where $\Sigma(f)$ vanishes 
yields $f_{\rm eq}$, 
which is determined by the tangent to $\Sigma(f)$ with the 
slope $\beta x=\beta_{c}$. The value $x=\beta_{c}/\beta$ 
is identical to the 
extremization of $f_{\rm 1RSB}(x)$ with respect to $x$. }
\label{fig2:REMcomplowT}
\end{center}
\end{minipage}
\end{tabular}
\end{figure}

In the microcanonical approach, the phase transition and 
free energy in each phase were easily derived by the 
pictorial interpretation of eq.\ (\ref{eq2:fmicREM}). 
We here reexamine this derivation 
from a view point of the complexity.
The form of $\Sigma(f)$ is identical to that of $s(u)$ given in fig.
\ref{fig2:REMent} by reading $u=f$ and $s(u)=\Sigma(f)$. 
There are two things which we should do. 
One is to derive the phase transition and 
the equilibrium value of the free energy 
$f_{\rm eq}\equiv - T (\log Z)/N $ in each phase.
The other is to elucidate relations between the derivation of 
$f_{\rm eq}$ and the conventional prescription of the 1RSB.
We start from the first point. 
The equilibrium free energy $f_{\rm eq}$ is obviously 
equal to $-g(1)/\beta$ from the definition. 
The relation between $g(x)$ and the complexity $\Sigma(f)$ 
is derived from eq.\ (\ref{eq2:gdef}) and (\ref{eq2:REMcomplexity}) as
\begin{equation} 
g(x)=\max_{f_{-}\leq f\leq f_{+}}\left\{-\beta x f+\Sigma(f) \right\}
=\max_{|f|<f_{0}}\left\{-\beta x f+\log 2-f^2 \right\},
\label{eq2:gREM}
\end{equation}
where $f_{0}$ is equal to $\sqrt{\log 2}$.
As mentioned in section \ref{sec2:purestat}, the complexity 
is the logarithm of the 
number of pure states and cannot be negative.  
Keeping $x=1$, for $T>T_{c}$ we can easily see that the maximum of 
eq.\ (\ref{eq2:gREM}) is given by $f^{*}(1)=-\beta/2$, which 
is identical to the abscissa of the curve $\Sigma(f)$ with slope $\beta$ 
as given in fig.\ \ref{fig2:REMcomphighT}.
For $T\leq T_{c}$, that abscissa $f^*(1)$ is out of the range $|f|<f_{0}$ 
and the maximum of eq.\ (\ref{eq2:gREM}) is given by 
$-f_{0}=-\sqrt{\log 2}$
as the microcanonical case, which leads to the correct result 
$f_{\rm eq}=-f_{0}$.

The free energies are correctly reproduced from the viewpoint of 
the complexity. Next, we see the relation with 
the conventional 1RSB prescription. 
The inequality $x=m\leq 1$ in the 1RSB construction is naturally understood 
from the specialty of $x=1$. For $x=1$, the generating function 
$g(1)$ directly relates to the correct $f_{\rm eq}$ 
unless the corresponding complexity becomes negative. 
Hence, in the usual case 
we should keep $x=1$ and calculate $g(1)$, which gives 
the correct solution for $T>T_{c}$. 
For $T\leq T_{c}$, however, a naive calculation of $g(1)$ leads to
the negative complexity, but this inconsistency is overcome  
in the 1RSB construction as follows. 
An important point is that the point $f=-f_{0}$ where the complexity 
vanishes is determined by a tangent to $\Sigma(f)$ with the slope 
$\beta_{c}=2\sqrt{\log 2}$.  
Hence, to select $f=-f_{0}$ below $T\leq T_{c}$, 
we can modify $x$ of $g(x)$ as $x=T/T_{c}$. 
This artificial prescription is just the extremization condition of 
$f_{\rm 1RSB}(x)=-g(x)/(\beta x)$ with respect to $x$. In fact,
\begin{eqnarray}
\hspace{-5mm}\Extr{x}\{  f_{\rm 1RSB}(x)\}
\Rightarrow
-\frac{1}{\beta}\Part{}{x}{}\frac{g(x)}{ x}=
\frac{1}{\beta x^2}\left(
g(x)-x\Part{g}{x}{}
\right)
=\frac{\Sigma(f(x))}{\beta x^2}=0,
\end{eqnarray}
which means that 
the extremization condition of $f_{\rm 1RSB}(x)$ with respect to 
$x$ leads to the point where the complexity vanishes.
The equilibrium free energy $f_{\rm eq}$ is expressed in the 
1RSB prescription as 
\begin{eqnarray}
f_{{\rm eq}}=f_{\rm 1RSB}\left(x=\frac{T}{T_{c}} \right)=
 \left.
-\frac{g\left(x\right)}{\beta x}
\right|_{x=\frac{T}{T_{c}}}
=
-\frac{
\max_{f}
\left \{ 
-\beta_{c} f + \Sigma(f)
\right \}
}{\beta_{c}}
=-f_{0}.
\end{eqnarray}
Hence, the correct solution is reproduced. 
These procedures are also illustrated in fig.\ \ref{fig2:REMcomplowT}.

\section{Viewpoint from the large deviation theory}\label{sec2:LDT}
In the previous sections, we gave the replica analysis with the RSB and its 
interpretation from the pure state statistics. In this section, 
we reexamine this result from the 
viewpoint of the large deviation theory, which 
treats asymptotic small probabilities of rare events in the large system limit.

For disordered systems, the
probability that the free energy, $-(1/N\beta)\log Z$, 
will take a certain value $f$, $P(f)$, 
fluctuates from sample to sample.
The large deviation theory tells us that,
 in most cases for large $N$,
this probability $P(f)$ can be scaled as
\begin{eqnarray}
P(f) \sim \exp \left \{ N R(f) \right \}, 
\label{rate_function}
\end{eqnarray}
where $R(f) (\le 0)$ is referred to as the rate function.
The rate function relates to the generating function  $\phi(n)$ 
as 
\begin{eqnarray}
&&e^{N\phi(n)}\equiv 
[Z^n]_{\V{J}}=\int d\V{J}P(\V{J})e^{-N\beta n f(\V{J})}
=\int df P(f)e^{-N\beta n f}
 \sim \int df e^{N\left( -\beta n f+R(f)\right)}\nonumber \\
&&\hspace{4cm} \Rightarrow \phi(n)=\max_{f_{-}\leq f\leq f_{+}}\{-\beta n f +R(f)\},\label{eq2:phi-R}
\end{eqnarray}
where the bounds $f_{+}$ and $f_{-}$ are required to satisfy the constraint 
$R(f)\leq 0$, as in the case of pure state statistics. 
If the rate function is a convex upward function, we can express 
the free energy and the rate function in forms parameterized by $n$
\begin{eqnarray}
-\beta f(n)=\frac{\partial \phi(n)}{\partial n}, \quad 
R(f(n))= \phi(n)-n\frac{\partial \phi(n)}{\partial n}. 
\label{eq2:R(f(n))}
\end{eqnarray}
This equation and the constraint $ R(f)\leq 0$ require that the following 
equation holds for $\forall n$:
\begin{equation}
\frac{\phi(n)}{n}\leq \Part{\phi(n)}{n}{}.\label{eq2:Rconst}
\end{equation}
Equations (\ref{eq2:phi-R}) and (\ref{eq2:R(f(n))})
indicate that the {\em typical} value of $f$, which is characterized
by the condition $R(f)=(1/N) \log P(f)=0$, can be evaluated as
\begin{eqnarray}
-\beta f_{\rm typ}=\frac{1}{N}[\log Z]_{\V{J}}=\lim_{n \to 0} \frac{\partial \phi(n)}{\partial n}, 
\label{standard_replica}
\end{eqnarray}
which is the basic identity of the replica method. 
The above discussion indicates that 
$n=1,2,\ldots > 0$ corresponds to {\em atypical} samples of 
$R(f) < 0$ representing a small probability. 
This means that the replica method can be regarded as a formula that infers
the behavior of typical samples by extrapolating the behavior for atypical samples. 

\subsection{Re-derivation of the 1RSB transition}\label{sec2:REMLDT}
Again we treat the REM as an example. 
As mentioned in section \ref{sec2:REM}, 
there are two solutions for the equations of states (\ref{eq2:q}) in the limit 
$p\to \infty$, namely the paramagnetic $q=0$ and spin-glass $q=1$ solutions.
Hence, there are corresponding two solutions for $\phi(n)$ as
\begin{eqnarray}
&&\phi_{\rm RS}(n;q=0)\equiv \phi_{\rm RS1}(n)=\frac{1}{4}\beta^2 n+n\log 2,\\
&&\phi_{\rm RS}(n;q=1)\equiv \phi_{\rm RS2}(n)=\frac{1}{4}\beta^2 n^2+\log 2.
\end{eqnarray}
The RS1 and RS2 solutions correspond to the paramagnetic and RS spin-glass 
solutions, $f_{P}$ and $f_{{\rm RS}SG}$ given in section \ref{sec2:REM}, 
respectively\footnote{Two limits $(q,\Wh{q})\to (1,\infty)$ and $n\to 0$ are not exchangeable and the values of free energy are different between 
$-\beta f_{{\rm RS}SG}$ and 
$\lim_{n\to 0}(\partial \phi_{\rm RS2}/\partial n)$.}.
If we follow the concept of the saddle-point method, 
the correct solution is given by larger $\phi(n)$, but naive application 
of this criterion yields incorrect results. To treat this problem 
precisely,
we should distinguish regions $T>T_{c}$ and $T\leq T_{c}$.

\begin{figure}[htbp]
\hspace{-7mm}
\begin{minipage}{0.32\hsize}
\begin{center}
   \includegraphics[height=40mm,width=50mm]{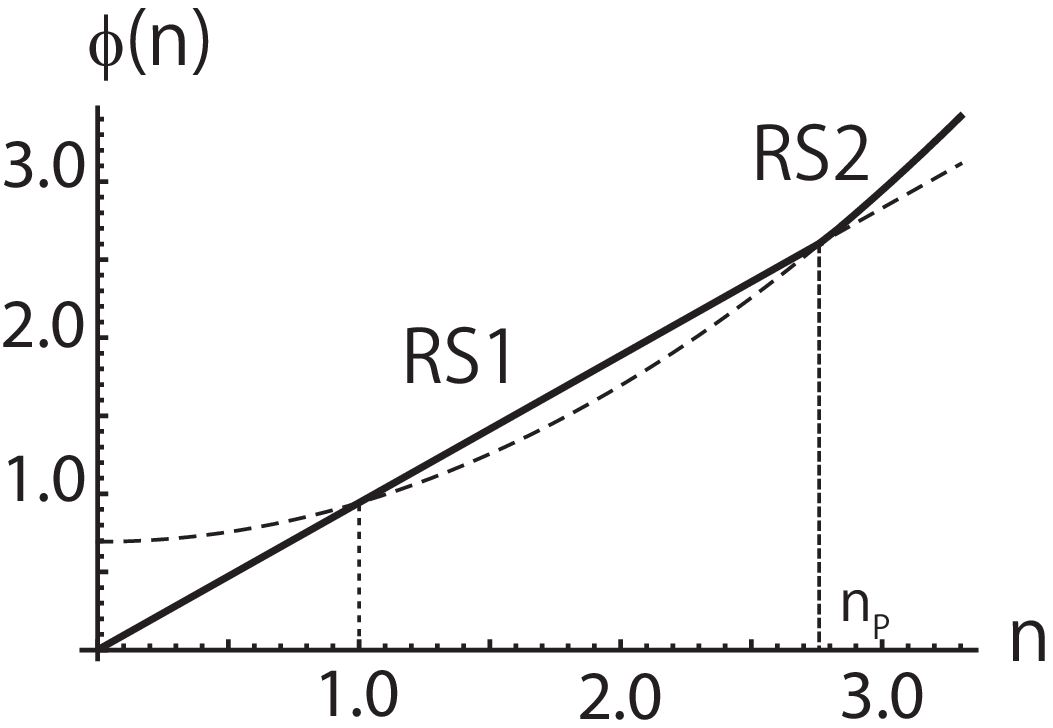}
\end{center}
\end{minipage}
\hspace{2mm}
 \begin{minipage}{0.32\hsize}
\begin{center}
   \includegraphics[height=40mm,width=50mm]{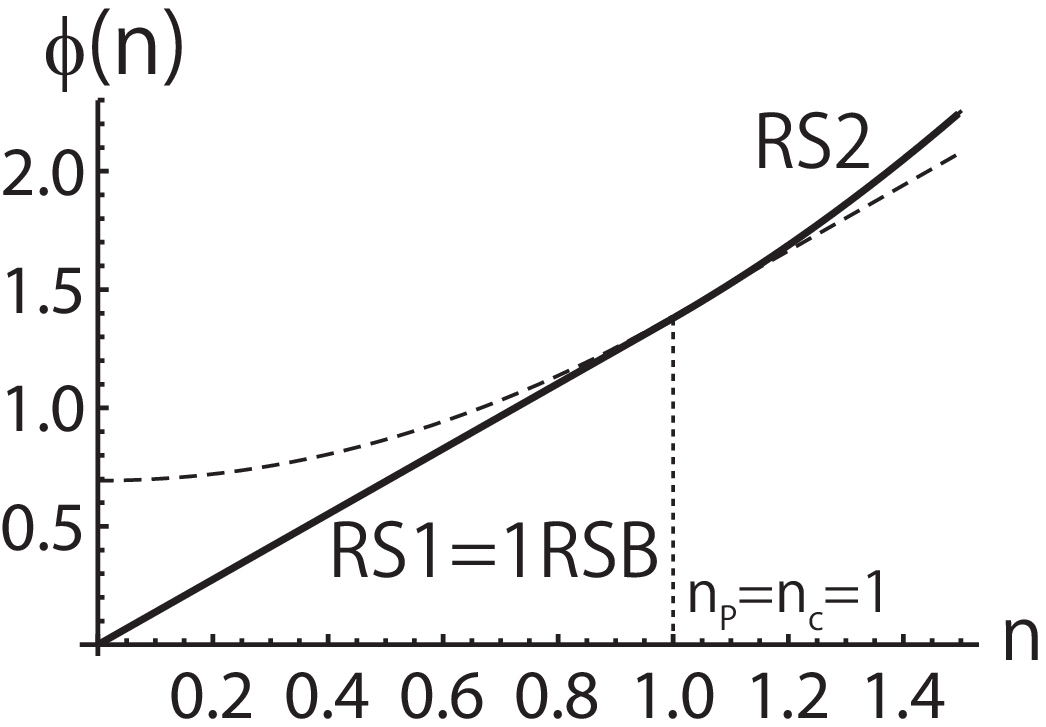}
\end{center}
 \end{minipage}
\hspace{2mm}
 \begin{minipage}{0.32\hsize}
\begin{center}
   \includegraphics[height=40mm,width=50mm]{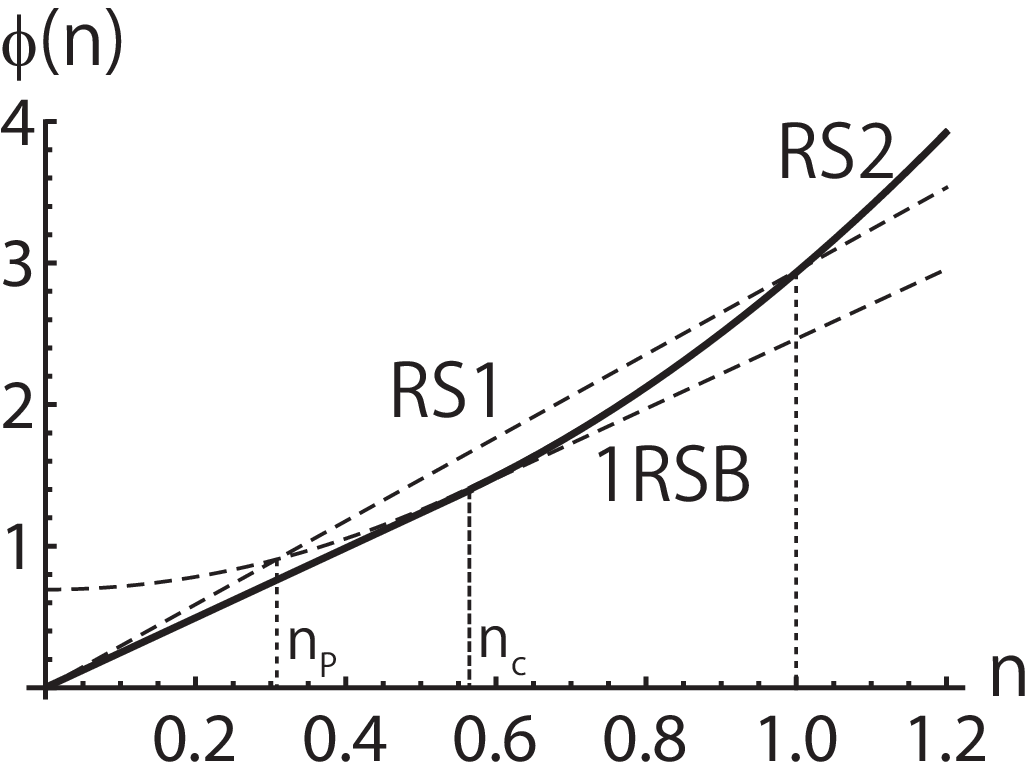}
\end{center}
 \end{minipage}
\caption{Behavior of $\phi(n)$. The solid lines represent the correct 
$\phi(n)$ and dotted lines are the RS and RSB branches. 
The corresponding values of temperature 
$T$ are $1,T_{c}\approx 0.6$ and 
$0.333$, from left to right.}
\label{fig2:phi}
\end{figure}

We first treat the high temperature region $T>T_{c}$.
The correct solution of $\phi(n)$ in this region is given by 
\begin{equation}
\phi(n)=
\left\{
\begin{array}{ll}
\phi_{\rm RS2}(n) & (n>n_{P}(T)\equiv (4\log 2)/\beta^2=\beta_{c}^2/\beta^2)\\
\phi_{\rm RS1}(n) & (n\leq n_{P}(T))
\end{array}
\right. ,\label{eq2:phi_highT}
\end{equation}
which gives the correct free energy 
$-\beta f_{P}=\lim_{n\to 0} 
\left( \partial \phi_{\rm RS1}(n)/\partial n \right)$.
Note that the boundary $n_{P}(T)>1$ is determined by equating 
two solutions  $\phi_{\rm RS1}$ and $\phi_{\rm RS2}$, which agrees with the 
concept of the saddle-point method, but 
another intersection at $n=1$ is 
ignored in the solution (\ref{eq2:phi_highT}) even though the RS2 solution 
exceeds the RS1 below $n=1$. 
The correctness of 
this prescription in the present case can be understood from the following 
criteria which should be satisfied by the correct $\phi(n)$:
\begin{enumerate}
\item{The generating function $\phi(n)$ 
is a convex downward function with respect to $n$.} 
\item{The rate function $R(f(n))=\phi(n)-n(\partial \phi(n)/\partial n)$ is nonpositive.}
\item{An equality 
$\lim_{n\to0 }\phi(n)=0$ should hold basically\footnote{In some cases, the identity $\lim_{n\to 0 }\phi(n)=0$ can be broken even though 
 the similar equality  for finite $N$,
$\phi_{N}(0)=0$, holds, due to taking the 
thermodynamic limit first. We will see such an example in chapter 3.}.
}
\item{The entropy $s=(\partial/\partial T)( \lim_{n\to 0}(
\partial \phi(n)/\partial n)/\beta)$ 
cannot be negative.}
\end{enumerate}
The RS2 solution clearly violates the third criterion which comes from 
a fact that for finite $N$ the generating function
$\phi_{N}(n)=\log[Z^n]_{\V{J}}/N$ necessarily becomes $0$ at $n=0$. 
Hence, we choose the RS1 solution as the correct one 
in the whole range of $n\leq 1$. This solution is illustrated 
in the left panel of fig.\ \ref{fig2:phi}.
Empirically, it is known that 
an intersection at $n=1$ should be neglected in most cases, but  
the general reason is not found and we only point out this fact here.

On the other hand, for the low temperature region $T\leq T_{c}$, 
we need to construct another solution for $\phi(n)$ since 
the solution (\ref{eq2:phi_highT}) gives incorrect results.
It is known that the correct solution is given by 
\begin{equation}
\phi(n)=
\left\{
\begin{array}{ll}
\phi_{\rm RS2}(n) & (n>n_{c}(T)\equiv (2\sqrt{\log 2})/\beta=
\beta_{c}/\beta)\\
\frac{\phi_{\rm RS2}(n_{c})}{n_{c}} n & (n\leq n_{c}(T))
\end{array}
\right. ,\label{eq2:phi_lowT}
\end{equation}
and is illustrated in the right panel of fig.\ \ref{fig2:phi}.
The content of this solution is as follows.
For large $n$, the larger RS solution, $\phi_{\rm RS2}(n)$, 
is correct as the high temperature region.
Decreasing $n$, 
we find an intersection of the RS1 and RS2 solutions 
at $n=1$, but as mentioned in the 
previous paragraph this intersection should be neglected. 
Actually, if we adopt 
the RS1 solution, we obtain the negative entropy at $n=0$, which violates 
the fourth criterion in the previous paragraph.
Proceeding to the range of $n < 1$, we can find that 
the rate function $R(f(n))$ becomes positive
below $n_{c}(T)$, which violates the constraint $R(f(n))\leq 0$.  
A natural prescription to avoid this inconsistency is keeping 
the value of $R(f(n))$ as $0$ below $n_{c}$, which leads to 
the solution (\ref{eq2:phi_lowT}).
This prescription can be proved to be correct 
in more mathematically proper procedures which focus on 
the convexity of $\phi(n)$ and the monotonicity of $\phi(n)/n$ 
\cite{Naka}. 

The solution (\ref{eq2:phi_lowT}) relates to the 1RSB solution. 
In fact, if we set $q_{0}=0$ in eq.\ (\ref{eq2:phi1RSB}), which 
is a necessary condition for the nontrivial 1RSB in the current case 
solution\footnote{The nontrivial solution of equations of states 
(\ref{eq2:q1})-(\ref{eq2:qhat}) is only given by $q_{1}>0$ and $q_{0}=0$ in 
the present model.}, 
the functional form of $\phi_{\rm 1RSB}(n,m)$ can be expressed by using 
the RS solution as\footnote{Another 1RSB order parameter $q_{1}$ is identified with the RS one $q$.}
\begin{equation}
\phi_{\rm 1RSB}(n,m)=\frac{n}{m}\phi_{\rm RS}(m)\label{eq2:1RSB-RS}.
\end{equation}
The interpretation of this equation is simple. The equality $q_{0}=0$ means 
that each pure state is perfectly uncorrelated. Thus, 
$\phi_{\rm 1RSB}(n,m)$ becomes proportional to $n/m$, as we can see in fig.\ 
\ref{fig2:1RSB}, which comes from the independent $y=n/m$ sets of replicas. 
In each set of the replicas, $x=m$ replicas are included and the overlap
$q_{1}$ takes a value, which is the same situation as the RS one. 
These considerations naturally lead to eq.\ (\ref{eq2:1RSB-RS}).
  
\begin{figure}[t]
\begin{tabular}{cc}
\hspace{-5mm}
\begin{minipage}[t]{0.50\hsize}
\begin{center}
 \includegraphics[height=50mm,width=60mm]{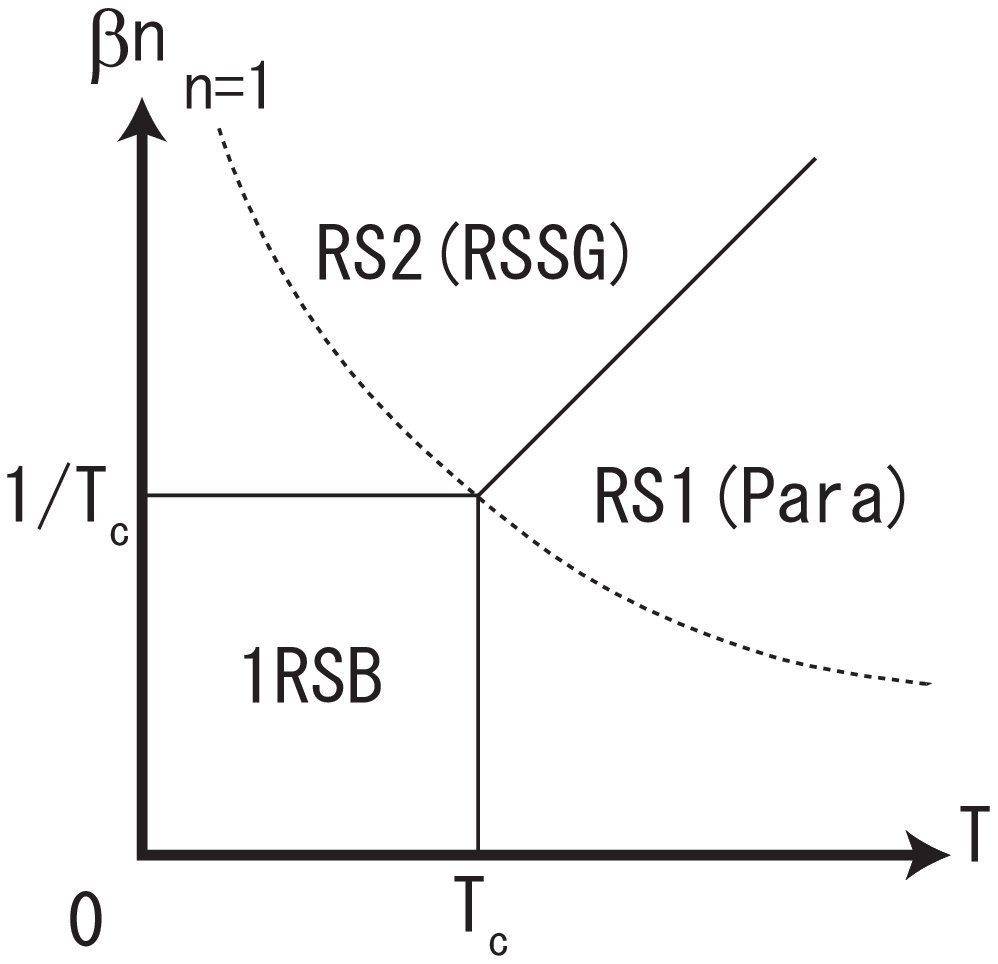}
 \caption{Phase diagram of the REM on the $T$-$\beta n$ plane.
The solid lines denote the phase boundaries and the dotted line represents 
the line $n=1$.}
\label{fig2:PD_REM}
\end{center}
\end{minipage}
\hspace{2mm}
 \begin{minipage}[t]{0.50\hsize}
\begin{center}
 \includegraphics[height=50mm,width=60mm]{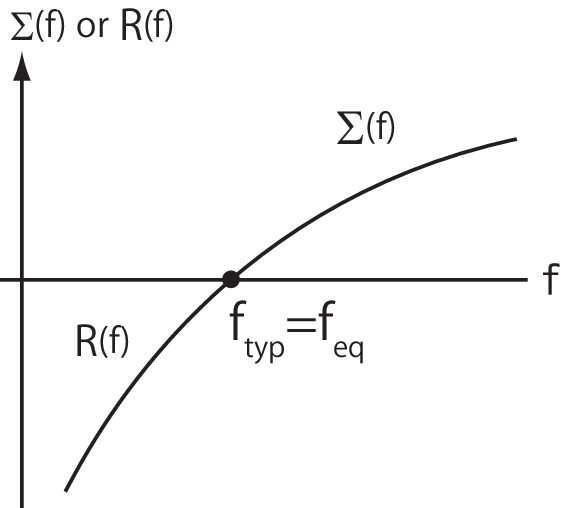}
 \caption{Schematic diagram of the complexity $\Sigma(f)$ 
and rate function $R(f)$. The complexity and rate function have 
disjointed domains of definition but their functional forms are identical, which leads to that the rate function is naturally 
continued to the complexity by using the 
RS generating function $\phi_{\rm RS}(n)$.}
 \label{fig2:Sigma-R}
\end{center}
\end{minipage}
\end{tabular}
\end{figure}

Taking the extremization condition of eq.\ (\ref{eq2:1RSB-RS}) with respect to 
$m$ with fixing $T<T_{c}$, 
we find that the breaking parameter $m$ becomes equal to 
$n_{c}$, and the solution (\ref{eq2:phi_lowT}) is reproduced.
Hence, the solution $\phi(n)=(\phi_{\rm RS2}(n_{c})/n_{c})n$ can be 
regarded as the 1RSB solution. Using eqs.\ (\ref{eq2:phi_highT}) and 
(\ref{eq2:phi_lowT}), 
we can depict a phase diagram of the REM on the $T$-$\beta n$ plane 
and the result is given in fig.\ \ref{fig2:PD_REM}.
If we see the complexity and rate function, the relation between 
the 1RSB construction and large deviations becomes clearer. 
Using eqs.\ (\ref{eq2:f(x)_2}) and (\ref{eq2:1RSB-RS}), 
we can express the complexity as 
\begin{equation}
\Sigma(f(m))=\phi_{\rm RS}(m)-m\Part{\phi_{\rm RS}(m)}{m}{},
\label{eq2:RScomp}
\end{equation}
the functional form of which is obviously 
identical to that of the rate function 
$R(f(n))=\phi_{\rm RS}(n)-n(\partial \phi_{\rm RS}(n)/\partial n)$, though 
their domains of definition are disjointed except for 
a point of typical value of 
free energy $f_{\rm typ}$ (which is identified with the equilibrium free energy $f_{\rm eq}$ in the context of pure state statistics), 
where the complexity and rate function become $0$. 
This is also illustrated in fig.\ \ref{fig2:Sigma-R}.

Summarizing the above discussions, 
we have seen the physical significance of the RSB 
from various perspectives by treating the REM as an example.  
In the remaining sections of this chapter, we treat the finite $p$ 
cases by utilizing the above discussions. 
In addition, 
we present some arguments about the microscopic description of 
pure states, which becomes important when we treat problems 
requiring microscopic information like the optimization problems.

\section{Finite $p$ cases}
In this section, we treat the finite $p$ cases of the fully-connected 
$p$-spin interacting model. Especially, we concentrate on the 
$p=2$ and $3$ cases, since it is known that the qualitative behavior 
for $p\geq 4$ is similar to either one of those cases. 
To treat the $p=2,3$ cases, 
we need to consider the transition to the FRSB.
Although the 1RSB transition can be observed by calculating the complexity
in the 1RSB framework, the FRSB is signaled by the local instability of the 
RS solution, which is called the de Almeida-Thouless (AT) instability 
\cite{Alme}. 
Hence, we first explain the AT instability, and then 
present the phase diagrams of the $p=2,3$ cases.

\subsection{de Almeida-Thouless instability}\label{sec2:AT}
Let us remember that the generating function $\phi(n)$ has 
the following form 
\begin{eqnarray}
\phi(n)=\Extr{
q^{\mu\nu},
\Wh{q}^{\mu\nu}
}
\left\{ \frac{\beta^2  }{2}\sum_{\mu < \nu}( q^{\mu\nu} )^p
      -\sum_{\mu<\nu}q^{\mu\nu}\Wh{q}^{\mu\nu} 
  +\frac{1}{4}\beta^2   n
  +\log \Tr{} 
  e^{\sum_{\mu<\nu}\Wh{q}^{\mu\nu}
 S^{\mu}S^{\nu}}
\right\}.\label{eq2:phi_again}
\end{eqnarray}
We consider the deviations of the order parameter $q^{\mu \nu}$ 
around the RS saddle-point
\begin{equation}
q^{\mu\nu}=q+\Delta^{\mu\nu},
\end{equation}
and expand $\phi(n)$ with respect to $\Delta^{\mu\nu}$ to second 
order. The conjugate order parameter $\Wh{q}^{\mu\nu}$ becomes
\begin{equation}
\Wh{q}^{\mu\nu}=\frac{1}{2}p\beta^2 (q^{\mu\nu})^{p-1}
\approx 
\frac{1}{2}p\beta^2
\left\{
q^{p-1}+(p-1)q^{p-2}\Delta^{\mu\nu}+\frac{1}{2}(p-1)(p-2)
\left(\Delta^{\mu\nu}\right)^2
\right\},
\end{equation}
hence the leading two terms in eq.\ (\ref{eq2:phi_again}) become
\begin{equation}
\frac{\beta^2  }{2}\sum_{\mu < \nu}( q^{\mu\nu} )^p
      -\sum_{\mu<\nu}q^{\mu\nu}\Wh{q}^{\mu\nu} 
\approx
-\frac{1}{4}\beta^2p(p-1)^2q^{p-2}\sum_{\mu<\nu}(\Delta^{\mu\nu})^2,
\end{equation}
where we leave only the second order terms because the zeroth order terms 
are irrelevant to the variation of $\phi(n)$, $\Delta\phi(n)$, and 
the first order terms become $0$ since we take the saddle-point condition 
with respect to $q$. 
Similarly, we can expand the last term of eq.\ (\ref{eq2:phi_again}) 
as 
\begin{eqnarray}
&&\hspace{-8mm}
\log \Tr{} 
  e^{\sum_{\mu<\nu}\Wh{q}^{\mu\nu}
 S^{\mu}S^{\nu}}
\approx
\frac{1}{4}p(p-1)(p-2)\beta^2 q^{p-3}
\sum_{\mu<\nu} (\Delta^{\mu\nu})^2 \Ave{S^{\mu}S^{\nu}}_{\rm RS}
\nonumber \\
&&\hspace{-8mm}+
\frac{1}{8} p^2 (p-1)^2 \beta^4 q^{2p-4}
\sum_{\mu<\nu} \sum_{\delta<\omega} \Delta^{\mu\nu}\Delta^{\delta\omega} 
\left(
\Ave{S^{\mu}S^{\nu}S^{\delta}S^{\omega}}_{{\rm RS}}
-\Ave{S^{\mu}S^{\nu}}_{{\rm RS}}
\Ave{S^{\delta}S^{\omega}}_{{\rm RS}}
\right),
\end{eqnarray}
where the brackets $ \Ave{ \cdots }_{{\rm RS}}$ denote the average over the 
RS weight $e^{ \Wh{q} \sum_{\mu<\nu} S^{\mu} S^{\nu} }$.
Regarding that an equality $\Ave{S^{\mu}S^{\nu}}_{{\rm RS}}=q$ holds, we 
eventually derive the variation of the generating function 
$\Delta \phi(n)$ as 
\begin{eqnarray}
&&
\Delta \phi(n)=-\frac{1}{4}\beta^2 p(p-1)q^{p-2}
\sum_{\mu<\nu} \sum_{\delta<\omega} \Delta^{\mu\nu}\Delta^{\delta\omega} 
\Biggl\{
\delta_{(\mu\nu),(\delta\omega)}
\nonumber \\
&&
\hspace{20mm}-\frac{1}{2}p(p-1) \beta^2 q^{p-2}
\left(
\Ave{S^{\mu}S^{\nu}S^{\delta}S^{\omega}}_{{\rm RS}}
-\Ave{S^{\mu}S^{\nu}}_{{\rm RS}}
\Ave{S^{\delta}S^{\omega}}_{{\rm RS}}
\right)
\Biggr\},
\end{eqnarray}
where $\delta_{(\mu\nu),(\delta\omega)}$ is the Kronecker delta function which 
yields $1$ when $(\mu\nu)=(\delta\omega)$ and $0$ otherwise.
We denote the matrix of coefficients of $\Delta^{\mu\nu}$ by $G$ 
which is called the {\it Hessian} matrix.
The local stability of the RS solution requires all the eigenvalues of $G$ 
be positive.
Here we classify the elements of $G$ by using the following notations
\begin{eqnarray}
&&G_{(\mu\nu)(\mu\nu)}=1-\frac{1}{2}p(p-1)\beta^2q^{p-2}(1-
\Ave{S^{\mu}S^{\nu}}_{{\rm RS}})
\equiv P,\label{eq:P}\\
&&G_{(\mu\nu)(\mu\delta)}=-\frac{1}{2}p(p-1)\beta^2q^{p-2} 
\left(
\Ave{S^{\nu}S^{\delta}}_{{\rm RS}}-\Ave{S^{\mu}S^{\nu}}_{{\rm RS}}^2
\right)\equiv Q
,\\
&&G_{(\mu\nu)(\delta\omega)}=-\frac{1}{2}p(p-1)\beta^2q^{p-2}
\left(
\Ave{S^{\mu}S^{\nu}S^{\delta}S^{\omega}}_{{\rm RS}}
-\Ave{S^{\mu}S^{\nu}}_{{\rm RS}}^2
\right)\equiv R.
\end{eqnarray}

\subsubsection{Eigenvalues of the Hessian and the AT condition}
Let us find the eigenvectors of $G$ 
by a heuristic approach found by Almeida and Thouless.
The first eigenvector $\V{s}_{1}$ is obtained by assuming
$\Delta^{\mu\nu}=a$ for any $\mu,\nu$.
In this condition, 
from the eigenvalue equation $G\V{s}_{1}=\lambda_{1}\V{s}_{1}$ 
we can easily derive the first eigenvalue $\lambda_{1}$ as 
\begin{equation}
\lambda_{1}=
P+2(n-2)Q+\frac{1}{2}(n-2)(n-3)R \label{eq2:lambda1}.
\end{equation}
The next type of solution $\V{s}_{2}$ 
is obtained by treating a replica $\theta$ 
as a special one.
We assume $\theta=1$ without loss of generality.
This solution $\V{s}_{2}$ has 
$\Delta^{\mu\nu}=b$ 
when $\mu$ or $\nu$ is equal to $1$, 
$\Delta^{\mu\nu}=c$ otherwise.
The first row of the eigenvalue equation 
$G\V{s}_{2}=\lambda_{2}\V{s}_{2}$
gives
\begin{equation}
(P+(n-2)Q)b+
\left(
(n-2)Q+  
\frac{1}{2}(n-2)(n-3)R
\right)c=\lambda_{2}b. \label{eq2:l2_1}
\end{equation}
We here impose the orthogonal condition 
of the solutions $\V{s}_{2}$ and $\V{s}_{1}$, which leads to
\begin{equation}
(n-1)b+\frac{1}{2}(n-1)(n-2)c=0\label{eq2:l2_2},
\end{equation}
Solving eqs.\ (\ref{eq2:l2_1}) and (\ref{eq2:l2_2})
under the condition $b,c \neq 0$, 
we get
\begin{eqnarray}
\lambda_{2}=P+(n-4)Q-(n-3)R.\label{eq2:lambda2}
\end{eqnarray}

The third mode $\V{s}_{3}$ is obtained by treating two replicas 
$\theta,\omega$ as special ones. 
This solution $\V{s}_{3}$ has 
$\Delta^{\theta\omega}=d$, 
$\Delta^{\theta\mu}=e$ when $\mu\neq \omega$ 
and
$\Delta^{\mu\nu}=f$ otherwise. 
Without loss of generality, we put $\theta=1$ and $\omega=2$.
The orthogonal condition of $\V{s}_{1}$ and $\V{s}_{3}$ yields
\begin{equation}
d+2(n-2)e+\frac{1}{2}(n-2)(n-3)f=0,\label{eq2:l3_1}
\end{equation}
and 
the orthogonal condition of $\V{s}_{2}$ and $\V{s}_{3}$ also gives
\begin{equation}
d+(n-2)e=0,\,\, e+\frac{1}{2}(n-3)f=0.\label{eq2:l3_2}
\end{equation}
The first row of the 
eigenvalue equation $G\V{s}_{3}=\lambda_{3}\V{s}_{3}$
leads to 
\begin{equation}
Pd+2(n-2)Qe+ \frac{1}{2}(n-2)(n-3)Rf=\lambda_{3}d. \label{eq2:l3_3},
\end{equation}
Solving eqs.\ (\ref{eq2:l3_1})-(\ref{eq2:l3_3}), 
we obtain
\begin{equation}
\lambda_{3}=P-2Q+R \label{eq:lambda3}.
\end{equation}
The eigenvectors $\V{s}_{1}$, $\V{s}_{2}$ and $\V{s}_{3}$ 
construct $n(n-1)/2$-dimensional space, hence all 
the eigenvalues are exhausted by these three modes.

For the stability of the saddle point, 
all of the eigenvalues must be non-negative. 
We hereafter concentrate on a region $n\leq 1$. 
In this region, from simple 
algebras we can see that inequalities 
$\lambda_{1},\lambda_{2}\geq \lambda_{3}$ hold, which 
means that the mode $\V{s}_{3}$ is the relevant mode for the stability
of the RS solution and is called the {\it replicon} mode.
To obtain an explicit form of $\lambda_{3}$,  
we calculate
the factor $r\equiv\Ave{S^{\mu}S^{\nu}S^{\delta}S^{\omega}}$ in $R$
and the result is  
\begin{equation}
r\equiv \Ave{S^{\mu}S^{\nu}S^{\delta}S^{\omega}}_{\rm RS}
=\frac{\int Dz \cosh^n \sqrt{\Wh{q}}z \tanh^4 \sqrt{\Wh{q}}z }{
\cosh^n \sqrt{\Wh{q}}z
}.\label{eq2:r} 
\end{equation}
Remembering $\Ave{S^{\mu}S^{\nu}}_{\rm RS}$ in $P,Q$ and $R$ 
is equal to $q$ given in eq.\ (\ref{eq2:q}),
we finally obtain 
\begin{equation}
\lambda_{3}=P-2Q+R=1-\frac{1}{2}p(p-1)\beta^2 q^{p-2}
\frac{
\GI{z}( \cosh \sqrt{ \Wh{q} } z )^{n-4}
}{ \GI{z} (\cosh \sqrt{ \Wh{q} } z )^n }
\geq 0,
\end{equation}
as the AT stability condition of the RS solution.
Once the AT stability breaks, it is believed 
that the FRSB solution is required 
to describe the behavior of the system. 

\subsubsection{Connection to the divergence of the spin-glass susceptibility}
As we can expect, the AT condition is related to the 
divergence of 
the spin-glass susceptibility $\chi_{SG}$ which is defined 
in section \ref{sec1:spinglass}.
To elucidate this point, we here calculate the spin-glass susceptibility 
for the fully-connected $p$-spin interacting model. 
The following calculations are based on those in reference \cite{ATSG}.

We first introduce the modified generating function $\Wt{\phi}(n,m,\V{F})$, 
in which we introduce the RS interaction $\V{F}$ among 
$m$ out of $n$ replicas,
\begin{eqnarray}
&&N \Wt{\phi}_{N}(n,m,\V{F})
=\log 
\left[
\left(
\Tr{}e^{-\beta H}
\right)^{n-m}
\left(
\Tr{}e^{\beta \sum_{a=1}^{m}H^{a}
+\sum_{i}^{N}F_{i}\sum_{a<b}S_{i}^{a}S_{i}^{b}}
\right)
\right]_{\V{J}} 
\nonumber \\
&&=
\log 
\left[
\left(
\Tr{}e^{\beta 
\sum_{i_{1}<\ldots <i_{p}} J_{i_{1}\ldots i_{p}} 
\sum_{\mu=1}^{n}
S_{i_{1}}^{\mu} \ldots S_{i_{p}}^{\mu}
+\sum_{i}^{N}F_{i}\sum_{a<b}S_{i}^{a}S_{i}^{b}}
\right)
\right]_{\V{J}}. \label{eq2:phi1}
\end{eqnarray}
We hereafter assume that the indices $a,b$ run only among $m$ replicas, 
while $\mu,\nu$ run all $n$ replicas.
Note that not only the limit $\V{F}\to 0$ but also 
the limit $m\to 1$ reproduce the normal generating function $\phi(n)$ 
from eq.\ (\ref{eq2:phi1}). 
The limit $m\to 1$ can be taken by using the analytic continuation from 
$m \in \mN$ to $\mC$ 
\begin{equation}
N \Wt{\phi}(n,m,\V{F})=
\log
\left[
\prod_{i=1}^{N}\GI{z_{i}}
\left(
\Tr{}e^{-\beta H+\sum_{i=1}^{N}\sqrt{F_{i}}z_{i}S_{i} }
\right)^{m}
\left(
 \Tr{}e^{-\beta H}
\right)^{n-m}
\right]_{\V{J}},
\end{equation}
where we have used the Hubbard-Stratonovich transformation. 
To see the stability of the system against introduction $\V{F}$, 
let us expand $\Wt{\phi}(n,m,\V{F})$ with respect to $\V{F}$. 
Denoting $Z(\V{J})=\Tr{}e^{-\beta H(\V{S}|\V{J})}$, 
we can express the first derivative as
\begin{equation}
N\Part{ \Wt{\phi} }{F_{k}}{}=
\frac{
\left[
Z^{n-m}(\V{J})\left(\Tr{}
\left(\sum_{a<b}S_{k}^{a}S_{k}^{b} \right)
e^{
\beta \sum_{a=1}^{m}H^{a}
+\sum_{i}^{N}F_{i}\sum_{a<b}S_{i}^{a}S_{i}^{b}
}
\right)
\right]_{\V{J}}
}{
\left[
Z^{n-m}(\V{J})\left(
\Tr{} 
e^{\beta \sum_{a=1}^{m}H^{a}
+\sum_{i}^{N}F_{i}\sum_{a<b} S_{i}^{a}S_{i}^{b}  } 
\right)
\right]_{\V{J}}
}=
\left[
\sum_{a<b}
\Ave{
S_{k}^{a}S_{k}^{b}
}_{m,\V{F}}
\right]_{n,\V{F}},
\end{equation}
where the brackets $\Ave{(\cdots)}_{m,\V{F}}$ 
and $[(\cdots)]_{n,\V{F}}$ denote the following 
averages
\begin{eqnarray}
&&\Ave{(\cdots)}_{m,\V{F}}=\frac{1}{Z^{m}(\V{J},\V{F})}
\Tr{}
\left(
\cdots 
\right)
e^{
\beta \sum_{a=1}^{m}H^{a}
+\sum_{i}^{N}F_{i}\sum_{a<b}S_{i}^{a}S_{i}^{b}
}\label{eq2:aveF}
\\
&&
[(\cdots)]_{n,\V{F}}=
\frac{
\left[
\left(
\cdots
\right)
Z^{n-m}(\V{J})Z^{m}(\V{J},\V{F})
\right]_{\V{J}}
}{
\left[
Z^{n-m}(\V{J})Z^{m}(\V{J},\V{F})
\right]_{\V{J}}
}\label{eq2:avenF}
\end{eqnarray}
where we define $Z(\V{J},\V{F})=\Tr{}
e^{
\beta \sum_{a=1}^{m}H^{a}
+\sum_{i}^{N}F_{i}\sum_{a<b}S_{i}^{a}S_{i}^{b}
}$, 
and the second derivative becomes
\begin{eqnarray}
\hspace{-5mm}
N\frac{\partial^{2} \Wt{\phi} }{\partial F_{k} \partial F_{l}}
=\left[
\sum_{a<b}\sum_{c<d} 
\Ave{S_{k}^{a}S_{k}^{b}S_{l}^{c}S_{l}^{d}}_{m,\V{F}}
\right]_{n,\V{F}}
-
\left[
\sum_{a<b}
\Ave{S_{k}^{a}S_{k}^{b}}_{m,\V{F}} 
\right]_{n,\V{F}}
\left[
\sum_{c<d} \Ave{S_{l}^{c}S_{l}^{d}}_{m,\V{F}} 
\right]_{n,\V{F}}\label{eq2:d2phit}.
\end{eqnarray}
The first term of this equation gives the following three contributions
\begin{eqnarray}
&&\frac{m(m-1)}{2}
\left[
\Ave{
S_{k}^{a} S_{l}^{a} S_{k}^{b} S_{l}^{b}
}_{m,\V{F}}
\right]_{n,\V{F}}
+
m(m-1)(m-2)\left[
\Ave{ 
S_{k}^{a}S_{l}^{a} 
S_{k}^{b} S_{l}^{d}
}_{m,\V{F}}
\right]_{n,\V{F}}
\nonumber
\\
&&+
\frac{m(m-1)(m-2)(m-3)}{4}
\left[
\Ave{
S_{k}^{a}
S_{l}^{b}
S_{k}^{c}
S_{l}^{d}
}_{m,\V{F}}
\right]_{ n,\V{F} },
\label{eq2:firstcont}
\end{eqnarray}
Hereafter, we concentrate on the case where 
$m$ is slightly larger than unity $m=1+\epsilon$. In this case, 
eq.\ (\ref{eq2:firstcont}) yields $O(m-1)$ terms. 
On the other hand, 
the second term in eq.\ (\ref{eq2:d2phit})
is negligible because the order is $O((m-1)^2)$. 
Taking the limit $\V{F}\to 0$, we can find that 
the $m$ replicas become completely independent and 
$\Ave{(\cdots)}_{m,\V{F}}$ is reduced to the thermal average 
$\Ave{(\cdots)}_{m}$ with respect to the $m$-replicated original Hamiltonian 
$\sum_{a=1}^{m}H^{a}$, which yields the following expressions of 
the derivatives up to the order $O(m-1)$,
\begin{eqnarray}
\left.
N\Part{ \Wt{\phi} }{F_{k}}{}\right|_{\V{F}=0}
\approx
\frac{(m-1)}{2}q,
\,\,\,
\left.
N\frac{\partial^{2} \Wt{\phi} }{\partial F_{k} \partial F_{l}}
\right|_{\V{F}=0}
\approx
\frac{m-1}{2}
\left[
\left(\Ave{S_{l}S_{k}}-\Ave{S_{l}}\Ave{S_{k}}\right)^{2}
\right]_{n},\label{eq2:phitpp1}
\end{eqnarray} 
where the brackets $[(\cdots)]_{n}$ denote 
$[(\cdots)]_{n}=\lim_{\V{F}\to 0}[(\cdots)]_{n,\V{F}}=[Z^n(\cdots)]_{\V{J}}/[Z^n]_{\V{J}}$.
Hence, we obtain the modified generating function as 
\begin{equation}
N\Wt{\phi}(n,1+\epsilon,\V{F})
\approx
N\phi(n)+\frac{\epsilon}{2}q\sum_{i}F_{i}
+\frac{\epsilon}{2}\V{F}^{\rm T}\Wh{\chi}_{SG}\V{F}
+O(\epsilon^2),
\label{eq2:phit_expand}
\end{equation}
where $\V{F}^{\rm T}$ denotes the transpose matrix of $\V{F}$ and 
$\Wh{\chi}_{SG}$ represents the spin-glass susceptibility matrix 
having the component $(\Wh{\chi}_{SG})_{lk}=\left[
\left(\Ave{S_{l}S_{k}}-\Ave{S_{l}}\Ave{S_{k}}\right)^{2}
\right]_{n}$.
Equation (\ref{eq2:phit_expand}) implies that 
the largest eigenvalue of the matrix $\Wh{\chi}_{SG}$ is related to 
the stability of the system against introduction $\V{F}$. 
We can expect that the system statistically has the translational invariance,
which implies that the eigenvector of the largest eigenvalue is 
given by $\V{1}=(1,1,\cdots,1)/\sqrt{N}$. 
Hence, the largest eigenvalue becomes
\begin{equation}
\frac{1}{N}\V{1}^{\rm T}\Wh{\chi}_{SG}\V{1}=\frac{1}{N}\sum_{l,k}
\left[
\left(\Ave{S_{l}S_{k}}-\Ave{S_{l}}\Ave{S_{k}}\right)^{2}
\right]_{n}=\chi_{SG}.\label{eq2:chiSG}
\end{equation}
This gives a natural extension of the spin-glass susceptibility
 to the finite replica case and implies the relation between 
the divergence of the spin-glass susceptibility and 
the instability against introduction of interactions between replicas.

Next, we express $\Wt{\phi}(n,m,\V{F})$ in a more tractable form.
Using the replica method and 
following the calculations in section \ref{sec2:RM}, 
we can obtain
\begin{eqnarray}
\Wt{\phi}(n)=\Extr{
q^{\mu\nu},
\Wh{q}^{\mu\nu}
}
\left\{ \frac{\beta^2  }{2}\sum_{\mu < \nu}( q^{\mu\nu} )^p
      -\sum_{\mu<\nu}q^{\mu\nu}\Wh{q}^{\mu\nu} 
  +\frac{1}{4}\beta^2   n
  +
\frac{1}{N}
\log 
\Tr{} 
  e^{L}
\right\},\label{eq2:phit_ext}
\end{eqnarray}
where 
\begin{equation}
L=
\sum_{i=1}^{N}
\left\{
\sum_{\mu<\nu}\Wh{q}^{\mu\nu}
 S_{i}^{\mu}S_{i}^{\nu}
+
F_{i}
\sum_{a<b}
 S_{i}^{a}S_{i}^{b}
\right\}.
\end{equation}
The interaction $\V{F}$ breaks the replica symmetry and we should 
choose the appropriate form of $q^{\mu\nu}$ to take the saddle point. 
Without loss of generality, we can assume that 
the RS interaction $\V{F}$ is introduced into
the first $m$ replicas, $(1,\cdots,m)$. 
A natural form of $q^{\mu\nu}$ is then expressed as 
\begin{equation}
(q^{\mu\nu},\Wh{q}^{\mu \nu})=
\left\{
\begin{array}{ll}
(q_{1},\Wh{q}_{1}) & 
(\, \mu\in (1,\cdots, m)\,\, {\rm and}\,\, \nu \in (1,\cdots, m)\,) 
\\
(q_{0},\Wh{q}_{0}) & (\, {\rm otherwise}\,)
\end{array}
\right. .
\end{equation}
Under this ansatz, we get 
\begin{eqnarray}
&&\Wt{\phi}(n,m,\V{F})
=\Extr{q^{1},q^{0},\Wh{q}^1,\Wh{q}^0}
\Biggl\{
\frac{1}{4}n\beta^{2}
+\frac{1}{2}(n(n-1) -m(m-1))
\left( 
-\Wh{q}_{0}q_{0}+\frac{1}{2}\beta^2  q_{0}^{p}
\right)
\nonumber \\
&&
+\frac{1}{2}m(m-1)
\left( 
-\Wh{q}_{1}q_{1}+\frac{1}{2}\beta^2  q_{1}^{p}
\right)+\frac{1}{N}\log\Tr{} e^{L_{0}}
\Biggr\},
\end{eqnarray}
where 
\begin{equation}
L_{0}=\sum_{i=1}^{N}\left\{
\Wh{q}_{0}\sum_{\mu<\nu}S_{i}^{\mu}S_{i}^{\nu}
+(\Wh{q}_{1}-\Wh{q}_{0}+F_{i})\sum_{a<b}S_{i}^{a}S_{i}^{b}
\right\}.
\end{equation}
The saddle-point condition gives 
\begin{eqnarray}
&&q_{1}=\frac{2}{m(m-1)}\frac{1}{N}\sum_{i=1}^{N}
\Ave{\sum_{a<b}S_{i}^{a}S_{i}^{b}}_{L_{0}}, 
\label{eq2:q1F}
\\
&&
q_{0}=\frac{1}{N}\sum_{i=1}^{N}\frac{ 
\Ave{\sum_{\mu<\nu}S_{i}^{\mu}S_{i}^{\nu} }_{ L_{0} }
-\Ave{ \sum_{a<b}S_{i}^{a}S_{i}^{b} }_{ L_{0} } 
}{n(n-1)/2-m(m-1)/2}
\\
&&
\Wh{q}_{1}=\frac{1}{2}p\beta^2  q_{1}^{p-1},
\,\,\,\, 
\Wh{q}_{0}=\frac{1}{2}p\beta^2  q_{0}^{p-1},
\end{eqnarray}
where 
\begin{equation}
\Ave{(\cdots)}_{L_{0}}=\frac{1}{\Tr{}e^{ L_{0} } }\Tr{}(\cdots)e^{ L_{0} }.
\end{equation}  
In this expression, the derivatives of $\Wt{\phi}$ 
are given by
\begin{equation}
\Part{\Wt{\phi} }{F_{l}}{}
=
\frac{1}{2N}m(m-1)q_{1},
\,\,\,
\frac{\partial^{2} \Wt{\phi} }{\partial F_{k} \partial F_{l}}
=
\frac{1}{2N}m(m-1)\Part{q_{1}}{F_{k}}{}
.\label{eq2:phitpp2}
\end{equation}
For obtaining the eigenvalue of the Hessian $
\partial^{2} \Wt{\phi}/(\partial F_{k} \partial F_{l})$, 
we should calculate 
$\partial q_{1}/\partial F_{k}$.   
The (statistically) translational invariance 
of the system enables us to simplify the calculation 
by putting $F_{l}=F$ for $\forall{l}$.  
Differentiating eq.\ (\ref{eq2:q1F}) with respect 
to $F$, we get
\begin{equation} 
\Part{q_{1}}{F}{}=C+\Part{\Wh{q}_{1}}{F}{}\frac{2}{m(m-1)}
\left\{ 
\Ave{\sum_{a<b}\sum_{c<d}S^{a}S^{b}S^{c}S^{d}}_{L_{0}}
-
\Ave{ 
\sum_{a<b}S^{a}S^{b}
}_{L_{0}}
\Ave{ 
\sum_{c<d}S^{c}S^{d}
}_{L_{0}}
\right\},\label{eq2:qF}
\end{equation}
where the particular form of the factor $C$ is irrelevant for our current 
purpose and is omitted here. 
The second term in the braces $\{\cdots\}$ 
is again negligible if we consider 
the case $m=1+\epsilon$ because it is $O(\epsilon^2)$. 
In addition, in the limit $F\to 0$ we can expect that 
$q_{1}$ and $q_{0}$ become the RS spin-glass order parameter $q$, 
which leads to 
that the average $\Ave{(\cdots)}_{L_{0}}$ becomes identical to 
the average $\Ave{(\cdots)}_{\rm RS}$ 
in the limit $F\to 0$. These conditions yield 
\begin{eqnarray}
&&\Ave{\sum_{a<b}\sum_{c<d}S^{a}S^{b}S^{c}S^{d}}_{L_{0}}
\to
\frac{m(m-1)}{2}
+
m(m-1)(m-2)
\Ave{ 
S^{a} S^{b}
}_{\rm RS}
\nonumber
\\
&&+
\frac{m(m-1)(m-2)(m-3)}{4}
\Ave{
S^{a}
S^{b}
S^{c}
S^{d}
}_{\rm RS}
\approx
\frac{\epsilon}{2}\left\{
1-2q+r
\right\},
\end{eqnarray}
where $q$ and $r$ are defined in eqs.\ (\ref{eq2:q}) and (\ref{eq2:r}), respectively. 
Substituting 
these conditions and 
$\partial \Wh{q}_{1}/\partial F =
p(p-1)\beta^2  q^{p-2}( \partial\Wh{q}_{1}/\partial F)/2$  
into eq.\ (\ref{eq2:qF}), we finally get 
\begin{equation}
\left. \Part{q_{1}}{F}{}\right|_{F=0}\propto 
\left(
1-\frac{1}{2}\beta^2  p(p-1) q^{p-2}(1-2q+r) 
\right)^{-1}.\label{eq2:qFlast}
\end{equation}
Using eqs.\ (\ref{eq2:q}) and (\ref{eq2:r}), 
we find that this condition is identical to the AT condition.
From (\ref{eq2:phitpp1}) and (\ref{eq2:phitpp2}), the spin-glass 
susceptibility is proportional to $\partial q_{1}/\partial F$ and 
hence the AT condition coincides with the divergence of the spin-glass
susceptibility.

\subsection{Phase diagrams for the $p=2,3$ cases }
\begin{figure}[htbp]
\hspace{-7mm}
\begin{minipage}{0.5\hsize}
\begin{center}
   \includegraphics[height=50mm,width=60mm]{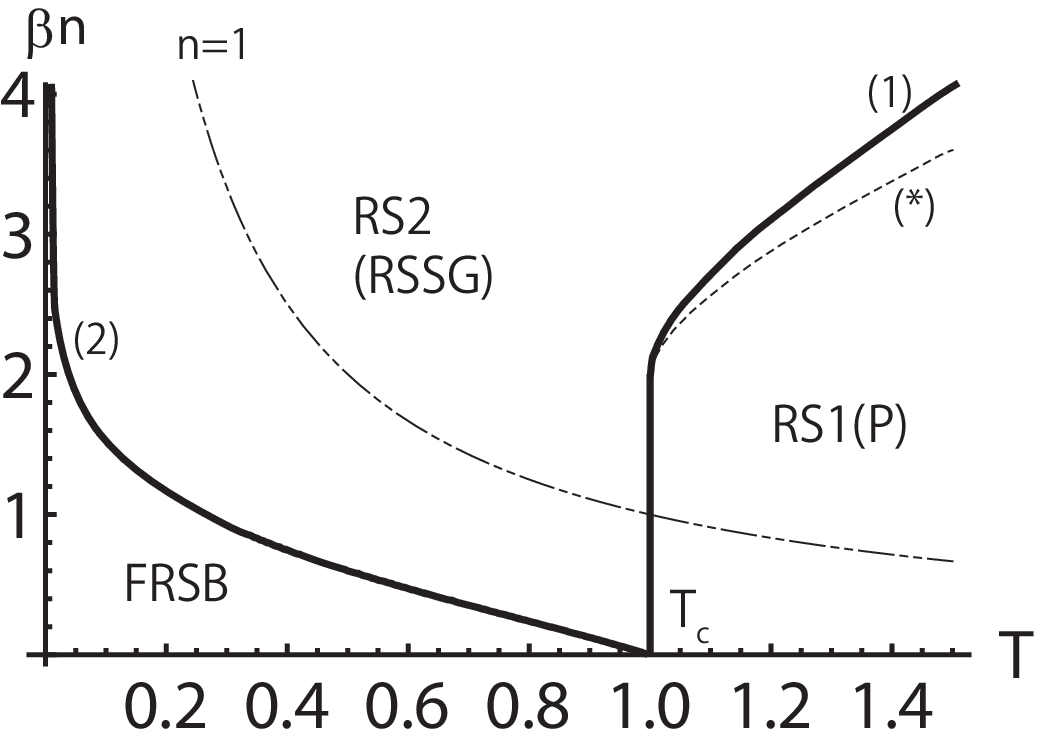}
 \caption{Phase diagram of the SK model $T$-$\beta n$ plane.  
The solid lines denote phase boundaries.  The boundary between RS2
 and FRSB phases is the AT line and is labeled $(2)$. 
The line $n=1$ is drawn dashed.}
 \label{fig2:2SKPD}
\end{center}
\end{minipage}
\hspace{2mm}
 \begin{minipage}{0.5\hsize}
\begin{center}
   \includegraphics[height=50mm,width=70mm]{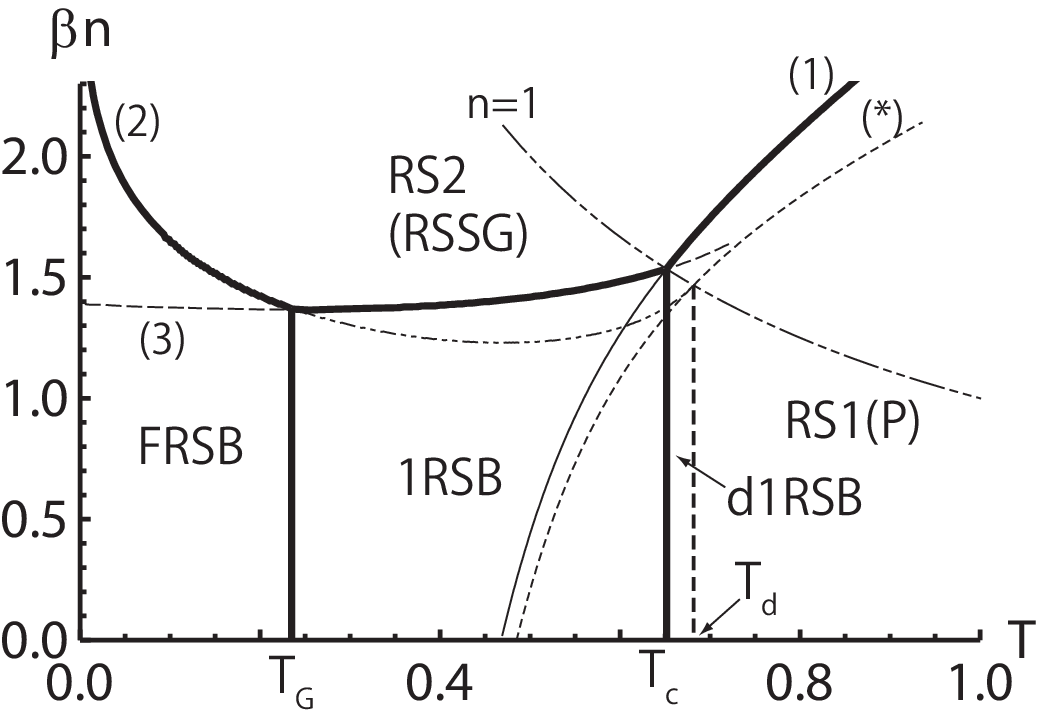}
   \caption{Phase diagram of the $p=3$ case on $T$-$\beta n$ plane.
The solid lines denote phase boundaries.
Above the line indicated by (*), there are two RS saddle points $(q=0,q>0)$. The dynamical transition temperature $T_{d}$ exists above the equilibrium transition $T_{c}$.}
 \label{fig2:3SKPD}
\end{center}
 \end{minipage}
\end{figure}

As mentioned in section \ref{sec2:REMLDT}, 
the 1RSB solution can be calculated by the RS solution as
$\phi_{\rm 1RSB}(n,m)=(n/m)\phi_{\rm RS}(m)$ 
in the present model\footnote{This is a particular property of models with 
$q_{0}=0$. If the external field is introduced into the present model, the equality $q_{0}= 0$ does not hold any more, and we must consider the 1RSB solution directly. We will treat such a model with $q_{0}\neq 0$ in chapter 3.}. The AT condition is also assessed by the RS solution and hence the RS solution is sufficient to derive 
phase diagrams for finite $p$.  
The resultant plots for $p=2$ and $3$ 
are given in figs.\ \ref{fig2:2SKPD} and 
\ref{fig2:3SKPD}, respectively. 
The solid lines in figs.\ \ref{fig2:2SKPD} and \ref{fig2:3SKPD} represent phase boundaries.
We see the contents of these phase diagrams in the following. 

First, let us consider the $p=2$ case.   
In this case, the transition temperature $T_{c}$ at $n=0$ 
is assessed by the 
perturbative approach because the transition is 
of second order.
We expand the right-hand side of 
eq.\ (\ref{eq2:q}) with respect to $\Wh{q}=\beta^2 q$ to second order
and get 
\begin{equation}
q\approx \beta^2 q+(n-2)\beta^4 q^2.
\end{equation}
This equation indicates that for $n\leq 2$ the transition is of second order 
and the transition temperature is $T_{c}=1$ and 
a phase boundary exists in a vertical form at $T_{c}$ 
from $n=0$ to $n=2$. 
For larger $n$, the transition is of first order
 and it is required to numerically solve eq.\ (\ref{eq2:q}).
Above the line labeled by $(*)$ in fig.\ \ref{fig2:2SKPD}, 
there are two stable solutions $q=0$ and $q>0$, and
we call them the RS1 and RS2 solutions, respectively, 
similarly to the REM case. 
For large $n$ the RS2 solution dominates the system in the whole range of 
temperature. 
For small $n$, 
at high temperatures $T>T_{c}$ 
we can draw a phase boundary between the RS1 and RS2 phases 
by equating $\phi_{\rm RS1}(n)$ and $\phi_{\rm RS2}(n)$
and the actual boundary is indicated by the line $(1)$ 
in fig.\ \ref{fig2:2SKPD}.
On the other hand, at low temperatures $T\leq T_{c}$, 
we must consider the RSB of the RS2 solution. 
The line where the AT stability breaks 
is the AT line which indicated by $(2)$ in fig.\ \ref{fig2:2SKPD}. 
Below the AT line, the system requires the FRSB solution. 
Although 
there exists a region where the rate function assessed by the RS2 solution 
becomes positive below the AT line, that region is irrelevant because the RS solution does not hold any meaning below the AT line.
In the low temperature limit $\beta \to \infty$, we can analytically derive 
the asymptotic behavior of the AT line. We here omit the detailed calculations 
and only give the result 
\begin{equation}
\beta n \approx \sqrt{
\frac{4}{p}
\log\left( \sqrt{ \frac{p}{2} }  \beta \right)
}.
\end{equation}
Hence the AT line diverges in the $T$-$\beta n$ plane. 

Next, we consider the $p=3$ case. 
In this case, transitions are of first order and we need numerical calculations to determine phase boundaries.
Unlike the SK $(p=2)$ case, there is a region where 
the 1RSB solution is relevant, which is signaled by the behavior of the 
rate function $R(f(n))$.  
The line indexed by $(3)$ denotes the critical condition $R(f(n))=0$ and below this line the rate function is naturally continued to the positive 
complexity, which means the emergence of the 1RSB phase.  
The meaning of other symbols in fig.\ \ref{fig2:3SKPD} 
are the same as in the $p=2$ case.
The equilibrium transition from the paramagnetic to 1RSB phases 
occurs at $T_{c}$ and the equilibrium 
free energy $f_{\rm eq}$ shows 
a singularity at this temperature.  
In addition, 
at lower temperatures the AT line $(2)$ exceeds the line $(3)$, which 
leads to another phase transition at $T_{G}$. The system 
is dominated by the FRSB solution below the AT line, 
and the transition from the 1RSB to FRSB is sometimes called the 
Gardner transition \cite{Gard}. 

Another important fact for the $p=3$ case is 
that a non-equilibrium transition, the so-called dynamical transition, 
occurs at $T_{d}$. At this temperature, the equilibrium free energy does not
show any singularity, but the phase structure of the system drastically 
changes. 
For $T>T_{d}$, only the paramagnetic solution $q=0$ exists 
for $n=1$\footnote{
Remember the 1RSB-RS correspondence (\ref{eq2:1RSB-RS}), 
i.e. the replica number $n$ of the RS solution corresponds to the 1RSB 
breaking parameter $m$, and the specialty of $m=1$.}, which means that the complexity takes 
$0$ at $f=f_{P}$ and $-\infty$ otherwise. 
On the other hand, for $T_{c}\leq T \leq T_{d}$, there are two solutions 
$q>0$ and $q=0$, which indicates that 
the nontrivial 1RSB saddle point exists and many pure states appear, 
which yields the complexity in a 
certain range of free energy. 
These pure states affect the dynamics of the system and the system cannot 
reach the equilibrium states, 
although the equilibrium value of free energy $f_{\rm eq}$ 
is completely the same as the paramagnetic one $f_{P} $ for $T>T_{d}$.
This transition is sometimes called the dynamical 1RSB transition (d1RSB).
In the REM, there exist two solutions $q=1$ and $q=0$ in the 
whole temperature region and we did not 
realize the existence of this dynamical transition\footnote{
We can also interpret as that 
the dynamical transition temperature of the REM 
is $T_{d}=\infty$.}. 
This dynamical transition revealed by the complexity analysis 
is also confirmed by some numerical approaches \cite{MontEPJ2003,MontPRB2004,BoucCugl} 
which directly investigate dynamical properties.   
The complexity is nowadays considered to be 
one of the most important concepts to 
understand not only dynamical behaviors of various systems
but also the replica method itself.

\section{Microscopic description of pure states}
In this final section of this chapter, we present the microscopic description
of pure states of spin glasses. 
A basis of this attempt is the so-called TAP equation 
which is a set of equations with respect to the local 
magnetizations $\{m_{i}\}$
 and is first derived by Thouless et al. \cite{TAP}.
\subsection{TAP method}
We here treat the SK model, the $p=2$ case, as an example. 
The TAP equation
for the SK model is given by
\begin{equation}
m_{i}=\tanh \beta
\left\{
\sum_{j\neq i}J_{ij}m_{j}+h_{i}
-\beta (1-q)m_{i}
\right\}.\label{eq2:TAP}
\end{equation}
For readers interested in the derivation of this equation, 
we refer to \cite{ADVA}.
The corresponding TAP free energy $f_{\rm TAP}(\{ m_{i} \})$ is 
expressed as
\begin{equation}
Nf_{\rm TAP}(\{ m_{i} \})=-\sum_{i<j}J_{ij}m_i m_j-T
\sum_{i}
s_{0}(m_i)
\end{equation}
where 
\begin{eqnarray}
&&s_{0}(q,m_i)=
-\frac{1+m_{i}}{2}\log\left( \frac{1+m_{i}}{2} \right)
-\frac{1-m_{i}}{2}\log\left( \frac{1-m_{i}}{2} \right)
+\frac{\beta^2}{4}(1-q)^2 \nonumber \\
&&=
-
\frac{1}{2}\log (1-m_{i}^2)-m_{i}\tanh^{-1}(m_i)+\log2
+\frac{\beta^2}{4}(1-q)^2.\label{eq2:s0}
\end{eqnarray}
where $q$ is the spin-glass order parameter and equals to  
$(\sum m_{i}^{2})/N$. 
Taking variation of $f_{\rm TAP}(\{ m_{i} \})$ with respect to $m_{i}$, we 
again obtain eq.\ (\ref{eq2:TAP}).

In the TAP context, pure states are identified with the 
solutions of eq.\ (\ref{eq2:TAP})\footnote{Historically, the notion of pure state was first founded through 
analyzing the TAP equation. }. The number of TAP solutions 
with the free energy value $f$, $\mathscr{N}_{\rm TAP}(f)$, 
is hence given by
\begin{equation}
\mathscr{N}_{\rm TAP}(f)=\int
\prod_{i}
\left\{
dm_{i}\delta
\left(
\Part{ f_{\rm TAP}(\{ m_{i} \})}{m_i}{} 
\right)
\right\} 
\Abs{\det{G}}\delta(f_{\rm TAP}(\{ m_{i} \})-f)\label{eq2:Nf},
\end{equation}
where $G$ is the Hessian of the TAP free energy 
\begin{equation}
G_{ij}=\frac{\partial^2 f_{\rm TAP}(\{ m_{i} \})}{\partial m_i \partial m_j}
=-J_{ij}+
\left(
\beta (1-q)+\frac{1}{\beta}\frac{1}{1-m_{i}^2}
\right)\delta_{ij}.
\end{equation}
The TAP complexity is defined by 
$\Sigma_{\rm TAP}(f)=(\log \mathscr{N}_{\rm TAP}(f))/N$ and is 
assumed to be self averaging, which leads to 
the quantity to be calculated as 
\begin{equation}
\Sigma_{\rm TAP}(f)=\frac{1}{N}[\log \mathscr{N}_{\rm TAP}(f)]_{\V{J}}.
\end{equation}
This was first formulated by Bray and Moore \cite{Bray} and 
is reexamined many times \cite{Mona1995,Tana,Cava1,CavaJPA2003}.
The calculations are straightforward but rather involved and
we only give a sketch of actual calculations and refer 
the results of \cite{CavaJPA2003}.
The evaluation procedure is as follows:
\begin{enumerate}
\item{Replace $[\log \mathscr{N}(f)]_{\V{J}}$  by 
$\log [\mathscr{N}^{n}(f)]_{\V{J}}$ by using the replica method.}
\item{The modulus of $\Abs{\det{G}}$ is removed to simplify the calculation (there is no {\it a priori} justification of this assumption) and $\det{G}$ is expressed by using Grassmann variables $\psi_{i}$ in an integral form $\det{G}=
\int \prod_{i} \left( d\psi_{i} d\bar{\psi}_{i} \right) 
\exp\left( \sum_{i,j} \psi_{i} \bar{\psi}_{j} G_{ij} \right)$.}
\item{The delta functions in eq.\ (\ref{eq2:Nf}) are transformed 
into the Fourier expressions as 
$\delta(f_{\rm TAP}(\{ m_{i} \})-f)
=\int du/(2\pi i) \exp 
\left\{ u(f_{\rm TAP}(\{ m_{i} \})-f)\right\}$
, and all the relevant factors 
are expressed in an exponential form $\exp(A)$ with an effective action $A$.}
\item{The configurational average is performed and then 
the RS and the Becchi-Rouet-Stora-Tyutin (BRST) supersymmetry are imposed on the effective action $A$.}
\item{Finally, the complexity is assessed by using the saddle-point method and 
taking the $n\to 0$ limit. }
\end{enumerate}
Note that it is known that the RS assumption usually yields the same result 
with the 
annealed approximation $[\log \mathscr{N}_{\rm TAP}(f)]_{\V{J}}\approx
\log[ \mathscr{N}_{\rm TAP}(f)]_{\V{J}}$, and actual calculations are also
performed in this approximation in \cite{CavaJPA2003} as other references 
\cite{Bray,Tana,Cava1}. 
Leaving a parameter $u$ which 
determines the value of TAP free energy $f_{\rm TAP}$,
we can express the TAP complexity as 
\begin{eqnarray}
&&\Sigma_{\rm TAP}(f;u)=
\frac{1}{4}\beta^2 u(-u-1)q^2+\frac{\beta^2}{2}uq-\frac{1}{4}\beta^2 u 
\nonumber \\
&&
\hspace{10mm}+\log \GI{z} \left( 2 \cosh \sqrt{\beta^2 q} z \right)^{-u}
-\beta u f \label{eq2:TAPcomp}
\end{eqnarray}
Taking the maximization with respect to $u$ gives the TAP complexity for 
a free energy value $f$, $\Sigma_{\rm TAP}(f)$. 
Comparing eqs.\ (\ref{eq2:gx_p}) with $p=2$ and (\ref{eq2:TAPcomp}), 
we can find a formal correspondence between the TAP complexity and 
 the generating function $g(x)$ assessed in the 1RSB level as 
\begin{equation}
\Sigma_{\rm TAP}(f;u=-x)=g(x)+\beta x f.
\end{equation}
Taking the minimization with respect to $x$ of the right-hand side
yields the complexity in the replica theory of the 1RSB level, hence the TAP and replica theories completely give the same result. 
This fact supports not only the consistency of different methods
 but also the correctness of microscopic description of pure states 
based on the TAP equation.

\subsection{Zero temperature limit}
In the previous discussion, 
we have seen that 
 pure states statistics based on the microscopic description, the TAP method, 
gives the same result as the replica one.
To investigate the microscopic description of pure states in more detail, 
we here concentrate on the zero temperature limit. 
In this limit, pure states are related to stable configurations 
against single spin flips. 

Although it is possible to propose some stabilities 
of a spin configuration at zero temperature,
the stability against single spin flip can be considered as 
the most natural one. 
Consider a spin $S_i$ at site $i$ and the effective field $h^{{\rm eff}}_{i}$ 
\begin{equation}
h^{{\rm eff}}_{i}=\sum_{j\neq i} J_{ij}S_{j}.
\end{equation}
The stability condition of $\{S_{i}\}$ 
against any single spin flip is that inequalities 
\begin{equation}
h^{{\rm eff}}_{i}S_{i}\geq 0 
\end{equation}
hold for all $i$. 
This is the same condition as equalities
\begin{equation}
S_{i}=\sgn{\sum_{j\neq i}J_{ij}S_{j}}\label{eq2:TAPT=0}
\end{equation} 
hold for all $i$.
Since $m_i$ can be identified with the spin $S_i$ in the limit 
$\beta \to \infty$, eq.\ (\ref{eq2:TAPT=0}) 
is nothing but the TAP equation at zero temperature with 
a condition\footnote{This condition has some delicate 
points as pointed out in \cite{CavaJPA2003}, but 
we do not treat this problem and only 
adopt the condition $\beta (1-q)\to 0$ in the limit $\beta\to \infty$.}
\begin{equation}
\beta (1-q)\to 0.
\end{equation}
Equation (\ref{eq2:TAPT=0}) and the discussion in the previous subsection 
imply that pure states at zero temperature are connected to stable spin 
configurations against any single spin flip. 

Note that this simple relation is, however, considered to be peculiar to 
the fully-connected models. 
This important fact was first pointed out 
by Biroli and Monasson \cite{Biro1}.
In order to follow their discussion,  
we here define a {\it $k$-stable} 
configuration as the configuration the energy 
of which cannot be decreased by flipping any subset of $k$ (or less than $k$) 
spins. 
In this context, pure states of fully-connected models at zero temperature 
correspond to 
$1$-stable spin configurations. 
Denoting the number of $k$-stable configurations with the energy density 
$e$ as $\mathscr{N}_{k}(e)$, 
we can define a kind of entropy of $k$-stable configurations,
 $s_{k}(e)$, as $s_{k}(e) \equiv (\log \mathscr{N}_{k}(e))/N$ in the 
$N\to \infty$ limit.  
Biroli and Monasson have shown that 
the TAP complexity $\Sigma_{\rm TAP}(e)$ is identical to 
$s_{\infty}(e)$ by treating a hybrid system of one dimensional and 
fully-connected models. 
They have also alleged that 
the $k$-stable entropy $s_{k}(e)$ does not depend on $k$ 
in the fully-connected models, which leads to an accidental correspondence 
between $\Sigma_{\rm TAP}(e)$ and $s_{1}(e)$. 
This peculiar property to the fully-connected models 
can be understood as follows:
Let us denote the energy deviation induced by flipping a spin $S_{i}$ as 
$\Delta E_{i}$. If we flip two spins $S_{i}$ and $S_{j}$, 
the energy deviation $\Delta E_{ij}$ equals to $\Delta E_{ij}=\Delta E_{i}+\Delta E_{j}-2J_{ij}S_{i}S_{j}$. This energy deviation $\Delta E_{ij}$ becomes, however, $\Delta E_{ij}=\Delta E_{i}+\Delta E_{j}$ in the $N\to \infty$ limit 
because $J_{ij}$ is scaled as $O(1/\sqrt{N})$, which means that 
a $1$-stable configuration is also 
$2$-stable, and the same holds for all $k$.
Therefore, $s_{k}(e)$ does not depend on $k$ and $s_{1}=\cdots =s_{\infty} =\Sigma_{\rm TAP}$. 

In conclusion, the microscopic description of pure states is given by the TAP 
equation. The TAP method gives a simple description of pure states for the 
fully-connected models, but for more realistic models there still remain some 
unclear points about the existence and the properties of pure states.

\section{Summary}
In this chapter, we reviewed the replica method in detail by treating 
the fully-connected $p$-spin interacting model as an example. 
The relation between the replica symmetry breaking (RSB) and the emergence of 
many pure states was closely explained and 
the distribution of pure states, 
the complexity $\Sigma(f)$, was calculated 
in the $1$-step RSB (1RSB) level. 
This description of the 1RSB was also reexamined from a viewpoint 
of the large deviation theory, and the 1RSB transition was interpreted from 
the rate function $R(f(n))$. In our particular model, 
the complexity and rate function share the same functional form, which 
implies a close relation between these quantities.
In addition, the microscopic description of pure states was presented from 
the TAP context. At zero temperature it was explained that 
stable spin configurations against 
$\infty$-spin flips correspond to pure states, but  
for the fully-connected models pure states are also identical to the 
stable spin configurations against single spin flip. 
It was shown from a simple discussion about the energy balance 
that this property is peculiar to the fully-connected models. 

Viewpoints from pure state statistics and the large deviation theory proposed 
in this chapter are rather recent results. 
In past days the RSB formulation was performed without realizing  
these backgrounds and some confusing discussions were proposed. 
This chapter is written to clarify these points 
by comparing conventional and modern operations of the RSB.

Pure states at zero temperature generally correspond to 
spin configurations stable against $\infty$-spin flips after taking the 
thermodynamic limit $N \to \infty$, but for the fully-connected models 
such configurations are the same as $1$-stable spin configurations.
This peculiar property is useful for 
identifying pure states to numerically evaluate the complexity.
In the next chapter, we will utilize this property and 
calculate the complexity numerically to verify the prediction of the replica 
result.

\chapter{Analyses of weight space structure and   
rare events in the Ising perceptron}
In the previous chapter, we introduced the replica method and 
also elucidated its aspects as a tool to obtain the pure state statistics 
and large deviations.
In this chapter, 
we reexamine these aspects extensively 
by treating a model of a neuron, namely the so-called Ising perceptron.

As seen in sec.\ \ref{sec2:LDT}, the rate function $R(f)$, which 
represents a small probability of atypical samples,
and the complexity $\Sigma(f)$, which represents a large number 
of pure states that appear for typical samples, share 
not only the similar mathematical structure 
but also the same functional form for the fully-connected 
$p$-spin interacting model.
This fact 
naturally motivates us to further explore more general relationships between 
$R(f)$, and $\Sigma(f)$, 
including cases for which 
the formal accordance of functional forms between $R(f)$ 
and $\Sigma(f)$ does not hold. 

The Ising perceptron considered here 
is a simple model of a neuron and
stores random input-output
patterns. 
There are two reasons for considering this system. 
First, the Ising perceptrons can be macroscopically characterized 
by a few sets of order parameters and 
are relatively easier to handle.
Despite the simplicity, it is known that 
this model exhibits rich behavior 
in the phase space involving nontrivial RSB phenomena \cite{GardREM,Krau} 
and is considered to show different functional forms of $R(f)$ and 
$\Sigma(f)$, 
which is highly suitable for our purpose. 
The second reason is that the meaning of complexity 
for the Ising perceptrons of finite size is clearer.
For the Ising perceptrons, 
a pure state at zero temperature can be identified with a stable cluster, 
the definition of which will be given in section \ref{sec3:numerical}, 
with respect to single spin flips \cite{Cocc,Biro1,Arde}. 
For samples of small systems, the size of the clusters can be numerically 
evaluated by exhaustive enumeration without any ambiguity. 
This property is extremely useful for justifying theoretical predictions 
through numerical experiments.  

Our investigations reveal that the generating function $g(x)$ of the complexity
$\Sigma(f)$
shows an extraordinary behavior, the origin of which is that 
$\Sigma(f)$ and $R(f)$ of the Ising perceptron are not 
convex upward functions, which is assumed in usual analyses. 
The results presented in this chapter are summarized in reference \cite{Obuchi:09-2}. 
\section{Model}
A simple perceptron is a model of a neuron, 
the function of which is a map from $\mR^N$ to $\{+1,-1\}$ as
\begin{equation}
y
= \left \{
\begin{array}{ll}
+1, & {\V{S}\cdot \V{x}}/{\sqrt{N}} > 0, \cr
-1, & {\V{S}\cdot \V{x}}/{\sqrt{N}} < 0, 
\end{array}
\label{eq3:perceptron}
\right . 
\end{equation}
where $\V{x} \in \mR^N $ is the input pattern and $y \in \{+1,-1\}$ 
is the output label.
The vector $\V{S}$ denotes the synaptic weights. 
\begin{figure}[htbp]
\begin{center}
   \includegraphics[height=50mm,width=70mm]{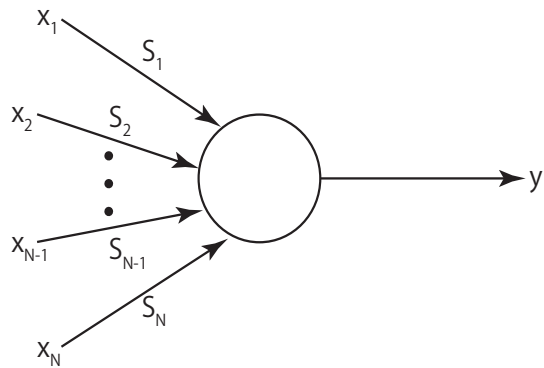}
 \caption{A conceptual picture of a simple perceptron.}
 \label{fig:Perceptron}
\end{center}
\end{figure}
We hereafter focus on the Ising weight case, in which each synaptic 
weight only takes $S_{i}=\pm 1$.
In a general scenario, the perceptron stores a given set of $M$ 
labeled patterns 
\begin{equation}
D^{M}=\{(\V{x}_{1},y_{1}),\cdots,(\V{x}_{M},y_{M})\}, \label{eq3:examples}
\end{equation}
by adjusting the weight $\V{S}$ so as to completely reproduce 
the given label $y_\mu$ for the input $\V{x}_\mu$ for $\mu=1,2,\ldots,M$. 

In the following, we consider a situation in which 
the patterns are independently and identically distributed (i.i.d.). 
In the conventional situations of the neural networks, a network consisting 
from Ising perceptrons is considered and 
 in that 
situation the input $\V{x}$ is a set of outputs of other perceptrons 
\cite{STAT}. 
This means that 
the input $\V{x}$ is usually chosen as random signs 
$x_{i}=\pm 1$, but in the thermodynamic limit 
we can regard $\V{S}\cdot \V{x}$ as multivariate Gaussian random variables
due to the central limit theorem.  
Hence, it is convenient to assume the following distributions 
\begin{eqnarray}
P(\V{x})=\left( \frac{1}{\sqrt{2\pi}} \right)^N \exp \left(-\frac{\V{x}^2}{2} \right),\label{eq3:dist^x}\\
P(y)=\frac{1}{2}\left( \delta(y-1)+\delta(y+1) \right),
\end{eqnarray} 
from the beginning of the analysis.
The question we address here 
is how the space of the weights that store $D^{M}$ 
is characterized macroscopically when the pattern ratio 
$\alpha=M/N \sim O(1)$ is fixed as $M$ and $N$ tend to infinity. 
In order to express this problem 
in the statistical mechanical context,
we define the Hamiltonian as
\begin{equation}
\mathscr{H}(\V{S}|D^{M})=\sum_{j=1}^{M}\Theta \left(
-y_{j}\frac{\V{S}\cdot\V{x}_{j}}{\sqrt{N}}
\right),\label{eq3:Hamiltonian}
\end{equation}
which counts the number of patterns which are 
incompatible with the weight $\V{S}$.
The function $\Theta(x)$ is the step function where 
$\Theta(x)=1 \, (x>0)$ and $\Theta(x)=0$ otherwise.   
Using the Hamiltonian we can define
the partition function 
\begin{equation}
Z(D^M)=\Tr{\V{S}}e^{-\beta \mathscr{H}(\V{S}|D^{M})}, 
\end{equation}
The quantity $Z(D^M)$ varies randomly depending on the quenched randomness
$D^M$, which naturally leads us to evaluate
the  generating function 
$\phi(n)=(1/N) \log \left [Z^n(D^M) \right ]_{D^M}$, 
where $\left [ \cdots \right ]_{D^M}$ represents the operation of 
averaging with respect to $D^M$.


\section{Formalism}\label{sec3:formulation}
The  generating function 
$\phi(n)$ can be evaluated by using the replica method.
Following the standard prescription of the replica method,  
we calculate the averaged $n$th moment in $n\in \mN$ 
\begin{eqnarray}
&&\left[Z^n \right]_{D^M}=\Tr{}
\left[
\exp
\left(
-\beta \sum_{\mu=1}^{n}\sum_{j=1}^{M}
\Theta 
\left(
-y_{j}\frac{ \V{S}^{\mu}\cdot\V{x}_{j}}{\sqrt{N}}
\right)
\right)
\right]_{D^M}
\end{eqnarray}
where the brackets $[\cdots]_{D^M}$ denote the average over the 
quenched randomness $D^M$.
According to eq.\ (\ref{eq3:dist^x}),
the variable 
$u^{\mu}_{j}=-y_{j}\V{S}^{\mu}\cdot\V{x}_{j}/\sqrt{N}$ 
$(\mu=1,2,\ldots,n;j=1,2,\ldots,M)$ can 
be regarded as multivariate Gaussian random variable, 
which is characterized as 
\begin{eqnarray}
\left[u^{\mu}_{j} \right]_{D^M}=0,\,\,
\left[u^{\mu}_{j}u^{\nu}_{k} \right]_{D^M}=
\delta_{j k} 
\left (\delta_{\mu \nu}+(1-\delta_{\mu \nu})q^{\mu \nu} \right) ,
\label{eq3:moments^u}
\end{eqnarray}
where $q^{\mu \nu}=(1/N) \sum_{i=1}^{N} S_i^\mu S_i^\nu$ ($\mu,\nu=1,2,\ldots,n$). 
This observation yields the following expression
\begin{eqnarray}
&&\left[
\exp
\left(
-\beta \sum_{\mu=1}^{n}\sum_{j=1}^{M}
\Theta 
\left(
-y_{j}\frac{ \V{S}^{\mu}\cdot\V{x}_{j}}{\sqrt{N}}
\right)
\right)
\right]_{D^M}\nonumber \\
&&=
\int\prod_{\mu<\nu}dq^{\mu \nu}
\delta\left(
\V{S}^{\mu}\V{S}^{\nu}-Nq^{\mu \nu}
\right)
\left(
\left[
\exp
\left(
-\beta \sum_{\mu=1}^{n}
\Theta 
\left(
u^{\mu}
\right)
\right)
\right]_{\V{u}}
\right)^M\label{eq3:Part}
\end{eqnarray}
where $\left [ \cdots \right ]_{\V{u}}$ denotes the average with 
respect to the multivariate Gaussian variables the moments of which are 
given by eq.\ (\ref{eq3:moments^u}), 
and the independency of $M$ examples is reflected in eq.\ (\ref{eq3:Part}).
Using the Fourier expression of the delta function, we get
\begin{eqnarray}
&&\hspace{0cm}\left[Z^n \right]_{D^M}
=
\int\prod_{\mu<\nu}\frac{dq^{\mu \nu}d\Wh{q}^{\mu \nu}}{2\pi}
\exp N 
\Biggl(
-\sum_{\mu<\nu} q^{\mu \nu} \Wh{q}^{\mu \nu}
+\log \Tr{} e^{ \sum_{ \mu<\nu }\Wh{q}^{\mu \nu}S^{\mu}S^{\nu} } 
\nonumber \\
&&\hspace{7cm}+
\alpha 
\log
\left[
e^{-\beta \sum_{\mu}\Theta(u^{\mu})}
\right]_{\V{u}}
\Biggr),\label{eq3:rawmoment} 
\end{eqnarray}
where we put $\alpha=M/N$. Using the saddle point method, we obtain 
the  generating function as 
\begin{equation}
\phi(n,\beta)=\Extr{q^{\mu\nu},\Wh{q}^{\mu\nu}}
\left\{
-\sum_{\mu<\nu} q^{\mu \nu} \Wh{q}^{\mu \nu}
+\log \Tr{} e^{ \sum_{ \mu<\nu }\Wh{q}^{\mu \nu}S^{\mu}S^{\nu} } 
+\alpha 
\log
\left[
e^{-\beta \sum_{\mu}\Theta(u^{\mu})}
\right]_{\V{u}}
\right\},\label{eq3:phi}
\end{equation}
where the symbol $\Extr{x}$ expresses to take the extremization condition 
with respect to $x$.
\subsection{The replica symmetric solution}
The replica symmetric (RS) ansatz gives
\begin{equation}
\Wh{q}^{\mu\nu}=\Wh{q},\,q^{\mu\nu}=q,
\end{equation}
which leads to
\begin{eqnarray}
&&\sum_{\mu<\nu} q^{\mu \nu} \Wh{q}^{\mu \nu}=\frac{1}{2}n(n-1)\Wh{q}q,\\ 
&&\Tr{} e^{ \sum_{ \mu<\nu }\Wh{q}^{\mu \nu}S^{\mu}S^{\nu} }
=e^{-\frac{1}{2}n\Wh{q}}\int Dz (2\cosh\sqrt{\Wh{q}}z)^n,
\end{eqnarray}
where $Dz=dze^{-z^2/2}/\sqrt{2\pi}$ and we assume that the domain of 
integration of $\int Dz$ is $]-\infty,\infty[$ if there is no 
explicit indication. 
Under the RS ansatz,
the Gaussian variable $u^{\mu}$ can be decomposed to
two independent Gaussian variables of zero mean and unit variance 
$x^{\mu}$ and $z$ as
\begin{equation}
u^{\mu}=\sqrt{1-q}x^{\mu}+\sqrt{q}z.
\end{equation}
It is easy to check that this assumption satisfies 
eq.\ (\ref{eq3:moments^u}).
Applying this expression, we get 
\begin{eqnarray}
&&\left[
e^{ -\beta \sum_{\alpha} \Theta(u^{\mu})} 
\right]_{\V{u}}
=\int Dz \int \prod_{\mu=1}^{n} Dx^{\mu} e^{ -\beta \Theta (u^{\mu}) }
=\int Dz \left( \int Dx e^{ -\beta \Theta (u)} \right)^n 
\nonumber\\
&&
=\int Dz 
\left( 
\int_{-\infty}^{-\sqrt{\frac{q}{1-q}}z} Dx 
+\int^{\infty}_{-\sqrt{\frac{q}{1-q}}z} Dx e^{-\beta}
\right)^n
\nonumber \\
&&=\int Dz 
\left(
e^{-\beta}+(1-e^{-\beta})
E
\left(
\sqrt{\frac{q}{1-q}}z
\right)
\right)^n,
\end{eqnarray}
where we define the complementary error function $E(x)$
\begin{equation}
E(x)=\int_{x}^{\infty}Dz.\label{eq3:H(x)}
\end{equation}
Note that the relation $E(x)+E(-x)=1$ holds.
Using the above expressions, we get the RS generating function as 
\begin{eqnarray}
&&\phi_{{\rm RS}}(n,\beta)=
\Extr{q,\Wh{q}}
\Biggl\{
-\frac{1}{2}n(n-1)\Wh{q}q-\frac{1}{2}n\Wh{q}+\log\int Dz (2\cosh\sqrt{q}z)^n
\nonumber
\\
&&+\alpha \log \int Dz 
\left(E_{\beta}
\left(
\sqrt{\frac{q}{1-q}}z
\right)
\right)^n
\Biggr\},\label{eq3:phiRS}
\end{eqnarray}
where 
\begin{equation}
E_{\beta}(x)=e^{-\beta}+(1-e^{-\beta})
E
\left(
x
\right).
\end{equation}
The saddle point conditions yield
\begin{eqnarray}
&&q=\frac{\int Dz \cosh^n \sqrt{\Wh{q}}z \tanh^2 \sqrt{\Wh{q}}z }{
\cosh^n \sqrt{\Wh{q}}z
},\label{eq3:q}  \\
&&
\Wh{q}=\frac{\alpha}{2\pi}\frac{(1-e^{-\beta})^2}{1-q}
\frac{
\int Dz e^{-\frac{q}{1-q}z^2}
\left(
E_{\beta}\left(
\sqrt{\frac{q}{1-q}}z
\right)
\right)^n
}
{
\int Dz
\left(
 E_{\beta}\left(
\sqrt{\frac{q}{1-q}}z
\right)
\right)^{n}
},\label{eq3:qhat}
\end{eqnarray}
the detailed derivation of which is given in appendix \ref{app:IP-saddle}.
Before discussing the significance of the RS solution,
we proceed to the replica symmetry breaking (RSB) solution. 
Especially, we utilize the 1RSB ansatz 
for investigating the complexity and the rate function 
in a unified framework.

\subsection{The 1RSB solution}
The 1RSB ansatz is expressed as
\begin{equation}
(q^{\mu\nu},\Wh{q}^{\mu\nu})=
\left\{
\begin{array}{ll}
(q_{1},\Wh{q}_{1}) & (\,\, {\rm in\,\, the\,\, same\,\, block} \,\,)\\
(q_{0},\Wh{q}_{0}) & (\,\, {\rm otherwise}\,\,)
\end{array}
\right.. 
\end{equation}
This assumption yields 
\begin{eqnarray}
&&\sum_{\mu<\nu} q^{\mu \nu} \Wh{q}^{\mu \nu}=\frac{1}{2}n(m-1)\Wh{q}_{1}q_{1}+
\frac{1}{2}n(n-m)\Wh{q}_{0}q_{0}
,\\ 
&&\Tr{} e^{ \sum_{ \mu<\nu }\Wh{q}^{\mu \nu}S^{\mu}S^{\nu} }
=e^{-\frac{1}{2}n\Wh{q}_{1}}
\int Dz_{0} 
\left(
\int Dz_{1}
(2\cosh h)^m
\right)^{n/m}
,
\end{eqnarray}
where $h=\sqrt{\Wh{q}_{1}-\Wh{q}_{0} }z_{1}+\sqrt{\Wh{q}_{0} }z_{0}$ and 
$m$ is the Parisi's breaking parameter and expresses the size of a block. 
In the 1RSB ansatz, the Gaussian variable $u^{\mu}$ is decomposed to
\begin{equation}
u^{\mu}=u^{l,\mu_{l}}=\sqrt{q_{1}-q_{0}}x_{l}+\sqrt{1-q_{1}}y_{\mu_{l}}
+\sqrt{q_{0}}z, 
\end{equation}
where the variables $x_{l},y_{\mu_{l}},z$ are drawn 
from the normal distributions and the index $l$ indicates a block 
and $\mu_{l}$ specifies a replica in the $l$th block. 
This transformation enables us to derive
\begin{eqnarray}
&&\left[
e^{-\beta \sum_{\mu}\Theta(u^{\mu})}
\right]_{ \V{u} }=
\left[
\prod_{l} \prod_{\mu_{l}}
e^{-\beta \Theta(\sqrt{q_{1}-q_{0}}x_{l}+\sqrt{1-q_{1}}y_{\mu_{l}}
+\sqrt{q_{0}}z)}
\right]_{\V{u}}\nonumber\\
&&=\int Dz 
\left(
\int Dx 
\left\{
E_{\beta}(y_{0}(z,x))
\right\}^m  
\right)^{n/m},
\end{eqnarray}
where 
\begin{equation}
y_{0}(z,x)=-\sqrt{\frac{q_{0}}{1-q_{1}}}z-\sqrt{\frac{q_{1}-q_{0}}{1-q_{1}}}x.
\end{equation}
Finally, we get 
\begin{eqnarray}
&&\phi_{{\rm 1RSB}}(n,m,\beta)
=
\Extr{q_{1},q_{0},\Wh{q}_{1},\Wh{q}_{0}}
\Biggl\{
-\frac{1}{2}n(m-1)\Wh{q}_{1}q_{1}-\frac{1}{2}n(n-m)\Wh{q}_{0}q_{0} \nonumber \\
&&
-\frac{1}{2}n\Wh{q}_{1}+
\log 
\int Dz_{0} 
\left(
\int Dz_{1} (2\cosh h)^m
\right)^{n/m} \nonumber \\
&&+\alpha \log
\int Dz_{0} 
\left(
\int Dz_{1} 
\left\{
E_{\beta}(y_{0}(z_{0},z_{1}))
\right\}^m  
\right)^{n/m}
\Biggr\}
.\label{eq3:phi1RSB}
\end{eqnarray}
The saddle point conditions yield the following equations of state
\begin{eqnarray}
q_{1}=I^{-1}
\int Dz_{0} 
\left(
\int Dz_{1} \cosh^m h
\right)^{n/m}
\frac{
\int Dz_{1} \cosh^m h \tanh^2 h
}{
\int Dz_{1} \cosh^m h
},\label{eq3:q1}\\
q_{0}=I^{-1}
\int Dz_{0} 
\left(
\int Dz_{1} \cosh^m h
\right)^{n/m}
\left(
\frac{
\int Dz_{1} \cosh^m h \tanh h
}{
\int Dz_{1} \cosh^m h
}
\right)^2
,\\
\Wh{q}_{1}=\frac{\alpha}{1-q_{1}}\Wh{I}^{-1}
\int Dz_{0} 
\left(
\int Dz_{1} E_{\beta}^m
\right)^{n/m}
\frac{
\int Dz_{1} E_{\beta}^m 
\left(
\frac{E_{\beta}'}{E_{\beta}}
\right)^2
}{
\int Dz_{1} E_{\beta}^m
},\\
\Wh{q}_{0}=\frac{\alpha}{1-q_{1}}\Wh{I}^{-1}
\int Dz_{0} 
\left(
\int Dz_{1} E_{\beta}^m
\right)^{n/m}
\left(
\frac{
\int Dz_{1} E_{\beta}^m 
\frac{E_{\beta}'}{E_{\beta}}
}{
\int Dz_{1} E_{\beta}^m
}
\right)^2,\label{eq3:q0hat}
\end{eqnarray}
where 
\begin{eqnarray}
E_{\beta}'=E_{\beta}'(y_{0}(z_{0},z_{1}))\equiv\left.\frac{dE_{\beta}(x)}{dx}\right|_{x=y_{0}}=-\frac{1-e^{-\beta}}{\sqrt{2\pi}}e^{-\frac{1}{2}y_{0}^2},\\
I=\int Dz_{0} 
\left(
\int Dz_{1} \cosh^m h
\right)^{n/m},\\
\Wh{I}=
\int Dz_{0} 
\left(
\int Dz_{1} E_{\beta}^m
\right)^{n/m}.
\end{eqnarray}
Note that we omit the arguments of $E_{\beta}(y_{0}(z_{0},z_{1}))$ and 
$E_{\beta}'(y_{0}(z_{0},z_{1}))$ in eqs.\ (\ref{eq3:q1})-(\ref{eq3:q0hat}). 
Before investigating the equations of state 
(\ref{eq3:q1})-(\ref{eq3:q0hat}), we again discuss the physical meaning 
of the 1RSB formulation for exploring the relation between 
the concepts of $\phi(n)$, $\Sigma(f)$, and $R(f)$.

\subsection{The complexity and 1RSB formulation 
revisited}\label{sec3:complexity}
\subsubsection{Two ways of the zero-temperature limit}
Let us focus on the $y\to 0$ (or $n\to 0$) limit as in 
section \ref{sec2:purestat}. Remember that 
the generating function $g(x|D^M)$ of the complexity 
 satisfies the following relation 
\begin{equation}
e^{Ng(x|D^M)}=\int df e^{N(-\beta x f+\Sigma(f) ) }.\label{eq3:g-Sigma}
\end{equation}
For the perceptron problems, we are mainly interested in the structure 
of the weight space at zero temperature since 
it directly relates to the learning process of the perceptrons. 

To investigate the weight space, there 
are two different ways for accessing the limit of $\beta \to \infty$.

One way is to take the $\beta \to \infty$ limit with keeping $m=x\sim O(1)$,
and we call this limit the {\it entropic} limit.
For our Hamiltonian (\ref{eq3:Hamiltonian}), the ground state 
energy $E_{GS}$ is given by $0$ or positive constant depending on 
the sample $D^M$. In such a situation, the $x$th moment of 
the partition function of pure state $\gamma$, $Z_{\gamma}^{x}= e^{-\beta x f_{\gamma}}$ 
becomes in the $\beta \to \infty$ limit
\begin{equation}
e^{-\beta x f_{\gamma}}=
\left\{
\begin{array}{ll}
e^{xs_{\gamma}} & (u_{GS}\equiv E_{GS}/N=0)\\
0 & (u_{GS}>0)
\end{array}
\right. ,
\end{equation}
where $s_{\gamma}$ is the entropy of the pure state $\gamma$.
Hence, this limit enables to investigate the weight space structure 
with $u_{GS}=0$ in detail, 
while it is ill defined for instances with $u_{GS}>0$.
The corresponding generating function becomes  
\begin{equation}
g(x|D^M)=\max_{s_{-}\leq s\leq s_{+}}\{xs+\Sigma(s) \}.
\end{equation}
This entropic limit is appropriate when samples with $u_{GS}=0$ are 
typically produced, which corresponds to the low $\alpha$ case.
Note that if we want to directly investigate this limit we can choose 
the Boltzmann factor $\eta(\V{S}|D^M)$ as
\begin{equation}
\eta(\V{S}|D^M)
=\prod_{\mu=1}^M 
\Theta \left (y_{\mu}\frac{\V{S}\cdot\V{x}_{\mu}}{\sqrt{N}}
\right),\label{eq3:BF}
\end{equation}
which is obtained by taking the limit $\beta \to \infty$ in $e^{-\beta \mathscr{H}(\V{S}|D^{M})}$ and is equal to the number of patterns that are incompatible 
with the weight $\V{S}$.

The other way is to take the $\beta \to \infty$ limit with the $m\to 0$ limit 
keeping $\beta m \to \xi \sim O(1)$, and we call this limit the 
{\it energetic} limit. 
Although the entropic limit cannot treat 
instances with $u_{GS}>0$, but the energetic 
limit works well for such instances. 
For a simple explanation,  
we here use two arguments for the complexity 
$\Sigma(u,s)$. In this notation, 
the generating function $g(x|J)$ becomes
\begin{equation}
g(x|D^M)=\max_{u,s} \{ x(s-\beta u)+\Sigma(u,s) \}.
\end{equation}
If we take the limit $\beta \to \infty$ under the condition $u>0$, 
the entropic term becomes irrelevant. After that, 
we take $x \to 0$ limit keeping $\beta x \to \xi$ 
and obtain the finite result 
\begin{equation}
g(\xi|D^M)=\max_{u_{-}\leq u \leq u_{+}}\{-\xi u+\Sigma_{u}(u) \},
\end{equation}
where 
\begin{equation}
\Sigma_{u}(u)=\max_{s_{-}\leq s \leq s_{+}}\Sigma(u,s).
\end{equation}
This limit is appropriate for the case that samples with $u_{GS}>0$ 
typically produced, which corresponds to the high $\alpha$ region.
In such a region, 
we can investigate the energy landscape by using the energetic limit.

The two limits, entropic and energetic, are in a complementary 
relation. 
Both limits can detect the critical capacity $\alpha_{s}$, 
below which typical samples are {\it separable} ($u_{GS}=0$) 
but not separable for $\alpha>\alpha_{s}$.
However, for our present purpose, i.e. to investigate the relation 
among $\phi(n)$, $\Sigma(f)$, and $R(f(n))$,
the entropic limit is 
more suitable. 
We can expect that for $\alpha >\alpha_{s}$ behaviors of separable samples 
which are atypically generated can be investigated by introducing the finite 
replica number $n>0$ into the entropic limit, which 
naturally extends the applicable range of this limit. 
On the other hand, for the energetic limit we do not have any clear reason
to introduce the finite replica. 
We can also expect that introducing $n>0$ into the energetic limit 
enables to investigate $u>0$ region for $\alpha<\alpha_{s}$, but 
the $u>0$ region can also be treated by introducing finite temperature 
$T>0$ for the typical case $n=0$. Thus, we cannot probably distinguish 
the effects of finite replica $n>0$ and of finite temperature $T>0$.
Hence, we hereafter concentrate on the entropic limit. 

\subsubsection{The RS and 1RSB solutions in the entropic limit}
The entropic limit is easily taken by performing the limit 
$\beta \to \infty$ in eq.\ (\ref{eq3:phiRS}). 
In this limit, 
there are two solutions for eqs.\ (\ref{eq3:q})-(\ref{eq3:qhat}): \\
\noindent{\bf RS1:} $0<q<1$ and $\widehat{q} < + \infty$. 
\begin{eqnarray}
\phi_{\rm RS1}(n)&=&
 -\frac{n(n-1)}{2}q\Wh{q}-\frac{1}{2}n\Wh{q} + \log  
\left (\int Dz
\left ( 2 \cosh 
\left (\sqrt{\widehat{q}} z 
\right )
\right )^n
\right ) \nonumber \\
&&+\alpha \log  \left (\int Dz
E^n\left (\sqrt{\frac{q}{1-q}}z \right ) \right ). 
\label{RS1}
\end{eqnarray}
\\
\noindent{\bf RS2:} $q=1$ and $\widehat{q} = + \infty$. 
\begin{eqnarray}
\phi_{\rm RS2}(n)=(1-\alpha) \log 2. 
\label{RS2}
\end{eqnarray} 
Similarly, we can find three solutions of the 1RSB saddle-point 
equations (\ref{eq3:q1})-(\ref{eq3:q0hat}): \\
\noindent{\bf 1RSB1:} $(q_1,q_0)=(1,q)$ and $(\widehat{q}_1, \widehat{q}_0)
=(+\infty, \widehat{q})$, where $q$ and $\widehat{q}$ take the same values as those for
$\phi_{\rm RS1}(n)$. 
\begin{eqnarray}
\phi_{\rm 1RSB1}(n,m)=\phi_{\rm RS1}\left (\frac{n}{m}\right ). 
\label{1RSB1}
\end{eqnarray}

\noindent{\bf 1RSB2:} $(q_1,q_0)=(q,q)$ and $(\widehat{q}_1, \widehat{q}_0)
=(\widehat{q},\widehat{q} )$, where $q$ and $\widehat{q}$ take the same values as those for 
$\phi_{\rm RS1}(n)$.
\begin{eqnarray}
\phi_{\rm 1RSB2}(n,m)=\phi_{\rm RS1}(n). 
\label{1RSB2}
\end{eqnarray}

\noindent{\bf 1RSB3:} $(q_1,q_0)=(1,1)$ and $(\widehat{q}_1, \widehat{q}_0)
=(+\infty, +\infty)$. 
\begin{eqnarray}
\phi_{\rm 1RSB3}(n,m)=\phi_{\rm RS2}(n)=(1-\alpha)\log 2. 
\label{1RSB3}
\end{eqnarray}

In usual analyses, Parisi's 1RSB parameter $m$ is determined 
by the extremum condition in evaluating 
$\phi(n)=\mathop{\rm Extr}_{m}\left \{\phi_{\rm 1RSB*}(n,m) \right \}$, 
where $\rm *=1,2$ and $\rm 3$. In addition, there might be no need
to classify ${\bf 1RSB2}$ and ${\bf 1RSB3}$ as 
1RSB solutions because ${\bf 1RSB2}$ and ${\bf 1RSB3}$ are completely reduced to 
${\bf RS1}$ and ${\bf RS2}$, respectively. However, handling these 
three solutions as 1RSB solutions with leaving the $m$-dependence 
of $\phi_{\rm 1RSB}(n,m)$ explicitly, is crucial
for the current purpose as we will see in section \ref{sec3:Tresult}.

The formulation relating $\phi(n)$ and $\Sigma(s)$ is easily 
obtained by following the same discussion 
as section \ref{sec2:purestat}.
Assuming that the complexity $\Sigma(s)$ is a convex upward function, 
we can get the parameterized forms 
\begin{eqnarray}
s(x)=\Part{}{x}{}
\left (x 
\left.
\Part{\phi_{\rm 1RSB}(n,x)}{n}{}
\right|_{n=0} 
\right ), 
\\
\Sigma(s(x))=
- x^2 
\left.
\frac{
\partial^ 2\phi_{\rm 1RSB}(n,x)}{
\partial x \partial n
} 
\right|_{n=0}. 
\label{phi_complexity}
\end{eqnarray}
On the other hand, the relation  
 $\phi(n)=\phi_{\rm 1RSB}(n,x)|_{x=1}$ also holds 
in the entropic limit. 
This means that the rate function can be assessed from $\phi_{\rm 1RSB}(n,m)$ 
as follows:
\begin{eqnarray}
&&s_{\rm eq}(n)=
\Part{\phi_{\rm 1RSB}(n,1)}{n}{},  \\
&&
R(s_{\rm eq}(n))=\phi_{\rm 1RSB}(n,1)-n \Part{\phi_{\rm 1RSB}(n,1)}{n}{},
\label{phi_rate}
\end{eqnarray}
where we define the equilibrium value of entropy 
\begin{equation}
s_{\rm eq}\equiv\lim_{N\to \infty}\lim_{\beta \to \infty}(1/N)\log Z
=\max_{s_{-}\leq s \leq s_{+}}\{s+\Sigma(s) \}
\end{equation}
which corresponds to 
the total number of weights that are compatible with $D^{M}$.
In eq.\ (\ref{phi_complexity}), the parameter $x$ can vary only in such a range 
that both $s(x) \ge 0$ and $\Sigma(s(x)) \ge 0$ hold. 
Similarly, the conditions $s_{\rm eq}(n) \ge 0$ and 
$R(s_{\rm eq}(n)) \le 0$ restrict
the range of $n$ in eq.\ (\ref{phi_rate}). 
These constitute the main result of this chapter. 

Here, three issues are noteworthy. First, for a class of disordered systems, 
including the $p$-spin interacting model without external fields, 
two equalities
$\phi_{\rm 1RSB}(n,m)=(n/m) \phi_{\rm RS}(m)$ 
and $\phi_{\rm 1RSB}(n,m=1)=\phi_{\rm RS}(n)$, hold 
in assessing the complexity and rate function,
respectively,  where $\phi_{\rm RS}(n)$ is an identical 
RS solution of the generating function $\phi(n)$. 
Inserting these functions into eqs.\ (\ref{phi_complexity}) and 
(\ref{phi_rate}) offers an identical functional form  
both for the complexity and the rate function, while their 
domains of definition are disjointed, 
except for a point of the typical value of 
free energy $f^*$ (or entropy $s^*$), as 
mentioned in section \ref{sec2:LDT}. 
The current system, however, does not possess 
this property because $\phi_{\rm 1RSB}(n,m)=(n/m) \phi_{\rm RS}(m)$ 
does not hold for {\bf 1RSB1}, {\bf 1RSB2}, or {\bf 1RSB3}
while $\phi_{\rm 1RSB}(n,m=1)=\phi_{\rm RS}(n)$ is always satisfied.
Second, eqs.\ (\ref{phi_complexity}) and (\ref{phi_rate}) are 
valid only when $\phi_{\rm 1RSB}(n,m)$ are stable against any perturbation 
for a further RSB. Fortunately, in the present problem, 
a stable solution against any known 
RSB instabilities can be constructed for $\forall{\alpha} > 0$ and $\forall{n}>0$.   
This implies that, in the present analysis, there is no need to consider further RSB. 
Finally, however, we have to keep in mind that eqs.\ (\ref{phi_complexity}) and (\ref{phi_rate}) depend on the assumptions that correct $\Sigma(s)$ and $R(s)$ are convex upward
functions, respectively. 
When the convex property does not hold, the estimates of 
eqs.\ (\ref{phi_complexity}) and (\ref{phi_rate}) 
represent not the correct solution, but rather its convex hull. 
The following analytical and experimental assessment indicates
that this is the case for $\Sigma(s)$ of sufficiently low $\alpha$
and $R(s_{\rm eq})$ of sufficiently high $\alpha$. 

\section{Theoretical predictions }\label{sec3:Tresult}
We are now ready to use the formalism developed above
to analyze the behavior of the weight space of the Ising perceptron. 

\subsection{Complexity for $\alpha< \alpha_{\rm s}=0.833 \ldots$}
In order to perform the analysis, it is necessary to select 
a certain solution (functional form) among the three candidates 
of {\bf 1RSB1}, {\bf 1RSB2}, and {\bf 1RSB3}. 
Analyticity and physical plausibility are two guidelines for this task. 

The replica method is a scheme to infer the properties for real replica 
numbers $n \in \mR$ by analytical continuation 
from those for natural numbers $n = 1,2,\ldots \in \mN$. 
This indicates that, for examining typical ($n\to 0$) behavior, 
it is plausible to select the solution 
of $\phi(n)$ that is dominant around $n \ge 1$, because 
unity is the natural number that is closest to zero. 
For $\alpha < \alpha_{\rm s}=0.833 \ldots$, this solution
is $\phi_{\rm RS1}(n)$. 
In addition, the relevant $\phi_{\rm 1RSB}(n,m)$ must agree with this 
solution at $m=1$. 
These considerations offer two candidates of $g(x)$ as
\begin{eqnarray}
g_{\rm 1RSB1}(x)=
x \frac{\partial}{\partial n}\phi_{\rm 1RSB1}(n,x)|_{n=0}
=\phi_{\rm RS1}^\prime(0), 
\label{g1RSB1}
\end{eqnarray}
and 
\begin{eqnarray}
g_{\rm 1RSB2}(x)=
x \frac{\partial}{\partial n}\phi_{\rm 1RSB2}(n,x)|_{n=0}
=x \phi_{\rm RS1}^\prime(0). 
\label{g1RSB2}
\end{eqnarray}

We combine these solutions to construct an entire functional 
form of $g(x)$ based on physical considerations. 
For $x \gg 1$, $g(x)$ should vary approximately linearly with respect to 
$x$, because a single pure state of the largest entropy 
typically dominates $\sum_\gamma Z_\gamma^x$. 
In addition, $s(x)=(\partial/\partial x) g(x)$ for $x \sim 0$
should be smaller than that for $x \gg 1$ because $s(x)$ should increase
monotonically with respect to $x$. 
Furthermore, $g(x)$ must be a continuous function. These considerations reasonably 
yield an entire functional form of $g(x)$ as
\begin{eqnarray}
g(x)=\left \{
\begin{array}{ll}
\phi_{\rm RS1}^\prime(0), & x \leq 1,\\
x\phi_{\rm RS1}^\prime(0), & x > 1,
\end{array}
\right .
\label{gx}
\end{eqnarray}
which yields the complexity as
\begin{eqnarray}
\Sigma(s)=\left \{
\begin{array}{ll}
\phi_{\rm RS1}^\prime(0)-s, & 0\leq s \leq \phi_{\rm RS1}^\prime(0), \\
-\infty, & \mbox{otherwise}.
\end{array}
\right .
\label{typicalcomplexity}
\end{eqnarray}

\begin{figure}[htbp]
\begin{center}
   \includegraphics[height=50mm,width=50mm]{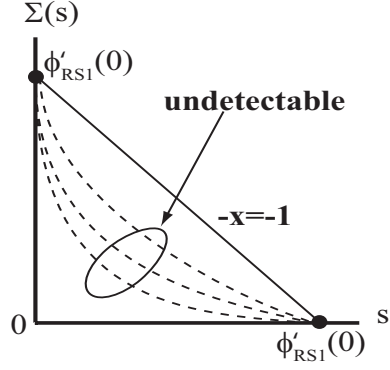}
 \caption{Schematic profile of complexity eq.\ (\ref{typicalcomplexity}). 
The characteristic exponent of the size distribution of
pure states cannot be correctly assessed in the current formalism if 
it is a convex downward function (dashed curves). 
In such cases, the complexity $\Sigma(s)$ assessed from $g(x)$ (solid line)
is the convex hull (black circle) of the correct exponent. 
}
\label{fig:typ-g(x)}
\end{center}
\end{figure}

The piecewise linear profile of eq.\ (\ref{gx}) is 
somewhat extraordinary. 
This is thought to be because the correct complexity 
is not convex upward in this system. 
When $\Sigma(s)$ is convex upward, 
the current formalism using the saddle-point method 
defines a one-to-one map between 
$g(x)$ and $\Sigma(s)$. However, if $\Sigma(s)$ is not
convex upward, the functional profile of a region in which  
the correct complexity is convex downward is lost and 
only the convex hull is obtained by the transformation from $g(x)$, 
as shown in figure \ref{fig:typ-g(x)}. 
The piecewise linear profile of $g(x)$ presumably signals that 
this actually occurs in the current problem. 
Similar behavior of the complexity could also be observed 
in a certain type of random energy models \cite{BoucMeza}.

The physical implication of eq.\ (\ref{typicalcomplexity}), 
the profile of which is obtained by connecting
two points $(s,\Sigma)=(0,\phi_{\rm RS1}^\prime(0))$ 
and $(\phi_{\rm RS1}^\prime(0), 0)$ with a straight line 
having a slope of $-x=-1$, is that the weight space is equally 
dominated by exponentially many clusters of vanishing entropy 
and a subexponential number of large clusters composed 
of exponentially many weights. 
The existence of large clusters may accord with 
an earlier study which reported that 
local search heuristics of a certain type 
manage to find a compatible weight efficiently
up to a considerably large value of $\alpha$ near to the capacity $\alpha_{s}$ \cite{Brau}.
On the other hand, the coexisting exponentially many 
small clusters may be a major origin of a known difficulty 
in finding compatible weights by Monte Carlo sampling schemes
\cite{Horner1,Horner2}. 

\subsection{Rate function for $\alpha> \alpha_{\rm s}=0.833 \ldots$\label{sec3:typ} 
and a transition at $\alpha_{\rm GD}=1.245 \ldots$}
For $\alpha_{\rm s} < \alpha$, eq.\ (\ref{typicalcomplexity}) becomes 
negative, which implies that there exist no compatible weights for 
typical samples of $D^M$. In such cases, the rate function $R(s)$, 
which characterizes a small probability that atypical samples 
that are compatible with the Ising perceptrons are generated, 
becomes relevant in the current analysis. Therefore, 
we focus on the assessment of this exponent for this region. 

\begin{figure}[htbp]
\hspace{-7mm}
\begin{minipage}{0.32\hsize}
\begin{center}
   \includegraphics[height=40mm,width=50mm]{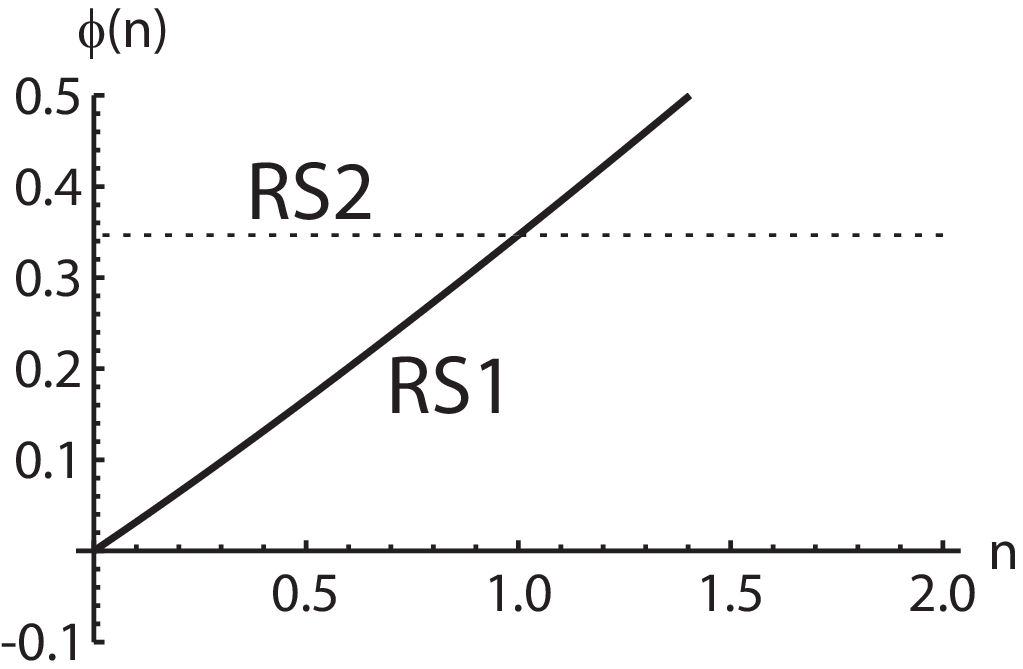}
\end{center}
\end{minipage}
\hspace{2mm}
 \begin{minipage}{0.32\hsize}
\begin{center}
   \includegraphics[height=40mm,width=50mm]{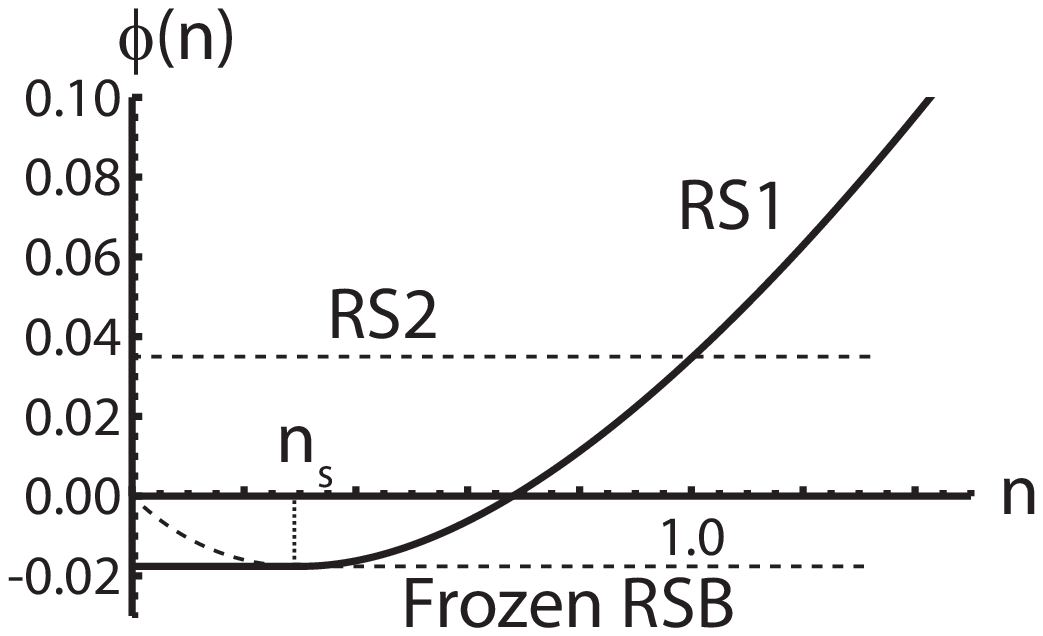}
\end{center}
 \end{minipage}
\hspace{2mm}
 \begin{minipage}{0.32\hsize}
\begin{center}
   \includegraphics[height=40mm,width=50mm]{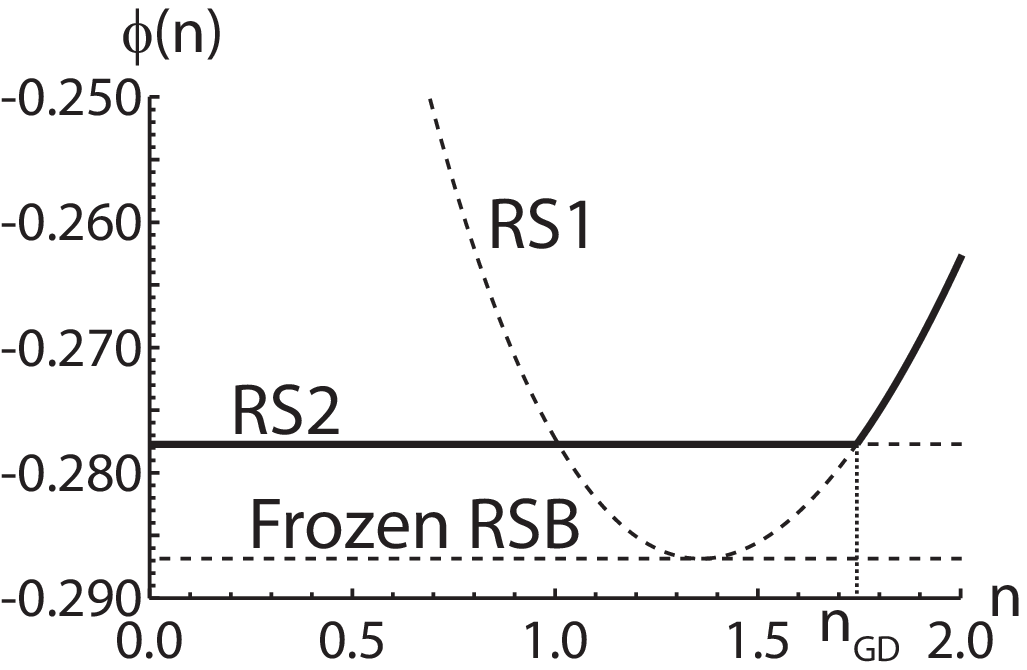}
\end{center}
 \end{minipage}
\caption{Behavior of $\phi(n)$. The solid lines denote 
the correct $\phi(n)$, and the 
dotted lines are the RS and frozen RSB branches. 
The corresponding values of 
the parameter $\alpha$ are $0.5,0.95$, and 
$1.4$, from left to right.}
\label{fig:phi2}
\end{figure}

For $\alpha_{\rm s} < \alpha < \alpha_{\rm GD}=1.245\ldots$, 
$\phi_{\rm RS1}(n)$ dominates the generating function $\phi(n)$
in the vicinity of $n \ge 1$ as for $\alpha < \alpha_{\rm s}$. 
This means that $\phi_{\rm 1RSB1}(n,m=1)=\phi_{\rm 1RSB2}(n,m=1)=
\phi_{\rm RS1}(n)$ should be used to assess $R(s)$ of 
relatively frequent events that correspond to $0<n<1$. 
However, this function is minimized to a negative value 
at a certain point at which $0<n_{\rm s}(\alpha)<1$, which implies
that assessment by na\"{i}vely using 
$\phi_{\rm RS1}(n)$ for $n<n_{\rm s}(\alpha)$ 
leads to incorrect results, which yield 
a negative equilibrium value of the entropy $s_{\rm eq}(n)
=(\partial/\partial n)\phi_{\rm RS}(n) < 0$. 
In order to avoid this inconsistency, we fix the value of
$\phi(n)$ to $\phi_{\rm RS1}(n_{\rm s}(\alpha))$, which is reduced to 
the conventional construction of a frozen RSB solution. 
In particular, this yields an assessment of 
\begin{eqnarray}
R(0)=\phi_{\rm RS1}(n_{\rm s}(\alpha))=
\mathop{\rm min}_{n} \{\phi_{\rm RS1}(n) \}, 
\label{separable_rate}
\end{eqnarray}
which has the physical meaning of a characteristic exponent of 
a small probability that a given sample set $D^M$ is 
separable by certain Ising perceptrons. 
For $\alpha \geq \alpha_{\rm GD}=1.245 \ldots$, 
on the other hand, 
the dominant solution of $\phi(n)$ in the vicinity of $n \ge 1$
is updated from $\phi_{\rm RS1}(n)$ to $\phi_{\rm RS2}(n)
=(1-\alpha)\log 2$, which yields 
\begin{eqnarray}
R(0)=\phi_{\rm RS2}(n)=(1-\alpha)\log 2. 
\label{separable_rate2}
\end{eqnarray}
In order to provide a visual representation of the above discussions, 
we depict the behaviors of $\phi(n)$ in figure \ref{fig:phi2}.

The difference in physical behavior 
between $\alpha_{\rm s} < \alpha < \alpha_{\rm GD}$ 
and $\alpha > \alpha_{\rm GD}$ is expected to be as follows. 
For $\alpha_{\rm s} < \alpha < \alpha_{\rm GD}$, the dominant solution  
around $n>n_{\rm s}(\alpha)$, $\phi(n)=\phi_{\rm RS1}(n)$, varies smoothly.
This leads to the following behavior of $R(s_{\rm eq})$ in the vicinity of $s =0$: 
\begin{eqnarray}
R(s_{\rm eq}) = R(0) -A  s^2_{\rm eq} +\ldots, 
\label{below_alpha_s}
\end{eqnarray}
where $A >0$ is a certain constant, 
which implies that large clusters can appear with a relatively 
large probability although typical samples of $D^M$ are not
separable by the Ising perceptrons. 
On the other hand, for $\alpha > \alpha_{\rm GD}$, 
$\phi(n)=\phi_{\rm RS2}(n)$ is constant for $n<n_{\rm GD}(\alpha)$,
which is characterized by $\phi_{\rm RS2}(n_{\rm GD}(\alpha))=
\phi_{\rm RS1}(n_{\rm GD}(\alpha))$ 
and $n_{\rm GD}(\alpha)>1$, and is switched to $\phi(n)=\phi_{\rm RS1}(n)$ 
for $n>n_{\rm GD}(\alpha)$ at $n=n_{\rm GD}(\alpha)$, which is accompanied by 
a jump in the first derivative. This indicates that the rate function 
$R(s_{\rm eq})$ is not convex upword 
in the region of $0 < s_{\rm eq} < (\partial /\partial n) 
\phi_{\rm RS1}(n_{\rm GD}(\alpha))$ as was mentioned for $\Sigma(s)$
in the previous subsection, which implies that the events of 
$s_{\rm eq}=0$ overwhelm those of $s_{\rm eq}>0$ in relative probabilities. 
Therefore, the generation of large clusters should be considerably 
rare for $\alpha$ of this region.

\subsection{Phase diagram on the $n$-$\alpha$ plane}
\begin{figure}[htbp]
\begin{center}
   \includegraphics[height=50mm,width=60mm]{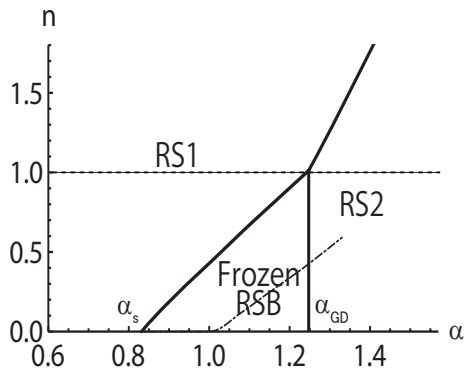}
 \caption{Phase diagram on the $n$-$\alpha$ plane. 
Solid lines are phase boundaries, and the dotted line denotes 
$n=1$. The dotted-dashed line expresses the AT line for the RS1 
solution, but is irrelevant. 
The AT line vanishes at a certain value of $\alpha$, 
because the solution for $0<q<1$ vanishes at this point.
The RS2 solution involves the AT instability only on the $n=0$ line, 
which is presumably of no relevance in the replica analysis. 
}
 \label{fig:PD}
\end{center}
\end{figure}

The above considerations are sufficient to draw a phase diagram 
on the $n$-$\alpha$ plane, which is depicted in figure \ref{fig:PD}. 

The value of the tricritical point 
$\alpha_{\rm GD}=1.245 \ldots$ is identical to the critical 
ratio of the perfect learning of the Ising perceptrons in the teacher-student 
scenario \cite{Gyor,Enge}. 
Formally, this agreement is explained as follows. 
The dominant solution for $n < 1$ is determined by 
whether $\phi_{\rm RS1}(n)$ or $\phi_{\rm RS2}(n)$ dominates
around $n \ge 1$. Since $\phi_{\rm RS1}(n=1)=\phi_{\rm RS2}(n=1)$
is always guaranteed, the critical condition is 
given as $(\partial /\partial n)\phi_{\rm RS1}(n)|_{n=1}=
(\partial /\partial n)\phi_{\rm RS2}(n)|_{n=1}=0$. 
On the other hand, $(\partial /\partial n)\phi_{\rm RS1}(n)|_{n=1}$ 
generally provides the equilibrium value of entropy after learning in the teacher-student scenario, 
the target of which can be dealt with as an $(n+1)$-replicated system, in which 
the teacher is handled as an extra replica. 
Therefore, the condition of perfect learning, which 
indicates that the weight of the student agrees perfectly 
with that of the teacher after learning, 
is identical to the vanishing entropy condition of the 
$(n+1)$-replicated system in the limit $n\to 0$, which 
agrees with $(\partial /\partial n)\phi_{\rm RS1}(n)|_{n=1}=0$, 
giving the critical value $\alpha_{\rm GD}$ in the current problem. 
Although the agreement is justified formally in this manner, 
its physical implication remains somewhat unclear. 
The line $n=1$, which passes through the tricritical point, 
may have an analogous relation to the 
concept of Nishimori's line in the theory of spin glasses \cite{STAT,Nish}. 

Finally, we mention the de 
Almeida-Thouless (AT) condition in this model \cite{Alme}. 
The AT (stability) condition of $\phi_{\rm RS}(n)$ with the 
order parameters $q$ and $\Wh{q}$
is expressed as follows:
\begin{eqnarray}
\frac{\alpha}{(1-q)^2}
\frac{
\int Dz 
E^{n}
\left(
\frac{E^{\prime \prime}}{E}
-
\left(
\frac{E^\prime}{E}
\right)^{2}
\right)^2
}
{
\int Dz 
E^{n}
}
\frac{\int Dz \cosh^{n-4} \sqrt{\Wh{q}}z }{
\int Dz \cosh^n \sqrt{\Wh{q}}z
}
\leq 1.\label{eq3:AT}
\end{eqnarray}
The derivation is given in appendix \ref{app:IP-AT}. 
This condition for $\phi_{\rm RS1}(n)$ 
is broken in a certain region on the $n$-$\alpha$ 
plane, but is irrelevant because the region is always 
included in $n < n_{s}(\alpha)$, for which the relevant solution is 
already switched to that of the frozen RSB. 
On the other hand, $\phi_{\rm RS2}$ is stable for $n>0$ 
but becomes unstable only on $n=0$, the deviation of which is also given 
in appendix \ref{app:IP-AT}, as reported 
in \cite{GardPer}.
The relevance of this instability for $\alpha\geq \alpha_{\rm GD}$ may 
require more detailed discussions,
but we assume herein that this instability can be ignored 
because only the asymptotic behavior of $\phi(n)$ in the 
limit $n\to 0$ is relevant in procedures of the replica method.

\section{Numerical validation}\label{sec3:numerical}
For validating the theoretical predictions obtained 
in the previous section, we carried out extensive numerical experiments. 

In describing the experiments, let us first define the cluster in the present system. 
The cluster is a set of spin configurations that are stable
with respect to single spin flips \cite{Cocc,Biro1}. 
Clusters have the following properties:
\begin{itemize}
\item{Any configuration belongs to a cluster.}
\item{
When a spin configuration ``A'' can be moved to another 
configuration ``B'' by a single spin flip without 
changing the number of incompatible patterns, 
``A'' and ``B'' belong to the same cluster.
}
\end{itemize}
In the following, we concentrate on vanishing energy clusters, which are 
composed of weights that are perfectly compatible with $D^M$. 

Before going into details, 
we elucidate the relation between the cluster and the pure state.
Identifying the microscopic description of a pure state is generally 
a delicate problem, but
in the Ising perceptron 
a pure state  
can be identified with a cluster,
as mentioned in the opening sentence in this chapter. 
There is no proof of this statement but 
it is naturally understood by 
considering the following aspects of the present problem: 
The Boltzmann weight of 
$\eta(\V{S}|D^M)$ in eq.\ (\ref{eq3:BF})
becomes completely zero if there is any incompatible pattern. 
This means that accessing from a cluster to a different cluster 
by single spin flips 
is impossible because those clusters are completely separated by states
with zero probability 
$\eta(\V{S}|D^M)=0$.
This naturally leads to identifying a cluster with a pure state, 
because a pure state is a set of configurations 
which cannot be accessed from other sets by natural dynamics.
Several earlier studies support this description 
\cite{Cocc,Biro1,Arde}, 
and we hereafter admit this 
assumption. 

Now let us return to the experiments.
We denote the size of a cluster 
as $Q$ and the number of size-$Q$ 
clusters for a sample $D^M$ as $C(Q|D^{M})$.  
The entropy of a cluster $s$ is considered to be identified by $s=(1/N)\log Q$, and the complexity
$\Sigma(s|D^{M})$ corresponds to $(1/N)\log C(Q|D^{M})$.
The clusters can be numerically evaluated, and hence 
we can construct the 1RSB generating function from the numerical data as
\begin{equation}
\phi_{\rm 1RSBnum}(n=xy,m=x)=\frac{1}{N}\log
\left[
\left(
\sum_{\gamma}Q_{\gamma}^{x}
\right)^y
\right],
\end{equation}
where $\left [ \ldots \right ]$ denotes the sample average operation
with respect to $D^M$. 
In the typical limit $y\to 0$, 
this yields the following expression:
\begin{equation}
g_{\rm num}(x)=
\lim_{y \to 0} \frac{\partial}{\partial y} \phi_{\rm 1RSBnum}(xy,x)
=
\frac{1}{N}
\frac{
\left[
\Theta
\left(
\sum_{\gamma}Q_{\gamma}^{x}
\right)
\log
\left(
\sum_{\gamma}Q_{\gamma}^{x}
\right)
\right]
}
{
\left[
\Theta
\left(
\sum_{\gamma}Q_{\gamma}^{x}
\right)
\right]
}
\label{eq3:gnum},
\end{equation}
where the step function $\Theta(x)$ comes from the differentiation of 
$\log\left[
\left(
\sum_{\gamma}Q_{\gamma}^{x}
\right)^y
\right]$ with respect to $y$. 
This means that if there is no cluster for a sample $D^M$, 
then the contribution of $\Theta
\left(
\sum_{\gamma}Q_{\gamma}^{x}
\right)
\log
\left(
\sum_{\gamma}Q_{\gamma}^{x}
\right)$ vanishes.

In order to examine the consistency with the replica analysis, 
we assess eq.\ (\ref{eq3:gnum}) based on data obtained in extensive 
numerical experiments. The function $g_{\rm num}(x)$ is evaluated by the exact 
enumeration of weights that are compatible with $D^M$, which are 
referred to hereinafter as {\em solutions}. 
The procedure is summarized as follows: 
\begin{enumerate}
\item{Generate $M$ examples $D^{M}=\{
(y_{1},\V{x}_{1})
\cdots (y_{M},\V{x}_{M})
\}$.} 
\item{
Enumerate all solutions.
} 
\item{
Partition the solutions into clusters, and calculate
$\sum_{\gamma}Q_{\gamma}^{x}$ for an appropriate set of $x$. 
We actually took $41$ points between $x=0$ and $2.0$. 
}
\item{
Repeat the above procedures until sufficient data are obtained and 
calculate 

$
\left[
\Theta
\left(
\sum_{\gamma}Q_{\gamma}^{x}
\right)
\log
\left(
\sum_{\gamma}Q_{\gamma}^{x}
\right)
\right]
$ by taking the sample average.  
}
\end{enumerate} 

\begin{figure}[htbp]
\begin{tabular}{cc}
\hspace{-2mm}
\begin{minipage}[t]{0.48\hsize}
\begin{center}
 \includegraphics[height=50mm,width=60mm]{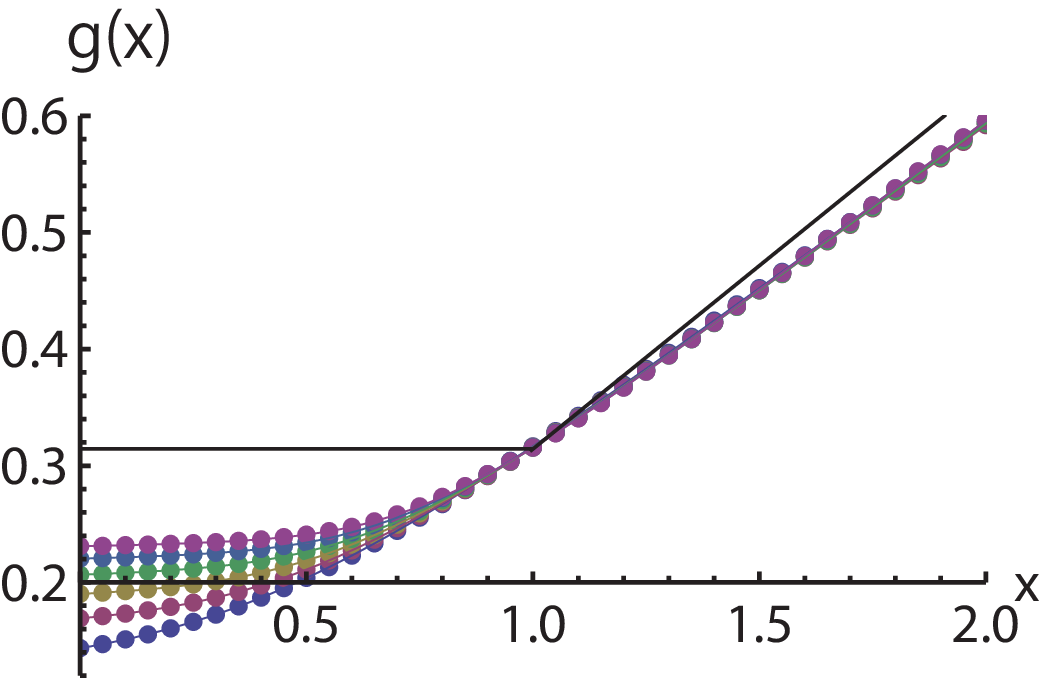}
 \caption{
Behavior of $g_{\rm num}(x)$
for $\alpha=0.5$. 
The system sizes are $N=14,16,18,20,22$, and $24$, 
from bottom to top. 
The solid lines denote $g(x)$ given by eq.\ (\ref{gx}).
The number of samples is $32,000$ for each $N$. 
Error bars are smaller than the size of markers.
As the system size grows,  
the profiles of $g_{\rm num}(x)$
approach the theoretical prediction. 
}\label{fig:gnum}
\end{center}
\end{minipage}
\hspace{2mm}
 \begin{minipage}[t]{0.48\hsize}
\begin{center}
\includegraphics[height=50mm,width=60mm]{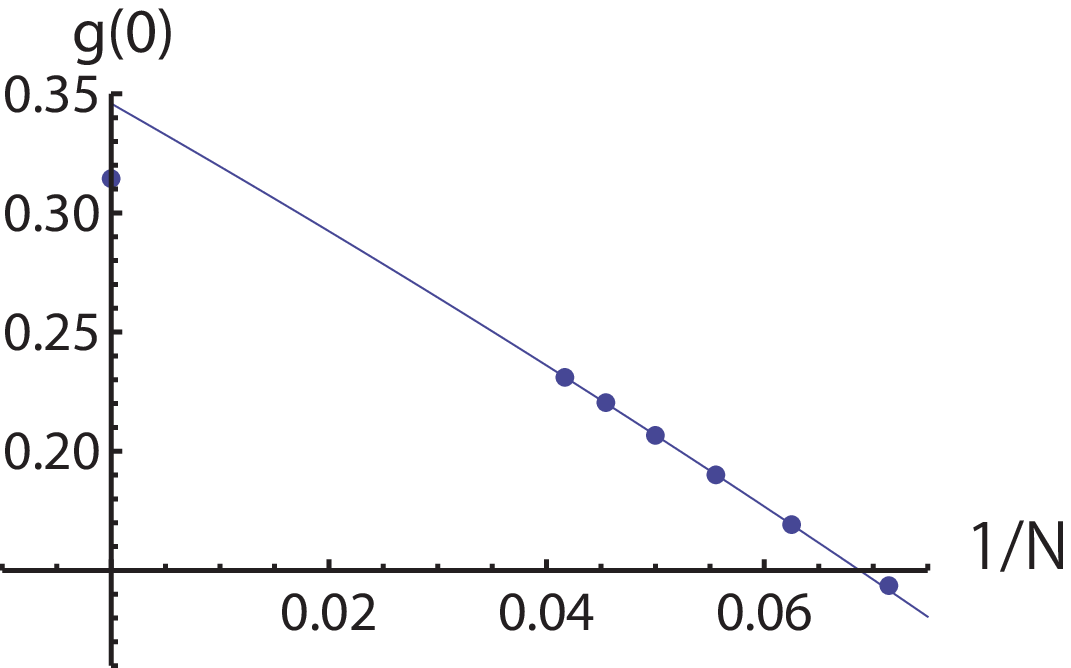}
 \caption{
Size dependence of $g_{\rm num}(0)$. The data are the same 
as fig.\ \ref{fig:gnum}.
The solid line is provided by quadratic fitting to the data. 
The point at $1/N=0$ is the theoretical value 
derived by the replica method.
The data tend to reach this theoretical value as the system size 
grows.  
}\label{fig:g0}
\end{center}
\end{minipage}
\end{tabular}
\end{figure}

The resultant plots of $g_{\rm num}(x)$ for $\alpha=0.5$ are 
shown in fig.\ \ref{fig:gnum}.
As the system size grows, 
the numerical data for $x\le 1$ exhibit 
flatter slopes approaching the theoretical 
prediction $g(x)=\phi^\prime_{\rm RS1}(0)$ for $x < 1$. 
This can also be seen in fig.\ \ref{fig:g0} 
as the systematic approaching of $g_{\rm num}(0)$ 
to the theoretical value of $g(0)=0.314 \ldots$ derived from 
the replica analysis. The difference between the numerical extrapolation 
and the analytical result at $1/N=0$ is considered to be the systematic error
due to 
higher order contributions of $1/N$. 
The profiles of $x>1$, on the other hand, are approximately  
straight lines, and the gradient of the slopes appear smaller
than that of the theoretical prediction $\phi_{\rm RS1}^\prime(0)$.  
However, the data still slowly move closer to 
$x \phi_{\rm RS1}^\prime(0)$ ($x>1$) as $N$ becomes larger as a whole, 
implying consistency with the theoretical prediction.

\begin{figure}[htbp]
\begin{center}
   \includegraphics[height=50mm,width=60mm]{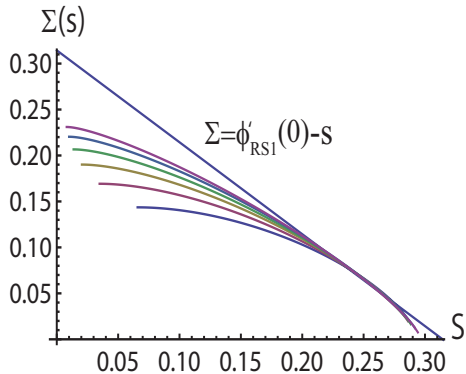}
 \caption{$\Sigma(s(x))$ obtained from $g_{\rm num}(x)$ 
using the relation (\ref{phi_complexity}) for 
$\alpha=0.5$. The system sizes increase 
from bottom to top. The solid line denotes the asymptotic 
line in the thermodynamic limit predicted by the replica analysis.}  
\label{fig:sSigmaB}
\end{center}
\end{figure}

Complexity $\Sigma(s)$ can also be assessed from the numerical data. 
One scheme for evaluating $\Sigma(s)$ is to use the 
relation (\ref{phi_complexity}) with a polynomial 
interpolation of the numerical data. 
We determined the order of the polynomial using Akaike's information 
criteria \cite{Akai} and eventually selected a $27$th degree polynomial, 
but the obtained results were not so sensitive to details of 
the choice of the polynomial. 
The assessed profiles of $\Sigma(s)$ are plotted in fig.\ \ref{fig:sSigmaB}.
The curves appear to approach the line predicted in the previous 
section as $N$ increases, which supports our replica analysis. 

\begin{figure}[htbp]
\begin{tabular}{cc}
\hspace{-2mm}
\begin{minipage}[t]{0.48\hsize}
\begin{center}
 \includegraphics[height=60mm,width=70mm]{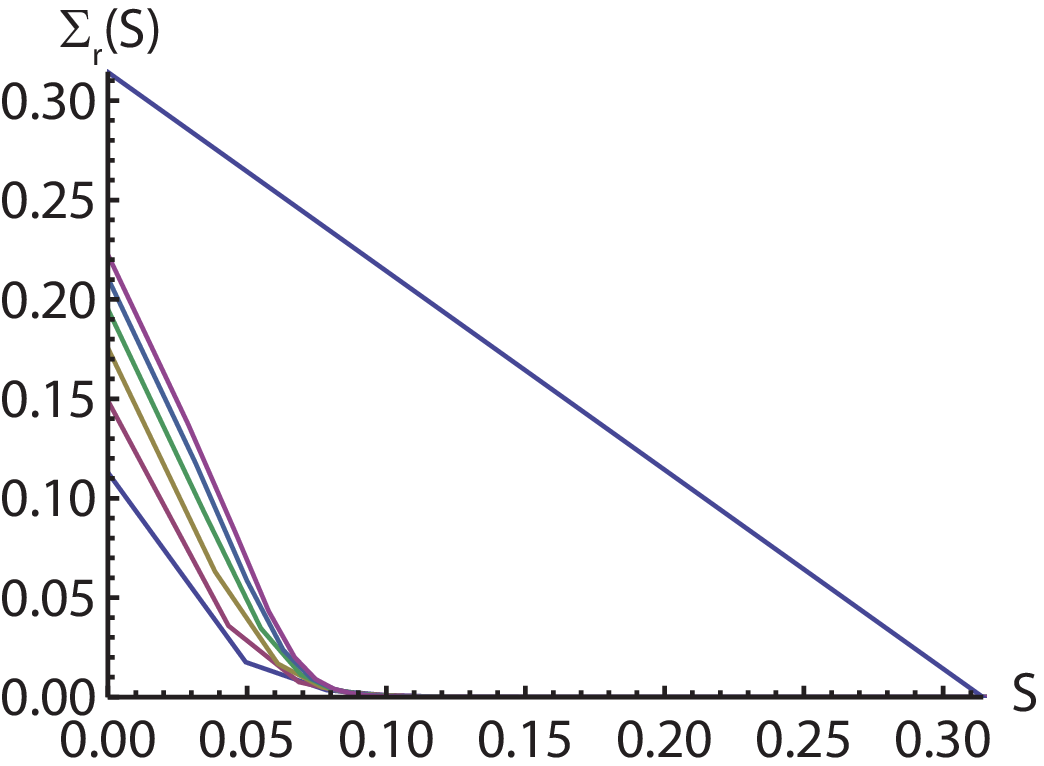}
 \caption{Plot of the raw complexity 
$\Sigma_{\rm r}(s)$ for $\alpha =0.5$. 
The system size increases from $N=14$ to $24$ in increments of 2, 
from bottom to top. The solid line is the same 
as that shown in fig.\ \ref{fig:sSigmaB}.  
}
\label{fig:rSigma}
\end{center}
\end{minipage}
\hspace{2mm}
 \begin{minipage}[t]{0.48\hsize}
\begin{center}
\includegraphics[height=60mm,width=70mm]{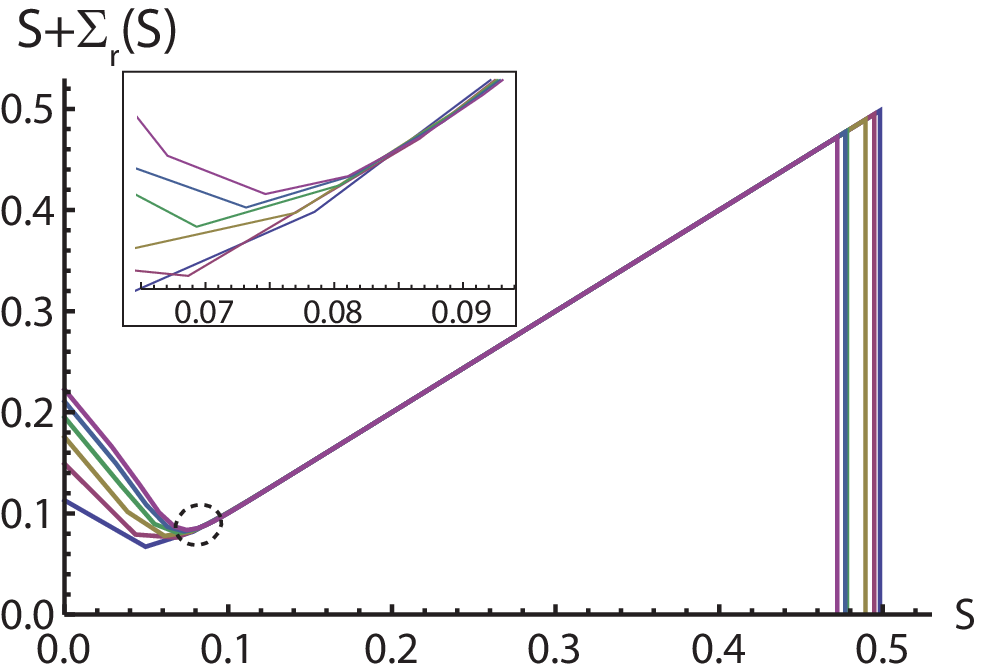}
\caption{Plot of 
$s+\Sigma_{\rm r}(s)$.  
As the system size increases, the curve appears to converge 
to a V-shape function, indicating that $\Sigma_{\rm r}(s)$ is 
convex downward. 
The inset shows a close-up of the region enclosed by the dotted ellipse. 
}
\label{fig:rspSigma}
\end{center}
\end{minipage}
\end{tabular}
\end{figure}

However, the complexity curve shown in fig.\ \ref{fig:sSigmaB} might 
lose the information about the correct distribution of the clusters, 
as mentioned in section \ref{sec3:typ}.
In order to examine this possibility, we directly evaluate 
the distribution of pure states in a rather naive manner. 
We refer to the result of this assessment as 
the raw complexity, which is defined as 
\begin{equation}
\Sigma_{\rm r}(s=(1/N)\log Q|D^{M})= 
\frac{1}{N}
\Theta
\left(
C(Q|D^{M})
\right)
\log
\left(
C(Q|D^{M})
\right).
\end{equation}
Taking the sample average yields the typical profile
of $\Sigma_{\rm r}( s|D^{M})$ as 
$\Sigma_{\rm r}(s)=
[\Sigma_{\rm r}( s|D^{M})]$, 
the result of which for $\alpha=0.5$ is shown in 
fig.\ \ref{fig:rSigma}. 
We took $32,000$ samples 
in the evaluation for each size and joined the plots to obtain 
smooth curves.
This figure indicates that 
$\Sigma_{\rm r}(0)$ approaches the value of the theoretical 
prediction $\phi_{\rm RS1}^\prime(0)|_{\alpha=0.5}=0.314\ldots$ 
from below as $N$ increases. 
However, $\Sigma_{\rm r}(x)$ for $x \ge 0.1$ appears to 
remain approximately constant at zero, indicating that $\Sigma_{\rm r}(x)$ converges to a convex 
downward function.
We also plot the function $s+\Sigma_{\rm r}(s)$ in figure 
\ref{fig:rspSigma}. 
This plot shows two peaks and one dip of $s+\Sigma_{\rm r} (s)$,
indicating that $\Sigma_{\rm r} (s)$ is convex downward. 
The position of the right-hand peak tends to move left 
to the right terminal point $s=0.314 \ldots$
of the theoretical prediction as the system size increases, while 
the dip appears to be bounded at the point $x=0.084$
 as shown in the inset. 
In conclusion, these figures indicate
that the exponent that characterizes the size distribution 
of the pure states, $\Sigma_{\rm r}(s)$, is not 
a convex upward function in this system and does not agree with 
$\Sigma(s)$, which is evaluated by the relation (\ref{eq2:f(x)}). 

Next, we assessed the rate function for the region of 
$\alpha>\alpha_{\rm s}$.
In this region, the generation of samples that are perfectly compatible with the Ising perceptrons rarely occurs and is dominated by $s=0$. 
Therefore, we numerically evaluated the probability that 
a given set of samples $D^M$ could be separated by the Ising 
perceptron, $P_{\rm sep}$, and estimated $R(0)$ as
$R(0)=(1/N) \log P_{\rm sep}$. 

\begin{figure}[htbp]
\begin{tabular}{cc}
\hspace{-2mm}
\begin{minipage}[t]{0.48\hsize}
\begin{center}
 \includegraphics[height=50mm,width=60mm]{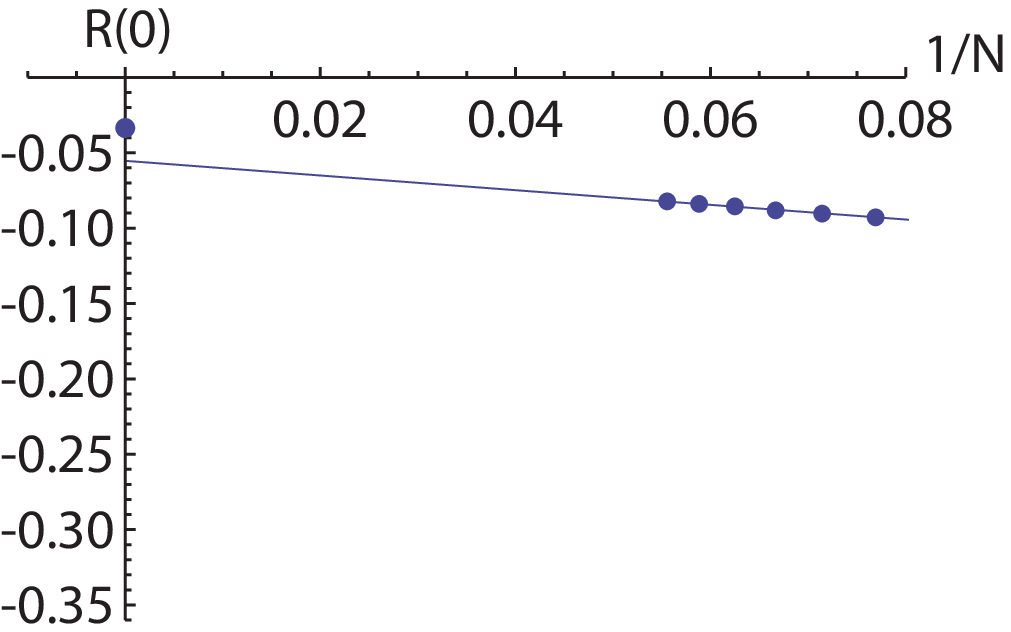}
 \caption{Size dependence of the rate function for $\alpha=1.0$. 
The point at $1/N=0$ is the value predicted by the frozen 
RSB solution. 
The system size increases from $N=12$ to $18$ in increments of $1$. 
The data from $320,000$ samples were evaluated for each $N$. 
The statistical errors are smaller than the markers.
}
\label{fig:Ra10}
\end{center}
\end{minipage}
\hspace{2mm}
 \begin{minipage}[t]{0.48\hsize}
\begin{center}
\includegraphics[height=50mm,width=60mm]{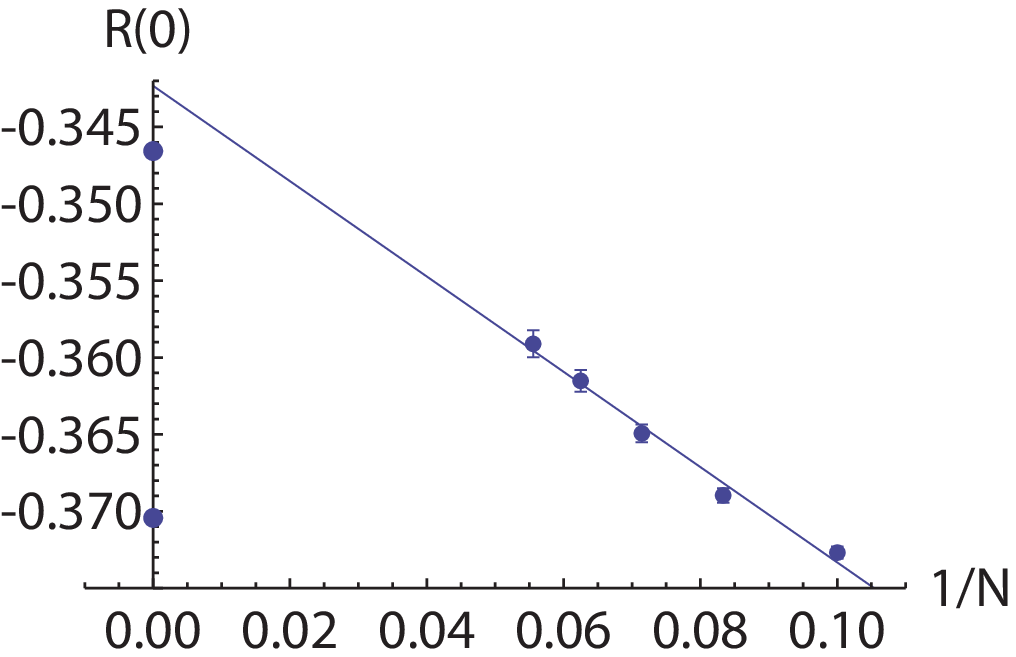}
\caption{Size dependence of the rate function for $\alpha=1.5$. 
The upper and lower plots at $1/N=0$ are given by the RS2 and 
frozen RSB solutions, respectively. 
The system size increases from $N=10$ to $18$ in increments of $2$. 
The data from $25,600,000$ samples were evaluated for each $N$. 
}
\label{fig:Ra20}
\end{center}
\end{minipage}
\end{tabular}
\end{figure}

The resultant plots are given in figures \ref{fig:Ra10} 
and \ref{fig:Ra20} for $\alpha=1.0$ and $1.5$, respectively.
The solid lines in these figures were obtained by the linear fitting 
for the numerical data. 
These figures show that the theoretical predictions
are reasonably consistent with the values of extrapolation of 
the numerical data. 
The statistical errors are sufficiently small, and hence 
the differences between the analytical and 
numerical results should be the systematic errors due to 
the nonlinearity of the Ising perceptron.
  
\section{Summary}
In this chapter, 
we investigated the structure of the weight space 
of Ising perceptrons in which a set of random patterns 
is stored using the derivatives of the generating function of 
the partition function. This was achieved by carrying out a finite-$n$ 
replica analysis under the assumption of one-step replica 
symmetry breaking (1RSB) handling Parisi's 1RSB parameter as a 
control parameter. 
To directly investigate the weight space structure, we took the 
zero-temperature limit and studied the entropy of the zero-energy 
weight.
For $\alpha < \alpha_{\rm s}=0.833 \ldots$, 
the analysis of $n \to 0$ indicates that 
the characteristic exponent of the size distribution of pure states 
is not convex upward, which implies that the weight space is equally 
dominated by a single large cluster of exponentially many 
weights and exponentially many clusters of a single weight. 
For $\alpha > \alpha_{\rm s}$, 
a set of random patterns is rarely compatible with the 
Ising perceptron. 
The $n\to 0$ analysis enables us to assess the rate function 
that characterizes a small probability that a cluster of 
a given entropy will emerge after the storage of random patterns.  
We found that a cluster of finite entropy is generated 
with a relatively high probability for $\alpha_{\rm s} 
< \alpha <\alpha_{\rm GD}=1.245 \ldots$, but this 
is very rare for $\alpha > \alpha_{\rm GD}$. 
These theoretical predictions have been validated by extensive 
numerical experiments. 
We also drew a complete phase diagram 
on the $n$-$\alpha$ plane, in which $(n,\alpha)=(1,\alpha_{\rm GD})$
becomes a tricritical point. The line $n=1$ that passes through 
the tricritical point is analogous to the Nishimori line in 
the theory of spin glasses.

\chapter{The replica zeros of $\pm J$ Ising spin glass at zero temperature}
In the previous chapters, we have seen that 
the generating function $\phi(n)$ 
shows the analyticity breaking leading to
phase transitions with respect to the replica number $n$ in several models.
Employing the Parisi scheme and some discussions based on the physical plausibility,
we have detected such transitions and 
constructed the reasonable solutions after the transitions.  
These schemes are very powerful and can give physically plausible predictions, but still involve some questions about the mathematical correctness, as pointed out in section \ref{sec1:replica}. 
The possible problems are summarized as follows:
\begin{enumerate}
\item{The uniqueness of 
the analytical continuation from natural to real (or complex)
numbers.}
\item{The consistency between the analytical continuation of $\phi(n)$ 
in the replica prescription 
and 
the possible analyticity breaking
of $\phi(n)=\lim_{N\to \infty}\phi_{N}(n)$.}
\end{enumerate}
 
The first problem claims that 
even if all the moments of $[Z^n]$ are given 
for $n \in \mN$, in general it is impossible to uniquely 
continue the analytical expressions for $n\in \mN$ to 
$n \in {\mR}$ (or $\mC$). 
The Carlson's theorem guarantees that
the analytical continuation from $n \in \mN$
to $n \in \mC$ is uniquely determined 
if $[Z^n]^{1/N}< O(e^{\pi |n| })$  
holds as ${\rm Re}(n)$ tends to infinity \cite{THEO}. 
Unfortunately, the moments of the SK model grow as
$[Z^n]^{1/N}< O(e^{C |n|^2})$, where $C$ is a constant,
and therefore this sufficient condition is not satisfied. 
van Hemmen and Palmer conjectured that 
the failure of the RS solution of the SK model 
might be related to this issue \cite{Hemm}, 
but as we have seen 
in chapter 2 this is not the case because $\phi(n)$  
actually loses its analyticity in a certain region 
of $n$ and $T$.

The second issue concerns the possible breaking of the analyticity 
of $\phi(n)$.  
The naive premise of the analytical continuation in the replica 
theory is the analyticity of $\phi_{N}(n)$ with respect to $n$. 
However,  
even if 
$\phi_N(n)$ is analytic
with respect to $n$ for finite $N$, 
the analyticity 
of $\phi(n)=\lim_{N \to \infty} \phi_N(n)$
can be broken, as we have seen in the previous chapters. 
Once the analyticity breaking occurs, in general we  
cannot depend on 
the analytical continuation from large $n$ to small $n$
to obtain $\phi(n)$; 
but 
the Parisi scheme somehow gives the 
correct behavior of $\phi(n)$ by partly 
utilizing the analytical continuation. 
The question is how the Parisi scheme overcomes 
the analyticity breaking and reaches the small $n$ region.

For considering this question, in this chapter,
we investigate the analyticity breaking of $\phi(n)$ 
with respect to $n$ in completely different ways. To this end, we 
propose a method based on the Lee-Yang theory \cite{LeeYang} of phase 
transitions.
In particular, we observe the zeros of $[Z^n]_{\V{J}}$ with respect to complex 
$n \in \mC$, which will 
be referred to as ``replica zeros'' (RZs), 
and examine how some sequences of the zeros approach the real axis
of $n$ as the system size $N$ grows.
For some discrete versions of the random energy models 
\cite{DerrREM,Mouk1,Mouk2,Ogur1}, 
this strategy is known to successfully characterize the 1RSB transition
accompanied by the singularity of the rate function $R(f)$ 
\cite{Ogur2}. 
We extend the scheme to be applicable to other tractable systems, 
$\pm J$ models with a symmetric distribution on two types of lattices, 
ladder systems and Cayley trees (CTs) 
with random fields on the boundary. 
There are some reasons for treating these models.
First, these models can be investigated 
in a feasible computational time by the cavity method 
\cite{MezaRevisit,Bowm}. 
Especially, at zero temperature this approach gives 
a simple iterative formula to yield the
partition function. 
Employing the replica method and the cavity method, 
we can perform symbolic calculations of  
the moment $[Z^n]_{\V{J}}$, 
which enables us 
to directly solve the
equation of RZs $[Z^n]_{\V{J}}=0$. The second reason is the
existence of the spin-glass phase. It is known that the spin-glass phase 
is present for CTs \cite{Chay,Mott,Carl,Lai} 
and is absent for ladder systems. Therefore, we can 
compare the behavior of RZs, which are considered to be 
dependent on the spin-glass ordering.  

Our result indicates that the RZs of 
CTs are relevant to neither 
1RSB nor FRSB. 
We consider some possible explanations about this fact, and 
also refer to some relatives of CTs, which are considered to show 
the RSB, to clarify the properties of the relatives from 
the RZs of CTs. 

The results in this chapter are written in reference \cite{Obuchi:09-1}.

\section{Fundamental knowledge about models and formulations}\label{sec4:FK}
We start from reviewing the models and formulations 
which will be needed in the following sections.
\subsection{Definition of the Cayley tree and the properties}
\begin{figure}[htbp]
\begin{center}
    \includegraphics[height=40mm,width=50mm]{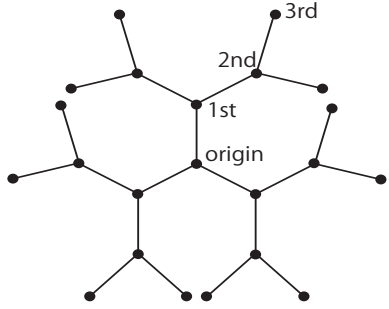}
 \caption{A picture of CT with coordination
 number $c=3$ and total generation $L=3$.}
 \label{fig:CT}
\end{center}
\end{figure}

The definition of the Cayley tree is summarized as follows:
\begin{itemize}
\item The Cayley tree (CT): A tree of finite size consisting of an origin 
      and its neighbors. The first generation is built from $c$ neighbors which
      are connected to the origin. Each site in the $g$th generation is
      connected to new $c-1$ sites without overlap 
      and all these new sites comprise the
      $g+1$th generation. Iterating this procedure to the $L$th
      generation, we obtain the CT, and the $L$th generation 
      becomes its boundary. 
\end{itemize} 
As an example, we give a picture of a CT in fig.\ \ref{fig:CT}.
Let us consider a general Ising system defined on the CT
\begin{equation}
\mathscr{H}=-\sum_{\Ave{i,j}}J_{ij} S_{i}S_{j}-\sum_{i}H_{i}S_{i}.
\end{equation} 
For CTs, it is possible to efficiently 
assess the partition function by an iterative method, 
{\em i.e.} the cavity method\footnote{The term `cavity method' is usually used in a different meaning and is considered to be 
a generalized Bethe approximation \cite{INFO,ADVA}. 
Mathematical and physical aspects 
about the cavity method in this sense are very fruitful, 
but 
in this thesis we only use 
this term to indicate 
the generic name of the iterative methods 
to calculate the partition function.}.
The basis of the cavity method for a CT 
is a formula for evaluating an effective field by a partial trace: 
\begin{equation}
\sum_{S_{j}} \exp\left\{
\beta( J_{ij} S_{i}S_{j}+H_{i}S_{i}+h_{j}S_{j})
\right\}=A\exp(\beta h_{i}S_{i}).
\label{eq4:basic}
\end{equation}
A simple algebra offers
\begin{equation}
h_{i}=H_{i}+{\widehat h}_{j}(J_{ij},h_{j}),\,\, 
A(J_{ij},h_{j})=\frac{2\cosh \beta J_{ij}\cosh \beta h_{j}}
{\cosh\beta {\widehat h}_{j}}, \label{eq4:A}
\end{equation}
where
\begin{equation}
\beta {\widehat h}_{j}(J_{ij},h_{j})
=\tanh^{-1}(\tanh \beta J_{ij} \tanh \beta h_{j} ) \label{eq4:bias}. 
\end{equation}
The fields $h_{j}$ and $\Wh{h}_{j}$ are sometimes termed the cavity field and 
cavity bias, respectively. 

\begin{figure}[htbp]
\begin{center}
    \includegraphics[height=40mm,width=50mm]{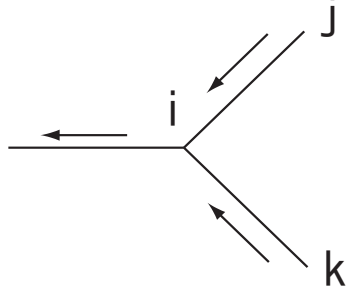}
 \caption{Local structure of a CT with coordination
 number $c=3$.}
 \label{fig:RBG}
\end{center}
\end{figure}

For CTs, 
iterating the above equations from the boundary gives 
the series of cavity fields and biases  
$\{ h_{j},\Wh{h}_{j}\}$.
In general, a cavity field becomes a summation of the
cavity biases from 
its $c-1$ descendants ($c$ is the coordination number): 
\begin{equation}
h_{i}=H_{i}+\sum_{j=1}^{c-1} {\widehat h}_{j}(J_{ij},h_{j}).
\label{eq4:bias_to_field}
\end{equation}
For simplicity, we mainly concentrate on the $c=3$ case, 
the local structure of which is given in fig.\ \ref{fig:RBG}, 
but the generalization to general $c$ is straightforward.
Let us denote the partition function 
in the absence of $i$'s ascendants as $Z_{i}$. 
Equations (\ref{eq4:basic})--(\ref{eq4:bias_to_field})
imply that the partition function is updated as 
\begin{equation}
Z_{i}=
\sum_{S_{i},S_{j},S_{k}}Z_{j}Z_{k}\exp\{-\beta (-H_{i}S_{i}+\Delta H_{ij}+\Delta H_{ik})\} 
\rho_{j}(S_{j})
\rho_{k}(S_{k}),\label{eq4:BPgeneral}
\end{equation}
where
\begin{equation}
\rho_{j}(S_{j})=\frac{\exp(\beta h_{j}S_{j})}{2\cosh \beta h_{j}},
\end{equation}
is the one-site marginal in the absence of $j$'s ascendants and
$\Delta H_{ij}=-J_{ij}S_{i}S_{j}$
is the bond Hamiltonian added by a propagation procedure.
As a final step, the contribution from the origin of 
the tree is calculated as  
\begin{eqnarray}
Z= 
Z_{1}Z_{2}Z_{3}\prod_{i=1}^{3}(2\cosh\beta J_{i}) \frac{
2\cosh\beta(H_{0}+\sum_{i=1}^{3}\Wh{h}_{i})
}{
\prod_{i=1}^{3}(2\cosh\beta \Wh{h}_{i})
}
, \label{eq4:BPcenter}
\end{eqnarray}
where $H_{0}$ is the external field on the origin 
and the whole partition function $Z$ is derived.

For the sake of simplicity, 
we consider pure ferromagnets $J_{ij}=J$ and $H_{i}=H$ for a while. 
In this case each cavity field depends only on the generation $g$ 
and the recursion relation becomes
\begin{equation}
h_{g}=H+(c-1)\Wh{h}_{g-1},\,\,\
\Wh{h}_{g-1}=\frac{1}{\beta}\tanh^{-1}(\tanh(\beta J)\tanh(\beta h_{g-1}) )
\label{eq4:h^pure}
\end{equation}
where $g$ is the index of the generation, but for simplicity of notation, 
we labeled the boundary as $g=0$ and the origin as $g=L$, 
which is in the inverse relation to fig.\ \ref{fig:CT}.
The partial partition function $Z_{g}$, which represents 
the partition function of a branch consisting from 
a $g$th-generation spin and its descendants, 
is also calculated by 
\begin{equation}
Z_{g}=Z_{g-1}^{c-1}(2\cosh\beta J)^{c-1}\frac{
2\cosh\beta h_{g}
}{
(2\cosh\beta \Wh{h}_{g-1})^{c-1}
}.\label{eq4:Z^pure}
\end{equation}
At the origin, we need to merge $c$ branches, which yields the 
whole partition function $Z$ 
\begin{equation}
Z=Z_{L-1}^{c}(2\cosh\beta J)^{c}\frac{
2\cosh\beta(H+c\Wh{h}_{L-1})
}{
(2\cosh\beta \Wh{h}_{L-1})^{c}
}.\label{eq4:Zc^pure}
\end{equation}
These formulas enable us to efficiently 
assess the partition function.
\subsection{Bulk properties of Cayley trees}\label{sec4:bulk} 
Usually, phase transitions are related to 
the bulk properties of the system.  
Contributions from the boundary condition are negligible 
and the inside part of the system dominates the whole behavior.
Unfortunately, this is not the case for a CT, 
because the CT is extraordinary 
in that the number of the boundary spins $N_{B}$ 
are comparable to the whole number of the spins $N$.
Due to this peculiarity, 
the CTs are known to  show some extraordinary critical behaviors 
\cite{MatsPTP1974,Mull1975,Mull1977}.
This problem may be also interesting 
but we are not involved in this topic. 

\begin{figure}[htbp]
\begin{tabular}{cc}
\hspace{-5mm}
 \begin{minipage}[t]{0.5\hsize}
\begin{center}
\includegraphics[height=40mm,width=50mm]{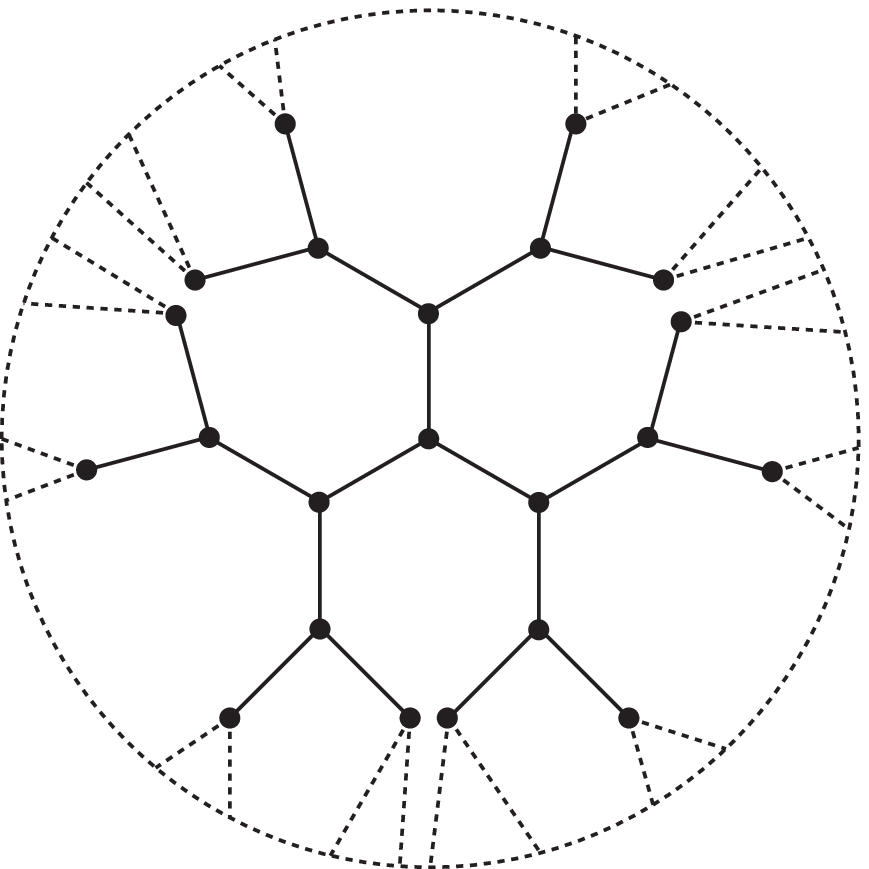}
 \caption{Schematic diagram of a BL with coordination number 
$c=3$.}
 \label{fig:BL}
\end{center}
\end{minipage}
\hspace{2mm}
\begin{minipage}[t]{0.5\hsize}
\begin{center}
\includegraphics[height=40mm,width=50mm]{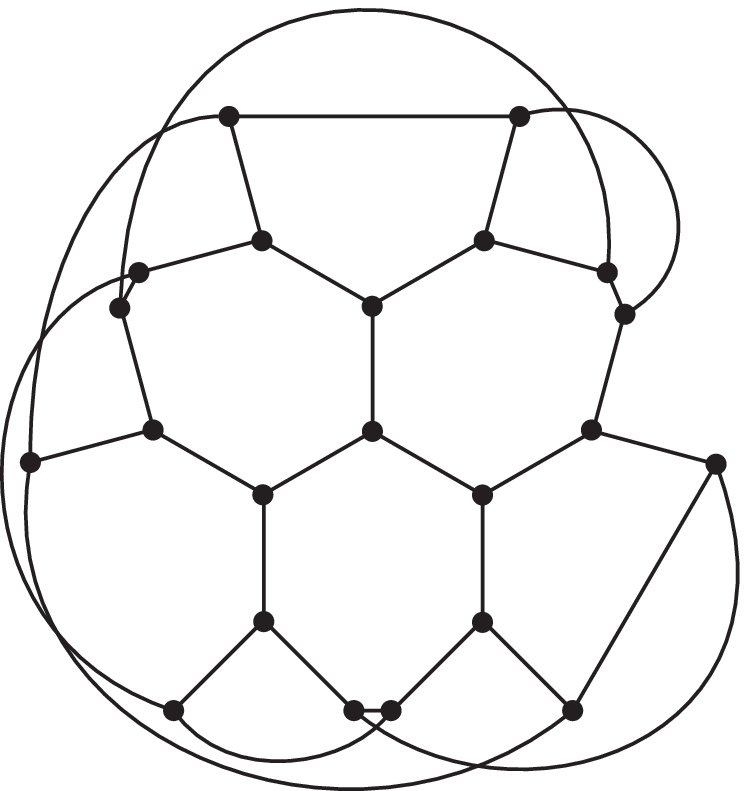}
 \caption{A sample of RRG with coordination number $c=3$ and $N=22$.}
 \label{fig:RRG}
\end{center}
\end{minipage}
\end{tabular}
\end{figure}

On the other hand, considering the bulk part of a CT is also 
meaningful.
For this, we define the following two systems which are closely related to 
CTs:
\begin{itemize}
\item {The Bethe lattice (BL): A lattice consisting of the first 
  $L^\prime$ generations
      of a CT, for which $L \to \infty$ is taken. 
      Alternatively, we can define a BL as
      a finite CT of $L^\prime$ generation, the boundary condition 
      of which is given by the convergent cavity field distribution 
      of the infinite CT. Unlike for a CT, the boundary condition depends on 
      the external parameters, e.g. the 
      temperature $T$ and the external field$H$.   
}

\item {The regular random graph (RRG): A randomly generated graph 
      under the constraint of a fixed connectivity $c$.
      There exist many cycles in this lattice, which makes it difficult 
      to calculate the partition function for finite $N$.
     In the limit $N \to \infty$ under appropriate conditions, 
      however, 
	the contribution coming from the loops becomes negligible
      and the RRG becomes an exactly solvable model. 
 Moreover, it is known that the RRG and the BL share many
      identical properties
      \cite{MezaRevisit,Wong,MezaZero}. 
       }
\end{itemize} 
As examples, we give diagrams of a BL and RRG in figs.\ \ref{fig:BL} and 
\ref{fig:RRG}, respectively.
The RRG is sometimes identified with the BL, but 
we here distinguish these two systems.  
That is because the main purpose of this chapter is to clarify 
the asymptotic properties of $\phi_{N}(n)$ 
from finite $N$ to infinite $N$, and our definition of 
the BL is useful to compare these limits. 
Here, the term ``RRG'' is used only 
to refer to systems of infinite size.

A BL is the deep inside part of a CT and the behavior is 
 determined by the limit $g\to \infty$ in eq.\ (\ref{eq4:h^pure}). 
In this limit, 
we can easily derive the self-consistent equations
satisfied by the converging cavity field $h^*$ and bias $\Wh{h}^{*}$
\begin{equation}
h^{*}=H+(c-1)\Wh{h}^{*},\,\,\
\Wh{h}^{*}=\frac{1}{\beta}\tanh^{-1}(\tanh(\beta J)\tanh(\beta h^{*}) ),
\label{eq4:h*^pure}
\end{equation}
Note that this equation is identical to the `Bethe approximation'.
The name of `Bethe lattice' comes from this fact that the Bethe approximation 
is exact on this lattice. 
When the external field is zero $H=0$, these self-consistent equations 
have the paramagnetic solution $h=0$ at high temperatures and the 
ferromagnetic solution $h>0$ at low temperatures. 
Expanding eq.\ (\ref{eq4:h*^pure}) with respect to $h$, 
we can obtain the following equation determining 
the critical temperature 
\begin{equation}
\tanh\beta J=\frac{1}{c-1}.
\end{equation}

\begin{figure}[htbp]
\begin{tabular}{cc}
\hspace{-5mm}
 \begin{minipage}[t]{0.5\hsize}
\begin{center}
\includegraphics[height=40mm,width=60mm]{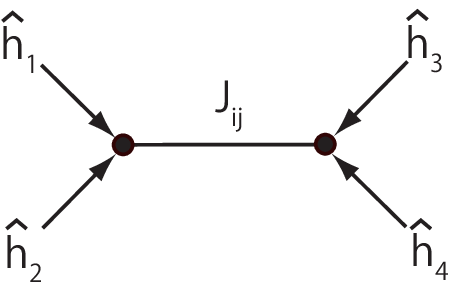}
 \caption{A pictorial representation of $f_{ij}$ for $c=3$.}
 \label{fig:bondcont}
\end{center}
\end{minipage}
\hspace{2mm}
\begin{minipage}[t]{0.5\hsize}
\begin{center}
\includegraphics[height=40mm,width=40mm]{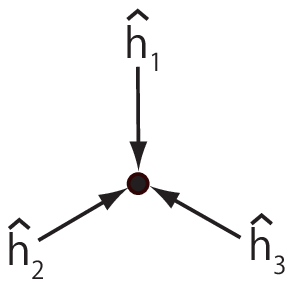}
 \caption{A pictorial representation of $f_{i}$ for $c=3$.}
 \label{fig:sitecont}
\end{center}
\end{minipage}
\end{tabular}
\end{figure}

For general tree systems, it is known that 
the free energy can be expressed as the 
following form \cite{MezaRevisit,Bowm,ADVA,Katsura1979,Nakanishi1981,Yedi}
\begin{equation}
F=\sum_{\Ave{i,j}}f_{ij}^{(2)}-\sum_{i}(c_{i}-1)f_{i}^{(1)},
\label{eq4:ftree}
\end{equation}
where $f^{(2)}_{ij}$ and $f^{(1)}_{i}$ is the bond and site contributions of the 
free energy, respectively
and 
$c_i$ is the number of bonds that site $i$ has. 
For regular CTs,
$c_i=c$ holds if $i$ is 
inside the tree, while $c_i=1$ for the boundary sites. 
The explicit forms of $f_{ij}$ and $f_{i}$ are
\begin{eqnarray}
&&\hspace{-15mm}-\beta f^{(2)}_{ij}=\log \sum_{S_{i},S_{j}} 
\exp \beta
\left\{
 J_{ij}S_{i}S_{j}+
\left(
H_{i}+\sum_{k\in \lambda_{i}-j}\Wh{h}_{k}
\right)S_{i} + 
\left(
H_{j}+\sum_{k\in \lambda_{j}-i}\Wh{h}_{k}
\right)S_{j}
\right\}
,\label{eq4:fij}
\\ 
&&
-\beta f^{(1)}_{i}=
\log \sum_{S_{i}} 
\exp
\left\{
\beta 
\left(
H_{i}+\sum_{k\in \lambda_{i}}\Wh{h}_{k}
\right)S_{i}
\right\},\label{eq4:fi}
\end{eqnarray}
where $H_{i}$ denotes the external field applied on site $i$ 
and $\Wh{h}_{k}$ denotes the cavity bias.
The symbol $\lambda_{i}$ denotes the neighbors of site $i$ 
and $\lambda_{i}-j$ represents the neighbors of site $i$ 
except for site $j$. These are expressed in the pictorial forms in figs.\ 
\ref{fig:bondcont} and \ref{fig:sitecont}.
For pure ferromagnets on CTs, the cavity bias takes different 
values depending on the generation, and we cannot obtain 
a simple form of the free energy.
On the other hand for BLs, the cavity bias takes an identical 
value $\Wh{h^*}$, which leads to the Bethe free energy 
$F/N=f^{\rm BL}$ as
\begin{equation}
f^{\rm BL}=\frac{r_{\rm I}+r_{\rm B}-1}{r_{\rm I}+r_{\rm B}}f^{(2)}-
r_{\rm I}(c-1)f^{(1)}=
r_{\rm I} f_{\rm I}
+r_{\rm B}f_{\rm B}. 
\label{eq4:BLfe}
\end{equation}
where
$r_{\rm I}=\left (1+c(c-2)^{-1}
\left ((c-1)^{L^\prime-1}-1 \right ) \right )/
\left (1+c(c-2)^{-1}
\left ((c-1)^{L^\prime}-1 \right ) \right )$ and 
$r_{\rm B}=1-r_{\rm I}$ represent 
the fractions of the number of sites inside the tree and 
on the boundary, respectively, and 
\begin{eqnarray}
f_{\rm I}=\frac{c}{2}f^{(2)}-(c-1)f^{(1)}, 
\label{eq4:fin}
\end{eqnarray}
and
\begin{eqnarray}
f_{\rm B}=\frac{1}{2}f^{(2)}. 
\label{eq4:fb}
\end{eqnarray}
The explicit expressions of bond and 
site contributions, $f^{(2)}$ and $f^{(1)}$, are 
derived from eqs.\ (\ref{eq4:fij}) and (\ref{eq4:fi}), respectively, 
by putting $\Wh{h}_{i}=\Wh{h}^{*}$, $J_{ij}=J$  and $H_{i}=H$.
These formulas give the free energy of the BL. 

Although the uniformness of the cavity fields is recovered in the BL 
by applying the convergent cavity field to the boundary spins, 
the BL still has a distinctive feature as a tree system, which 
appears in the existence of the boundary free energy (\ref{eq4:fb}).
Nevertheless, the complete separation of contributions between 
the inside and the boundary in the equation (\ref{eq4:BLfe})
implies that it is physically plausible to use
$f_{\rm I}$, instead of $f^{\rm BL}$,
in handling problems concerning the bulk part of the CT. 
In general, $f_{\rm I}$ agrees with the free energy of an RRG which 
can be derived by using the replica method and the derivation is given in 
appendix \ref{app:RRG}.
This fact provides 
the basis of the correspondence between BLs and RRGs. 

Using the relations between CTs, BLs, and RRGs discussed so far, 
we compare the analytical results of RRGs with 
the numerical plots of RZs of CTs 
for exploring possible links between them. 
The BL will be also useful to connect those distinctive results. 
According to this advantage,  
we clearly 
distinguish the three systems, CTs, BLs, and  RRGs, throughout this 
thesis.


\subsection{Zeros and singularities}
Phase transitions are generally reflected as the singularities of 
the free energy $-\beta f=(1/N)\log Z$. 
Lee and Yang pointed out that those singularities are related to the zeros 
of the partition function with respect to the complex parameters 
\cite{LeeYang}. 
For simplicity of explanation, 
let us consider an Ising system with an complex external 
field $H$. 
The partition function for finite $N$ becomes the following polynomial 
\begin{equation}
Z\left( H \right) = e^{N\beta H}\sum_{k=0}^{N}\Omega(k)Y^{k} 
,\label{eq4:Z-pori}
\end{equation}
where $\Omega(k)$ is the partition function of the system with the fixed magnetization $M=\sum_{i}S_{i}=N-2k$, and $Y$ equals to $e^{-2\beta H}$.
The partition function is the polynomial of $Y$ and has its zeros. 
Of course, these zeros are distributed in the complex $H$ plane and never on the real axis of $H$ for finite $N$.
However, as the system size $N$ grows, in several cases some sequences of the
zeros approaches the real axis, and finally a part of the zeros 
touch the real axis in the limit $N \to \infty$.
Roughly speaking, Lee and Yang proved that these zeros on the real axis 
are directly related to the phase transition of the system, and
that if the zeros never touch the real axis, the system never shows the 
phase transition. 
They also showed that for several models 
the existing region of the zeros is strongly limited and 
the phase transitions of those systems are well observed by the 
behavior of the zeros of $Z$ for finite $N$.

The discussion by Lee and Yang that the zeros of the partition function 
are closely related to the singularities of $\log Z$ is quite general and we 
can apply this to our problem, which motivates us to investigate the RZs 
given by the equation $[Z^n]_{\V{J}}=0$ for studying the singularities of $\phi(n)=\lim_{N \to \infty}(1/N)\log [Z^n]_{\V{J}}$ 

\subsection{Energetic zero-temperature limit for replica zeros}
Solving 
\begin{equation}
[Z^n]_{\V{J}}=0 \label{eq4:RZs}
\end{equation}
with respect to $n$ is our main objective. 
Unfortunately, this is, in general, a hard task even by 
numerical methods because eq.\ (\ref{eq4:RZs}) is 
a transcendental equation unlike eq.\ (\ref{eq4:Z-pori}), 
and becomes highly complicated as
the system size $N$ grows.
In the $T\rightarrow 0$ limit, however, the main contributions 
to the partition function only come
from the ground state and eq.\ (\ref{eq4:RZs}) becomes
\begin{equation}
[Z^n]_{\V{J}}\approx[d_{ {\rm g} }^{n}e^{-\beta n E_{ {\rm g} }} ]_{\V{J}}=0,
\label{eq4:RZs2}
\end{equation}
where $E_{{\rm g}}$ is the
energy of the ground state and $d_{{\rm g}}$ is the degeneracy. 
If $n$ is finite when $\beta \rightarrow \infty$, the term 
$e^{-\beta n E_{ {\rm g} }}$ diverges or vanishes and there is
no meaningful result. Therefore, we suppose that non-trivial
solutions exist only in the limit 
$n \rightarrow 0,\beta \rightarrow \infty,$ and 
$ y=\beta n \sim O(1)$. 
This is the same limit as the energetic zero temperature 
limit referred to in section \ref{sec3:complexity}.
Under this condition, eq.\ (\ref{eq4:RZs2}) becomes 
\begin{equation}
[e^{-y E_{ {\rm g} }} ]_{\V{J}}=0.
\label{eq4:RZs3}
\end{equation}
In the following, 
we focus on the $\pm J$ model whose Hamiltonian is given by
\begin{equation}
\mathscr{H}=-\sum_{\Ave{i,j}}J_{ij}S_{i}S_{j},
\end{equation}
and the distribution of interactions is  
\begin{equation}
P(J_{ij})=\frac{1}{2}\delta(J_{ij}-1)+\frac{1}{2}\delta(J_{ij}+1), 
\label{eq4:dist_int}
\end{equation}
assuming that the total number, $N_{B}$, of interacting spin pairs
$\Ave{i,j}$ is proportional to $N$, which is 
the case for ladder systems and CTs. 
This limitation restricts the energy of any state to an integer value.
As a result, eq.\ (\ref{eq4:RZs3}) can always
be expressed as a polynomial of $x=e^{y}$, 
which significantly reduces the numerical cost for 
searching for RZs. 

\section{General formula for ladders and Cayley trees}\label{sec4:GF}
\subsection{The cavity method}\label{sec4:cavity}
We have already demonstrated the cavity method for pure ferromagnetic systems 
on CTs in 
section \ref{sec4:FK}, 
and 
we here only give a simple explanation of its generalization to treat 
 random systems, $k$-body interactions, and ladder systems. 

\begin{figure}[htbp]
\begin{tabular}{ccc}
\hspace{-5mm}
 \begin{minipage}[t]{0.48\hsize}
\begin{center}
\includegraphics[height=40mm,width=50mm]{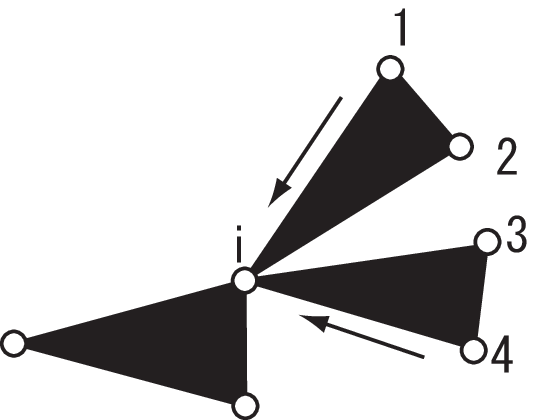}
 \caption{Local structure of a $3$-CT with coordination
 number $c=3$.}
 \label{fig:3bRBG}
\end{center}
\end{minipage}
\hspace{2mm}
\begin{minipage}[t]{0.48\hsize}
\begin{center}
\includegraphics[height=40mm,width=50mm]{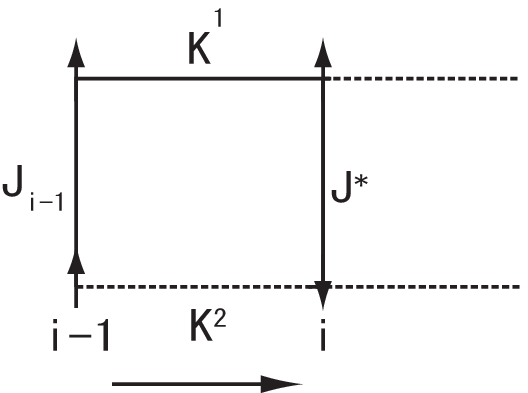}
 \caption{Local structure of a ladder with the width $2$.}
 \label{fig:w2RL}
\end{center}
\end{minipage}
\end{tabular}
\end{figure}

Equations (\ref{eq4:basic})--(\ref{eq4:bias_to_field}) 
are applicable to random systems 
in the unchanged forms.
Generalizing these equations to $k$-spin interacting CTs ($k$-CTs)
is straightforward;
the only necessity is to replace the partial trace (\ref{eq4:basic}) 
with that for a $k$-spin interaction, as 
\begin{equation}
\sum_{\{S_{j}\}} \exp
\left\{
\beta\left( S_{i}J_{l}\prod_{j=1}^{k-1}S_{j}+\sum_{j=1}^{k-1} h_{j}S_{j}
+H_{i}S_{i}
\right)
\right\}
=A\exp(\beta h_{i}S_{i}),
\label{eq4:3bbasic}
\end{equation}
where
\begin{equation}
h_{i}=H_{i}+\Wh{h}_{l},\,\, A=\frac{2^{k-1}\cosh\beta J_{l} \prod_{j=1}^{k-1}\cosh\beta h_{j}}{\cosh \beta \Wh{h}_{l}},
\end{equation}
\begin{equation}
\Wh{h}_{l}=\frac{1}{\beta}\tanh^{-1}\left(
\tanh \beta J_{l} \prod_{j=1}^{k-1}\tanh \beta h_{j}
\right).
\end{equation}
The $3$-CT's local 
structure with $c=3$ is shown in fig.\ \ref{fig:3bRBG} as an 
example.

For ladder systems, 
 similar relations can be derived with some slight modifications. 
We treat the width $2$ ladder as an example.
Summation over the two spins 
of $i$'s previous layer as in fig.\ \ref{fig:w2RL}, 
we can obtain an expression corresponding to
eq.\ (\ref{eq4:basic}) of the CT as 
\begin{equation}
\sum_{S_{i-1}^{1,2}}\exp\{\beta(J_{i-1}S_{i-1}^{1}S_{i-1}^{2}
+K^{1}S_{i-1}^{1}S_{i}^{1}+K^{2}S_{i-1}^{2}S_{i}^{2}
+J^{*} S_{i}^{1}S_{i}^{2}  ) \}=A\exp
\left\{\beta
(J^{*}+\Wh{J}_{i-1})S_{i}^{1}S_{i}^{2}  
\right\}.
\end{equation}
From simple algebras we get
\begin{eqnarray}
&&\Wh{J}_{i-1}(J_{i-1},K^1,K^2)=\frac{1}{\beta}\tanh^{-1}(\tanh\beta K^{1} \tanh\beta K^{2} 
\tanh\beta J_{i-1}), \\
&& A(J_{i-1},K^1,K^2)=4\frac{\cosh\beta J_{i-1}\cosh\beta K^{1}\cosh\beta K^{2}}{\cosh\beta \Wh{J}_{i-1}(J_{i-1},K^1,K^2) },
\end{eqnarray}
which means that the effective bonds $J_{i}$ between $S_{i}^{1}$ and
$S_{i}^{2}$ becomes
\begin{equation}
J_{i}=J^{*}+\Wh{J}_{i-1}(J_{i-1},K^1,K^2).\label{eq4:J}
\end{equation}
As eq.\ (\ref{eq4:BPgeneral}) in the CT case,
the partition function is also updated as 
\begin{equation}
Z_{i}=\sum_{S_{i-1}^{1,2},S_{i}^{1,2}}Z_{i-1}e^{-\beta \Delta H_{i,i-1} } 
\rho(S_{i-1}^{1},S_{i-1}^{2})
\end{equation}
where 
\begin{equation}
\rho(S_{i-1}^{1},S_{i-1}^{2})=\frac{e^{\beta J_{i-1}S_{i-1}^{1}S_{i-1}^{2} }}{4\cosh \beta J_{i-1}}
\end{equation}
and 
$\Delta H_{i,i-1}=-K^{1}S_{i-1}^{1}S_{i}^{1}-K^{2}S_{i-1}^{2}S_{i}^{2}
-J^{*} S_{i}^{1}S_{i}^{2}$ is the partial Hamiltonian added by an 
iterative procedure.
The above formulas for the CTs and ladders enable us to assess 
the partition function of a given bond configuration $\{ J_{ij}\}$ 
in a feasible computational cost.

Taking the $T \to 0$ limit yields the ground-state
energy for a given bond configuration. 
For the $c=3$ CT, eqs.\ (\ref{eq4:bias}) and 
(\ref{eq4:BPgeneral})
become
\begin{equation}
{\widehat h}_{j} \rightarrow \sgn{J_{ij}h_{j}}\min(|J_{ij}|,|h_{j}|), \label{eq4:T0bias}
\end{equation}
and 
\begin{eqnarray}
&&E_{i} =E_{j}+E_{k}-J_{ij}-J_{ik}\nonumber \\
&&
\hspace{10mm}+\left\{
\begin{array}{ll}
0 & \, (\,\,\sgn{J_{ij}J_{ik}h_{j}h_{k}} \geq 0\,\,)\\
2\min(|J_{ij}|,|J_{ik}|,|h_{j}|,|h_{k}|) & \, (\,\,{\rm otherwise}\,\,)\,\, 
\end{array}
\right.
.
\end{eqnarray}
where we put the ground-state energy as 
$E_{i}=\lim_{\beta \to \infty} -(1/\beta)\log Z_{i}$ 
and assume $\sgn{0}\equiv 0$. The extension to general $c$ and 
$k$-body interactions is straightforward and we here omit. 
Similarly, for the ladder, we can derive 
\begin{equation}
{\widehat J}_{i-1} 
\rightarrow \sgn{J_{i-1}K^1 K^2}
\min(|J_{i-1}|,|K^1|,|K^2|), \label{eq4:T0bias^J}
\end{equation}
and 
\begin{eqnarray}
&&E_{i} =E_{i-1}-K^1-K^2-J^{*} \nonumber \\
&&\hspace{10mm}+
\left\{
\begin{array}{ll}
0 & \,\, (\,\,\sgn{J_{i-1}J^{*} K^1 K^2} \geq 0\,\,) \,\,\\
2\min(|J_{i-1}|,|J^{*}|,|K^1|,|K^2|) & \,\, (\,\,{\rm otherwise}\,\,)\,\, 
\end{array}
\right. .
\end{eqnarray}
For a given single sample of interactions and boundary conditions, 
a simple application of the cavity method enables us to 
evaluate the partition function in a feasible computational time. 
Unfortunately, this does not fully 
resolve the problem of the computational 
cost for assessing the moments (\ref{eq4:RZs}) since 
the cost for evaluating the average over all possible samples of 
interactions and boundary conditions grows exponentially with 
respect to the number of spins.
 
\subsection{Use of the replica method}
To overcome the difficulty mentioned immediately above, 
we perform the configurational average 
for $n \in \mN$  
and use the analytical continuation of the obtained 
expressions to $n \in \mC$ in the level of the cavity recursions, 
the method of which may be considered as 
a generalization of the replica method. 

For this purpose, we first evaluate the $n$th moment of the 
partition function $Z$ 
\begin{eqnarray}
&&\Xi(n)\equiv [Z^n]_{\V{J}}=\Tr{} \prod_{\Ave{i,j} } \left[
\exp
\left(
\beta J_{ij}\sum_{\mu=1}^{n}
S^{\mu}_{i}S^{\mu}_{j}  
\right) 
\right]_{\V{J}}, \label{eq4:effZ}
\end{eqnarray}
for $n \in \mN$, where $\mu$ is the replica index.  
Let us denote the effective Hamiltonian as
\begin{equation}
H_{{\rm eff}}
=\sum_{\Ave{i,j}} H_{ij}
=\sum_{\Ave{i,j}}-\frac{1}{\beta}\log 
\left[ 
\exp
\left(
\beta J_{ij} \sum_{\mu=1}S_{i}^{\mu}S_{j}^{\mu}
\right)
\right]_{\V{J}},\label{eq4:repH}
\end{equation}
where $[\cdots]_{\V{J}}$ stands for the configurational 
average with respect to 
the interactions $\{J_{ij}\}$. 
This means that eq.\ (\ref{eq4:effZ}) is simply 
the partition function of an $n$-replicated system,
which is defined on the same lattice as the original system, 
and is free from 
quenched randomness. 
Therefore, we can again use the cavity method to calculate $\Xi(n)$.

\subsubsection{The case of the CT} 
For the $c=3$ CT, we can derive the following recursion relation 
for $\Xi(n)$ as
\begin{eqnarray}
&&\hspace{-15mm} \Xi_{i}(n)=
\sum_{\V{S}_{i}, \V{S}_{j}, \V{S}_{k}} 
\left[\exp\left\{\beta 
\left(J_{ij}\sum_{\mu=1}^{n}
S^{\mu}_{i}S^{\mu}_{j} +J_{ik}\sum_{\mu=1}^{n}
S^{\mu}_{i}S^{\mu}_{k}
\right)\right\}
\right]_{\V{J}}
\rho_{j} \rho_{k }  \Xi_{j}(n) \Xi_{k}(n)  \label{eq4:SPijk}\\
&&\hspace{-15mm} 
=\sum_{\V{S}_{i}} 
\rho_{i}(\V{S}_{i})  \Xi_{i}(n),\label{eq4:SPi} 
\end{eqnarray}
where $\rho_{i}$ is the one-site marginal distribution 
of site $i$. 
The expressions (\ref{eq4:SPijk}) and (\ref{eq4:SPi})
define the updating rules of $\rho_{i}$ and $\Xi_{i}(n)$. 

So far, we have made no assumptions or approximations
and therefore eq.\ (\ref{eq4:SPi}) yields exact assessments 
for $n \in \mN$, but  
to practically calculate $\Xi(n)$, 
we need to specify the form of $\rho_{i}$.
In the present case, the form of $\rho_{i}$ can be determined as
follows.
First, we assume the RS which requires that 
the correlation functions are independent of combinations of replica
indices
\begin{equation}
\Ave{S^{1}_{i}S^{2}_{i}}_{\rho_{i}}=\Ave{S^{\mu}_{i}S^{\nu}_{i}}_{\rho_{i}}=q^{(2)},
\end{equation}
where the brackets $\Ave{(\cdots)}_{\rho}$ 
denote the average over the distribution $\rho_{i}$. 
This assumption is correct for $n\in \mN$ because 
the replicated Hamiltonian eq.\ (\ref{eq4:repH})
 is invariant under the permutation of the
replica indices. 
Moreover, 
in our case as long as the
number of spins $N$ is finite, the RS is exact even for real $n \in \mR$. 
This is because the $\pm J$ model satisfies the necessary 
conditions of the Carlson's theorem \cite{THEO}, as explained in section \ref{sec4:unique}.
With the RS, the expression of $\rho_{i}$ is uniquely
determined as
\begin{equation}
\rho_{i}(\V{S_{i}})=
\frac{
\sum_{l=0}^{n}q^{(l)}_{i}
\sum_{ \{ a_{1} \ldots a_{l} \} } S^{a_{1}}_{i} \cdots S^{ a_{l} }_{i}   }
{2^n},\label{eq4:rho}
\end{equation}
where the symbol $\sum_{ \{ a_{1} \ldots a_{l} \} }$ stands for a
summation over all combinations of $l$ choices from $n$ replicas and
$q^{(l)}$ denotes the correlation functions among $l$ replica spins.
It is easy to check that the normalization condition  
is satisfied and all the moments are recovered in this expression. 
Generally,
eq.\ (\ref{eq4:rho}) can be transformed as follows
\begin{equation}
\rho_{i}(\V{S}_{i})=
\int P_{i}(x)\prod_{\mu=1}^{n}\left(\frac{1+x S^{\mu } }{2}  
\right)dx,
\end{equation}
where the distribution function $P_{i}(x)$ must be defined to satisfy 
relations
\begin{equation}
\int x^{l}P_{i}(x)dx =q^{(l)}_{i}.
\end{equation}
If $|x|$ is bounded by $1$, the nonnegativity of the distribution
$\rho_{i}$ is satisfied.   
Under this assumption, 
we can perform a variable transformation $x=\tanh \beta h$ as
\begin{equation}
\rho_{i}=\int \pi_{i}(h)\prod_{\mu=1}^{n}\left(\frac{1+\tanh(\beta h) S^{\mu }_{i} }{2}  
\right)dh
=\int \pi_{i}(h)
\frac{e^{\beta h \sum_{\mu} S^{\mu}_{i}  }  }{(2\cosh \beta h)^n}  
dh.\label{eq4:pifi}
\end{equation}
This expression can be interpreted as that 
the cavity field $h$ fluctuates by
the randomness and has the distribution $\pi_{i}(h)$.

Inserting eq.\ (\ref{eq4:pifi}) into eq.\ (\ref{eq4:SPi})
and performing some simple algebraic steps give 
\begin{eqnarray}
&& \Xi_{i}=\sum_{ \V{S}_{i}}
\Xi_{j}\Xi_{k}
( 2\cosh\beta)^{2n}\int dh_{i}
\frac{e^{\beta h_{i} \sum_{\mu} S^{\mu}_{i}}}{(2\cosh \beta h_{i})^n} 
\Bigg\{ 
\int \!\!\! \int \pi_{j}(h_{j}) \pi_{k}(h_{k}) \cr 
&& \hspace{-0cm}\times \left[
\delta(h_{i}-{\widehat h}_{j}-{\widehat h}_{k} ) 
\left(
\frac{2\cosh \beta h_{i}}{2\cosh\beta {\widehat h}_{j}2\cosh\beta {\widehat h}_{k} }
\right)^n
\right]_{\V{J}}dh_{j}dh_{k} 
\Bigg\} \label{eq4:rhoijZij} \\
&&=
\sum_{ \V{S}_{i}}
\int dh_{i}
\pi_{i}(h_{i}) 
\frac{e^{\beta h_{i} \sum_{\mu} S^{\mu}_{i}}}{(2\cosh \beta h_{i})^n} 
\Xi_{i}. 
\label{eq4:rhoiZi}
\end{eqnarray}
Equations (\ref{eq4:rhoijZij}) and (\ref{eq4:rhoiZi}) provide the 
cavity equation of $\pi_{i}(h_{i})$: 
\begin{eqnarray}
\hspace{-7mm} \pi_{i}(h_{i})
\propto
\int \!\!\! \int \pi_{j}(h_{j}) \pi_{k}(h_{k})\left[
\delta(h_{i}-{\widehat h}_{j}-{\widehat h}_{k} ) 
\left(
\frac{2\cosh \beta h_{i}}{2\cosh\beta {\widehat h}_{j}2\cosh\beta {\widehat h}_{k} }
\right)^n
\right]_{\V{J}}dh_{j}dh_{k}\label{eq4:pih1},  
\end{eqnarray}
\begin{eqnarray}
\hspace{-7mm} \Xi_i=\Xi_{j}\Xi_{k}( 2\cosh\beta)^{2n}
\int \!\!\! \int dh_{j}dh_{k}
\, \pi_{j}(h_{j})\pi_{k}(h_{k}) 
\left[
\left(
\frac{2 \cosh \beta (\widehat{h}_j +\widehat{h}_k)}{
2\cosh\beta {\widehat h}_{j}2\cosh\beta {\widehat h}_{k} }
\right) ^n
\right]_{\V{J}},\label{eq4:Zbranch}
\end{eqnarray}
which are applicable to $\forall{n} \in \mC$. 
When the algorithm reaches the origin of the CT, 
the moment of eq.\ (\ref{eq4:effZ}) is assessed as 
\begin{eqnarray}
&&\Xi(n)=[Z^n]_{\V{J}}
=\Xi_{1}\Xi_{2}\Xi_{3}(2\cosh\beta )^{3n}
\int \!\!\! \int\!\!\! \int dh_{1}dh_{2}dh_{3}
\, \pi_{1}(h_{1})\pi_{2}(h_{2})\pi_{3}(h_{3})  \cr 
&& \times\left[
\left(
\frac{1+\tanh\beta J_{1}\tanh\beta J_{2}\tanh\beta h_{1}\tanh\beta h_{2}
+R }{4}
\right) ^n
\right]_{\V{J}},\label{eq4:Zfinal}
\end{eqnarray}
where $R$ is two terms with the indices $1,2,3$ rotated.

\subsubsection{The case of the ladder}
For the width-$2$ ladder, the moment $\Xi(n)$ is updated as
\begin{eqnarray}
&&\Xi_{i}=\sum_{\V{S}_{i-1}^{1,2},\V{S}_{i}^{1,2}}
\left[
\exp
\left\{
\beta\sum_{\mu=1}^{n}
\left(
J_{i-1}S_{i-1}^{1,\mu}S_{i-1}^{2,\mu}
+K^{1}S_{i-1}^{1,\mu}S_{i}^{1,\mu}+K^{2}S_{i-1}^{2,\mu}S_{i}^{2,\mu}
+J^{*} S_{i}^{1,\mu}S_{i}^{2,\mu} 
\right)  
\right\}
\right]_{\V{J}} \nonumber \\
&&\hspace{15mm}
\times \Xi_{i-1} \rho_{i-1}(\V{S}_{i-1}^{1},\V{S}_{i-1}^{2})
\label{eq4:i-1}
\\
&&
=\sum_{\V{S}_{i}^{1,2}}
\Xi_{i} \rho_{i}(\V{S}_{i}^{1},\V{S}_{i}^{2})
\label{eq4:i}
\end{eqnarray}
These equations also define the recursion rule of the two-site marginal 
distribution $\rho_{i}(\V{S}_{i}^{1},\V{S}_{i}^{2})$.
As in the CT case, we can
specify the form of $\rho_{i}(\V{S}_{i}^{1},\V{S}_{i}^{2})$ 
from symmetries of the present problem as
\begin{equation}
\rho_{i}(\V{S}_{i}^{1},\V{S}_{i}^{2})=\int dJ_{i} \pi_{i}(J_{i}) \frac{
e^{\beta J_{i} \sum_{\mu} S_{i}^{1, \mu }S_{i}^{2,\mu}}  
}{(4\cosh \beta J_{i})^n},\label{eq4:rho^J}
\end{equation}
which can be interpreted as that the effective bond 
fluctuates by quenched randomness. 
The equation of the moment $\Xi_{i}$ then becomes
\begin{eqnarray}
\hspace{-4mm}\Xi_{i}=\Xi_{i-1}\int dJ_{i-1}\pi_{i-1}(J_{i-1})\left[
(4\cosh\beta K^{1}\cosh\beta K^{2})^{n}
\left(\frac{\cosh\beta (J^{*}+\Wh{J}_{i-1})}{\cosh \beta \Wh{J}_{i-1}}\right)^n
\right]_{\V{J}}.
\label{eq4:Xi}
\end{eqnarray}
Equations (\ref{eq4:i-1})-(\ref{eq4:rho^J}) imply that 
the recursion relation of $\pi_{i}(J_{i})$ becomes
\begin{equation}
\pi_{i}(J_{i})\propto \int dJ_{i-1} \pi_{i-1}(J_{i-1})
\left[
\delta(J_{i}-J^{*}-\Wh{J}_{i-1})
\left(\frac{\cosh\beta (J^{*}+\Wh{J}_{i-1})}{\cosh \beta \Wh{J}_{i-1}}\right)^n
\right]_{\V{J}}.\label{eq4:piJ}
\end{equation}
These relations yield the algorithm to calculate $\Xi(n)$ for the width-$2$
ladder.
\subsection{Energetic zero-temperature limit in the cavity recursions}
Let us take the energetic zero-temperature limit 
$\beta \to \infty$, $n \to 0$ keeping $y=\beta n \approx O(1)$  
in the iterative procedures of the cavity method. 

We first focus on the width-$2$ ladder. 
In our present model, the interaction $J_{ij}$ takes $\pm 1$ with 
an equal probability $1/2$, 
which means that the initial condition of the recursion relation 
(\ref{eq4:piJ}), $\pi_{0}(J_{0})$, takes the following form
\begin{equation}
\pi_{0}(J_{0})=\frac{1}{2}(\delta(J_{0}-1)+\delta(J_{0}+1)),
\end{equation}
and the initial condition of the moment $\Xi_{0}$ becomes 
in the energetic zero-temperature limit as 
\begin{equation}
\Xi_{0}(n)=(2\cosh \beta)^n \to e^{y}. 
\end{equation}

The relation (\ref{eq4:T0bias^J}) and the fact that 
$|J_{ij}|$ equals to $1$ mean that 
for $\forall i$ 
$J_{i}$ only takes five integers $0,\pm 1,\pm 2$
at zero temperature.
This condition considerably simplify the current problem.
The cavity equation (\ref{eq4:piJ}) becomes
\begin{eqnarray}
&&\pi_{i}(J_{i})
\propto 
\int dJ_{i-1} 
\pi_{i-1}(J_{i-1})
\Biggl[
\delta(J_{i}-J^{*}-\sgn{K^1 K^2 J_{i-1}})
\nonumber
\\
&&
\hspace{50mm}
\times 
e^{y
\left(
\left|
J^{*}+\sgn{K^1 K^2 J_{i-1}}
\right|-
\left|
\sgn{K^1 K^2 J_{i-1}}
\right|
\right)}
\Biggr]_{\V{J}}
,\label{eq4:piJT0}
\end{eqnarray}
and 
the distribution $\pi_{i}(J_{i})$ 
takes the following form 
\begin{equation}
\pi_{i}(J_{i})=p_{i;0}\delta(J_{i})
+\sum_{f=1}^{2} p_{i;f}(\delta(J_{i}-f)+\delta(J_{i}+f)),
\label{eq4:piJex}
\end{equation}
where $\V{p}_{i}=(p_{i;0},p_{i;1},p_{i;2})$ represents a probability vector
satisfying $p_{i;0}+2\sum_{f} p_{i;f}=1$ and $p_{i;f}\geq 0$. 
The symmetry $p_{i;f}=p_{i;-f}$ comes from the symmetric distribution 
of $\{ J_{ij}\}$.
Equations (\ref{eq4:piJT0}) and 
(\ref{eq4:piJex}) offer the simple recursion equations as follows; 
\begin{equation}
(p_{i;2},p_{i;1},p_{i;0})
\propto
\left(
\frac{(1-p_{i-1;0})}{2},\,\,
p_{i-1;0},\,\,
(1-p_{i-1;0})e^{-2y}
\right),\label{eq4:p^ladder}
\end{equation}
and the recursion relation of the moment $\Xi_{i}$ (\ref{eq4:i-1})
is also simplified as
\begin{equation}
\Xi_{i}=\Xi_{i-1}e^{3y}
\left\{
p_{i-1;0}+\frac{1}{2}(1+e^{-2y})(1-p_{i-1;0})
\right\}.\label{eq4:Xi^ladder}
\end{equation}
Equations (\ref{eq4:p^ladder}) and (\ref{eq4:Xi^ladder}) 
indicate that only $p_{i;0}$ is relevant for the evaluation of the 
moment. This is a peculiar property to the symmetric distribution of 
$J_{ij}=\pm 1$. We here write down the explicit form of the cavity 
equation of $p_{i;0}$
\begin{eqnarray}
p_{i;0}=\frac{1-p_{i-1;0}}{1-p_{i-1;0}-(1+p_{i-1;0})e^{2y}}.\label{eq4:T0piJ}
\end{eqnarray}
The formulas (\ref{eq4:Xi^ladder}) and (\ref{eq4:T0piJ}) 
enable us to assess the moment $\Xi(y)$ 
in a feasible computational time and 
therefore offer a useful scheme for examining the RZs.

Next, we proceed to the CT case. 
Unlike the ladder case, for the CT 
we should choose  
the boundary conditions appropriately.  
This is because the CT has no cycle in the graph and hence the frustration 
cannot be introduced into the system unless the boundary condition is 
appropriately chosen.
 
The boundary condition we here treat  
is  
the random external field $h_i=\pm 1$, 
the sign of which is determined with an equal probability of $1/2$. 
As the result of this boundary condition, 
the cavity field distribution and the moment at the boundary, 
$\pi_{0}(h_{0})$ and $\Xi_{0}$, become
\begin{eqnarray}
\pi_0(h_0)=\frac{1}{2} \left ( \delta(h_0-1)+\delta(h_0+1) \right )
\label{eq4:boundary_dist}
\end{eqnarray}
and 
\begin{eqnarray}
\Xi_{0}=(2 \cosh \beta)^n \to e^{y}, 
\label{eq4:boundary_Z}
\end{eqnarray}
respectively, which yield the initial conditions of the cavity recursions.
The relevance of the boundary condition to 
the current objective systems is discussed later. 

Equation (\ref{eq4:boundary_dist}) in conjunction with 
the property $|J_{ij}|=1$ again restrict the 
functional form of 
$\pi_i(h_i)$ in eq.\ (\ref{eq4:pih1}) 
as
\begin{equation}
\pi_{i}(h_{i})=
p_{i;0}\delta(h_i)+
\sum_{f=1}^{c-1}
p_{i;f}\left (\delta(h_{i}-f)+\delta(h_{i}+f) \right ).
\label{eq4:symmdist}
\end{equation}
The symmetry $\pi_{i}(h_{i})=\pi_{i}(-h_{i})$ comes 
from the symmetry of 
the boundary-field distribution (\ref{eq4:boundary_dist}) 
as in the ladder case.

After the configurational average is performed, the cavity-field 
distribution $\pi_{i}(h_{i})$ depends only on the 
distance, $g$, from the boundary. Therefore, we hereafter denote 
$\pi_{i}(h_{i})$ as $\pi_{g}(h_{i})$ and represent 
the distance of the origin from the boundary as $g=L$. 
The current scheme assesses $\V{p}_{g+1}$
using its descendents $\V{p}_{g}$
For the $c=3$ case, 
the actual formula is derived from eq.\ (\ref{eq4:pih1}) as
\begin{equation}
(p_{g+1;2},p_{g+1;1},p_{g+1;0})
\propto
\left(
\left(
\frac{(1-p_{g;0})}{2}
\right)^2
,\,\,
p_{g;0}(1-p_{g;0})
,\,\,
p_{g;0}^2+
2
\left(
\frac{1-p_{g;0}}{2}
\right)^2
e^{-2y}
\right).\label{eq4:p^CT}
\end{equation}
Similarly to the ladder case,  
the relevant part to the assessment of $\Xi(n)$ is 
only for $p_{g;0}$, the precise recursion relation of which is given by
\begin{equation}
p_{g+1;0}=\frac{p_{g;0}^2+2\left( \frac{1-p_{g;0}}{2} \right)^2 
e^{-2y}}{1-2(1-e^{-2y})\left( \frac{1-p_{g;0}}{2}\right)^2}, 
\label{eq4:p0precise}
\end{equation}
being accompanied by an update of the moment 
\begin{eqnarray}
\Xi_{g+1}=\Xi_{g}^2e^{2y}
\left\{
1-
2(1-e^{-2y})\left(
\frac{1-p_{g;0}}{2}
\right)^2
\right\}\label{eq4:T0Zh1}. 
\end{eqnarray}
After evaluating $p_{g;0}$ and $\Xi_g$ using this algorithm 
up to $g=L-1$, the full moment, $\Xi(y)$, in the current limit
$n \to 0$ and $\beta \to \infty$ keeping $y=n\beta \sim O(1)$
is finally assessed as
\begin{eqnarray}
\Xi(y)=\Xi_{L-1}^{3}e^{3y}
\left\{
1-3(1-e^{-2y})\left( \frac{1-p_{L-1;0}}{2} \right)^2(1+p_{L-1;0})
\right\}. 
\label{eq4:T0Zh2}
\end{eqnarray}
As eqs.\ (\ref{eq4:p^ladder}) and (\ref{eq4:Xi^ladder}) for 
the ladder case, 
eqs.\ (\ref{eq4:p0precise})--(\ref{eq4:T0Zh2}) 
enable us to calculate the moment $\Xi(y)$ in a feasible time, 
which constitute the main result of this chapter.  

\subsection{Uniqueness of the analytic continuation}\label{sec4:unique}
The analytical continuation 
from $n \in \mN$ to $n \in \mC$ cannot be determined uniquely
in general systems.  
However, in the present system, we can show the uniqueness 
of the continuation. Therefore, the RS solution assumed above
is correct. 

For this, let us consider the modified moment
$[(Z e^{-\beta N_{B}})^n]_{\V{J}}^{1/N}$, where $N_{B}$ is the total number of
bonds. 
This quantity satisfies the inequality 
\begin{eqnarray}
&&\left|
\left[
\left(
Z e^{-\beta N_{B}}
\right)^n
\right]_{\V{J}}^{1/N}
\right|
\leq
\left[
\left(
Z e^{-\beta N_{B}}
\right)^{{\rm Re}(n)}
\right]_{\V{J}}^{1/N} \cr
&& \leq 
\left[
\left(\Tr{}1
\right)^{{\rm Re}(n)}
\right]_{\V{J}}^{1/N}
=2^{{\rm Re}(n)}< O(e^{\pi |n| }),
\end{eqnarray}   
for finite $N$.
Suppose that we have an analytic
function $\psi(n;N)$ that satisfies the condition 
$|\psi (n;N)|< O(e^{\pi |n|})$. 
The Carlson's theorem guarantees that if the equality 
$|\psi (n;N)-[(Z e^{-\beta N_{B}})^n]_{\V{J}}^{1/N}|=0$ holds 
for $\forall{n}\in \mN$,
$\psi(n;N)$ is identical to $[(Z e^{-\beta N_{B}})^n]_{\V{J}}^{1/N}$
for $\forall{n} \in \mC$.   
Because $e^{-\beta N_{B}}$ is a non-vanishing constant, this means
that the analytic continuation of $[Z^{n}]_{\V{J}}^{1/N}$ is uniquely
determined. This indicates that 
expressions in the above subsections, 
which analytically continued under the RS ansatz, 
are correct for finite $N$ (or equivalently, finite $L$).
Hence, in the current problem,
the possible problem of the analytic continuation of 
$\phi(n)=\lim_{N\to \infty}(1/N)\log [Z^n]_{\V{J}}$ 
is summarized into the point whether 
the analyticity breaking on the real axis 
occurs or not in the limit $N \to \infty$. 

\section{Results for the ladders}\label{sec4:RZ^ladder}
\begin{figure}[t]
\begin{tabular}{cc}
\hspace{-5mm}
\begin{minipage}[t]{0.5\hsize}
\begin{center}
   \includegraphics[height=60mm,width=75mm]{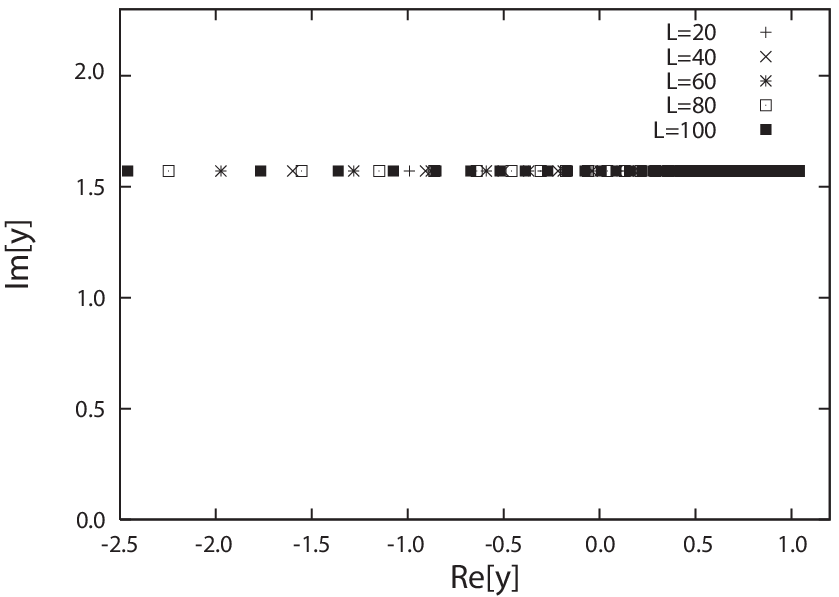}
 \caption{RZs for ladders with $2\times L$. All
 the zeros lie on ${\rm Im} (y)=\pi/2$ and never reach the 
real axis of $y=\beta n$. The inequality ${\rm Re}(y)\leq \log 2\sqrt{2}$
 holds, as shown in appendix \ref{app:proof}.}
 \label{fig:ladderzeros}
\end{center}
\end{minipage}
\hspace{2mm}
 \begin{minipage}[t]{0.5\hsize}
\begin{center}
\includegraphics[height=60mm,width=75mm]{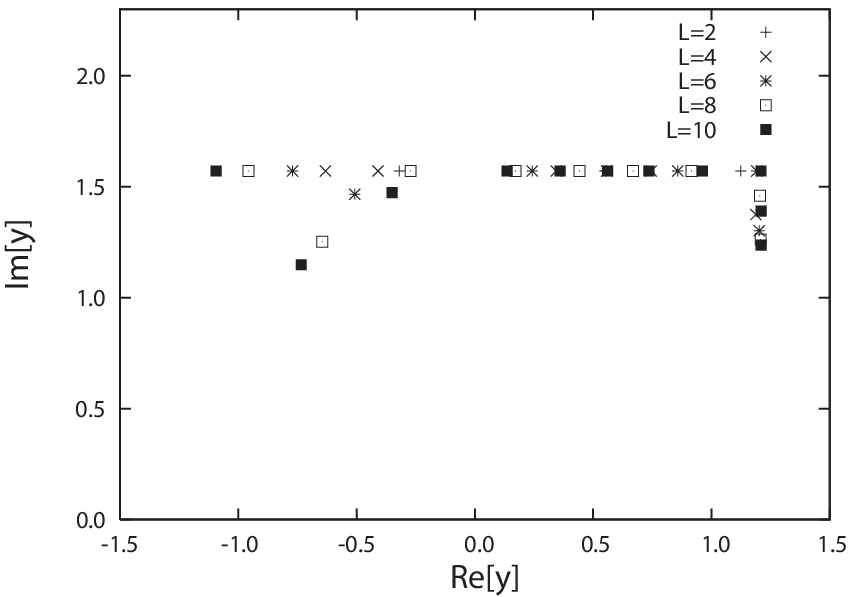}
\caption{Zeros of width-$4$ ladders. Some 
of the zeros approach  the real axis around ${\rm Re} (y)\approx 1.2$ and $-0.8$, but the
rate of approach rapidly decreases as $L$ grows.}
\label{fig:w4ladderzeros}
\end{center}
\end{minipage}
\end{tabular}
\end{figure}

The concrete procedure to obtain RZs  
for the width-$2$ ladder with the length $L$ 
is summarized as follows:
\begin{enumerate}
\item{To obtain a series of $p_{i;0}$, eq.\ (\ref{eq4:T0piJ}) 
is recursively applied under the initial condition $p_{0;0}=0$ until 
$i$ reaches $L$. This can be symbolically performed by using 
computer algebra systems such as {\em Mathematica}.} 
\item{Using the series $\{ p_{i;0} \}$, 
the moment $\Xi_{i}$ is recursively 
calculated by using eq.\ (\ref{eq4:Xi^ladder}) under the initial 
condition $\Xi_{0}=e^{y}$ until $i$ becomes $L$. 
Then, the full moment $\Xi(y)=\Xi_{L}$ 
is obtained.}
\item{Solving $\Xi_{L}=0$ with respect to $x=e^{y}$ numerically.}
\end{enumerate}
Although the right hand side of eq.\ (\ref{eq4:T0piJ}) 
is expressed as a rational function,
the moment $\Xi(y)$ 
is guaranteed to be certain polynomials of $x$ since 
the contribution from the denominator is canceled 
in each step of eqs.\ (\ref{eq4:Xi^ladder}) and (\ref{eq4:T0piJ}). 
The procedures of 1, 2 and 3 can be performed in a polynomial 
time with respect to the number of spins. 
Figure \ref{fig:ladderzeros} shows the plot for 
a $2 \times L$ ladder. 
Notice that all RZs lie on a line ${\rm Im}(y)=\pi/2$.
This fact can be mathematically proven, 
as detailed in \ref{app:proof}. 
The physical significance of this behavior is
that the generating function $g_{N}(n)$ is analytic 
with respect to
real $y$ even for the $N \rightarrow \infty$ limit. 

We also investigated larger-width ladders. 
However, the larger-width systems requires 
many-body interactions to calculate 
the marginal distribution $\rho$.
It makes the problem complicated and  
simple relations like (\ref{eq4:rho^J}) cannot
be obtained. 
Hence, when treating larger-width ladders,
we directly use the cavity method for a given sample $\{J_{ij}\}$ 
to obtain the ground state energy \cite{Kado}, which is just 
the use of the cavity method we mentioned in the end of 
section \ref{sec4:cavity}.
In this prescription, 
we numerically 
assess the distribution of the ground state energies $P(E_{g})$
by enumerating all the bond configurations $\{J_{ij}\}$. 
Using the distribution $P(E_{g})$, the RZs equation is derived as 
\begin{equation}
\sum_{E_{g}} P(E_{g}) e^{-y E_{g}}=0.
\end{equation}
This equation is solved numerically in the same way as the width-$2$ case.
Note that computational times required in the counting process to obtain 
$P(E_{g})$ exponentially increases as $L$ grows, 
which makes it infeasible to obtain $\Xi_{L}$ for large $L$, unlike 
the width-$2$ case. 

The RZs for $3 \times L$
ladder were calculated and 
found qualitatively similar results as for the width-2 case.   
For a width-$4$ ladder,
the RZ plot is given in fig.\ \ref{fig:w4ladderzeros}. 
We can observe that some zeros approach the real axis around ${\rm Re}
(y)\approx 1.2$ and $-0.8$, 
but the rate of approach decreases rapidly 
as $L$ grows. This implies that the RZs do not reach the
real axis, which agrees with a naive
speculation that ladders are essentially one-dimensional systems and 
therefore do not involve any phase transitions as long as the 
width is kept finite. 
These results indicate that the RZs for 
the ladders do not reach the real axis 
of $y$ in the limit $N\to \infty$. 
This means that the analyticity breaking of $\phi(n)$ does not 
occur for the ladder systems, 
which accords with an intuition that the ladders are essentially 
one-dimensional systems in the thermodynamic limit and should not 
exhibit the spin-glass ordering.

\section{Results for the CTs}
As for the ladder case, we first summarize 
the concrete procedure to obtain RZs for the CT:
\begin{enumerate}
\item{Recursively applying (\ref{eq4:p0precise}) 
under the initial condition $p_{0;0}=0$ until 
$g$ reaches $L-1$ to obtain the series$\{ p_{g;0} \}$. } 
\item{The moment $\Xi_{g}$ is recursively assessed from
eq.\ (\ref{eq4:T0Zh1}) by using the series $\{ p_{g;0} \}$
 until $g$ becomes $L-1$. }
\item{The full moment $\Xi(y)=\Xi_{L}$ 
is derived from eq.\ (\ref{eq4:T0Zh2}) using $\Xi_{L-1}$ and $p_{L-1;0}$}
\item{Solving $\Xi_{L}=0$ with respect to $x=e^{y}$ numerically.}
\end{enumerate} 
The procedures 1--4 can be performed in a polynomial 
time with respect to the number of spins, which is the 
same as the ladder case. 

On the other hand, 
for CTs the number of spins and the degree of the polynomial 
$\Xi_{L}$ increase exponentially as $O\left (((k-1)(c-1))^L \right )$ 
as $L$ grows, 
which makes it infeasible to solve $\Xi_{L}=0$ for large $L$. 
For instance, it is computationally difficult
to evaluate RZs beyond $L =7$ for $(k,c)=(2,3)$ and $L=4$ for $(k,c)=(3,4)$
by use of today's computers of reasonable performance. 
This prevents us from accurately examining the convergence of RZs to the real 
axis in the limit $L \to \infty$ by means of numerical methods. 

However, the data of small $L$ still strongly indicate that 
the qualitative behavior of RZs
can be classified distinctly 
depending on whether certain bifurcations
 occur for the cavity field distribution in the 
limit of $L \to \infty$, which 
implies that the RZs of the CTs reflect 
the phase transitions with respect to $y$.
Unfortunately, we also clarify that the transitions 
are related to no RSB by some analytical discussions.
In the following subsections, we give detailed discussions to lead 
this conclusion presenting plots of RZs. 

\subsection{Plots of the replica zeros for CTs}\label{sec:plots}
We here present the results for CTs. 
\begin{figure}[t]
\begin{tabular}{cc}
\hspace{-5mm}
\begin{minipage}[t]{0.5\hsize}
\begin{center}
 \includegraphics[height=60mm,width=75mm]{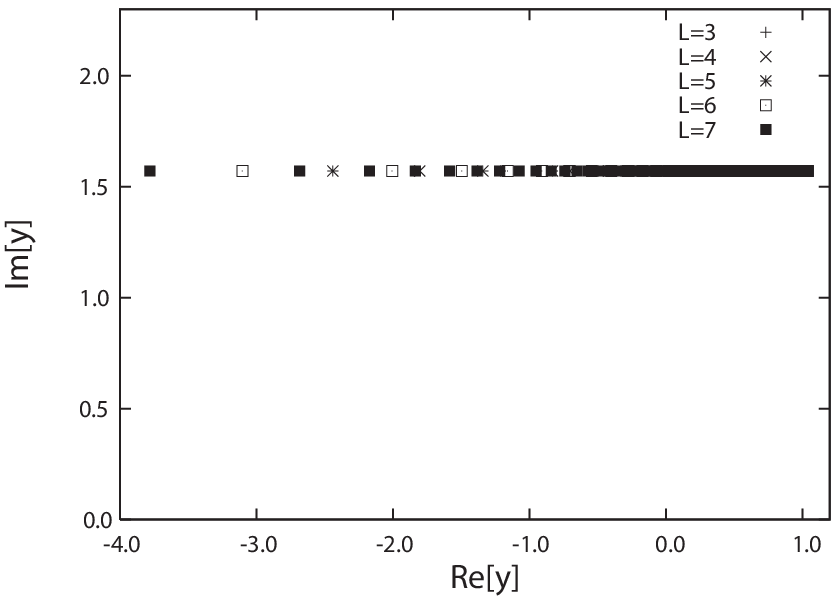}
 \caption{Plot of RZs for a
 CT with $c=3$. All the zeros 
lie on the line ${\rm Im} (y)=\pi/2$, as for a $2 \times L$ ladder. }
\label{fig:RBGzeros}
\end{center}
\end{minipage}
\hspace{2mm}
 \begin{minipage}[t]{0.5\hsize}
\begin{center}
\includegraphics[height=60mm,width=75mm]{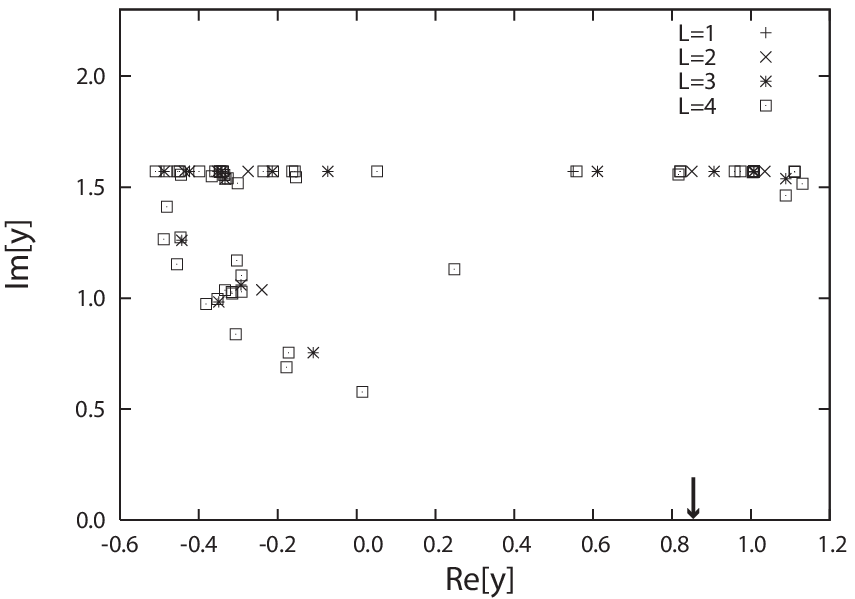}
\caption{RZs plot for a $3$-CT with $c=3$. 
A sequence of zeros approaches 
 the real axis as the number of generations $L$ increases. The arrow
 indicates the collision point expected from the study of the
 $L \to \infty$ limit in section \ref{sec:thermo}.}
\label{fig:3bRBGzeros}
\end{center}
\end{minipage}
\end{tabular}
\end{figure}
The plots for a CT and for a $3$-CT with $c=3$ are shown 
in figs.\ \ref{fig:RBGzeros} and \ref{fig:3bRBGzeros},
respectively. 
It is known that the FRSB and 1RSB transitions occur in RRGs 
with $(k,c)=(2,3)$ and $(k,c)=(3,3)$, respectively 
\cite{MontEPJ2003,MezaRevisit,MezaZero,Thou}.

Figure \ref{fig:RBGzeros} shows that 
RZs of the $c=3$ CT lie on a line ${\rm Im}(y)=\pi/2$.
Interestingly, this behavior is the same as 
the $2\times L$ ladder case in fig.\ \ref{fig:ladderzeros}.
This result indicates that 
 there is no phase transition or breaking 
of analyticity of $\phi(n)$ with respect to real $y$.
This strongly suggests 
that the RZs of CTs cannot detect the FRSB transitions 
in RRGs.

On the other hand, for the $3$-CT case in fig.\ 
\ref{fig:3bRBGzeros}, 
a sequence of RZs approaches 
a point $y_c$ on the real axis from the line 
${\rm Im}(y)=\pi/2$ as the number
of generations $L$ increases, although the value of $y_{c}$ is 
far from $y_{s}=\infty$ where the corresponding RRG occurs 
the 1RSB transition. 
The arrow in fig.\ \ref{fig:3bRBGzeros} represents the speculated 
value of $y_{c}$ from the analysis of the corresponding BL, the detail
of which is given in the next subsection.

\begin{figure}[t]
\begin{tabular}{cc}
\hspace{-5mm}
\begin{minipage}[t]{0.5\hsize}
\begin{center}
\includegraphics[height=60mm,width=75mm]{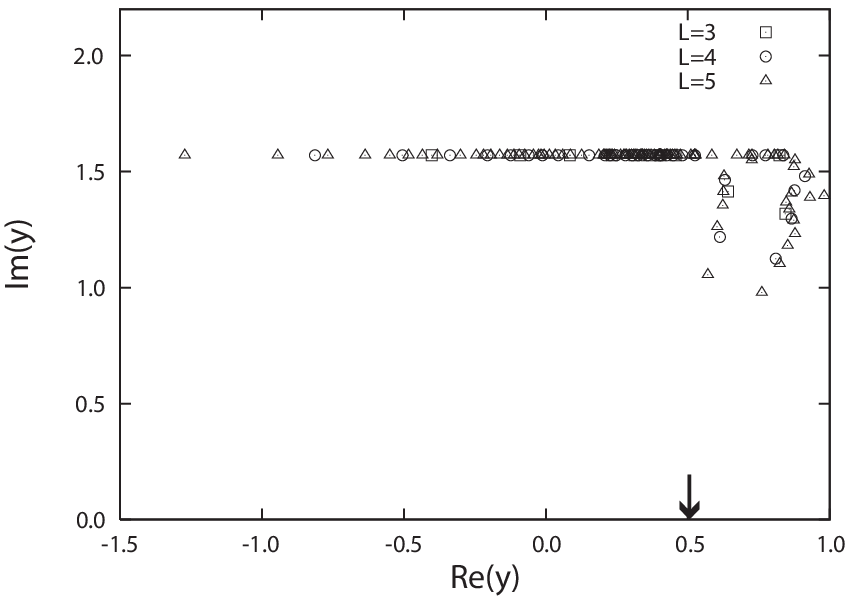}
 \caption{RZs of a CT with 
 $c=4$. We consider only an $L$-generation branch  in this case
 because of computational limits.
RZs approach the
 real axis as $L$ increases around $y_c \approx 0.5$. The arrow indicates the location of 
the singularity of the
 cavity-field distribution.}
 \label{fig:RBGzerosc4}
\end{center}
\end{minipage}
\hspace{2mm}
\begin{minipage}[t]{0.5\hsize}
\begin{center}
\includegraphics[height=60mm,width=75mm]{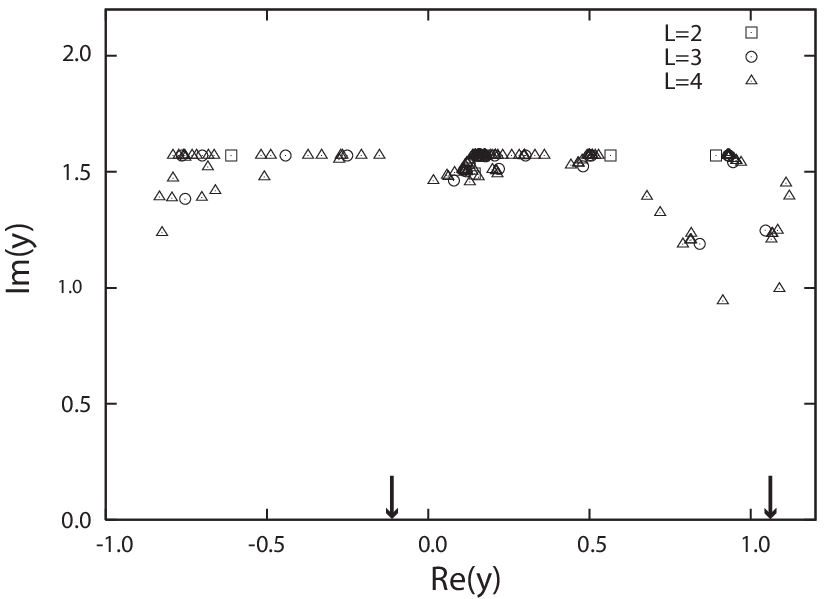}
   \caption{RZs of a $3$-CT with $c=4$. We consider only an $L$-generation branch. 
The zeros approach the real axis around $y_{c}\approx 1.1$. 
There are two singular points of the cavity field distribution in this
 case, both of which are indicated by arrows.}
   \label{fig:b3c4RBGzeros}
\end{center}
 \end{minipage}
\end{tabular}
\end{figure}

A similar tendency is also observed
for a CT and $3$-CT with $c=4$, plots of which are presented in
figs.\ \ref{fig:RBGzerosc4} and \ref{fig:b3c4RBGzeros}, respectively.
The 1RSB critical values are $y_{s}=0.38926$ for the CT
and $y_{s}=1.41152$ for the 3-CT. 
Again, these values are far from the values
of $y_{c}$, which can be observed 
in figs.\ \ref{fig:RBGzerosc4} and \ref{fig:b3c4RBGzeros}.  

These results 
indicate that certain phase transitions occur for some CTs,
although they are irrelevant to 1RSB and FRSB. 
It is difficult to identify the critical value $y_{c}$ 
from the plots because of the computational limits. 
Instead, 
in the following subsection we investigate the
$L \to \infty$ limit of these models. 
The arrows in figs.\ \ref{fig:3bRBGzeros}--\ref{fig:b3c4RBGzeros} 
represent the transition
points $y_{c}$ determined by this investigation.

\subsection{The thermodynamic limit and phase transitions of 
BLs}\label{sec:thermo}
In order to identify the value of $y_c$, 
we here consider the thermodynamic limit. 
Direct investigation of CTs in this limit is difficult, 
and we employ BLs to speculate the behavior of CTs.

\subsubsection{Transitions of the convergent cavity-field distribution}
\begin{figure}[t]
\hspace{-5mm}
\begin{tabular}{cc}
\begin{minipage}[t]{0.5\hsize}
\begin{center}
   \includegraphics[height=50mm,width=60mm]{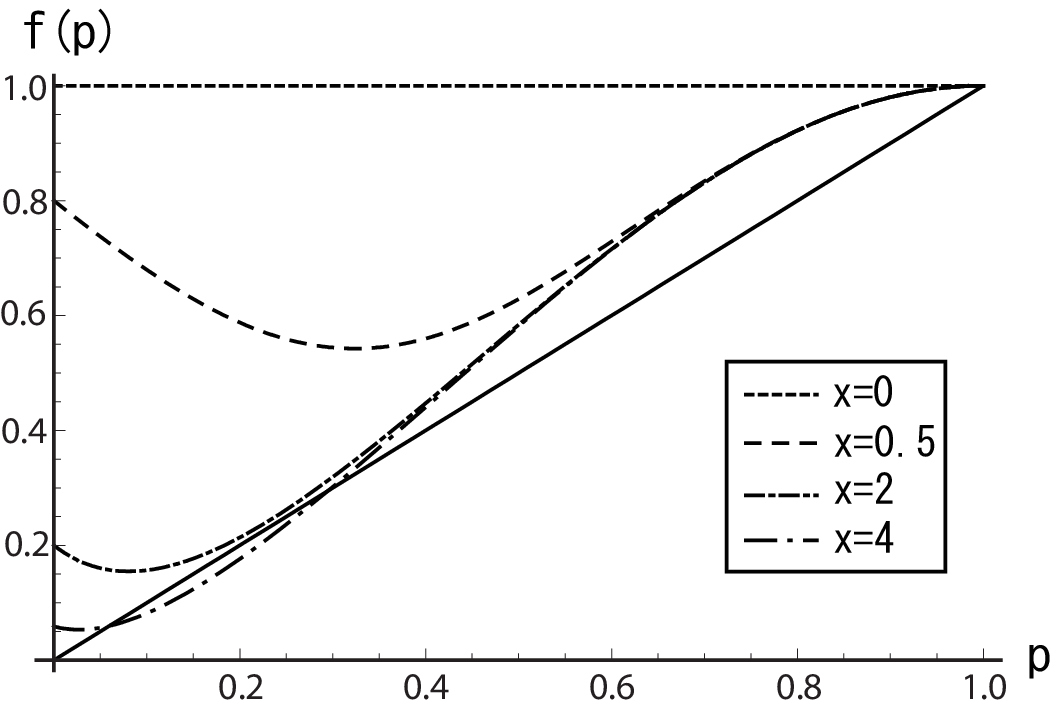}
 \caption{Return map of a $3$-CT with $c=3$. The
 convergent point of the recursion discontinuously changes depending on
 $x$. The solid line represents the function $f(p)=p$.}
 \label{fig:rmb3c3}
\end{center}
\end{minipage}
\hspace{2mm}
\begin{minipage}[t]{0.5\hsize}
\begin{center}
   \includegraphics[height=50mm,width=60mm]{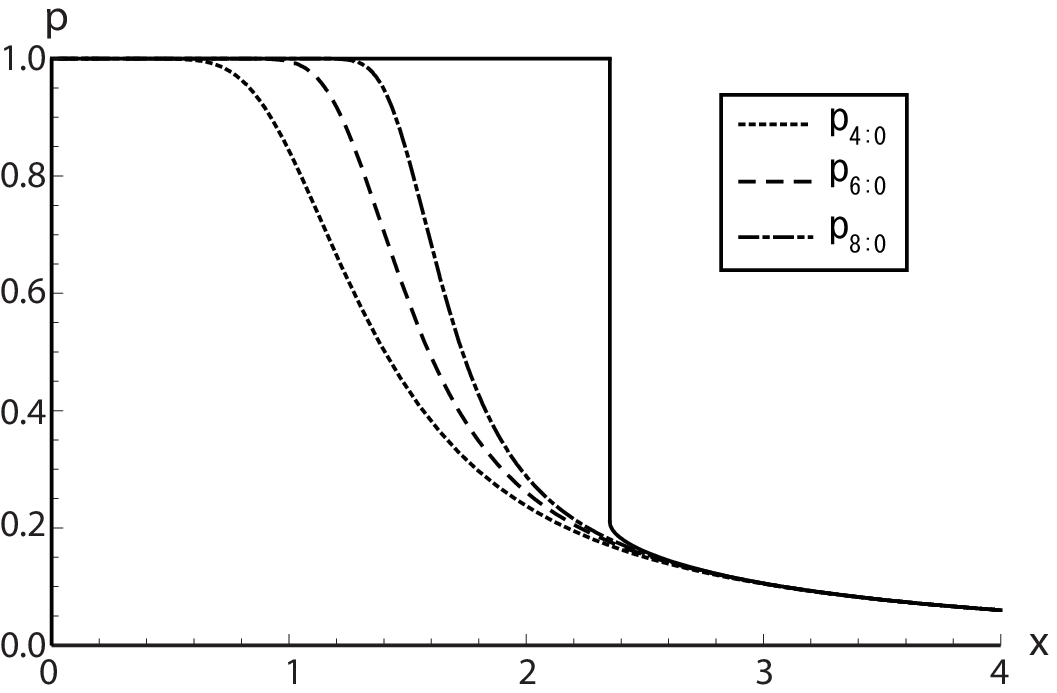}
 \caption{Asymptotic behavior of $p_{g;0}$ of a $3$-CT with
 $c=3$. A finite jump of $p$ occurs at $x\approx 2.35$. 
The solid line denotes the $\L \to \infty $ solution $p_{*}$.}
 \label{fig:apb3c3}
\end{center}
\end{minipage}
\end{tabular}
\end{figure}

\begin{figure}[htbp]
\begin{center}
    \includegraphics[height=50mm,width=70mm]{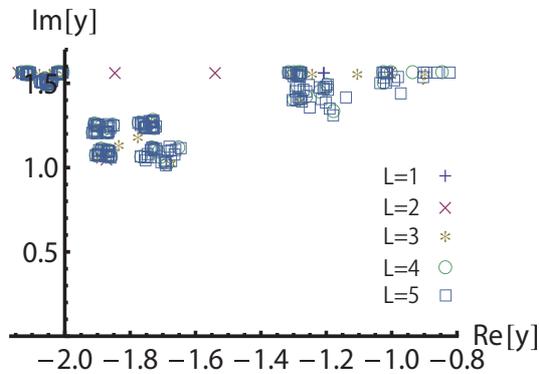}
 \caption{Plot of RZs of a $c=3$ $3$-CT with the initial condition 
$p_{0;0}=1/2$. The RZs do not approach the real axis even as $L$ grows.}
 \label{fig:k3RZp05}
\end{center}
\end{figure}

The first step is to take the $L\to \infty$ limit  
by equating 
$p_{g+1;0}$ and $p_{g;0}$ in the iterative equation of $p_{g;0}$, 
which
yields the boundary condition $p_*$ of the BL.  
For
a $c=3$ 3-CT, the iterative equation 
is given by
\begin{equation}
p_{g+1;0}=\frac{\left\{p_{g;0}^2+2p_{g;0}(1-p_{g;0})\right\}^2+\frac{1}{2}e^{-2y}(1-p_{g;0})^4}{1-\frac{1}{2}(1-p_{g;0})^4(1-e^{-2y})}.
\label{eq4:33CFD}
\end{equation} 
A return map of the recursion of $p_{g;0}$ and the convergent solution 
$p_{*}$ in the initial condition $p_{0;0}=0$ 
are presented in 
figs.\ \ref{fig:rmb3c3} and \ref{fig:apb3c3}, respectively.
The return map shows that 
there are three fixed points 
for $x \gsim 2.35$, while 
$p=1$ is the only fixed point for $x \llsim 2.35$. 
This situation is in contrast to the $c=3$ CT case, in which 
the cavity-field distribution uniformly 
converges to an analytic function: 
\begin{equation}
p_{*}=\frac{2+x^2-\sqrt{x^4+8x^2}}{2(1-x^2)} \label{eq4:pb2c3},
\end{equation}
which can be derived from eq.\ (\ref{eq4:p0precise}). 
This implies that when eq.\ (\ref{eq4:boundary_dist})
is put on the boundary of the CT, 
the boundary condition of the BL, which was obtained by 
an infinite number of recursions $L-L^\prime \to \infty$,
exhibits a discontinuous transition from $p_* < 1$ 
to $p_*=1$ at $x \approx 2.35 \Leftrightarrow y_c \approx 0.86$
as $y$ is reduced from the above.  
Actually, in fig.\ \ref{fig:3bRBGzeros}, 
RZs of the $c=3$ 3-CT seem to approach $y_c \approx 0.86$, 
marked by an arrow. 
This indicates that RZs obtained by our framework are 
relevant to the phase transition of the boundary of a BL, 
which is not related to 1RSB.    
This can be also confirmed from the initial condition dependence of the RZs 
behavior. Equation (\ref{eq4:33CFD}) 
implies that the fixed point
 $p=1$ is always stable, which can be seen from fig.\ \ref{fig:rmb3c3}, 
and a sufficiently 
large initial condition $p_{0;0}>0$ leads to the convergence to this fixed 
point. In that case, it is reasonable that RZs do not approach the real axis 
of $y$. The actual plot of RZs for a $c=3$ $3$-CT with the initial condition 
$p_{0;0}=1/2$ is given in fig.\ \ref{fig:k3RZp05}. 
This figure clearly shows that 
the RZs do not approach the real axis. 
It can be also observed that as the initial condition $p_{0;0}$ 
becomes close to $1$, 
RZs tend to go to the limit ${\rm Re}(y) \to -\infty$  
and ${\rm Im}(y)=\pi/2$, and become 
to never approach the real axis. 
These observations support 
the present description that the RZs of CTs 
correspond to the transition at the boundary of BLs. 

\begin{figure}[t]
\hspace{-5mm}
\begin{tabular}{cc}
\begin{minipage}[t]{0.5\hsize}
\begin{center}
   \includegraphics[height=50mm,width=60mm]{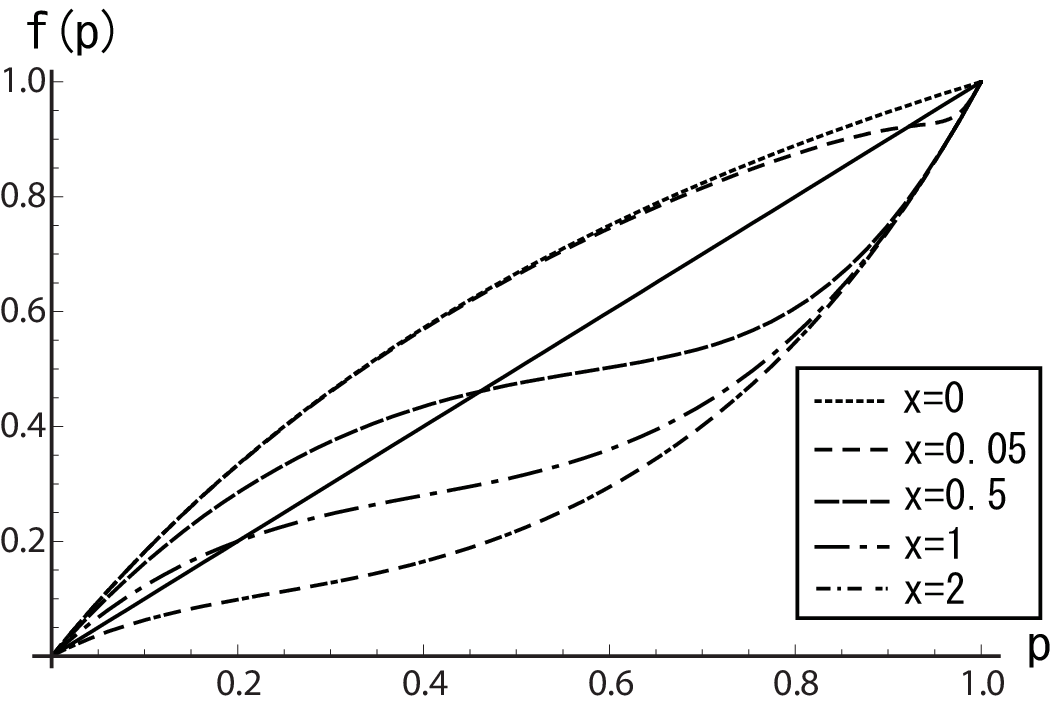}
 \caption{Return map of a CT with $c=4$. The
 stable fixed point is unique but shows a singularity at 
$x=e^{y} =\sqrt{3}$.}
 \label{fig:rmb2c4}
\end{center}
\end{minipage}
\hspace{2mm}
\begin{minipage}[t]{0.5\hsize}
\begin{center}
   \includegraphics[height=50mm,width=60mm]{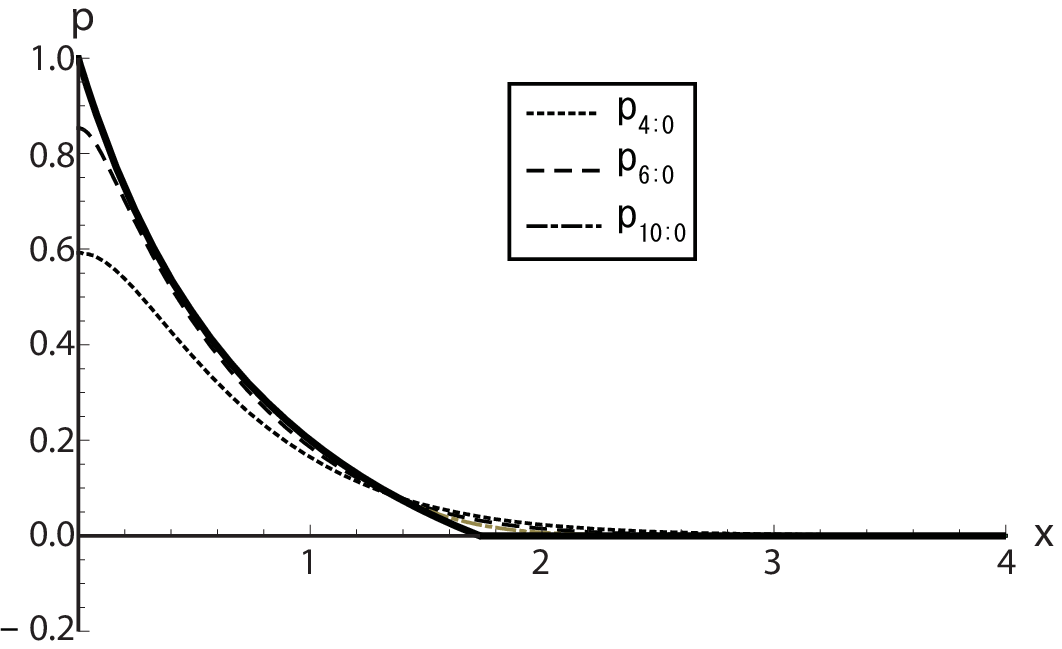}
 \caption{Asymptotic behavior of $p_{g;0}$ of a CT with $c=4$. In the thermodynamic limit, $p_{g;0}$ is continuous
 but the derivative becomes discontinuous at $x =\sqrt{3}$.}
 \label{fig:apb2c4}
\end{center}
\end{minipage}
\end{tabular}
\end{figure}

The same analysis for a $c=4$ CT shows that bifurcation 
of another type can occur for even $c$. 
For this model, 
the recursive equation of $p_{g;0}$ 
has a trivial solution $p_*=0$ for $\forall{x}$, 
which is always the case when $c-1$ is odd. 
The return map and plots of $p_*$ 
are shown in figs.\ \ref{fig:rmb2c4} and \ref{fig:apb2c4}, respectively. 
These figures indicate that  
there exists a continuous transition from $p_*=0$ to
$p_* >0$ at a certain value of $x$, 
which can be assessed as $x_{c}=\sqrt{3} \Rightarrow y_{c}\approx 0.5$.   
This is consistent with  
a certain sequence of RZs approaching 
the real axis around $y_c \approx 0.5$ in fig.\ \ref{fig:RBGzerosc4},
 which supports 
the analytical assessment of the critical points. 

In general, the discontinuous transition appears for 
cases of $k \ge 3$ spin interactions and 
the continuous transition occurs when $c$ is even. 
Actually, for a $c=4$ 3-CT, 
both discontinuous and continuous transitions
occur at 
$x \approx 0.86 \Rightarrow y \approx -0.15$ 
and $x=3 \Rightarrow y \approx 1.1$, respectively.
Figure \ref{fig:b3c4RBGzeros} shows 
a sequence of RZs approaching $y \approx 1.1$, 
while it is difficult to clearly identify a sequence converging to 
the other critical point $y \approx -0.15$. 
We consider that this is because the system size 
is not large enough, since a portion of the RZs in the left 
shows a tendency to approach the real axis,
though further increase of the system size 
is practically unfeasible due to the limitations of current 
computational resources. 

In conclusion, the RZs of CTs 
are related to the phase transitions on the boundary 
of the corresponding BLs. Regardless of the type of transition, 
a sequence of RZs approaches a critical point on the real 
axis when the BL provided from a CT 
in the limit $L \to \infty$ exhibits a phase transition 
on the boundary. 

\subsubsection{Relevance of the boundary condition and 
bulk properties of CTs}\label{sec4:BC} 
Equations (\ref{eq4:p0precise})--(\ref{eq4:T0Zh2}) imply that 
the generating function 
$\phi_N(y)=N^{-1} \log \Xi(y) $ for CTs
can also be separated into the bond and site contributions 
as eq.\ (\ref{eq4:ftree}) in the case of free energy 
\begin{eqnarray}
\phi_N(y)=\frac{1}{N} 
\sum_{\left \langle ij \right \rangle}
\phi_{\left \langle ij \right \rangle}^{(2)}(y) 
-\frac{1}{N}\sum_{i}(c_i-1) \phi_{i}^{(1)}(y)
. 
\label{eq4:decomposition}
\end{eqnarray}
As discussed in section \ref{sec4:bulk},
for a BL, the boundary condition given by 
the convergent solution of eq.\ (\ref{eq4:p0precise}), 
$p_*$, which becomes a function of $y$, 
simplifies the expression of 
eq.\ (\ref{eq4:decomposition}) as
\begin{eqnarray}
\phi_N^{\rm BL}(y)=r_{\rm I} \phi_{\rm I}(y)
+r_{\rm B}\phi_{\rm B}(y). 
\label{eq4:BLdecomposition}
\end{eqnarray}
and
\begin{eqnarray}
&&\phi_{\rm I}(y)=\frac{c}{2}\phi^{(2)}(y)-(c-1)\phi^{(1)}(y), 
\label{eq4:g2g1}
\\
&&\phi_{\rm B}(y)=\frac{1}{2}\phi^{(2)}(y), 
\label{eq4:g2}
\end{eqnarray}
where $\phi_{\rm I}(y)$ and $\phi_{\rm B}(y)$ represent 
contributions from a site  
inside the tree and on the boundary, respectively. 
In general, 
$\phi^{(2)}(y)$ and $\phi^{(1)}(y)$ are expressed as 
\begin{eqnarray}
&&\phi^{(2)}(y)=\log \left\{ \Tr{}
\left[\Wh{\rho}(\V{S}_{1})^{c-1}\Wh{\rho}(\V{S}_{2})^{c-1}
e^{\beta J\sum_{\mu}S_{1}^{\mu}S_{2}^{\mu}} 
\right]_{\V{J}}
\right\},\\
&&\phi^{(1)}(y)=\log \left\{ \Tr{}\Wh{\rho}(\V{S})^{c} \right\},
\end{eqnarray}
where
\begin{equation}
\Wh{\rho}( \V{S} )=\int d \Wh{h } \Wh{ \pi }( \Wh{h} )
\frac{ e^{\beta \Wh{h} \sum_{\mu} S_{\mu}  }   }
{ (2\cosh \beta \Wh{h})^n }
\end{equation}
and $\Wh{\pi}(\Wh{h})$ is the distribution of the cavity bias, which is
related to $\pi(h)$ as
\begin{equation}
\Wh{\pi}(\Wh{h})=\int dh \pi(h)\left[\delta\left(
\Wh{h}-\frac{1}{\beta}
\tanh^{-1}(\tanh \beta J \tanh \beta h) 
\right)\right]_{\V{J}}.
\end{equation}
These can be interpreted that cavity biases fluctuate due to the average 
over the quenched randomness.
These formulas are generalizations of eqs.\ (\ref{eq4:ftree})--(\ref{eq4:fi})  
to the generating function $\phi(n)$. 
For $c=3$ in the limit $\beta n \to y$, we have 
\begin{eqnarray}
&&\phi^{(2)}(y)
=\log e^{y} \left (1-\frac{1}{2}(1-e^{-2y})(1-p_*)^2 \right )^3,
\label{eq4:phi2}
\\
&&\phi^{(1)}(y)=\log \left (1-\frac{3}{4}(1-e^{-2y})(1-p_*)^2(1+p_*)\right )
\label{eq4:phi1}.
\end{eqnarray}
In general, $\phi_{\rm I}(y)$ agrees with $\phi(y)$ of an RRG, 
which is the same as the free energy case.

In spin-glass problems on cycle-free
graphs,
the replacement of $\phi_{N}^{\rm BL}(y)$ with 
$\phi_{\rm I}(y)$ is crucial. 
To see this, let us investigate the rate function 
$R(f)$ as in section \ref{sec2:LDT}.
We denote the boundary condition as 
$P_{\rm B}(\V{h})=\prod_{i \in {\rm boundary}} \pi_i(h_i)$. 
Since
CTs and BLs have no cycle, 
for the $\pm J$ model defined on those lattices
the boundary condition is  
the only relevant factor determining the 
property of the model.
Combining this fact with eq.\ (\ref{eq4:RZs3}), 
we can express the moment $\Xi(y)$ of CTs and BLs as 
\begin{equation} 
\Xi(y)=\int d\V{h}P_{\rm B}(\V{h})\exp\left (-y E_g(\V{h}) \right ),
\end{equation}
where $E_g(\V{h})$ is 
the ground state energy 
when $\V{h}$ is imposed on the boundary. 
This yields the following equation 
with respect to the rate function for finite $N$, $R_{N}(y)$, 
\begin{equation}
-R_{N}(y)=y^2 \Part{}{y}{}\left (\frac{\phi_N(y)}{y} \right )
=
\frac{1}{N}\int dh 
\widetilde{P}_{\rm B}(\V{h})
\log \frac{\widetilde{P}_{\rm B}(\V{h})}{P_{\rm B}(\V{h})}
\equiv
\frac{1}{N} D(\widetilde{P}_{\rm B} |P_{\rm B}) \ge 0,\label{eq4:KL}
\end{equation} 
where we define the probability 
distribution $\widetilde{P}_{\rm B}(\V{h})$ 
as 
$\widetilde{P}_{\rm B}(\V{h}) \equiv P_{\rm B}(\V{h})
\exp\left (-y E_g(\V{h}) \right )/\Xi(y)$ 
and introduce the so-called Kullback--Leibler (KL) 
divergence
$D(\widetilde{P}_{\rm B} |P_{\rm B})$ 
between $\widetilde{P}_{\rm B}(\V{h})$ 
and ${P}_{\rm B}(\V{h})$, which takes a positive value unless 
$\widetilde{P}_{\rm B}(\V{h})$ becomes identical to 
$P_{\rm B}(\V{h})$. 
This equation 
guarantees the non-positivity of the rate function  
$R_{N}(y)$.
The constraint $R_{N}(f)\leq 0$ is always satisfied 
even when $N \to \infty$. However, as discussed in section 
\ref{sec2:LDT},
this is not necessarily the case 
when we first take the thermodynamic limit 
$\lim_{N \rightarrow \infty}\phi_{N}(y)=\phi(y)$ and 
then
calculate the rate function by using the analytically continued 
$\phi(y)$ as 
$R(y)=-y^2 (\partial /\partial y)\left (y^{-1} \phi(y) \right )$.
This function $R(y)$ can be positive, and    
 the condition
$R(y_{s})=0$ signals the onset of $1$RSB as shown in section 
\ref{sec2:LDT}. 

The condition $R(y_{s})=0$ has already been investigated
for RRGs and indicates that 1RSB transitions occur for some types
of RRGs \cite{MontEPJ2003,MezaZero}. However, 
it is considered that such a symmetry breaking cannot be detected by 
an investigation based on eqs.\
(\ref{eq4:p0precise})--(\ref{eq4:T0Zh2})
because 
the boundary contribution is inevitably
taken into account for a BL as well as for a CT. 

\begin{figure}[htbp]
\hspace{-5mm}
\begin{tabular}{cc}
\begin{minipage}[t]{0.5\hsize}
\begin{center}
\includegraphics[height=50mm,width=70mm]{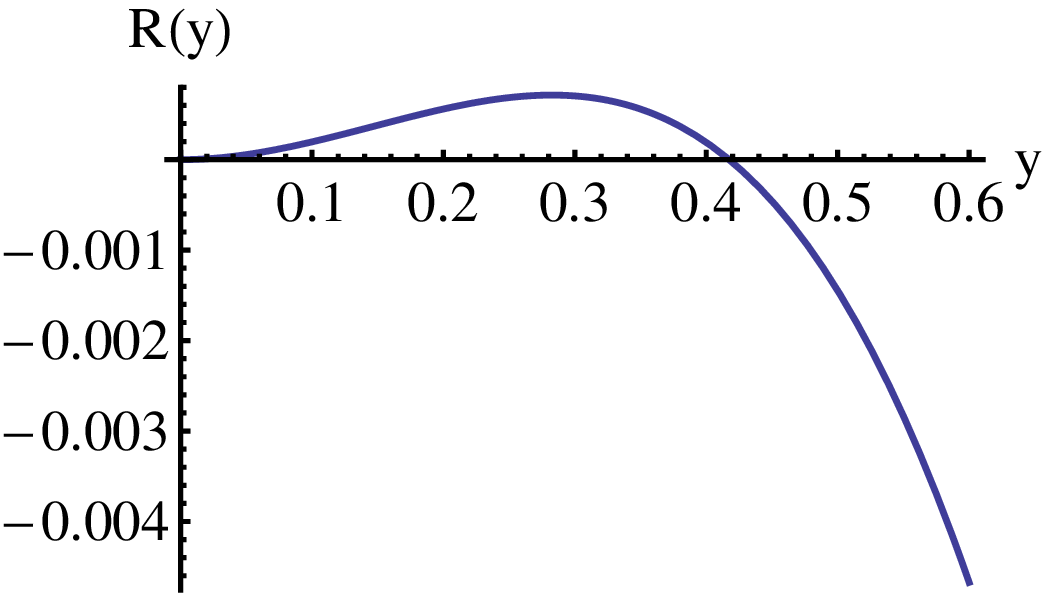}
 \caption{The rate function of a $c=3$ RRG. The 1RSB critical value
is  $y_{s}=0.41741$.}
 \label{fig:rateRRG}
\end{center}
\end{minipage}
\hspace{2mm}
\begin{minipage}[t]{0.5\hsize}
\begin{center}
   \includegraphics[height=50mm,width=60mm]{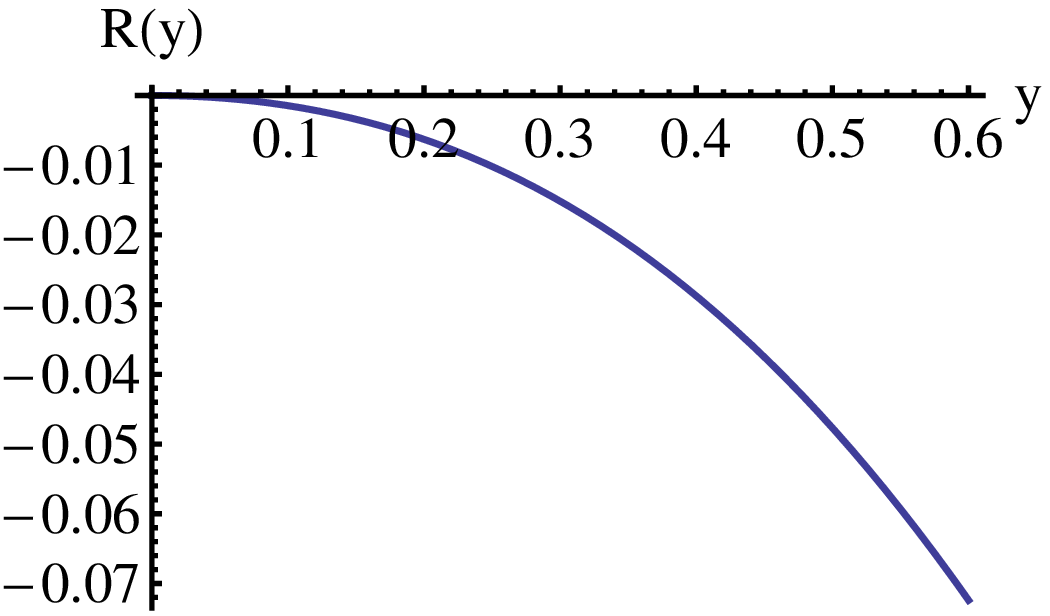}
 \caption{The rate function of a $c=3$ BL in the limit $L\to \infty$. 
The constraint $R(f)\leq 0$ is completely held.}
 \label{fig:rateBL}
\end{center}
\end{minipage}
\end{tabular}
\end{figure}

An example will clearly illustrate this point. 
Let us calculate the generating function of the $c=3$ BL and RRG. 
The generating function of the RRG (assessed only in the 
thermodynamic limit) is given by 
substituting eqs.\ (\ref{eq4:p0precise}), (\ref{eq4:phi2}), 
and (\ref{eq4:phi1}) into $\phi_{\rm I}(y)$, and the 
plot is given in fig.\ \ref{fig:rateRRG}.
This shows a 1RSB critical value $R(y_{s})=0$ at $y_{s}=0.41741$ and 
demonstrates the failure of the analytic continuation of $\phi(y)$. 
On the other hand,
the generating function of the BL for finite $N$ is given by
eq.\ (\ref{eq4:BLdecomposition}) with the conditions 
 (\ref{eq4:p0precise}), (\ref{eq4:phi2}), and (\ref{eq4:phi1}).
Clearly, the $N$-dependence of $\phi^{BL}_{N}(y)$ only appears in 
two factors $r_{\rm I}=\left (1+c(c-2)^{-1}
\left ((c-1)^{L^\prime-1}-1 \right ) \right )/
\left (1+c(c-2)^{-1}
\left ((c-1)^{L^\prime}-1 \right ) \right )$ and $r_{\rm B}=1-r_{\rm I}$
through the total generation $L$. 
These factors are analytic with respect to $L$ even in the limit 
$L\to \infty$, which guarantees 
the uniform convergence of $\phi^{\rm BL}_{N}(y)$ to 
$\lim_{L\to \infty} \phi^{\rm BL}_{L}(y)=\phi^{\rm BL}(y)$. 
Hence, the  
analytic continuation of $\phi(y)$ is successfully performed 
and the resultant rate function $R(f)$ 
never violates $R(f)\leq 0$, which can be actually seen in 
fig.\ \ref{fig:rateBL}.
Also, we can prove the absence of the 1RSB transition for CTs 
by showing the uniform convergence of $\phi_{N}(y)$. 
We succeeded to prove this for a $c=3$ CT and the details are 
shown in appendix \ref{app:rate}.

In summary, the RZs of CTs provided by the 
current scheme are irrelevant to the 1RSB 
because it does not appear 
in BLs and CTs due to the huge boundary effect. 
However, this does not mean the absence of another RSG, FRSB, in 
BLs and CTs. Next we examine this point.

\subsection{Possibility of the full-step replica symmetry breaking}
\label{sec4:AT}
The AT
condition, which is critical 
for the FRSB, 
has not yet been characterized
for sparsely connected spin models \cite{Thou,MezaPhysique1985,PariEPJB2001}. 
In fact, a previous research has found that 
critical values of the continuous transitions
from $p_*=0$ to $p_* > 0$ 
are candidates for those of the AT condition for systems 
of even $c$ \cite{MontEPJ2003}. 
This motivates us to further 
explore a possible link between RZs and the AT instability. 

As discussed in section \ref{sec2:AT}, 
the AT condition is identical to 
the divergence of the spin-glass susceptibility $\chi_{SG}$ 
in the 
fully-connected $p$-spin interacting model. 
Hence, it is reasonable to adopt the divergence of $\chi_{SG}$ 
 as the critical condition of 
the FRSB for BLs and RRGs \cite{Rivo,Krza,Mart}. 
Using the (statistical) uniformness of BLs and RRGs, 
it is sufficient 
to see the susceptibility of only the root site $0$. 
The spin-glass susceptibility of BLs 
for finite replica $n$ is generally written as
\begin{equation}
\chi_{SG} =\sum_{i}\left [
\left(\Part{\Ave{S_{0}}}{h_{i}}{} \right)^2 \right ]_n.
\label{eq4:chiSG}
\end{equation} 
where $\left [ (\cdots )\right ]_n$ means 
an average with respect to a modified distribution of coupling and 
boundary field 
\begin{equation}
P_n(  \{J_{ij}\},\{h_{i}\}  )=
\frac{
P(  \{J_{ij}\},\{h_{i}\}   )Z^n(\{J_{ij}\},\{h_{i}\} )
}
{
\sum_{\{J_{ij}\}}
P(\{J_{ij}\},\{h_{i}\})Z^n(\{J_{ij}\},\{h_{i}\} )
}. 
\end{equation}

In a cycle-free graph, an arbitrary pair of nodes is connected 
by a single path. Let us assign node indices from the origin 
of the graph $0$ to a node of distance $G$ along the path 
as $g=1,2,\ldots,G$. 
For a fixed set of couplings and boundary 
fields, the chain rule of the derivative operation 
indicates that
\begin{eqnarray}
\Part{\Ave{S_{0}}}{h_{G}}{}&=&\Part{\Ave{S_{0}}}{h_{0}}{}
\Part{h_0}{\Wh{h}_0}{}\Part{\Wh{h}_0}{{h}_1}{}
\cdots \Part{h_G}{\Wh{h}_{G}}{}
=\Part{\Ave{S_{0}}}{h_{0}}{}
\Part{h_0}{\Wh{h}_0}{} \prod_{g=1}^G
\Part{\Wh{h}_{g-1}}{h_{g}}{}
\Part{{h}_g}{\Wh{h}_g}{} \cr
&=& \Part{\Ave{S_{0}}}{\Wh{h}_0}{}
\prod_{g=1}^G
\Part{\Wh{h}_{g-1}}{\Wh{h}_g}{},
\label{chainrule}
\end{eqnarray}
as $h_g$ depends linearly on $\Wh{h}_g$ as 
$h_g=\Wh{h}_g+r_g$, where $r_g$ represents a sum of 
the cavity biases from other branches that flow into
node $g$. For a BL of $(k,c)=(2,3)$, the cavity equation yields 
an evolution equation of the cavity bias
\begin{eqnarray}
&&\Wh{h}_{g-1}=\frac{1}{\beta} 
\tanh^{-1}\left (\tanh(\beta J_g) \tanh(\beta (\Wh{h}_g+r_g))\right )\cr 
&&\to  \left \{
\begin{array}{ll}
{\rm sgn}\left (J_g (\Wh{h}_g+r_g) \right ) &(\,\, |\Wh{h}_g+r_g| \ge 1 \,\,) \cr
J_g (\Wh{h}_g+r_g)  &(\,\, \mbox{otherwise} \,\,) 
\end{array} \right .
(\,\,\beta\to \infty\,\,), 
\label{BPzerotemp}
\end{eqnarray}
where $J_g$ denotes the coupling between nodes $g-1$ and $g$, 
and similarly for other cases. 

To assess eq.\ (\ref{eq4:chiSG}), we take an average of 
the square of eq.\ (\ref{chainrule}) with respect to the 
modified distribution $P_n(\{J_{ij}\},\{h_{i}\})$. 
Here, $r_g$ can be regarded as a sample of a
stationary distribution determined by the convergent 
solution of eq.\ (\ref{eq4:p0precise}) for the BL. 
As $r_g$ is limited to being an integer and $|J_g|=1$, 
eq.\ (\ref{BPzerotemp}) gives 
\begin{eqnarray}
\left |
\frac{\partial \Wh{h}_{g-1}}{
\partial \Wh{h}_g}
\right |=
\left \{
\begin{array}{ll}
0 & (\,\, |\Wh{h}_{g}+r_g|> 1 \,\,)\\
0 \mbox{ or $1$} & (\,\, |\Wh{h}_{g}+r_g|= 1\,\,) \\
1 & (\,\, {\rm otherwise}\,\,)
\end{array}
\right.,  \label{eq4:transition}
\end{eqnarray}
where the value of $0$ or $1$ for the case of $|\Wh{h}_{g}+r_g|= 1$
is determined depending on the value of $\Wh{h}_{g}$.
When $\Wh{h}_{g}$ eqauls $0$ (and $|r_g|=1$ ), 
the values $0$
and $1$ are chosen with equal probability $1/2$ since
the sign of the infinitesimal fluctuation of $\Wh{h}_{g}$, 
$\delta \Wh{h}_{g}$, 
is determined in an unbiased manner due to the mirror symmetry 
of the distribution of couplings. 
On the other hand, under the condition of $\prod_{k=g+1}^{G} \left |
\partial \Wh{h}_{k-1}/\partial \Wh{h}_{k} \right | \ne 0 $, 
the case of $|\Wh{h}_g|=1$ (and $r_g=0$) always yields $\left |
\partial \Wh{h}_{g-1}/\partial \Wh{h}_g \right |=1$. 
This is because $\Wh{h}_g \delta \Wh{h}_g <0$ is 
guaranteed for $|\Wh{h}_g|=1$ under this condition. 

Equation (\ref{eq4:transition}) indicates that 
the assessment of eq.\ (\ref{chainrule}) 
is analogous to an analysis of a random-walk 
which is bounded by absorbing walls. 
We denote by $P_{(G \to 0)}$ 
the probability that 
$\left |\partial \Wh{h}_{g-1}/\partial \Wh{h}_g\right |$ 
never vanishes during the walk from $G$ to $0$ and 
the value of 
$\prod_{g=1}^G
|\partial \Wh{h}_{g-1}/\partial \Wh{h}_g|$ is kept to unity. 
This indicates that 
\begin{eqnarray}
\left [\left (\Part{\Ave{S_{0}}}{h_{G}}{} \right )^2 \right ]_n
\propto 
P_{(G \to 0)}
\label{survingprob}
\end{eqnarray}
holds. Summing all contributions up to the boundary of the BL
yields the expression
\begin{equation}
\chi_{SG}  \propto 
\sum_{G=0}^{L^\prime} (k-1)^{G} (c-1)^{G} P_{(G \rightarrow 0)}. 
\label{expression_of_chiSG}
\end{equation}
The critical condition for convergence of eq.\ (\ref{expression_of_chiSG})
in the limit $L^\prime \to \infty$ is
\begin{eqnarray}
\log \left ((k-1)(c-1) \right )+\lim_{G \to \infty} \frac{1}{G} 
\log P_{(G \rightarrow 0)} =0.
\label{eq4:AT}
\end{eqnarray}
This serves as the ``AT'' condition in the current framework. 

For a BL, eq.\ (\ref{eq4:AT}) can be assessed by analyzing the 
random walk problem of eq.\ (\ref{eq4:transition}), as
shown in appendix \ref{app:AT^BL}. 
We evaluated the critical $y_{AT}$ values of eq.\ (\ref{eq4:AT}) 
for several pairs of $(k,c)$, 
shown in Table \ref{tab:yvalue} along with other critical values. 
\begin{table}[hbt]
    \begin{center}
\begin{tabular}{c|c|c|c}
\noalign{\hrule height 0.8pt}
$(k,c)$ & $y_{AT}$ & $y_{c}$ & $y_{s}$  \\
\hline
$(2,3)$ & 0.54397 & none & 0.41741 \\
\hline
$(2,4)$ & 0.54931 & $\log \sqrt{3}\approx 0.54931$ & $0.38926$  \\
\hline
$(3,3)$ & 1.51641  & 0.85545 & $\infty$ \\
\hline
$(3,4)$ & 1.09861  & $-$0.15082, $\log 3\approx 1.09861$ & 1.41152 \\
\noalign{\hrule height 0.8pt}
\end{tabular}
\end{center}
\caption{Relevant values of $y$. Note that each kind of $y$ is
 calculated using different models. 
The 1RSB transition point $y_{s}$ is for
 RRGs and $y_{AT}$ is for RRGs or BLs. The
 singularity of the cavity-field distribution $y_{c}$ is common for all the models.
Note that 
the value of $y_{AT}$ in the case $(k,c)=(3,4)$ is the same as 
the one in \cite{MontEPJ2003}. 
} \label{tab:yvalue}
\end{table}
These results show that the values of $y_c$, 
which signal the phase transitions of the boundary condition 
of the BL, agree with neither $y_{AT}$ or $y_{s}$ for $c=3$ 
implying irrelevance of RZs of CTs 
to the replica symmetry breaking, 
while for $c=4$ the values $y_{AT}$ and $y_{c}$ agree each other.
We consider that the correspondence for the $c=4$ cases is accidental 
since the case $(k,c)=(2,3)$ offers strong evidence that 
the AT instability cannot be detected by the current RZs.
This point may require more detailed discussions but we 
here leave it as a future work.

To investigate the irrelevance of RZs to the AT instability further, 
let us examine the singularities of RZs of a $(k,c)=(2,3)$ BL. 
The moment of the BL for finite generation $L'$, $\Xi_{L'}(y)$, 
is easily derived from eqs.\ 
(\ref{eq4:T0Zh1}), (\ref{eq4:T0Zh2}) and (\ref{eq4:pb2c3}), 
and the result is
\begin{equation}
\Xi_{L'}(y)=\Xi_{0}^{3\cdot 2^{L'-1}}
\left(
x^2-\frac{1}{2}(x^2-1)(1-p_{*}) 
\right)^{3\cdot (2^{L'-1}-1)}
\left(
x^3+\frac{3}{4}(1-x^2)(1-p_{*})^{2} (1+p_{*})
\right),
\end{equation}
where $\Xi_{0}=p_{*}+x p^{1}_{*}+x^2 p^{2}_{*}$ is the 
moment of a boundary spin, and the 
quantities $p^{1}_{*}$ and $p^{2}_{*}$ represent 
the probabilities that the convergent cavity field 
takes $\pm 1$ and 
$\pm 2$, respectively. 
The explicit forms of $p^{2}_{*}$ and $p^{1}_{*}$ 
are obtained from eq.\ (\ref{eq4:p^CT})
\begin{equation}
(p^{2}_{*},p^{1}_{*})=
\frac{2}{ 2-(1-p_{*})^{2} ( 1-x^{-2} ) }
\left(
\left(
\frac{ 1-p_{*} }{2} 
\right)^{2}
, 
p_{*}(1-p_{*})
\right).
\end{equation}
The RZs of the BL is easily derived as
\begin{equation}
\Xi_{L}(y)=0 \Rightarrow y= \pm \frac{\pi}{2}i,\,\,\, 
0.369207 \pm 1.02419 i,\,\,\, 0.23664 \pm 1.80591 i.
\end{equation}
This result implies 
that RZs of BLs are also irrelevant to the FRSB, because there is no RZ
on the real axis of $y$.

The irrelevance of RZs to the AT instability may be interpreted as follows. 
Remember the discussions given in section \ref{sec2:AT}. 
We can link the spin-glass susceptibility 
to $\phi_{N}(n)$ by considering the 
modified generating function 
\begin{eqnarray}  
&&N \Wt{\phi}_{N}(n,m, \V{F}) =\cr
&&\log \left[\left( \Tr{} \exp\left( -\beta\sum_{a=1}^{m}H({S^{a}})
+\sum_{l=1}^{N} F_{l}\sum_{a<b}S_{l}^{a}S_{l}^{b}  
\right)\right)\left(\Tr{} e^{-\beta H}\right)^{n-m} \right]_{n}, 
\label{eq4:phitilda}
\end{eqnarray}
in which the interaction
$\V{F}=(F_1,F_2,\ldots,F_N)$ are introduced 
among
$m$ out of $n$ replica systems.  
Analytically continuing eq.\ (\ref{eq4:phitilda})
to $n,m \in \mR$ and 
expanding the obtained expression around $\V{F} = 0$ for
$m \simeq 1$ yield
\begin{equation}
N\Wt{\phi}_{N}( \V{F};m,n)\approx N \phi_{N}(n)+
\frac{m-1}{2}q\sum_{i}F_{i}
+
\frac{m-1}{2}\V{F}^{T}\Wh{\chi}_{SG}\V{F}
+({\rm  higher\,\, orders}),
\label{eq4:phitildaexpansion}
\end{equation}
where $\Wh{\chi}_{SG}=\left (
\left [\left(\Ave{S_{l}S_{k}}-\Ave{S_{l}} \Ave{S_{k}}
\right)^2 \right ]_{n} \right )$ represents 
the spin-glass susceptibility matrix. 
Equation (\ref{eq4:phitildaexpansion}) implies that 
the divergence of the spin-glass susceptibility is linked to 
analytical singularities of $\lim_{N \to \infty} \Wt{\phi}_{N}(n,m,\V{F})$ 
for $m \ne 1$. However, for $m=1$, which corresponds to $\phi_N(n)$
examined in this chapter, it is 
difficult to detect the singularity because 
the factor $m-1$ with $\V{F}^{T}\Wh{\chi}_{SG}\V{F}$ 
makes the divergence of the spin-glass susceptibility 
irrelevant to the analyticity breaking of 
$\phi(n)=\lim_{N \to \infty}\phi_N(n)$.
This explains 
the irrelevance of the RZs to the AT instability 
in the current framework.
Considering systems of $m \ne 1$ may enable us to detect the 
AT instability by the RZs formula,
but such an investigation is beyond the current purpose and 
we leave it as a future work.

\subsection{Physical implications of the obtained solutions}
We concluded that bifurcations of the fixed point solutions
of the cavity-field equation correspond to phase transitions
of the boundary condition of a BL and are not relevant to either 1RSB or FRSB. 
Before closing this section, we discuss the physical 
implications of the obtained solutions.

A naive consideration finds that 
the solution of $p_{\infty;0}=p_{*}=1$ corresponds to a paramagnetic phase 
implying that any cavity fields vanish and therefore
all spin configurations are equally generated. 
Note that this phase is of the ground states 
in the limit $\beta \rightarrow \infty$ and is different from the usual
temperature-induced phase.
 
For finite $p_{*}<1$, relevant fractions of the spins can take any direction
without energy cost because the cavity field on the site is $0$. 
This implies that the ground state energy is highly degenerate, 
which means that this solution describes a RS spin-glass phase. 
Actually, it is easy to confirm that the following equality holds:
\begin{eqnarray}
q_{\mu \nu}=\frac{ [\Tr{} S^{\mu}_{g} S^{\nu}_{g} e^{-\beta \sum_{\mu} H^{\mu }}]_{\V{J}} }{[Z^n]_{\V{J}}}=\Tr{} S^{\mu}_{g}S^{\nu}_{g}\rho_{g}(\V{S})=1-p_{g;0}.
\end{eqnarray}  
Hence, the singularity of the cavity-field distribution in the limit 
$g\rightarrow \infty$ can be regarded as the transition of the spin-glass
order-parameter. A finite jump of $p_{*}$ for the $k=3$ case is
the first-order transition from the RS spin glass to paramagnetic phases, and such
a transition is also observed in the mean-field models.  
The transitions from $p_{*}=0$ to finite-$p_{*}$ 
for the $c=4$
case can be regarded as a saturation of $q$ to $q_{EA}=1$. 
We infer that these are the transitions from RS to RS phases. 
Notice that such a transition has not been observed for 
infinite-range models.
Our results indicate that this $q=1$ phase appears 
only when $c$ is even.
This means that such a phase is highly sensitive to
the geometry of the objective lattice. This may be 
a reason why such a transition has not been observed in other models.

\section{RZs for RRGs}
In section \ref{sec4:AT}, 
we have investigated the AT instability for BLs and RRGs 
and concluded that the instability cannot be detected by RZs 
of CTs and BLs. We have also inferred that 
the AT instability is difficult to observe in the systems 
without interactions among replicas. 
To examine this proposition,
 we present RZs for $c=3$ RRGs in this section.

Constructing the RZ equation $[Z^n]_{\V{J}}=0$ for a RRG is a
rather involved work because the average $[(\cdots)]_{\V{J}}$ 
requires not only 
the calculations of configurations of $J_{ij}=\pm 1$ but also 
the geometrical structure of the graph.
To overcome this difficulty,  
we fix a geometrical structure and 
only treat the configurations of the bond value $J_{ij}=\pm 1$ 
when constructing the RZ equation $[Z^n]_{\V{J}}=0$.
This prescription is not faithful to the definition of the RRG but 
we can expect that gaps coming from the difference of the definitions 
vanish in the thermodynamic limit. 

Actual procedures we performed for calculating the RZs of RRGs 
are rather simple:  
\begin{enumerate}
\item{Generate a $c=3$ RRG with a size $N$.} 
\item{
Set a bond configuration $\V{J}$
and calculate the ground-state energy by exploring all 
the spin configurations. 
} 
\item{
Continue the procedure $2$ until all the bond configurations are completed 
and 
make the distribution of the ground-state energy $P(E_{g})$.
}
\item{
Construct the RZ equation $\sum_{E_{g}}P(E_{g})e^{-y E_g}=0$ and solve it.
}
\item{Generate other RRGs with the same size $N$ and compare the results 
(actually we generate $16$ samples for a size).}
\end{enumerate} 
Clearly, these procedures require huge computational times, 
which makes 
it infeasible to calculate for $N >16$. 

\begin{figure}[t]
\hspace{-5mm}
\begin{tabular}{cc}
\begin{minipage}[t]{0.5\hsize}
\begin{center}
\includegraphics[height=50mm,width=60mm]{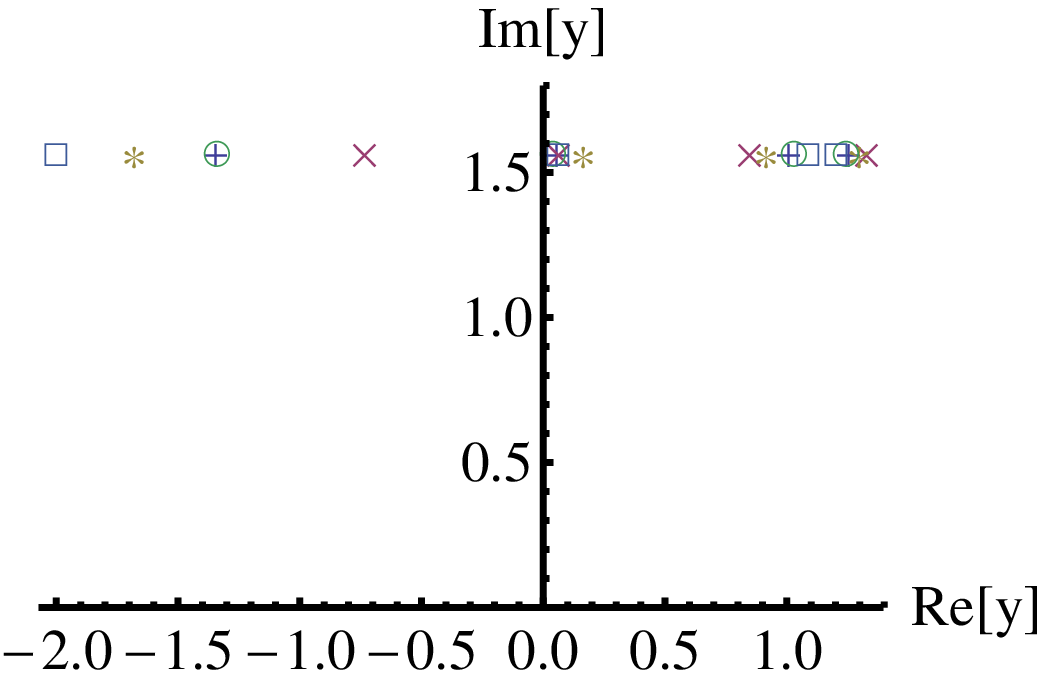}
 \caption{Plot of RZs for five samples of $c=3$ RRG of the size $N=14$. 
All of the RZs lie on the line ${\rm Im}(y)=i\pi/2$.}
 \label{fig:RZ^RRGN14}
\end{center}
\end{minipage}
\hspace{2mm}
\begin{minipage}[t]{0.5\hsize}
\begin{center}
   \includegraphics[height=50mm,width=60mm]{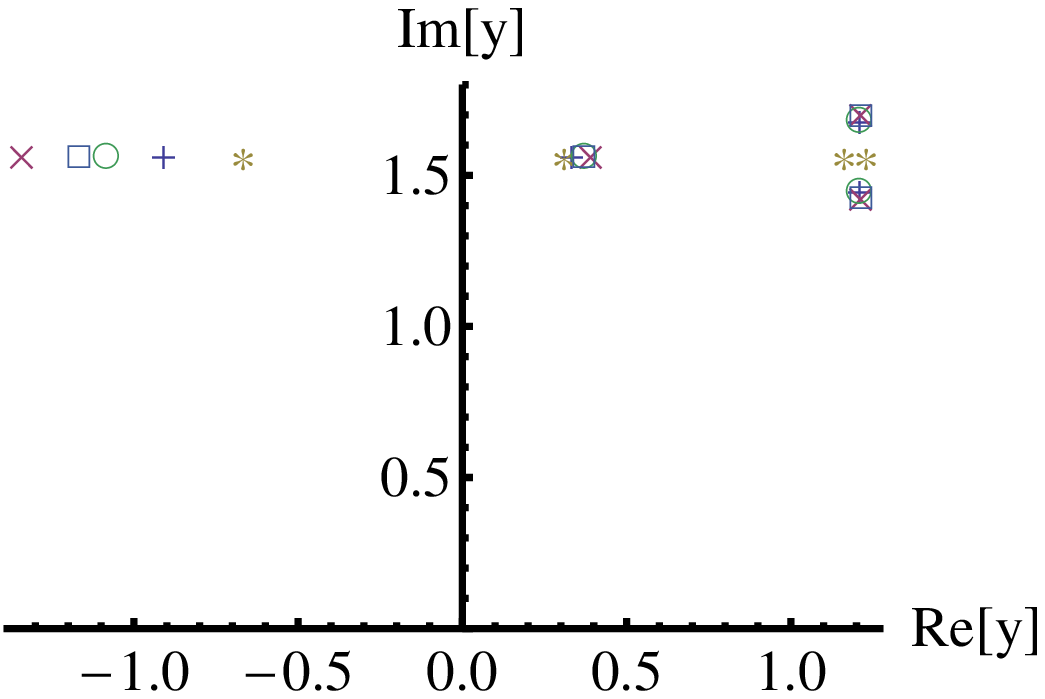}
 \caption{Plot of RZs for five samples of $c=3$ RRG of the size $N=16$. 
For some samples, some sequences of the zeros exist apart from
 the line ${\rm Im}(y)=i\pi/2$.}
 \label{fig:RZ^RRGN16}
\end{center}
\end{minipage}
\end{tabular}
\end{figure}

Some typical behavior of RZs of RRGs for the sizes $N=14$ and $16$ 
are given in figs.\ \ref{fig:RZ^RRGN14} and \ref{fig:RZ^RRGN16}, 
respectively.
We can see that the RZs lie on the line ${\rm Im}(y)=i\pi/2$ for 
$N=14$. 
Actually for $N\leq 14$,   
we could not find any samples which show 
RZs approaching to the real axis.
For the $N=16$ case,
some sequences of the zeros exist apart from the line ${\rm Im}(y)=i\pi/2$.
This result implies that RZs of RRGs
approach the real axis as the system size grows, which 
may signal the AT instability of the RRGs. 
However, it is difficult to draw 
any distinct conclusion 
from figs.\ \ref{fig:RZ^RRGN14} and \ref{fig:RZ^RRGN16} 
because the system sizes are not sufficiently large.
For further investigation,
we need some other idea to overcome the computational difficulty of 
large systems, and leave this problem as a future work.

\section{Summary}
In this chapter, 
we have investigated RZs for CTs and
ladders in the limit $T,n\rightarrow 0,\,\, 
\beta n \rightarrow y \sim O(1)$. 
Most of the
zeros exist near the line ${\rm Im}(y)=\pi/2$ in all cases investigated; 
in particular, for the $(k,c)=(2,3)$
CT and the width-$2$ ladder all the zeros lie on this
line. 
For the width-$2$ ladder we have proved that 
the generating function is 
analytic with respect to $y$ in this model. 
On the other hand, for some CTs, a relevant fraction
of the RZs spreads away from the line ${\rm Im}(y)=\pi/2$ and
approaches the real axis as the generation number $L$ grows. 
This implies that $\phi(n)$ has a singularity at a finite real $y$ in
the thermodynamic limit.  
A naive observation finds that the RZs collision points 
correspond to phase transitions of the boundary condition
of the BL. We have compared them with known 
critical conditions of the 1RSB and FRSB
and concluded that these conditions are irrelevant to the 
behavior of RZs. 
This is consistent with  
the absence of RSB in CTs
reported in some earlier studies \cite{Chay,Mott,Carl,Lai}.

The irrelevance of RZs to the 1RSB and FRSB has been considered 
in many aspects. 
By considering the boundary effect of CTs and BLs, 
we have concluded that the 1RSB cannot be observed 
by the current framework. 
On the other hand, the AT instability, the critical condition 
of the FRSB, has required 
some delicate discussions. 
We have inferred that 
the AT instability cannot be detected by the current formula 
without introducing the interaction among several replicas 
and the peculiarity of CTs and BLs is not essential.  
We have examined this statement by considering RZs of $c=3$ RRG but any 
distinct conclusion could not be obtained because the system sizes 
are not sufficiently large, which is caused by the computational difficulty 
of the moment $[Z^n]_{\V{J}}$ of RRGs.
Overcoming this difficulty needs some more idea and is left 
as a future work.
Another possible way to examine the statement is 
considering the replicated systems 
with interactions among replicas. 
This strategy seems to be more hopeful but 
here we leave this problem as a future work too.

\chapter{The replica symmetry breaking and 
the partition function zeros for Bethe lattice spin glasses}
In the previous chapter, we have investigated replica zeros of Cayley trees and discussed 
the possible links to the RSB. 
Our conclusion is that 
the replica zeros of Cayley trees and Bethe lattices cannot detect any RSB.
According to some discussions about the boundary condition of Bethe lattices 
we have clarified that 
the peculiarity of Cayley trees and Bethe lattices 
makes it difficult to identify the 1RSB transition by the replica zeros. 
On the other hand, the FRSB has required 
more delicate discussions 
and we have finally inferred that 
the replica zeros are essentially not able to detect the FRSB, not due to the 
peculiarity of Cayley trees and Bethe lattices.
To examine this conclusion, 
we investigate the partition function zeros of Bethe lattices with respect to the 
external field $H$ and temperature $T$ in this chapter. 

As mentioned in chapter 1, 
the spin-glass susceptibility is directly related to the nonlinear 
susceptibility as eq.\ (\ref{eq1:chinl}). 
This fact implies that 
singularities with respect to the external field $H$ well 
signal the AT condition, which is 
expected to be the same as the divergence of 
the spin-glass susceptibility. 

Utilizing the cavity method, we can accurately calculate the partition 
function zeros of Bethe lattices of infinite sizes. 
 Our result implies that the FRSB is related 
to the continuous zeros of the partition function and can be 
detected in Bethe lattices despite the peculiarity, which supports our 
conclusion in the previous chapter.
In addition, this result directly confirms a common belief 
that the AT instability leads to 
the continuous singular behavior of the system \cite{Bray1979}.

The results given in this chapter is in reference \cite{Matsuda:10}. 
Note that most of numerical
calculations in this chapter was done by Yoshiki Matsuda and he has the priority of
those results. The zeros formulation given in section \ref{sec5:formula} was mainly investigated by
Antonello Scardicchio and Markus M\"uller. The contributions of the author of this thesis
to the results are mainly on investigating the formulation of the AT instability based on the 
divergence of the
spin-glass susceptibility. Especially, the zero-temperature analysis of the AT instability
given in appendix \ref{app:AT^BL2} was performed exclusively by the present 
author.

\section{Phase boundaries and critical properties of Bethe lattice 
spin glasses}
\begin{figure}[htbp]
\begin{center}
   \includegraphics[height=50mm,width=70mm]{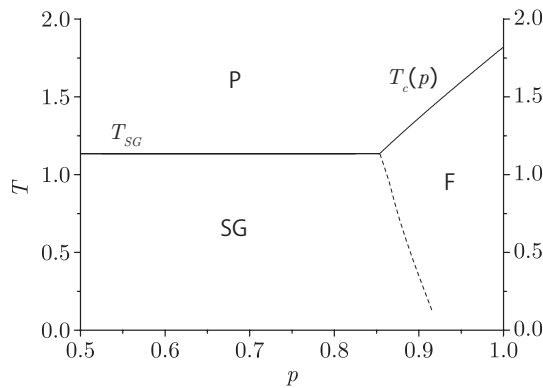}
 \caption{A $T$-$p$ phase diagram of the $\pm J$ model without the external 
field on a $c=3$ Bethe lattice. 
Symbols P, F, SG in the figure represent the paramagnetic, ferromagnetic, and 
spin-glass phases, respectively. The dashed line is the AT line. 
The boundary between the ferromagnetic and spin-glass phases 
is not determined. 
The critical temperature between the paramagnetic and spin-glass phases 
is given by $T_{SG}=1/\tanh^{-1}\left[{ {1/ \sqrt{2} } }\right]\simeq 1.13$.
The critical line $T = T_{c}(p)$ 
denotes the boundary between the paramagnetic and
ferromagnetic phases.
}
 \label{fig:PDofBL}
\end{center}
\end{figure}

Before proceeding to the formulation of the partition function zeros, 
we first review some
critical properties of Bethe lattice spin glasses in this section.
When the external field $H$ is absent, 
the critical conditions of Bethe lattice spin glasses 
are given by 
\begin{eqnarray}
(c-1)[\tanh(\beta J)]_{\V{J}}=1,\label{eq5:para_to_ferro}\\
(c-1)[\tanh^2(\beta J)]_{\V{J}}=1.\label{eq5:para_to_SG}
\end{eqnarray}
The first relation gives the boundary 
between the paramagnetic and ferromagnetic phases,
and the latter corresponds to the transition from  
the paramagnetic to spin-glass phases.
We here derive these relations by using the cavity method.
Taking the $n\to 0$ limit in eq.\ (\ref{eq4:pih1}),
we can find the equation of the cavity-field distribution (CFD)
$P_{i}(h)=\lim_{n\to 0}\pi_{i}(h)$ as
\begin{equation}
P_{i}(h)=\int 
\prod_{j=1}^{c-1}
\left(
dh_{j}P_{j}(h_{j})
\right)
\left[
\delta\left(
h-H-\sum_{j=1}^{c-1}\Wh{h}_{j}(J_{ij},h_{j})
\right)
\right]_{\V{J}}. \label{eq5:cfd}
\end{equation}
If the external field $H$ is absent,
this equation always has the paramagnetic solution 
$P(h)=\delta(h)$ . 
Instability of this paramagnetic solution signals phase transitions to
other phases. To perform the stability analysis,
we expand the relation (\ref{eq4:bias_to_field}) with respect to $h_{i}$
and get
\begin{equation}
h_{i}=\sum_{j=1}^{c-1}t_{ij}(h_{j}-\frac{1}{3}(1-t_{ij}^2)h_{j}^{3}),
\label{eq5:moment} 
\end{equation}
where $t_{ij}=\tanh(\beta J_{ij})$. Near the phase boundary, 
we can expect that the moments $\Ave{h^n}$ 
decrease rapidly as $n$ grows, which validates to ignore higher order terms. 
Here, we treat up to the third order. Taking powers of eq. (\ref{eq5:moment})
and averaging with respect to $P_{i}$ which is denoted by the brackets 
$\Ave{\cdots}$, we get
\begin{eqnarray}
&&\Ave{h}=
(c-1)(2p-1)\left(\Ave{h}-\frac{1}{3}(1-t^2)\Ave{h^3}\right),\label{eq5:1moment}
\\
&&\Ave{h^2}=(c-1)t^2\Ave{h^2}+(c-1)(c-2)(2p-1)^2t^2\Ave{h}^2,\\
&&\hspace{-15mm}\Ave{h^3}=(c-1)(2p-1)t^3
\left\{\Ave{h^3}+3(c-2)\Ave{h^2}\Ave{h}+(c-2)(c-3)(2p-1)^2\Ave{h}^3\right\}
\label{eq5:3moment},
\end{eqnarray}
where we assume the $\pm J$ model 
\begin{equation}
P(J_{i})=p\delta(J_{i}-J)+(1-p)\delta(J_{i}+J),\label{eq5:P(J)}
\end{equation}
and denote $t=\tanh(\beta J)$. 
To investigate the stability of the paramagnetic solution $\Ave{h}=\Ave{h^2}=\Ave{h^3}=0$, we linearize eqs.\ (\ref{eq5:1moment})--(\ref{eq5:3moment}) 
to yield
\begin{equation}
\left(      
   \begin{array}{c}
    \Ave{h} \\
    \Ave{h^2} \\
    \Ave{h^3} 
\end{array}
 \right)
=
 \left(      
   \begin{array}{ccc}
    (c-1)(2p-1)t & 0 & -\frac{1}{3}(c-1)(2p-1)t(1-t^2) \\
    0 & (c-1)t^2 & 0 \\
    0 & 0 & (c-1)(2p-1)t^3 
	 \end{array}
 \right)
\left(      
   \begin{array}{c}
    \Ave{h} \\
    \Ave{h^2} \\
    \Ave{h^3} 
\end{array}
 \right)
.\label{eq5:linearizedmoment}
\end{equation}  
This matrix has three eigenvalues $((c-1)(2p-1)t,(c-1)t^2,(c-1)(2p-1)t^3)$.
It is quite straightforward to 
rewrite these eigenvalues for general distribution $P(J)$, and the explicit form is $((c-1)[\tanh(\beta J)]_{\V{J}},(c-1)[\tanh^2(\beta J)]_{\V{J}},
(c-1)[\tanh^3(\beta J)]_{\V{J}})$.  
When an eigenvalue exceeds unity, the paramagnetic solution becomes unstable, which yields the critical conditions (\ref{eq5:para_to_ferro}) and (\ref{eq5:para_to_SG}).
In this chapter, 
we treat the $\pm J$ model with $J=1$ on 
a $c=3$ Bethe lattice, 
and here present the phase diagram in fig.\ \ref{fig:PDofBL}.
Symbols P, F, SG in fig.\ \ref{fig:PDofBL} 
represent the paramagnetic, ferromagnetic, and 
spin-glass phases, respectively. 
The boundary between the ferromagnetic and spin-glass phases  
is not given in this figure because 
there is no known method to calculate this boundary.
Although we did not discuss the AT condition so far, 
we provide the 
AT line as the dashed line in fig.\ \ref{fig:PDofBL}. 
The detailed derivation of the AT line will be 
given in section \ref{sec5:AT}.

From eqs.\ (\ref{eq5:1moment})--(\ref{eq5:3moment}), we can also obtain the critical exponents. Near the phase boundary between the paramagnetic and ferromagnetic phases, we can define a small parameter $\epsilon$ by $(c-1)[\tanh\beta J]_{\V{J}}=(c-1)(2p-1)t=1+\epsilon$. For $\epsilon<0$, the paramagnetic solution is stable but for $\epsilon>0$ it becomes unstable and the ferromagnetic solution emerges. 
For the lowest order with respect to $\epsilon$, we find the ferromagnetic solution as follows;
\begin{subequations}
\begin{align}
\Ave{h}=\sqrt{3}\sqrt{\frac{(c-1)(1-(c-1)t^2)}{(c-2) \{c-3-t^2(c(c-7)+9)\} }}\epsilon^{1/2},\\
\Ave{h^2}=\frac{3}{c-3-t^2(c(c-7)+9)}\epsilon,\\
\Ave{h^3}=\frac{3\sqrt{3}}{1-t^2}\sqrt{\frac{(c-1)(1-(c-1)t^2)}{(c-2)
\{c-3-t^2(c(c-7)+9)\}}}\epsilon^{3/2}.
\end{align}\label{eq5:ferro}
\end{subequations}
Equations (\ref{eq5:ferro}) indicate that 
the critical exponent $\beta$, which yields
the temperature dependence of spontaneous magnetization $m \sim (T-T_{c})^\beta$, equals to $1/2$. 
Near the P-SG boundary, similar calculations are possible. In this case, we need the moments up to the fourth order. Assuming the absence of the ferromagnetic order $\Ave{h^1}=\Ave{h^3}=0$, we get 
\begin{subequations}
\begin{align}
\Ave{h^2}=(c-1)t^2(\Ave{h^2}-\frac{2}{3}(1-t^2)\Ave{h^4}),\\
\Ave{h^4}=(c-1)t^4\Ave{h^4}+3(c-1)(c-2)t^4\Ave{h^2}^2.
\end{align}\label{eq5:SGeq}
\end{subequations}
Putting $(c-1)t^2=1+\epsilon$, we find the solution of eqs. (\ref{eq5:SGeq}) 
in powers of $\epsilon$
\begin{subequations}
\begin{align}
\Ave{h^2}=\frac{c-1}{2(c-2)}\epsilon+\frac{(c-1)(2t^2-3)}{2(c-2)(1-t^2)}\epsilon^2+O(\epsilon^3),\\
\Ave{h^4}=\frac{3(c-1)}{4(c-2)(1-t^2)}\epsilon^2+O(\epsilon^3).
\end{align}
\end{subequations}
This implies that the spin-glass order parameter $q$ is proportional to 
$\epsilon \propto T-T_{SG}$, which is known to be 
the same property as the SK model. 

\section{Formulations}
Next, we present the formulation to yield the zeros of the partition function. The cavity
method again plays a key role for this purpose. Antonello Scardicchio and Markus M\"uller
mainly contributed to the derivation of the zeros formula given in section \ref{sec5:formula}. Actual
implementation of the formula was done by Yoshiki Matsuda and the author of this thesis.
\subsection{The partition function zeros of Bethe lattices}\label{sec5:formula}
Let us first consider the zeros of the partition function with respect to 
the uniform external field $H$, the Lee-Yang zeros \cite{LeeYang}. 
The Lee-Yang zeros are known to be related to 
the magnetization of the system, and here 
we start from illustrating this point.
In general, the partition function of an Ising system 
in the external field $H$ is expressed as 
degree $N$ polynomial with respect to $e^{-2 \beta H}$
as
\begin{equation}
Z\left( H \right) = 
e^{N\beta H} 
\sum_{k = 0}^N { {\Omega \left( {k} \right)
 e^{-2k\beta H} } } 
=\xi e^{ N\beta H} \prod\limits_{i}^N {\left( {e^{-2\beta H}  - e^{-2\beta H_i} } \right)} ,\label{eq5:Z(H)}
\end{equation}
where $N$ is the system size and $\Omega(k)$ 
is the partition function for a 
fixed magnetization $M=N-2k=\sum_{i}S_{i}$. 
In the last equation, we factorize the polynomial and introduce the 
zeros $e^{-2\beta H_{i}}$. The leading factor $\xi$ is an 
irrelevant constant and hereafter omitted. 
Taking the logarithm of eq.\ (\ref{eq5:Z(H)}) and dividing by $N$, 
we get
\begin{equation}
\frac{1}{N}\log Z(H)= - \beta f\left( H \right) = \beta H + \int\!\!\!\int {d^2 H'g_{z} \left( {H'} \right)\log \left( {e^{ - 2\beta H}  - e^{ - 2\beta H' } } \right)}.\label{eq5:f(H)}
\end{equation}
where $g_{z}(H)$ is the density of zeros on the complex $H$ plane. 
The magnetization $m\left( H \right) = -\partial f/\partial H$ 
is then expressed as
\begin{equation}\label{mag(H)}
m\left( H \right) = 1 + \int\!\!\!\int {d^2 H' g_{z} \left( {H'} \right){2 \over {e^{2\beta \left( {H - H'} \right)}  - 1}}} .
\end{equation}
Taking an infinitesimal closed line integral with respect to complex $H$, 
we can derive $g_{z}(H)$ from the magnetization as
\begin{equation}
\label{g2(H)}
g_{z}(H)= \frac{\beta}{2\pi i} \lim_{r\to 0} \frac{1}{\pi r^2}\oint_{|H-H'|=r} m(H') dH'.
\end{equation}
This relation enables us to evaluate the density of zeros from the 
magnetization.

Next let us turn to discuss the properties of Bethe lattices. 
We have defined a Bethe lattice as an interior part of an infinitely large Cayley tree. 
This means that 
a Bethe lattice is statistically uniform and the typical magnetization  
is represented by the magnetization of the central spin, 
which is expressed by 
using the CFD of the central spin, $P^{(c)}(h^{(c)})$, as 
\begin{equation}\label{mcav}
m = \int {d^{2}h^{(c)} P^{(c)}\left( {h^{(c)}} \right)\tanh \left( \beta h^{(c)} \right) }.
\end{equation}
Note that the domain of integration is two-dimensional 
(as $d^{2}h^{(c)}$) because we are now
treating the complex cavity field.
The central CFD $P^{(c)}(h^{(c)})$ is derived from 
the convergent CFD $P(h)=\lim_{g \to \infty}P_{g}(h)$\footnote{The convergence of the CFD is shown rigorously in \cite{Chay} for the real cavity-field case. 
Although
this proof is not directly applicable to our case of complex cavity field, we numerically observed the
convergence for $p<1$ and hereafter assume it.}, 
where $P_{g}(h)$ denotes the CFD of the $g$th generation of a Cayley tree, 
as    
\begin{equation}\label{fdcent}
P^{(c)} \left( {h^{(c)}} \right) =  {\int {\left[
\delta \left(
h^{(c)} - H - \sum\limits_{j = 1}^{c} {\Wh{h}_j\left(J_{ij}, h_j \right) } 
\right) \prod\limits_{j = 1}^{c} {P \left( {h_j } \right)d^{2}h_j }\right]_{\V{J}} } }  .
\end{equation}
Substituting eq.\ (\ref{mcav}) to eq.\ (\ref{g2(H)}), 
we obtain
\begin{eqnarray}
\label{gh}
g_{z}(H) &=& \frac{\beta}{2\pi i}
\lim_{r\to 0} 
\frac{1}{\pi r^2}\oint_{|H-H'|=r} dH' \int {d^2 h^{(c)} P^{(c)}\left( {h^{(c) }} \right)\tanh \left( \beta h^{(c)}\left( H' \right) \right) } \nonumber\\
&=& 
\lim_{r\to 0} 
\frac{1}{\pi r^2}\int_{|H-H'| \le r} d^2 H' \int {d^2 h^{(c)} P^{(c)}\left( {h^{(c) }} \right) \sum\limits_{n \in \mathbb{N}	} {\delta \left( {\beta h^{\left( c \right)} \left( {H'} \right) - {{2n - 1} \over 2}\pi i } \right)}  }\nonumber\\
&=& 
\lim_{r\to 0} 
\frac{1}{\pi r^2}\int_{|H-H'|\leq r}\!\!\!\!\!\!\!\!\!\! d^2H' \sum\limits_{n \in \mathbb{N}} {P^{(c)}\left( \left( {2n - 1} \right)\pi i /2\beta  \right)}  \nonumber\\
&=& \sum\limits_{n \in \mathbb{N}} {P^{(c)}\left( \left( {2n - 1} \right)\pi i /2\beta  \right)}.
\end{eqnarray}
In the second line, 
we used the residue theorem under the condition that the radius $r$ 
is sufficiently small.
Equation (\ref{gh}) enables us to calculate the density of zeros from 
the central CFD $P^{(c)}(h^{(c)})$.


For a Bethe lattice, eq.\ (\ref{gh}) can also be interpreted as follows. 
Remember that the partition function of the whole system of a 
Bethe lattice is given by
\begin{equation}\label{pfcent}
Z= 2\cosh \left( {\beta h^{(c)} } \right)
\prod\limits_{j = 1}^c {\left( 
{{{\cosh \left( {\beta J_{ij} } \right)} \over {\cosh \left( {\beta \Wh{h}_j } \right)}}Z_j } \right)
}. 
\end{equation} 
This implies that the equation of zeros $Z=0$ 
becomes identical to\footnote{Note that eq.\ (\ref{pfcent}) seems to diverge when 
the factor $\cosh{\beta \Wh{h}_j }$ becomes $0$, 
but this is not the case. 
That is because the condition 
$\cosh{\beta \Wh{h}_j }=0$ always involves $Z_{j}=0$ and  
the factor $(\cosh{\beta \Wh{h}_j })^{-1}Z_{j}$ yields a finite value.} 
\begin{equation}
2\cosh \left( {\beta h^{(c)} }\right)=0 
\Rightarrow
\beta h^{(c)}=\frac{2n-1}{2}\pi i. \,\,\,\,(\forall{n}\in \mN). \label{cosh=0}
\end{equation}
Equations (\ref{gh}) and (\ref{cosh=0}) provide a simple interpretation of 
the zeros of the Bethe lattice. 
That is 
the probability distribution that   
the partition function of a central spin becomes $0$ 
is identical to the distribution of zeros of the whole Bethe lattice.
This interpretation is useful for evaluating the density of zeros 
with respect to complex temperature.
What we should do is to evaluate $P^{(c)}(h^{(c)})$ 
at the temperature $T=1/\beta$ fixed on a complex value 
and to calculate $g_{z}(\beta)$ using the relation 
(\ref{gh}). 
This prescription is one of the advantages 
by considering Bethe lattices and constitutes 
the main result of this chapter.

\subsection{Population method to evaluate the CFD}\label{sec5:remark}
In the previous subsection, we have derived the relation to obtain the distribution of zeros from the central CFD $P^{(c)}(h)$. 
The next step is to actually calculate the CFD. 
A widely-used numerical method for this purpose is the ``population method'', 
the main idea of which is to represent the distribution by a population of 
$N_{\rm pop}$ cavity fields $h_{i}$. 
The actual procedure is as follows:
\begin{enumerate}
\item{Set the initial population of $N_{\rm pop}$ cavity fields.}
\item
{Choose $c-1$ fields randomly and compute a new field according to 
eqs.\ (\ref{eq4:bias}) and (\ref{eq4:bias_to_field}). 
} 
\item{Replace a randomly chosen field with the new field.}
\end{enumerate} 
These procedures define a Markov chain on the space of the 
$N_{\rm pop}$ fields.  
Usually, it is known that 
this chain has a stationary distribution and 
the stationary distribution satisfies the self-consistent 
equation of the convergent CFD $P(h)$ in the limit $N_{\rm pop}\to \infty$ 
\cite{MezaRevisit}. 
This population method is usually employed to evaluate 
the real CFD but we here apply this to the complex CFD case. 

We should here notice that there are two delicate points 
in applying the population method to complex $P(h)$.

One is related to the relevance of the 
boundary condition. Even in the real parameter case,  
the boundary condition is very important to correctly 
consider the spin-glass problem on Cayley trees and Bethe lattices, as 
discussed in the previous chapter. 
The selection of the boundary condition is directly related to the 
selection of the convergent CFD or the selection of the 
saddle-point of the corresponding regular random graph. 
Moreover for the complex parameter case, 
there is a possibility that other additional 
solutions of $P(h)$ appear depending on the boundary condition.
To avoid confusions coming from such a situation, 
we choose the boundary condition
by referring to
an ``adiabatic'' criterion, which 
is based on an intuition that 
the physically relevant complex $P(h)$ 
should be smoothly continued to the real solution.
This means that in principle 
the correct complex $P(h)$ is derived 
by gradually increasing the imaginary part of the parameter 
during the cavity update. 
As one of the simplest choices according to this adiabatic criterion, 
we choose the fixed boundary condition, 
which yields the initial cavity bias as 
$\Wh{h}_{i}=J_{i}$ where $J_{i}$ is the bond between the 
outermost site $i$ and its ascendant.

The other is the dimensionality of the zeros 
distribution $g_{z}$. 
Hereafter let us focus on the $\pm J$ model, the bond 
distribution of which is given by eq.\ (\ref{eq5:P(J)}).
For the $p=1$ ferromagnetic case, 
the Lee-Yang 
zeros are distributed only in a one-dimensional subspace of the complex 
field plane, which is also the case for 
the zeros with respect to the complex temperature \cite{LeeYang,Fisher:64}. 
On the other hand, for several spin-glass systems the zeros 
are known to two-dimensionally spread in 
the complex plane 
\cite{Mouk1,Mouk2,Ozeki:88,Matsuda:08,Bhanot:93,Saul:93,
Damgaard:95,Derrida:91}.
These facts imply that 
the existence region of the zeros gradually spread 
as the ferromagnetic bias $p$ decreases
and 
the zeros distribution $g_{z}$ has its one and two-dimensional 
parts, $g_{z1}$ and $g_{z2}$, respectively, in a certain range of $p$.
The problem is 
the one-dimensional part is difficult to detect by the 
population method because we use the two-dimensionally distributed 
population.  
For the Lee-Yang zeros case, however, 
we can expect that $g_{z1}$ takes non-zero values 
only on the imaginary axis, 
which enables us to calculate $g_{z1}(H=i\theta)$ from the 
the following relation\footnote{This relation is easily understood by regarding the relation (\ref{eq5:f(H)}) between the free energy and the 
zeros distribution 
is the same as the one between 
the electrostatic potential and the charge distribution in a 
two-dimensional space. In this context, 
the magnetization corresponds 
to the electric field, which has 
the discontinuous jump on the line density of the charge. 
This analogy leads to eq.\ (\ref{eq5:gz1}). }\cite{Fisher:64} 
\begin{equation}
{\rm Re}\left(
\lim_{H_{r}\to 0}m(H=H_{r}+i\theta)
\right)=2\pi g_{z1}(\theta).
\label{eq5:gz1}
\end{equation}
In general, 
if the region where $g_{z1}$ takes non-zero values  
is exactly known, the evaluation of  
$g_{z1}$ is possible 
by using a similar relation to eq.\ (\ref{eq5:gz1}).
Unfortunately, however, 
the exact location of such region of $g_{z1}$ 
is not known in general situations.
In the remainder of this thesis,  
we calculate $g_{z1}$ only on 
the imaginary axis of the external field $H=i\theta$, 
and ignore other possible region of $g_{z1}$.

\subsection{Numerical procedures}
In the previous subsections we present the conceptual explanation 
of the procedures to obtain $g_{z}$ and the convergent CFD $P(h)$. 
However, 
the implementation of those procedures is also involved  
and we here summarize actual numerical procedures.

To this end, we first write down the update rule of the 
cavity bias $\Wh{h}_{i}$, which is more numerically tractable 
than the cavity field, as
\begin{equation}\label{recursion2}
\Wh{h}_i  = {1 \over \beta }
\tanh ^{ - 1} 
\left( {\tanh \left( {\beta J_{i} } \right)
\tanh \left( 
{\beta \left( {H + \sum_{j = 1}^{c - 1} \Wh{h}_j  
} \right)} \right)} \right),
\end{equation}
where $J_{i}$ is the bond between the site $i$ and 
its ascendant.
We hereafter assume that the
function $\tanh^{-1}$ 
takes the principal branch, which restricts the value of the imaginary
part of the bias to a range $[-\pi/\beta,\pi/\beta]$.
The cavity bias distribution (CBD) for the $g$th generation 
$\Wh{P}_{g}(\Wh{h})$ 
is represented by 
a population 
and 
is updated by eq.\ (\ref{recursion2}) 
until $\Wh{P}_{g}(\Wh{h})$ converges.

After the convergence, the central CFD 
$P^{(c)}(h^{(c)})$ is calculated from 
the convergent CBD 
$\Wh{P}(\Wh{h})=\lim_{g\to \infty}\Wh{P}_{g}(\Wh{h})$ as 
\begin{equation}\label{recursion3}
P^{(c)}(h^{(c)})=\int \prod_{j=1}^{c}
d^{2}\Wh{h}_{j} \Wh{P}(\Wh{h}_{j})
\delta\left( 
h^{(c)}-H-\sum_{j=1}^{c}\Wh{h}_{j}
\right).
\end{equation}
and $g_{z}$ is calculated from $P^{(c)}(h^{(c)})$ by using
eqs. (\ref{gh}) and (\ref{eq5:gz1}).

Before closing this subsection, 
we summarize the actual steps of the algorithm:
\begin{enumerate}
 \item{
Generate the initial population $\left\{\Wh{h}_{i}\right\}_{i=1}^{N_{\rm pot}}$ as $\Wh{h}_{i} =J_{i}=\pm 1$ according to the distribution (\ref{eq5:P(J)}).
}
 \item{
Update the population by the relation 
(\ref{recursion2}) until $\Wh{P}_{g}(\Wh{h})$ 
converges.
}
\item{
Calculate the central CFD $P^{(c)}(h^{(c)})$ by 
eq.\ (\ref{recursion3}).
}
 \item{
Estimate the two-dimensional part of the zeros density $g_{z2}$ 
by
\begin{eqnarray}
g_{z2}(H,T) &=& 
\sum_{ n \in \mathbb{N} } 
P^{(c)} 
\left( 
h^{(c)} = \left( 2n-1 \right)\pi i 
\right), 
 \label{density}
 \end{eqnarray}
  for a given $H$ and $T$.
\begin{enumerate}
\item{
When the external field is pure imaginary $H=i\theta$, 
estimate $g_{z1}(H=i \theta,T)$ separately by 
\begin{eqnarray}
\hspace{-20mm}g_{z1}(H=i\theta,T) 
&=& {1 \over {2\pi }} {\rm Re}\left( 
m( H = i \theta ) 
\right)
\nonumber\\
&=& {1 \over{2\pi } } 
{\rm Re}
\left( 
\lim_{ H_R  \rightarrow 0} 
\left. \int 
d^{2} h^{(c)} 
P^{(c)}\left( h^{(c)} \right)
\tanh \left( h^{(c)} \right)  
\right|_{H = i \theta + H_R}  
\right).
\label{densityg1}
 \end{eqnarray}
}
\end{enumerate}
}
\end{enumerate}
In the remainder of this thesis, 
we assume the conditions $c=3$ and $N_{\rm pop}=10^{6}$.
We performed at least $5000 N_{\rm pop}$ cavity iterations until the population converged. 
After the convergence, we took the average of the objective quantities 
($g_{z2}$ and $m(H=i\theta)$) over 
additional $5000 N_{\rm pop}$ steps to avoid possible fluctuation due to 
the finiteness of the population.

\section{Result}
In this section, using the formula given above, we derive the distributions of the zeros of the
presented model for several values of temperature $T$, external field $H$, and ferromagnetic
bias $p$. The physical significance of the zeros distribution is discussed and the phase
diagram of this model is re-derived in terms of the zeros. The relevance of the zeros
distribution to the AT instability is also argued, which leads to a conclusion that the AT
instability corresponds to the continuous distribution of zeros on the real axes of $T$ and
$H$.

\subsection{Zeros on the complex field plane}\label{oncomplezfp}
First, we show the density of zeros on the complex field  
plane, the Lee-Yang zeros, for fixed real temperatures. 
For simplicity of calculations, we rescale the complex field 
$H$ as $2\beta H$ when analyzing the Lee-Yang zeros hereafter.
\begin{figure}[htbp]
\begin{minipage}{0.5\hsize}
\includegraphics[width=1.00\linewidth]{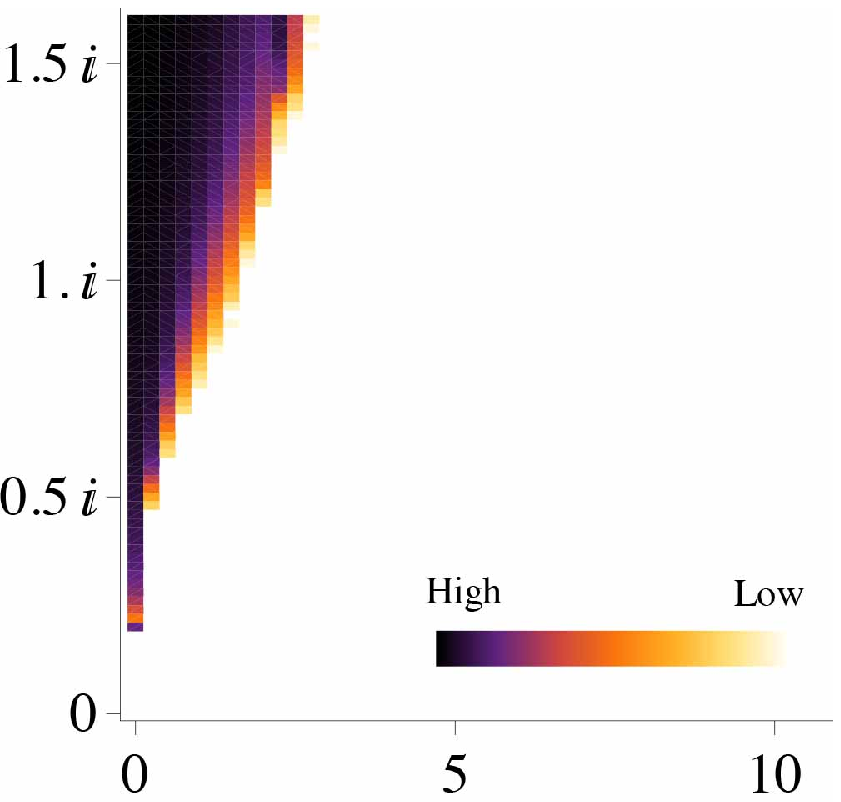}
\end{minipage}
\begin{minipage}{0.5\hsize}
\includegraphics[width=1.00\linewidth]{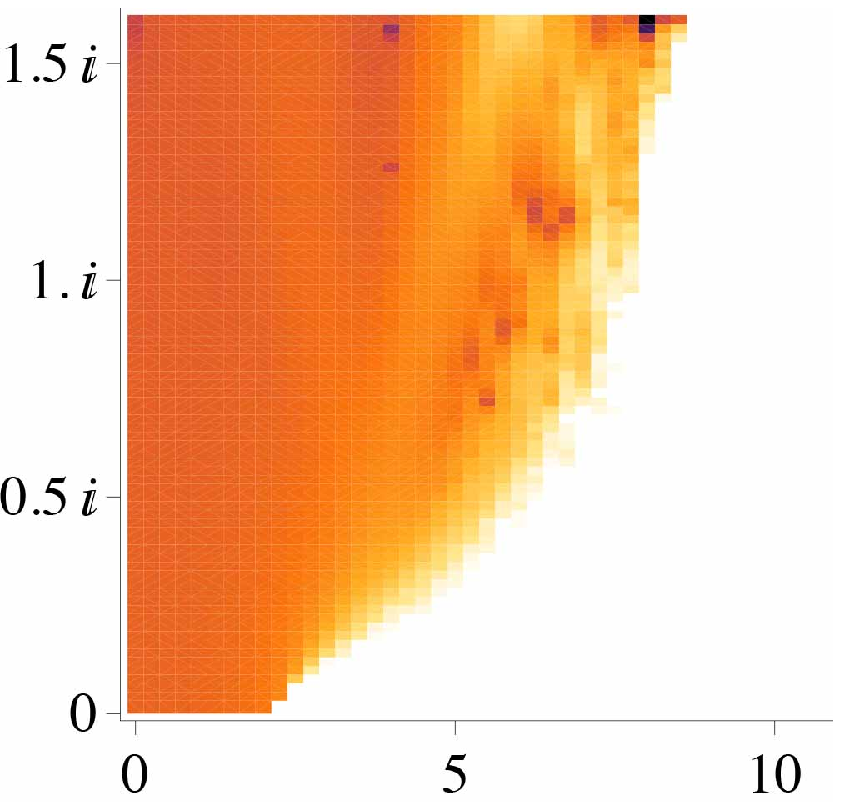}
\end{minipage}
\caption{
Distribution of zeros on the complex $2\beta H$ plane with $p=0.5$ at $T = 1.43$ (left) and $0.5$ (right). Densities are colored in logarithmic scale. Zeros with finite real part approach the real axis at low temperature, whereas $g_{z1}$ is numerically zero.
\label{fig:p05parasg}}
\end{figure}
Figure \ref{fig:p05parasg} is the distributions of zeros on the complex $2\beta H $ plane with $p = 0.5$ and $T = 1.43 > T_{SG} = 1/\tanh ^{ - 1} \left( {1/\sqrt {c - 1} } \right) $ (left) and 
$0.5 < T_{SG}$ (right). 
To draw this figure, we changed the value of ${\rm Re}( 2\beta H )$ from $0$ to $12$ with the increment $0.25$ and of 
${\rm Im} \:( 2\beta H )$
from $0.02$ to $\pi/2$ with the increment of $0.02$. 
Each evaluated point locates at the center of each cell in 
fig.\ \ref{fig:p05parasg} 
and the corresponding value 
of $g_{z}$ represents the value of the cell.
The density out of the range is omitted, since the range is sufficient to see zeros near the real axis which are relevant to 
the critical phenomena. 
Both $g_{z1}$ and $g_{z2}$ are plotted in the same figure 
and colored in logarithmic scale; the black dots show a very high density.

The left panel of fig.\ \ref{fig:p05parasg} ($T=1.43$) is for the paramagnetic phase. In both $g_{z1}$ and $g_{z2}$ parts, the zeros do not reach the real axis, which means there is no phase transition. 
On the other hand, the right panel of fig.\ \ref{fig:p05parasg} is in the spin-glass phase ($T=0.5$) and 
the zeros reach the real axis at a certain value $0<H_{SG} \in \mR$. 
This means that there is a phase transition at ${\rm Re}(H)=H_{SG}$ in the Bethe lattice. 
Besides, below the critical field $H_{SG}$, zeros continuously touch the real axis, which indicates that the phase transitions continuously occur 
in the range of $|{\rm Re}(H)|<H_{SG}$. 
This behavior is quite different from usual phase transitions, and 
seems to be related to the chaos effect of a small change of field.
The chaos effect is considered to be the result of the instability of 
randomly frozen spin configurations in the spin-glass phase and 
to be one of the characteristic properties of spin glasses.
Note that  
the one-dimensional density $g_{z1}(H=i\theta)$ is numerically zero at all 
the range of $\theta$ in this $p=1/2$ case, which implies the 
absence of singularities of the magnetization. 
This fact suggests that the Griffiths singularity \cite{Griffiths:69}, 
which is the essential singularity of the free energy with respect 
to the external field $H$ at $H=0$ and is considered to appear 
in some diluted ferromagnets and spin glasses, 
is absent in the current Bethe lattice spin glass without the ferromagnetic 
bias ($p=1/2$). 

\begin{figure}[htbp]
\begin{minipage}{0.5\hsize}
\includegraphics[width=1.00\linewidth]{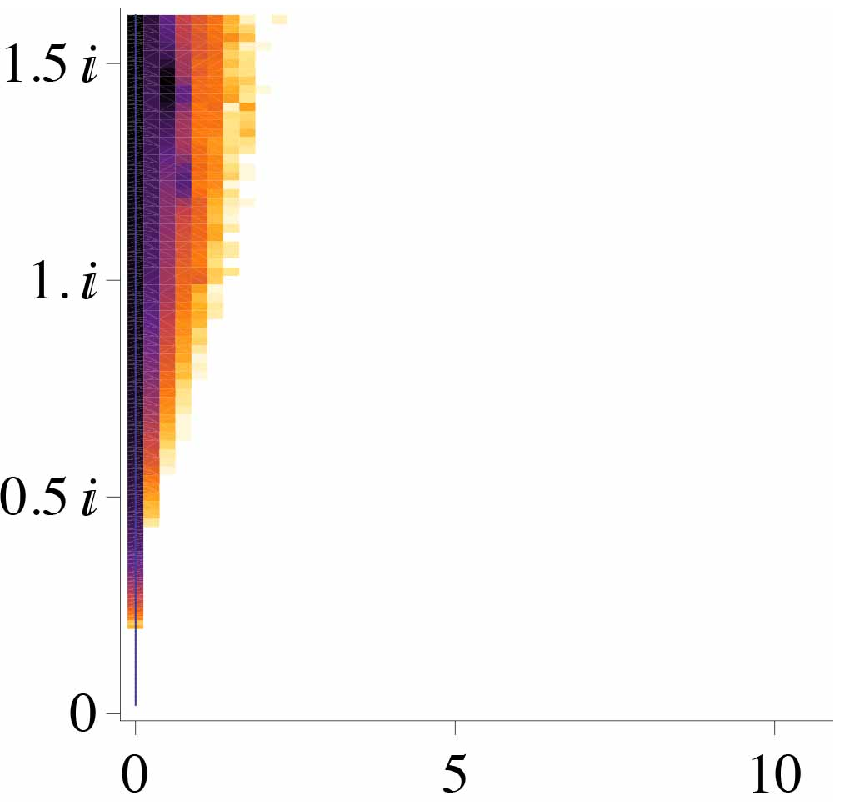}
\end{minipage}
\begin{minipage}{0.5\hsize}
\includegraphics[width=1.00\linewidth]{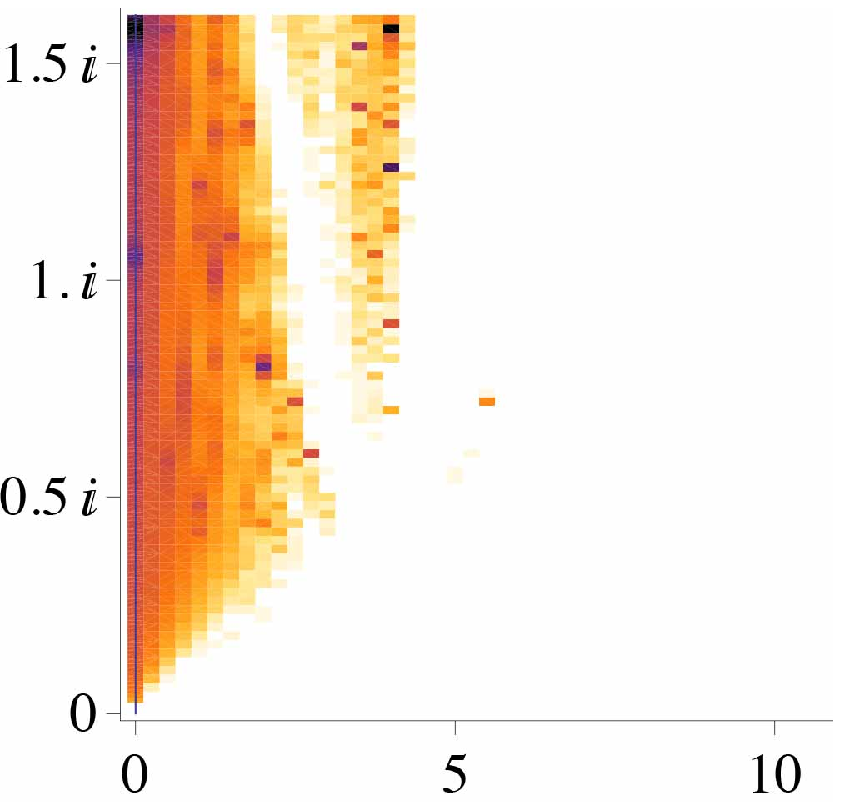}
\end{minipage}
\caption{
Distribution of zeros on the complex $2\beta H$ plane with $p=0.9$ at $T = 1.5$ (left: para) and $0.5$ (right: ferro). Only the one-dimensional part of the 
zeros on the imaginary axis (thin lines) approaches 
the origin at the low temperature. 
\label{fig:p09paraferro}}
\end{figure}

As the ferromagnetic bias $p$ increases, 
the ferromagnetic phase appears and the situation changes.
Figure \ref{fig:p09paraferro} is of the case $p=0.9$.
The right panel of fig.\ \ref{fig:p09paraferro} 
is in the ferromagnetic phase, in which $g_{z2}$ does not reach the real axis 
but only $g_{z1}$ touches the origin. 
Besides, comparing to the left panel of fig.\ \ref{fig:p05parasg}, 
we can find that $g_{z1}$ has a finite value on the imaginary axis away from the origin. 
These observations indicate that 
the ferromagnetic ordering is characterized by $g_{z1}$, while 
$g_{z2}$ signals the spin-glass ordering. 
According to this distinction of roles of $g_{z1}$ and $g_{z2}$, 
we can draw the phase diagram of the current system and will 
present the result in section \ref{sec5:PD}.

\subsection{Zeros on the complex temperature plane}
Assuming the temperature is complex, we can also obtain the density of zeros on the complex temperature plane at a fixed $H \in \mC$.

\begin{figure}[htbp]
\begin{minipage}{0.5\hsize}
\includegraphics[width=1.00\linewidth]{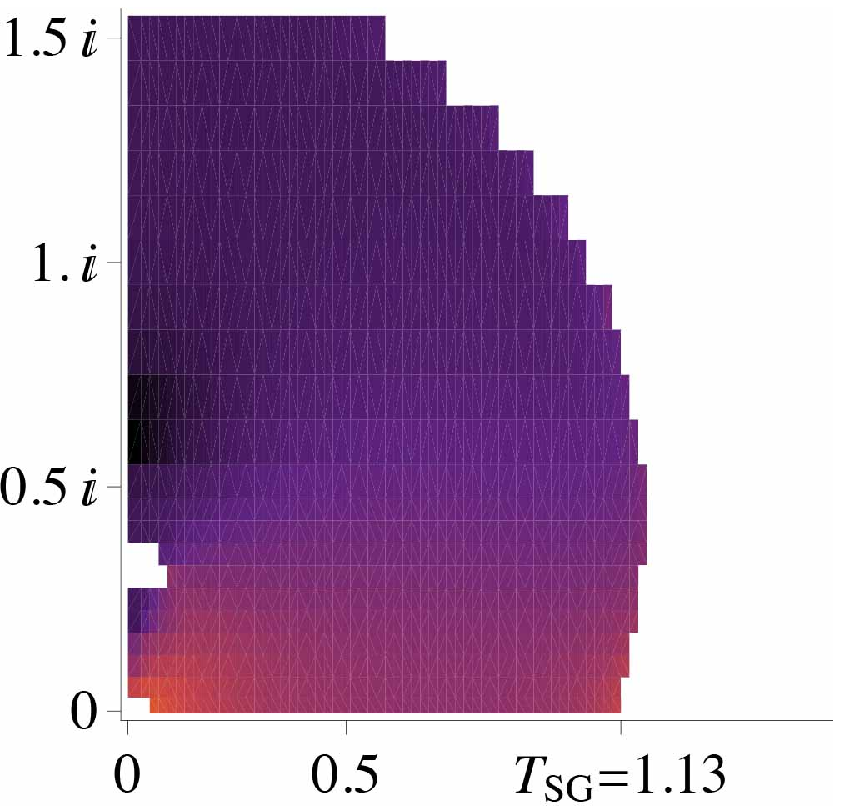}
\end{minipage}
\begin{minipage}{0.5\hsize}
\includegraphics[width=1.00\linewidth]{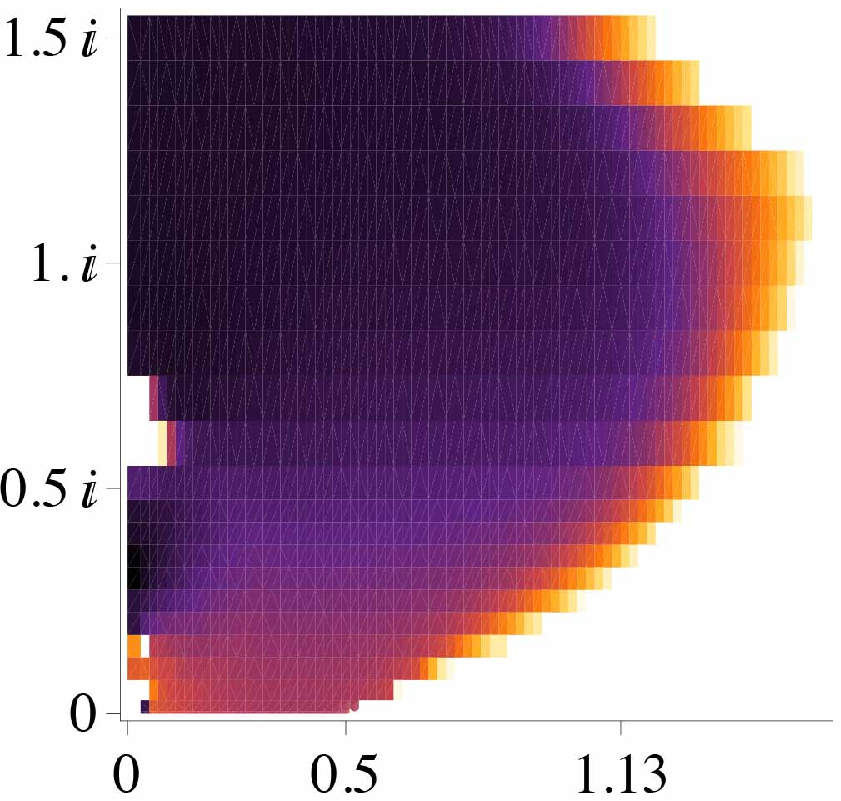}
\end{minipage}
\caption{
Distribution of zeros on the complex $T$ plane with $p=0.5$ and $H=0$ (left) and $H=0.5$ (right). Zeros reach the real axis below the critical temperature $T_{SG} \simeq 1.13$ (left) and $0.5$ (right). The apparent absence of zeros near the origin may be due to numerical rounding errors of $\tanh \beta$.
\label{fig:p05parasg_tp}}
\end{figure}

Figure \ref{fig:p05parasg_tp} shows the zeros on the temperature plane at $p=0.5$ with $H = 0$ (left) and $H = 0.5$ (right). 
All the (two-dimensional) zeros below the spin-glass transition temperature $T_{SG} \simeq 1.13$ also reach the real temperature axis. This result reconfirms that the spin-glass transition differs from usual phase transitions where zeros touch the real axis only at the transition temperature. The continuous singularities in 
the range of ${\rm Re}(T)\leq T_{SG}$ again implies 
the chaos effect against temperature deviation. 
The right panel of fig.\ \ref{fig:p05parasg_tp} shows that the spin-glass phase is stable under a weak field ($H=0.5$), which is in accordance with the Lee-Yang zeros result (the right panel of fig.\ \ref{fig:p05parasg}). 

\begin{figure}[htbp]
\begin{minipage}{0.5\hsize}
\includegraphics[width=1.00\linewidth]{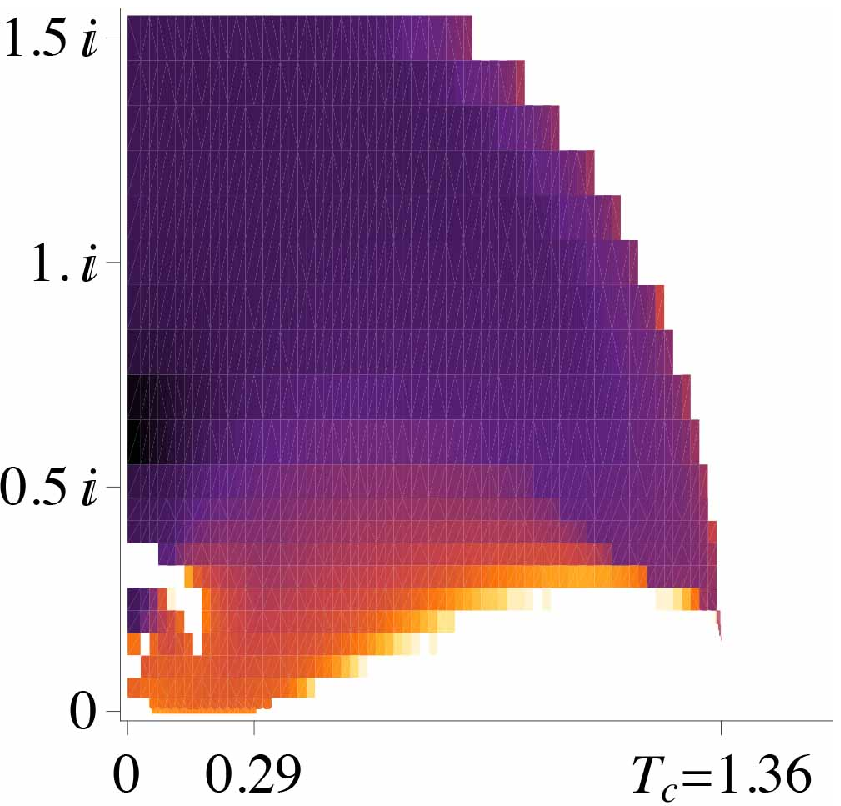}
\end{minipage}
\begin{minipage}{0.5\hsize}
\includegraphics[width=1.00\linewidth]{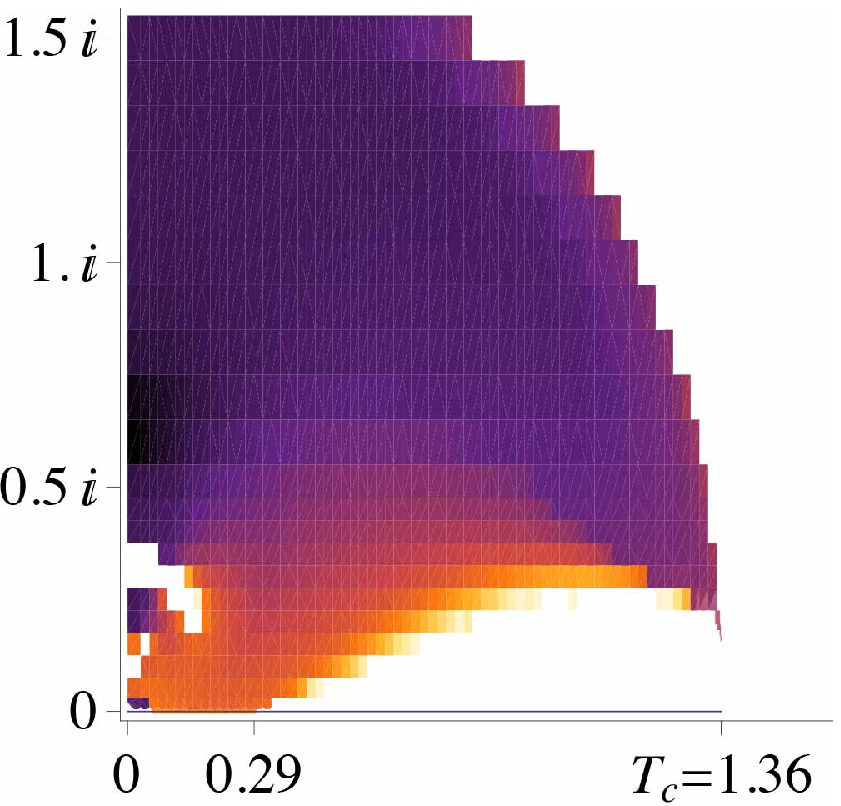}
\end{minipage}
\caption{
Distribution of zeros on the complex $T$ plane with $p=0.9$ and $2\beta H=0$ (left) and $2\beta H=10^{-4}i$ (right). The zeros tend to approach the real axis 
near $T=T_c$, where $T_c \simeq 1.36$ is the ferromagnetic critical temperature. It can be also observed that  
zeros reach the real axis at and below another lower critical temperature. 
This second critical temperature corresponds to the AT line 
as shown in fig.\ \ref{fig:pdpt}. 
The one-dimensional part of the zeros is revealed by introducing a small imaginary field and lies on the real axis in the right panel.
\label{fig:p09parasg_tp}}
\end{figure}

The distributions at high $p$ are also interesting. Figure \ref{fig:p09parasg_tp} shows the density at $p=0.9$ with $2\beta H=0$ (left) and $2\beta H=10^{-4}i$ (right). 
The critical temperature between the 
paramagnetic and ferromagnetic phases is known as 
$T_{c} \simeq 1.36$ at $p=0.9$. 
We can find that the zeros protrude and 
tend to distribute in a sharp curve 
near this critical temperature $T_{c}$, which  
implies the existence of the one-dimensional part of the zeros density 
near $T_{c}$. 
As mentioned in section \ref{sec5:remark}, however, 
we cannot directly detect this 
one-dimensional part by the current formalism using the two-dimensionally 
distributed population.  
Instead, we introduce a small pure imaginary field $2\beta H=10^{-4}i$ and 
investigate $g_{z1}$, the result of which is in the right panel of fig.\ \ref{fig:p09parasg_tp}. This figure clearly shows the presence of a critical temperature near ${\rm Re}(T)\simeq 1.36$; zeros have finite densities on the real axis below ${\rm Re}(T)\simeq 1.36$, which reflects that $g_{z1}(H=i\theta,T)$ 
is finite for $\theta \simeq 0$ in the ferromagnetic phase. On the other hand, the two-dimensional part of the zeros distribution, $g_{z2}$, also touch the real axis at and below ${\rm Re}(T) \simeq 0.29$. This situation is the same as fig.\ \ref{fig:p05parasg_tp} and indicates that the system is 
in the spin-glass phase in this region. 
As shown in later, this critical temperature quantitatively 
accords with the AT line, which we identify with the 
divergence point of the spin-glass susceptibility as the chapter $4$. 
This fact strongly supports our speculation that 
the two-dimensional part of the zeros distribution, $g_{z2}$, 
dominates the spin-glass behavior of the system.

\subsection{Phase diagram}\label{sec5:PD}
\begin{figure}[htbp]
\hspace{-5mm}
\begin{minipage}{0.5\hsize}
\includegraphics[width=1.0\linewidth]{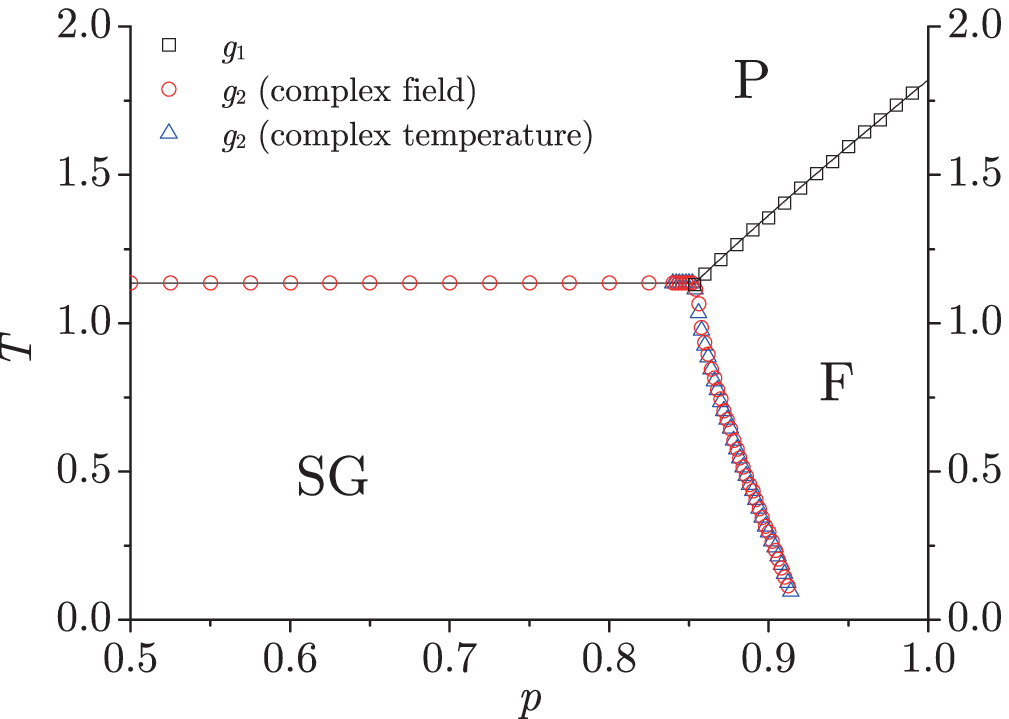}
\end{minipage}
\begin{minipage}{0.5\hsize}
\includegraphics[width=1.1\linewidth]{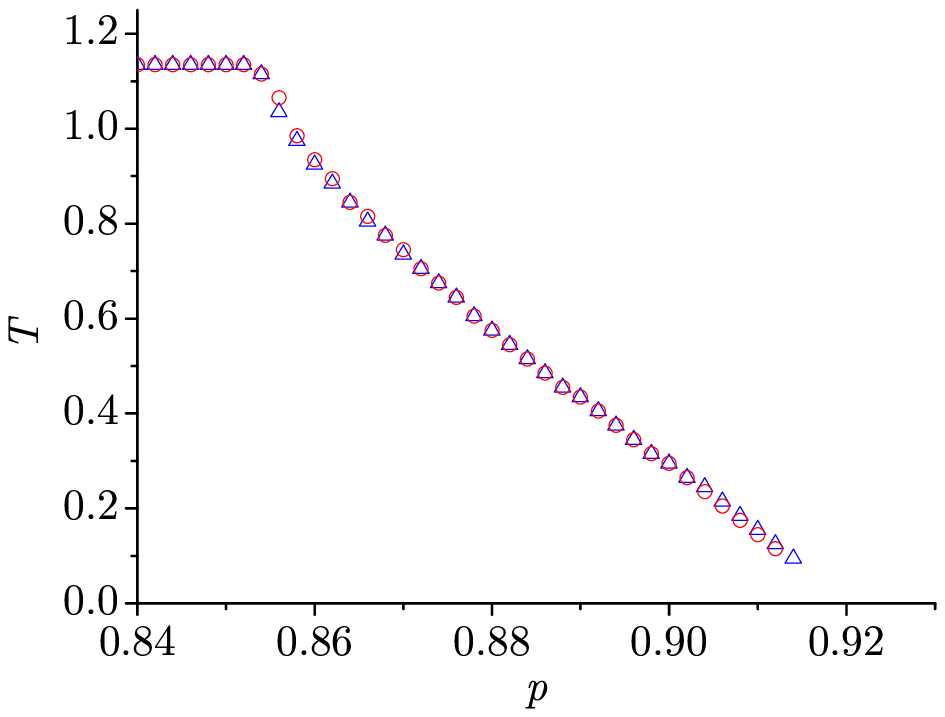}
\end{minipage}
\caption{
Phase diagrams on the $p$-$T$ plane written by the density of zeros. The solid lines are analytically given as $T_c = 1/\tanh^{-1}\left[{ {1/ \left( 4p-2\right) } }\right]$ (para and ferro) and $T_{SG} = 1/\tanh^{-1}\left[{ {1/ \sqrt{2} } }\right]$ (para and spin glass). Left: Circles and triangles are determined by two-dimensional density of zeros $g_{z2}$ on the complex field and temperature planes, respectively, which corresponds to the AT line as shown in section \ref{sec5:AT}. Squares are phase boundaries calculated by one-dimensional density of zeros $g_{z1}$ indicating the ferromagnetic order. Note that our result does not immediately conclude the phase boundary between mixed and spin-glass phases. Right: Blow up of the critical temperatures derived from $g_{z2}$ on the left panel.
\label{fig:pdpt}}
\end{figure}

The previous observations lead to an ansatz that two types of transitions are classified by one- and two-dimensional parts of the zeros distribution; 
the one-dimensional part $g_{z1}$ detects the ferromagnetic phase transition 
and the two-dimensional part $g_{z2}$ signals the emergence of the spin-glass
ordering. 
According to this ansatz, we can plot the $p$-$T$ phase diagram and 
the result is given in fig.\ \ref{fig:pdpt}.
Note that the $T$-$H$ phase diagram is also drawn by the same procedure. 
That diagram has also some interesting features but we here omit it 
because the $p$-$T$ phase diagram is sufficient for the current purpose 
of examining the AT instability of the Bethe lattice spin glass.
 
The resulting phase boundaries in fig.\ \ref{fig:pdpt} well 
agree with the analytical ones, $T_c = 1/\tanh^{-1}\left[{ {1/ \left( 4p-2\right) } }\right]$ (para and ferro) and $T_{SG} = 1/\tanh^{-1}\left[{ {1/ \sqrt{2} } }\right]$ (para and spin glass). 
This fact supports our ansatz that $g_{z1}$ and $g_{z2}$ correspond to the 
ferromagnetic and spin-glass ordering, respectively. 
To examine the relation between $g_{z2}$ and the AT instability further, 
in the next subsection 
we investigate the relation between 
the boundary obtained from $g_{z2}$ in fig.\ \ref{fig:pdpt} and 
the AT line, which we identify with the divergence of the spin-glass 
susceptibility.

\subsection{The divergence of the spin-glass susceptibility}\label{sec5:AT}
Let us remember the discussions in section \ref{sec4:AT} and 
develop a similar argument in the limit $n\to 0$. 
The spin-glass susceptibility is then given by 
\begin{equation}
\chi_{SG} =\frac{1}{N}\sum_{i,j}\left [
\left(\Part{\Ave{S_{i}}}{h_{j}}{} \right)^2 \right ]_{\V{J}}
=\sum_{j}
\left [
\left(\Part{\Ave{S_{0}}}{h_{j}}{} \right)^2 \right ]_{\V{J}}
\label{eq5:chiSG},
\end{equation}
Considering the fact that on a Bethe lattice 
an arbitrary pair of sites is connected by a single path 
and using the chain rule of the derivative operation, 
we can rewrite this quantity as
\begin{equation}
\chi_{SG}=\sum_{G=1}^{\infty}c(c-1)^{G-1}\left[ \left(\Part{\Ave{S_{0}}}{h_{G}}{}\right)^2
\right]_{\V{J}}
\propto 
\sum_{G}^{\infty}
(c-1)^{G}
\left[
\prod_{g=1}^G
\left(
\Part{\Wh{h}_{g-1}}{\Wh{h}_g}{}
\right)^2
\right]_{\V{J}}
,
\end{equation}
where the factor $c(c-1)^{G}$ denotes the number of sites of distance $G$ 
from the central site $0$.
This yields 
the divergence condition of $\chi_{SG}$ as 
\begin{equation}
\log(c-1)+\lim_{G\to \infty} \frac{1}{G}\log\left(
\left[
\prod_{g=1}^G
\left(
\Part{\Wh{h}_{g-1}}{\Wh{h}_g}{}
\right)^2
\right]_{\V{J}}
\right)=0.\label{eq5:AT}
\end{equation}

In order to estimate the divergence points of the spin-glass susceptibility at
finite temperatures, 
we numerically implement the calculation of 
the factor 
\begin{equation}
\left[\prod_{g=1}^G
\left(
\Part{\Wh{h}_{g-1}}{\Wh{h}_{g}}{}
\right)^2
\right]_{\V{J}}
=
\left[\left( 
\Part{\Wh{h}_{0}}{\Wh{h}_{G}}{}
\right)^2\right]_{\V{J}}.
\end{equation}
A naive consideration that 
the derivative can be estimated by the difference coming from a small deviation
yields
\begin{eqnarray}
\Part{\Wh{h}_0}{\Wh{h}_G}{} &\approx&  
\frac{{\Wh{h}_0 \left( {\Wh{h}_G  + \Delta \Wh{h}_G } \right) - \Wh{h}_0
 \left( {\Wh{h}_G } \right)}} {\Delta \Wh{h}_G }.
\label{eq:ATnum}
\end{eqnarray}
The procedure to evaluate this equation is as follows. 
We arrange two replicas of 
an identical population $\{\Wh{h}_{i}\}_{i=1}^{N_{\rm pop}}$ 
expressing the convergent CBD $\Wh{P}(\Wh{h})$.
In addition, we introduce 
a uniform perturbation ($\Delta \Wh{h}_{G}=10^{-4}$) 
into only one of two replicas, 
and then observe the square average of the variation 
$(1/N_{\rm pop})\sum_{i=1}^{N_{\rm pop}}(\Wh{h}_{i}(\Wh{h}_G  + \Delta \Wh{h}_G) - \Wh{h}_{i}(\Wh{h}_G))^2$
after a certain number of the cavity updates.

\begin{figure}[htbp]
\begin{minipage}{0.5\hsize}
\includegraphics[width=1\linewidth]{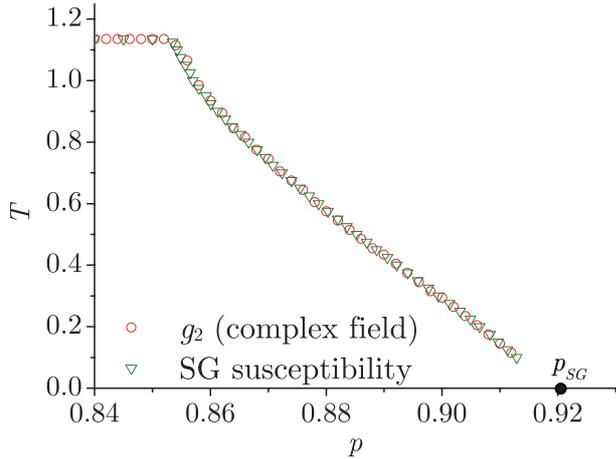}
\end{minipage}
\begin{minipage}{0.5\hsize}
\caption{
The divergence points of spin-glass susceptibility, which is denoted by 
the inverted triangles. 
The circles represent the phase boundary estimated by the two-dimensional distribution of zeros (fig.\ \ref{fig:pdpt}). They agree well each other.
The critical
value of the ferromagnetic bias $p_{SG} =
0.92067$ at zero temperature is indicated by
the colored circle.
\label{fig:sgsus}}
\end{minipage}
\end{figure}

In particular, 
we update two populations by $5000N_{\rm pop}$ iterations with the same set of $J_{ij}$. A critical line of the divergence of the spin-glass susceptibility is 
determined whether the square average is numerically zero or much larger than the original perturbation. The result is shown in fig.\ \ref{fig:sgsus} where the spin-glass susceptibility diverges below this line. This result well agrees  with the phase boundary drawn by $g_{z2}$ in fig.\ \ref{fig:pdpt}, 
which implies that the continuous distribution of $g_{z2}$ on the real 
axis corresponds to the divergence of the spin-glass 
susceptibility. 
Combining the behavior of $g_{z2}$ and 
 the discussions about the divergence of the spin-glass susceptibility 
in section \ref{sec2:AT}, 
we can expect that the current results reveal 
a physical aspect of the AT instability or the FRSB.
Namely,
the FRSB is related to the continuous singularities 
of the partition function, which is the common belief about the 
AT instability as mentioned in the beginning of this chapter. 
Consequently, we can reasonably conclude that the presented results 
leads to the AT instability of the Bethe lattice spin glass 
as the continuous singularities of the partition function 
and meets the purpose of this chapter.

Note that at zero temperature, evaluating the condition (\ref{eq5:AT}) can 
be performed in a more analytical manner. 
The result gives the critical value of the ferromagnetic bias
$p_{SG} = 0.92067$ which well 
agrees the estimated value by the finite-temperature results
derived by both the zeros and spin-glass susceptibility analyses. The details of the zero temperature
analysis are rather involved and presented in appendix \ref{app:AT^BL2}.

\section{Summary}
In this chapter, we founded the formulation to investigate the 
partition function zeros of Bethe lattices by utilizing the cavity method, 
for the purpose of investigating the AT instability of the Bethe lattice spin glasses. 
In the formulation, 
the cavity field distribution, which can be evaluated by using the 
population method, is directly linked to the density of zeros.
This enables us to assess the zeros density in infinite-size Bethe lattices 
by using the cavity method.
 
Using this formulation, 
we investigated the $\pm J$ model on the coordination $c=3$ Bethe lattice. 
The density of zeros consisted of two parts: One-dimensional and two-dimensional densities. 
We found that the two-dimensional density is related to the 
spin-glass ordering, while the one-dimensional part on the imaginary axis
of the external field corresponds to 
the ferromagnetic transition. 
These observations were confirmed by comparison with 
the conventional critical conditions.
The resultant $p$-$T$ phase diagram was drawn by those two different 
ways, and the
accordance between them is excellent. 

Our result about the zeros indicates that the system is 
singular everywhere in the spin-glass phase. 
This observation and the relation between the AT instability 
and the divergence of the spin-glass susceptibility enable us 
to reasonably conclude that the continuous singularities reveals 
a physical aspect of 
the AT instability and the FRSB phase is present in the Bethe 
lattice spin glass, which meets the purpose of this chapter.

Throughout this thesis, we consider the 1RSB and  
FRSB transitions separately. 
A possible further work is to clarify the behavior of zeros of the systems 
with the 1RSB. In the random energy model, it is known that 
the zeros touch the real axis at only the transition point 
\cite{Mouk1, Derrida:91}, which is in contrast to our result of 
the FRSB case.  
The current formalism can treat any RSB in the same framework and
can be expected to give useful information about other types of the RSB. 
Investigation along this line is quite hopeful and is currently under way.

\chapter{Conclusion}
In this thesis, the replica method was reviewed and examined 
from several viewpoints. 
The replica symmetry breaking (RSB) was the central concept about 
this issue 
and was investigated by several ways. 
The generating function of the partition function $Z$, 
$\phi(n)=(1/N)\log [Z^n]$ appearing in the 
replica method, played a key role in the investigation.

In chapter 2, we presented a review of the replica method in a form 
including recent descriptions of the method. 
The important concepts for this purpose are 
the probability distribution of the free energy and 
the pure states which are 
 disjointed sets of configurations in the phase space.  
Two characteristic exponents, the complexity characterizing the number of 
pure states and the rate function characterizing a small probability 
of the free energy, are naturally embedded in the replica procedures.
Practical calculations of those quantities were demonstrated on  
the fully-connected $p$-spin interacting model for several values of $p$. 
Especially, since the limit $p\to \infty$ makes the calculations 
easier,  
many aspects of the RSB in the one-step level (1RSB)
were elucidated in this limit. 
As a result, we found that 
the transition to the 1RSB phase
occurs due to the crisis of the complexity. 
This consequence was re-derived from the large deviations of the 
free energy, and it was revealed that 
the 1RSB is also signaled by the rate function of the probability distribution 
of the free energy. 
We also pointed out that the behavior of the rate function 
is closely related to the analytical 
properties of the generating function $\phi(n)$ 
and can be understood by some necessary conditions which 
should be satisfied by the correct $\phi(n)$.

On the other hand, another RSB, 
the full-step RSB (FRSB), is characterized by the de Almeida-Thouless (AT)
condition which is the local stability condition of the 
replica symmetric (RS) saddle-point. We demonstrated this point 
for $p=2$ and $3$ cases. The relation between the AT instability 
and the divergence of the spin-glass susceptibility was also discussed.  
In addition, for $p=3$, we pointed out that the dynamical 1RSB transition 
exists, which cannot be observed by the RS ansatz and require the 1RSB 
prescription with the complexity analysis. 
As a consequence of the above considerations, the 
phase diagrams for $p=2,3$ and $\infty$ on the $T$-$\beta n$ plane 
were presented. 

Besides, in the end of chapter 2, 
the microscopic description of the pure states were presented. 
To this end, 
the formulation by Thouless, Anderson and Palmer (TAP) was discussed 
and 
the equivalence with the replica result 
was shown; 
the characteristic exponent of the number of solutions 
of the TAP equation becomes 
identical to that of the complexity. 
We also showed that the solution of the TAP equation at zero temperature 
can be identified with the spin configurations stable against any single spin flip, 
which provides a simple interpretation of the pure state and is useful 
for identifying the pure states by numerical methods.

In chapter 3, we analyzed the Ising perceptron, which is a model of a neuron, 
by the replica method. 
To study the structure of the synaptic weight space, we intensively used the concept of the pure states and calculated 
the complexity in the 1RSB level. 
The complexity analysis revealed that the weight space 
consists of many clusters (pure states) for $\alpha\leq \alpha_{s}=0.833...$ 
where $\alpha=M/N$ is the ratio of 
the number of the embedded patterns $M$ to the system size $N$.
Besides, it was shown that the weight space is equally dominated by a single large cluster of exponentially many weights and exponentially many small clusters of a single weight. 
This fact means that any information about the middle-size clusters cannot be obtained in the replica method due to the convex downward form of the 
cluster-size distribution. 
On the other hand, for $\alpha\geq \alpha_{s}$, we 
calculated the rate function to evaluate the small probability that a given
set of random patterns is atypically separable by the Ising perceptron. 
The result showed that 
for $\alpha_{s}\leq \alpha\leq \alpha_{GD}=1.245...$ 
the rate function takes the minimum for $n_{s}>0$ and the system freezes out 
below $n_{s}$, which is a new type of frozen RSB phases and is not 
directly related the conventional 1RSB ones.
For $n \geq n_{GD}$, it was shown that the analyticity of the rate function drastically changes, which implies that the dominant configuration of the atypically separable patterns exhibits a phase transition at this critical ratio. 
Extensive numerical experiments were also conducted and supported 
the above theoretical predictions.

In chapter 4, zeros of the $n$th moment of the partition function $[Z^n]$ were 
investigated to directly reveal the analyticity breaking of $\phi(n)$
in a vanishing temperature limit $\beta \to \infty$, $n\to 0$ keeping $y=\beta n\sim O(1)$ by combining the replica and cavity methods.
In this limit, the
moment parameterized by $y$ 
characterizes the distribution of the ground-state energy.
We numerically investigated the zeros for $\pm J$ 
Ising spin glass models defined on several Cayley trees and
ladders. 
For several Cayley trees,  we
found that the zeros tend to approach the real axis of $y$ 
in the thermodynamic limit, which implies 
that the moment cannot be described by a single analytic function of 
$y$ as the system size tends to infinity. 
To explore the possible links of those analyticity breaking to the RSB, 
we examined the analytical properties of $\phi(n)$ and 
assessed the spin-glass susceptibility which can be 
identified with the AT instability. 
To this end we employed two relatives of Cayley trees, Bethe lattices and 
regular random graphs.
The result revealed that the analyticity breaking indicated by the zeros 
of Cayley trees is relevant to neither the 1RSB and FRSB.
We concluded that the peculiarity of Cayley trees and Bethe lattices 
makes the 1RSB not appear in those systems.
On the other hand,   
the FRSB required some delicate discussions.   
According to the analysis of the modified generating function introduced 
interactions between replicas, 
we finally inferred that 
the AT instability cannot be detected by the current formula. 

In chapter 5, to remove a possibility that the FRSB does not occur on Cayley 
trees and Bethe lattices considered in chapter 4, 
we investigated the zeros of the partition function of the $\pm J$ model with respect to the external field $H$ and temperature $T$ on a Bethe lattice 
with the coordination number $c=3$, by utilizing the cavity method.
The result 
indicates that once 
the spin-glass phase emerges, the two-dimensional part of the zeros 
continuously tough the real axis, which means that the system is 
singular everywhere in the spin-glass phase. 
Combining this observation and the relation between the AT instability 
and the divergence of the spin-glass susceptibility, 
we could reasonably conclude 
that the continuous singularities of the partition function 
reveals a physical aspect of 
the AT instability and the FRSB phase is present in the Bethe 
lattice spin glass.
\\

Main results of this thesis are summarized as follows:
\begin{enumerate}
\item{Clarifying the behavior of the generating function 
when the complexity and rate function are not convex upward (chapter 3).}
\item{Comparing the analytical value of the rate function with the numerical one qualitatively
and supporting the validity of the replica method (chapter 3).}
\item{Proposing a new method to investigate zeros of the moment $[Z^n]$ with respect to the replica number $n$ and 
finding the analyticity breaking of $\phi(n)$ by using the zeros (chapter 4).}
\item{Clarifying the relation among Cayley trees, Bethe lattices, and regular random graphs in the context of the RSB (chapter 4).}
\item{
Drawing a phase diagram of a Bethe lattice spin glass in terms of the partition function zeros and 
reasonably identifying the FRSB phase 
by the continuous singularities of the free energy (chapter 5).}
\end{enumerate}

Although the spin-glass theory has its long history, 
the essential comprehension of
physics of spin glasses is still lacking. One of the main controversial points in the theory
is whether the mean-field description is applicable to finite-dimensional systems or not.
To obtain the solution to this problem, much effort have been made not only on directly
studying finite-dimensional systems but also on improving the mean-field theory itself.
The presented researches in this thesis can be regarded as one of such improvements in
that our results provide some new aspects of the replica method being one of the central
methods of the mean-field theory of spin glasses.

Besides, our results suggest further possibilities. Deeper understanding about pure
states can provide some insights on several other problems, e.g. structural glasses, optimization
problems and so on. The zeros formulations given in chapter 4 and 5 are in
principle applicable to finite-dimensional systems, which can lead to further comprehension
of finite-dimensional spin glasses. Also, our results of the Lee-Yang zeros in chapter
5 can provide a possibility of detecting the RSB in real experiments by observing the
response of the system to the external field, like the Griffiths singularities \cite{Chan:06}.

In the light of the above possibilities, we have several hopeful future works.

Lessons from chapter 3 tell us that in other research fields there probably 
exist many other
systems exhibiting extraordinary behavior of the generating function, complexity and rate
function. This implies that investigation of such systems can reveal not only the properties
of the systems but also the remaining mysteries of the replica theory. Researches along
this line are already one of the rapidly-expanding mainstreams of the current spin-glass
theory and will continue to develop at least for the near future.

The replica-zeros formulation in chapter 4 should be improved to be a more tractable
form. In the current formulation, the replica zeros require huge computational time
which prevents us from treating systems of large sizes. Moreover, we should clarify the
relation between the replica zeros and the AT instability to investigate whether the AT
instability occurs in finite-dimensional systems or not. The formulation of the inter-replicas
interactions may be useful for this purpose and should be developed in the future.

The Lee-Yang zeros in chapter 5 may be most hopeful in clarifying the properties of
real spin glasses, since they are directly related to the external field being controllable
in real experiments. To give concise predictions for real experiments, we should obtain
more detailed information about the distribution of the Lee-Yang zeros. The precise
functional form of the distribution may enable to distinguish the RSB response from the
normal one in the scaling form of the magnetization with respect to the external field.
Obviously, we should find out the behavior of the zeros distribution in different types of
RSB. Investigation of this point is a promising research and is currently underway.

It would be expected that the future direction of the spin-glass theory is classified
broadly into two lines. One is the mathematical foundation of the mean-field theory in a
more rigorous way. The other is the application of the spin-glass theory to a wide variety
of different systems; real spin glasses, structural glasses, information processing tasks, and
so on. The results presented in this thesis probably contribute to both of the two directions.
This is because the mathematical structure of the replica method is closely related to both of
the mathematics and physics of spin glasses. We hope that the reviews and results in
this thesis become a basis to researches in such directions and inspire further study and
understanding of spin glasses and beyond.

\appendix
\chapter{Calculations for chapter 3}\label{app:IP}
\section{Derivation of the RS saddle-point equations and the limit $q\to 1$. }\label{app:IP-saddle}
Differentiation of eq.\ (\ref{eq3:phiRS}) with respect to $q$ yields
\begin{equation}
\Part{\phi}{q}{}=-\frac{1}{2}n(n-1)\Wh{q}+\alpha 
\frac{
\int Dz \Part{}{q}{}E_{\beta}^{n}
\left(
\sqrt{\frac{q}{1-q}}z
\right)
 }{\int Dz E_{\beta}^{n}
\left(
\sqrt{\frac{q}{1-q}}z
\right)
}=0.\label{eq3:diff.q}
\end{equation}
The following integration requires some algebras
\begin{equation}
\int Dz \Part{}{q}{}E_{\beta}^{n}
\left(
\sqrt{\frac{q}{1-q}}z
\right)=
n
\int Dz E_{\beta}^{n-1}\Part{E_{\beta}}{q}{}.\label{eq3:intE}
\end{equation}
The differentiation of $E_{\beta}$ with respect to $q$ gives
\begin{equation}
\Part{
}{q}{}E_{\beta}
\left(
\sqrt{\frac{q}{1-q}}z
\right)
=
-\frac{1-e^{-\beta}}{2\sqrt{2\pi}(1-q)^2}
\sqrt{
\frac{1-q}{q}
}
z e^{
-\frac{1}{2}\frac{q}{1-q}z^2
}
\end{equation}
Inserting this expression into eq.\ (\ref{eq3:intE}) and integrating by parts 
with respect to $z$ give
\begin{eqnarray}
&&\int Dz E_{\beta}^{n-1}ze^{-\frac{1}{2}\frac{q}{1-q}z^2} =(1-q)(n-1)\int \frac{dz}{\sqrt{2\pi}}
e^{-\frac{1}{2}\frac{1}{1-q}z^2}
E_{\beta}^{n-2}\Part{E_{\beta}}{z}{}\nonumber \\
&&=
(1-q)(n-1)
 \int \frac{dz}{\sqrt{2\pi}}
e^{-\frac{1}{2}\frac{1}{1-q}z^2}
E_{\beta}^{n-2}
\frac{-(1-e^{-\beta})}{\sqrt{2\pi}}
\sqrt{
\frac{q}{1-q}
}
e^{-\frac{1}{2}\frac{q}{1-q}z^2}.
\end{eqnarray}
Substituting this equation into eq.\ (\ref{eq3:diff.q}) and reducing some factors, we finally get the saddle-point equation (\ref{eq3:qhat}).
The other saddle-point equation (\ref{eq3:q}) is also derived in a similar 
way.

Next we take the limit $q\to 1$ in eqs.\ (\ref{eq3:q}-\ref{eq3:qhat}). 
In this limit, the argument of $E$, $z\sqrt{q/1-q}$, 
diverges to $\infty$ or $-\infty$ depending on the sign of $z$ and 
we need the asymptotic behavior of $E(x)$
\begin{equation}
E(x)\to \left\{
 \begin{array}{cc}
\hspace{-3mm} \frac{1}{\sqrt{2\pi}}
\left(\frac{1}{x}-\frac{1}{2x^3} \right)e^{-\frac{1}{2}x^2}\to 0,\, (x\to \infty) \\
	\hspace{17mm}   1, \hspace{24mm} (x\to-\infty)
		    \end{array}
			  \right.\label{eq3:asymptotic^H}
,
\end{equation}
which leads to $E_{\beta}(x)\to e^{-\beta}$ for $x\to \infty$ and $1$ for 
$x\to -\infty$.
Hence, we can calculate the asymptotic behavior of $\Wh{q}$ as
\begin{eqnarray}
\Wh{q}\to
\frac{ \alpha }{ 2\pi }
\frac{ ( 1-e^{-\beta} )^2 }{ 1-q }
\frac{
\int_{-\infty}^{0} 
Dz 
e^{ -\frac{1}{1-q}z^2} 
+
\int_{0}^{\infty} Dz 
e^{ -\frac{1}{1-q} z^2 } 
e^{-\beta n}
}
{
\int_{-\infty}^{0} Dz 
+
\int_{0}^{\infty} Dz 
e^{-\beta n}
}\nonumber \\
=
\frac{ \alpha }{ 2\pi }
\frac{ ( 1-e^{-\beta} )^2 }{ 1-q }
\frac{\sqrt{\frac{1-q}{1+q}}(1+e^{\beta n})/2 }{ (1+e^{\beta n})/2 }
\to\frac{ \alpha }{ 4\pi }
\frac{ ( 1-e^{-\beta} )^2 }{ \sqrt{1-q} }\to \infty.\label{eq3:RS2qhat}
\end{eqnarray}
This condition $\Wh{q}\to\infty$ enables us to calculate 
the following integration
\begin{eqnarray}
&&\int Dz (2\cosh \sqrt{\Wh{q}}z)^n
\to 
\int Dz e^{n\sqrt{\Wh{q}}|z|}
=\frac{2}{\sqrt{2\pi}} 
\int_{0}^{\infty} 
e^{-\frac{1}{2}z^2+n\sqrt{\Wh{q}}z}
=\nonumber \\
&&\frac{2}{\sqrt{2\pi}} 
\int_{0}^{\infty} dz
e^{-\frac{1}{2}\left(z-n\sqrt{\Wh{q}} \right)^2}
e^{\frac{1}{2}n^2\Wh{q}}
\to 2e^{\frac{1}{2}n^2\Wh{q}}.\label{eq3:asymptotic^ch}
\end{eqnarray}
Substituting these results into eq.\ (\ref{eq3:phiRS}), we obtain the RS2 
solution (\ref{RS2}).
 
\section{Derivation of the AT condition}\label{app:IP-AT}
We consider small deviations of eq.\ (\ref{eq3:phi}) 
around the RS saddle points.
We use the notation 
$q^{\mu\nu}=q+y^{\mu\nu}$, $\Wh{q}^{\mu\nu}=\Wh{q}+\Wh{y}^{\mu\nu}$ and expand
eq.\ (\ref{eq3:phi}) with respect to the perturbations 
$y^{\mu\nu}$ and $\Wh{y}^{\mu\nu}$. 
The terms of first order of $y^{\mu\nu}$ and $\Wh{y}^{\mu\nu}$ 
becomes automatically $0$ because 
we impose the saddle-point conditions with respect to $q$ and $\Wh{q}$. 
Hence, the relevant terms in the first term of eq.\ (\ref{eq3:phi}) 
become
\begin{eqnarray}
-\sum_{\mu<\nu}
q^{\mu\nu}\Wh{q}^{\mu\nu}
\approx
-\sum_{\mu<\nu}
y^{\mu\nu}\Wh{y}^{\mu\nu}.
\end{eqnarray}
Similarly, for the second term of eq.\ (\ref{eq3:phi}), 
\begin{eqnarray}
\log \Tr{\{S^{\mu}\} } e^{
\sum_{\mu<\nu}
\Wh{q}^{\mu\nu}S^{\mu}S^{\nu}
}
=\log\Tr{\{S^{\mu}\} } e^{
\Wh{q}\sum_{\mu<\nu}S^{\mu}S^{\nu}
+
\sum_{\mu<\nu}\Wh{y}^{\mu\nu}S^{\mu}S^{\nu}
}\nonumber\\
\approx
\sum_{\mu<\nu}
\sum_{\delta<\omega}
\Wh{y}^{\mu\nu}
\Wh{y}^{\delta\omega}
\left(
\Ave{S^{\mu}S^{\nu}S^{\delta}S^{\omega}}
-
\Ave{S^{\mu}S^{\nu}}\Ave{S^{\delta}S^{\omega}}
\right),
\end{eqnarray}
where the brackets $\Ave{(\cdots)}$ denote 
the following average 
\begin{equation}
\Ave{(\cdots)}=\frac{
\Tr{\{S^{\mu}\} }(\cdots) e^{\Wh{q}\sum_{\mu<\nu}S^{\mu}S^{\nu}}
}{
\Tr{\{S^{\mu}\} } e^{\Wh{q}\sum_{\mu<\nu}S^{\mu}S^{\nu}}
}.
\end{equation}
To expand the third term of eq.\ (\ref{eq3:phi}), we need some more algebras. 
First, we write down the general form of the probability distribution of 
$[f(\V{u})]_{\V{u}}$
\begin{equation}
[f(\V{u})]_{\V{u}}
=\sqrt{
\frac{\det{(Q^{-1})}}{(2\pi)^N}
}
\int 
\left(
\prod_{\mu=1}^{n}du^{\mu}
\right)
e^{-\frac{1}{2}\bra{u}Q^{-1}\ket{u}}
f(\V{u})
\end{equation}
where $\ket{u}$ expresses 
the $u$ vector whose $\mu$th component is $u^{\mu}$ and
$Q$ is the matrix whose elements are spin-glass order parameters 
\begin{equation}
Q_{\mu\nu}=q^{\mu\nu}.
\end{equation}
This distribution satisfies relations (\ref{eq3:moments^u}).
We here employ the following identity 
\begin{equation}
e^{
-\frac{1}{2}\bra{u}Q^{-1}\ket{u}
}
=
\sqrt{
\frac{\det{Q}}{(2\pi)^N}
}
\int 
\left(
\prod_{\mu=1}^{n} dv^{\mu}
\right)
e^{
-\frac{1}{2} \bra{v} Q\ket{v}
-
i\braket{v}{u}
}.
\end{equation}
Using these formulas, we get
\begin{eqnarray}
&&\log [f(\V{u})]_{\V{u}}
=\log
\frac{1}{(2\pi)^N}
\int
\left(
\prod_{\mu=1}^{n} du^{\mu}dv^{\mu}
\right)
e^{
-\frac{1}{2} \bra{v} Q\ket{v}
-i\braket{v}{u}
}f(\V{u})\nonumber \\
&&=
\log
\frac{1}{(2\pi)^N}
\int
\left(
\prod_{\mu=1}^{n} du^{\mu}dv^{\mu}
\right)
e^{
-\frac{1}{2} \bra{v}Q_{\rm RS}\ket{v}
-\sum_{\mu<\nu}y^{\mu\nu}v^{\mu}v^{\nu}
-i\braket{v}{u}
}f(\V{u})\nonumber \\
&&
\approx
\sum_{\mu<\nu}
\sum_{\delta<\omega}
y^{\mu\nu}y^{\delta\omega}
\left(
[v^{\mu}v^{\nu}v^{\delta}v^{\omega}]_{v}
-
[v^{\mu}v^{\nu}]_{v}[v^{\delta}v^{\omega}]_{v}
\right)
,
\end{eqnarray}
where the brackets $[(\cdots)]_{v}$ denote
the following average
\begin{equation}
[(\cdots)]_{v}=
\frac{
\int
\left(
\prod_{\mu=1}^{n} du^{\mu}dv^{\mu}
\right)
(\cdots)
e^{
-\frac{1}{2} \bra{v}Q_{\rm RS}\ket{v}
-i\braket{v}{u}
}f(\V{u})
}{
\int
\left(
\prod_{\mu=1}^{n} du^{\mu}dv^{\mu}
\right)
e^{
-\frac{1}{2} \bra{v}Q_{\rm RS}\ket{v}
-i\braket{v}{u}
}f(\V{u})
}.
\end{equation}
This average with respect to $v$ 
can be transformed to the average with respect to $u$ by using the integral 
by parts
\begin{eqnarray}
&&\int
\left(
\prod_{\mu=1}^{n} du^{\mu}dv^{\mu}
\right)
v^{\mu}
e^{
-\frac{1}{2} \bra{v}Q_{\rm RS}\ket{v}
-i\braket{v}{u}
}f(\V{u})\nonumber \\
&&=\frac{1}{i}
\int
\left(
\prod_{\mu=1}^{n} du^{\mu}dv^{\mu}
\right)
e^{
-\frac{1}{2} \bra{v}Q_{\rm RS}\ket{v}
-i\braket{v}{u}
}
\Part{ f(\V{u}) }{ {u^{\mu}} }{}.
\end{eqnarray}
Hence, we get
\begin{eqnarray}
&&\sum_{\mu<\nu}
\sum_{\delta<\omega}
y^{\mu\nu}y^{\delta\omega}
\left(
[v^{\mu}v^{\nu}v^{\delta}v^{\omega}]_{v}
-
[v^{\mu}v^{\nu}]_{v}[v^{\delta}v^{\omega}]_{v}
\right)
\nonumber \\
&&=
\sum_{\mu<\nu}
\sum_{\delta<\omega}
\left(
y^{\mu\nu}y^{\delta\omega}
\frac
{
[\partial_{\mu\nu\delta\omega} f]_{\V{u}}
}
{
[f]_{\V{u}}
}
-
\frac{
[\partial_{\mu\nu} f]_{\V{u}}
}{
[f]_{\V{u}}
}
\frac{
[\partial_{\delta\omega} f]_{ \V{u} }
}{
[f]_{\V{u}}
}
\right)
,
\end{eqnarray}
where the symbol $\partial_{x_{1}^{n_{1}}\cdots x_{l}^{n_{l}}} $ means
\begin{equation}
\partial_{x_{1}^{n_{1}}\cdots x_{l}^{n_{l}}}
=
\prod_{i=1}^{l}\Part{}{x_{i}}{n_{i}}
\end{equation}
and note that 
$[(\cdots)]_{\V{u}}$ is the same average as defined in Sec. 
\ref{sec3:formulation}.
Summarizing above results, 
we can derive the explicit form of the Hessian $G$ 
around the RS ansatz 
\begin{eqnarray}
&&G=
\sum_{\mu<\nu}
\sum_{\delta<\omega}
\Biggl(
-
\delta_{\mu\nu,\delta\omega}
\Wh{y}^{\mu\nu}y^{\mu\nu}
+
\Wh{y}^{\mu\nu}
\Wh{y}^{\delta\omega}
\left(
\Ave{S^{\mu}S^{\nu}S^{\delta}S^{\omega}}
-
\Ave{S^{\mu}S^{\nu}}\Ave{S^{\delta}S^{\omega}}
\right)\nonumber \\
&&+
\alpha 
y^{\mu\nu}y^{\delta\omega}
\left(
\frac{
[\partial_{\mu\nu\delta\omega} f]_{\V{u}}
}{
[f]_{\V{u}}
}
-
\frac{
[\partial_{\mu\nu} f]_{\V{u}}
}{
[f]_{\V{u}}
}
\frac{
[\partial_{\delta\omega} f]_{ \{u^{\mu}\} }
}{
[f]_{\V{u}}
}
\right)
\Biggr)
\end{eqnarray}
Here we classify the elements of $G$ by using the following notations
\begin{eqnarray}
P= \alpha
\left(
[\partial_{\mu^2\nu^2} f]_{\V{u}}/
[f]_{\V{u}}
-
([\partial_{\mu\nu} f]_{\V{u}}/
[f]_{\V{u}}
)^2
\right)
, \label{eq3:P}\\
Q=\alpha
\left(
[\partial_{\mu^2\nu\delta} f]_{\V{u}}/
[f]_{\V{u}}
-
[\partial_{\mu\nu} f]_{\V{u}}
[\partial_{\mu\delta} f]_{\V{u}}/
[f]_{\V{u}}^2
\right)
,\\
R=\alpha
\left(
[\partial_{\mu\nu\delta\omega} f]_{\V{u}}/
[f]_{\V{u}}
-
[\partial_{\mu\nu} f]_{\V{u}}
[\partial_{\delta\omega} f]_{\V{u}}/
[f]_{\V{u}}^2
\right)
,\\
\Wh{P}=1-\Ave{S^{\mu}S^{\nu}}^2=1-q^2, \\
\Wh{Q}=\Ave{S^{\mu}S^{\nu}}-\Ave{S^{\mu}S^{\nu}}\Ave{S^{\mu}S^{\delta}}
=q-q^2,\\
\Wh{R}=\Ave{S^{\mu}S^{\nu}S^{\delta}S^{\omega}}
-\Ave{S^{\mu}S^{\nu}}\Ave{S^{\delta}S^{\omega}}
=r-q^2,\label{eq3:Rhat}
\end{eqnarray}
where $r=\Ave{S^{\mu}S^{\nu}S^{\delta}S^{\omega}}$.
Using these notations, 
we can express the Hessian $G$ as 
\begin{equation}
G=
 \left(\begin{array}{c|c}	      
   \begin{array}{ccc}
    P & Q\cdots Q & R\cdots R \\
  & \ddots & \\
 Q\cdots Q  & R\cdots R & P
\end{array}                      & I \\ \hline
I & 	\begin{array}{ccc}
    \Wh{P} & \Wh{Q}\cdots \Wh{Q} & \Wh{R}\cdots \Wh{R} \\
  & \ddots &  \\
 \Wh{Q} \cdots \Wh{Q}  & \Wh{R}\cdots \Wh{R} & \Wh{P}
 \end{array}
\end{array}
 \right), \label{eq3:G}
\end{equation}
where $I$ is the unit matrix of rank $n(n-1)/2$.

Let us find the eigenvectors of this matrix 
by a heuristic approach found by de Almeida and Thouless.
The first eigenvector $\V{s}_{1}$ is obtained by assuming
$y^{\mu\nu}=a$ and $\Wh{y}^{\mu\nu}=\Wh{a}$
for any $\mu,\nu$
.
The upper half of the eigenvalue equation 
$G\V{s}_{1}=\lambda_{1}\V{s}_{1}$
 gives
\begin{equation}
\left(
P+2(n-2)Q+\frac{1}{2}(n-2)(n-3)R
\right)a+\Wh{a}=\lambda_{1}a,
\end{equation}
and the lower half yields
\begin{equation}
a+\left(
\Wh{P}+2(n-2)\Wh{Q}+\frac{1}{2}(n-2)(n-3)\Wh{R}
\right)\Wh{a}=\lambda_{1}\Wh{a}.
\end{equation}
These equations have the eigenvalue 
$\lambda_{1}$ with non-vanishing $a,\Wh{a}$, 
which must satisfies the following relation
\begin{equation}
\lambda_{1}^2-(X_{1}+\Wh{X}_{1})\lambda_{1}+(X_{1}\Wh{X}_{1}-1)=0\label{eq3:lambda1},
\end{equation}
where $X_{1}=P+2(n-2)Q+(n-2)(n-3)R/2$ and 
$\Wh{X}_{1}=\Wh{P}+2(n-2)\Wh{Q}+(n-2)(n-3)\Wh{R}/2$.
This equation says that this mode constructs a two-dimensional space.

The next type of solution $\V{s}_{2}$ 
is obtained by treating a replica $\theta$ 
as a special one.
We assume $\theta=1$ without loss of generality.
This solution $\V{s}_{2}$ has 
$y^{\mu\nu}=b$ and $\Wh{y}^{\mu\nu}=\Wh{b}$
when $\mu$ or $\nu$ is equal to $1$, 
$y^{\mu\nu}=c$ and $\Wh{y}^{\mu\nu}=\Wh{c}$ 
otherwise.
The first low of the eigenvalue equation 
$G\V{s}_{2}=\lambda_{2}\V{s}_{2}$
gives
\begin{equation}
(P+(n-2)Q)b+
\left(
(n-2)Q  
\frac{1}{2}(n-2)(n-3)R
\right)c+\Wh{b}=\lambda_{2}b,
\end{equation}
and the first low of the lower half yields
\begin{equation}
b+\left(
\Wh{P}+(n-2)\Wh{Q}
\right)\Wh{b}
+
\left(
(n-2)\Wh{Q}
+\frac{1}{2}(n-2)(n-3)\Wh{R}
\right)\Wh{c}=\lambda_{2}\Wh{b}.
\end{equation}
We here use the orthogonal condition 
of the solutions $\V{s}_{2}$ and $\V{s}_{1}$.
We expect that 
the upper half of $\V{s}_{2}$ 
is orthogonal with that of $\V{s}_{1}$ 
independent of the lower half space.
This leads to
\begin{equation}
(n-1)b+\frac{1}{2}(n-1)(n-2)c=0,
\end{equation}
and the same relation holds for $\Wh{b}$ and $\Wh{c}$.
Substituting these conditions, we get
\begin{eqnarray}
&&\left(
P+(n-4)Q-(n-3)R
\right)b
+\Wh{b}=\lambda_{2}b,\\
&&\left(
\Wh{P}+(n-4)\Wh{Q}-(n-3)\Wh{R}
\right)\Wh{b}
+b=\lambda_{2}\Wh{b},
\end{eqnarray}
and with non-vanishing $b$ and $\Wh{b}$, 
this leads to
\begin{equation}
\lambda_{2}^2-(X_{2}+\Wh{X}_{2})\lambda_{2}+(X_{2}\Wh{X}_{2}-1)=0
\label{eq3:lambda2},
\end{equation}
where $X_{2}=P+(n-4)Q-(n-3)R$ and 
$\Wh{X}_{2}=\Wh{P}+(n-4)\Wh{Q}-(n-3)\Wh{R}$.
There are $n$ possible choices of the special replica $\theta$
and 
there are two eigenvalues for the solutions $\V{s}_{1}$
and $\V{s}_{2}$, which means that $\V{s}_{1}$
and $\V{s}_{2}$ construct $2n$-dimensional space.

The third mode $\V{s}_{3}$ is obtained by treating two replicas 
$\theta,\omega$ as special ones. 
This solution $\V{s}_{3}$ has 
$y^{\theta\omega}=d$ and $\Wh{y}^{\theta\omega}=\Wh{d}$,
$y^{\theta\mu}=e$ and $\Wh{y}^{\omega\mu}=\Wh{e}$,
and 
$y^{\mu\nu}=f$ and $\Wh{y}^{\mu\nu}=\Wh{f}$
otherwise.
Following the similar argument to the case of $\V{s}_{2}$, 
we obtain
\begin{equation}
\lambda_{3}^2-(X_{3}+\Wh{X}_{3})\lambda_{3}+(X_{3}\Wh{X}_{3}-1)=0
\label{eq3:lambda3},
\end{equation}
where $X_{3}=P-2Q+R$ and 
$\Wh{X}_{2}=\Wh{P}-2\Wh{Q}+\Wh{R}$.
This solution constructs $n(n-3)$-dimensional space and all 
the eigenvalues are exhausted by these three modes.

For the stability of the saddle point, 
all of the eigenvalues must be non-negative. 
This condition corresponds to 
\begin{equation}
\forall{i},\hspace{5mm}X_{i}\Wh{X}_{i}\leq 1.\label{eq3:positivity}
\end{equation}
Note that the desired condition is $X_{i}\Wh{X}_{i}\leq 1$ 
but not $X_{i}\Wh{X}_{i}\geq 1$.
That is 
because $\Wh{q}^{\mu\nu}$ is originally 
pure imaginary variable, which means that 
$\delta q^{\mu\nu}\delta \Wh{q}^{\mu\nu}$ is  
associated with a factor $i$ and
$\delta \Wh{q}^{\mu\nu}\delta \Wh{q}^{\mu\nu}$
is with a factor $-1$. Hence, if we change the variable 
from $\Wh{q}^{\mu\nu}$ to $i\Wh{q}^{\mu\nu}$, 
the diagonal block of lower half of $G$ yields a factor $-1$
and the off-diagonal part becomes $iI$, which leads to 
the positivity condition of the eigenvalues as 
eq.\ (\ref{eq3:positivity}). 

The most relevant mode is
$\V{s}_{3}$ and to check this condition we next 
calculate eqs.\ (\ref{eq3:P}-\ref{eq3:Rhat}).
The order parameter $q$ is already given by eq.\ (\ref{eq3:q}) 
and the quantity 
$\Ave{S^{\mu}S^{\nu}S^{\omega}S^{\delta}}=r$ 
is calculated in a similar way to 
the derivation of eq.\ (\ref{eq3:q}). The result is 
\begin{equation}
r
=\frac{\int Dz \cosh^n \sqrt{\Wh{q}}z \tanh^4 \sqrt{\Wh{q}}z }{
\cosh^n \sqrt{\Wh{q}}z
}.\label{eq3:r} 
\end{equation}
The factor $P-2Q+R$ requires to assess 
$[\partial_{\theta}f]_{\V{u}}$.
This factor is calculated as
\begin{eqnarray}
&&[\partial_{\theta}f]_{\V{u}}=
\int Dz 
\int 
\prod_{\mu=1}^{n} Dx^{\mu} 
\Part{}{{u^{\theta}}}{}
f(
\V{u}
)
\nonumber \\
&&=
\int Dz 
\prod_{\mu\neq\theta}
\left(
\int Dx^{\mu} 
f(
u^{\mu}
)
\right)
\left(\int Dx^{\theta}
\Part{}{{u^{\theta}}}{}
f(
u^{\theta}=\sqrt{1-q}x^{\theta}+\sqrt{q}z
)
\right)
\nonumber \\
&&=
\int Dz 
\left(
E_{\beta}
\left(
\sqrt{\frac{q}{1-q}}z
\right)
\right)^{n-1}
\left(
\frac{1}{\sqrt{q}}
\Part{}{z}{}
E_{\beta}
\left(
\sqrt{\frac{q}{1-q}}z
\right)
\right)\nonumber\\
&&=
\frac{1}{\sqrt{1-q}}
\int Dz 
E_{\beta}^{n}\left(
\sqrt{\frac{q}{1-q}}z
\right)
\frac{E_{\beta}'\left(
\sqrt{\frac{q}{1-q}}z
\right)}{E_{\beta}\left(
\sqrt{\frac{q}{1-q}}z
\right)},
\end{eqnarray}
where we used the explicit form of the function 
$f(\V{u})=\exp(-\beta\sum_{\mu=1}^{n}\theta(u^{\mu}))$
and denoted $E_{\beta}'(x)=dE_{\beta}(x)/dx$.
All the derivatives in the factor $P-2Q+R$ can be similarly calculated. 
The result is 
\begin{equation}
P-2Q+R
=\frac{\alpha}{(1-q)^2}
\frac{
\int Dz 
E_{\beta}^{n}
\left(
\frac{E_{\beta}''}{E_{\beta}}
-
\left(
\frac{E_{\beta}'}{E_{\beta}}
\right)^{2}
\right)^2
}
{
\int Dz 
E_{\beta}^{n}
}
,
\end{equation}
where we omit the argument $\sqrt{q/(1-q)}z$ of the functions $E_{\beta},E_{\beta}'$ and $E_{\beta}''$,
and the explicit forms of $E_{\beta}'$ and $E_{\beta}''$ are
\begin{eqnarray}
E_{\beta}'(x)=-\frac{1-e^{-\beta}}{\sqrt{2\pi}}e^{-\frac{1}{2}x^2},
\hspace{2mm}
E_{\beta}''(x)\equiv\frac{d^2E_{\beta}(x)}{dx^2}=\frac{1-e^{-\beta}}{\sqrt{2\pi}}xe^{-\frac{1}{2}x^2}.\label{eq3:E's}
\end{eqnarray}
Finally we obtain the AT condition as
\begin{eqnarray}
X_{3}\Wh{X}_{3}=\frac{\alpha}{(1-q)^2}
\frac{
\int Dz 
E_{\beta}^{n}
\left(
\frac{E_{\beta}''}{E_{\beta}}
-
\left(
\frac{E_{\beta}'}{E_{\beta}}
\right)^{2}
\right)^2
}
{
\int Dz 
E_{\beta}^{n}
}
\frac{\int Dz \cosh^{n-4} \sqrt{\Wh{q}}z }{
\int Dz \cosh^n \sqrt{\Wh{q}}z
}
\leq1.
\end{eqnarray}

\subsection{AT instability of the RS2 solution}
The AT condition for the RS2 saddle point is calculated by taking 
the limits $q\to 1$ and $\Wh{q}\to \infty$. 
In the limit $\beta \to \infty$, the function $E_{\beta}(x)$ is reduced to the 
complementary error function $E(x)$ defined in eq.\ (\ref{eq3:H(x)}),
and we need to calculate 
the following integral to obtain the explicit form of 
the AT condition of the RS2 solution
\begin{eqnarray}
\int Dz 
E^{n}
\left(
\frac{E''}{E}
-
\left(
\frac{E'}{E}
\right)^{2}
\right)^2.\label{eq3:H^AT}
\end{eqnarray}
Remembering the argument of $E$ in the AT condition is $\sqrt{q/(1-q)}z$
and using eq.\ (\ref{eq3:asymptotic^H}), 
we find that for $z<0$ the function $E$ goes to $1$ 
and the integrand of eq.\ (\ref{eq3:H^AT}) becomes
\begin{equation}
E^{n}
\left(
\frac{E''}{E}
-
\left(
\frac{E'}{E}
\right)^{2}
\right)^2
\to
\left(\frac{1}{\sqrt{2\pi}}
\sqrt{\frac{q}{1-q}}z e^{-\frac{1}{2}\frac{q}{1-q}z^2}
-
\frac{1}{2\pi}
 e^{-\frac{q}{1-q}z^2}
\right)^2,
\end{equation}
This contribution vanishes in a rapid manner for any $n$.
Meanwhile, for $z>0$ eq.\ (\ref{eq3:H^AT}) becomes 
\begin{equation}
4\int_{0}^{\infty} Dz 
\left(
\sqrt{ \frac{1-q}{q} } \frac{1}{z}
e^{-\frac{1}{2}\frac{q}{1-q}z^2}
 \right)^n,
\end{equation}
which vanishes for $n>0$ but a constant $2$ remains if $n=0$. 
Assuming $0<n<4$, we obtain the AT condition in this limit as
\begin{equation}
\frac{\alpha}{(1-q)^2}
\frac{
4\int_{0}^{\infty} Dz 
\left(
\sqrt{ \frac{1-q}{q} } \frac{1}{z}
e^{-\frac{1}{2}\frac{q}{1-q}z^2}
 \right)^n
}
{
1/2
}
\frac{
	\sqrt{\frac{2}{\pi}}\frac{1}{(4-n)\sqrt{\Wh{q}}} 
}
{
2e^{\frac{1}{2}n^2\Wh{q}}
}\leq 1.
\end{equation}
Recalling eq.\ (\ref{eq3:RS2qhat}) indicating $\Wh{q}\propto 1/\sqrt{1-q}$, 
we can see that this condition is always satisfied for $n>0$ but is violated 
on $n=0$, which was already pointed out by Gardner \cite{GardPer}.

\chapter{Calculations for chapter 4}
\section{Derivation of the generating function $\phi(n)$ and free energy 
of regular random graphs}\label{app:RRG}
At first, we give the explicit form of 
the generating function $\phi(n)$ for the $k$-spin
interacting regular random graph
with coordination number $c$, since the derivation requires 
rather involved calculations.
The generating function $\phi(n)$ is given by
\begin{equation}
\phi^{\rm RRG}(n)=\log[Z^n]_{\V{J}}/N=\frac{c}{k}\log I_{1}-c\log I_{2} +\log I_{3},\label{eq4:betheg}
\end{equation}
  where 
\begin{eqnarray}
&&I_{1}=\left[
(2\cosh \beta J)^n\int \prod_{i=1}^{k} (dh_{i}\pi(h_{i}))
\left( 
\frac{1+\tanh \beta J \prod_{i}^{k}\tanh \beta h_{i} }{2^{k+1}}
\right)^n
\right]_{\V{J}},\label{eq4:I1}\\
&&I_{2}=\int d\Wh{h}dh \Wh{\pi}(\Wh{h})  \pi(h)
\left(
\frac{1+\tanh \beta \Wh{h}  \tanh \beta h}{4} \right)^n , \label{eq4:I2}\\
&&I_{3}=\int \prod_{i=1}^{c}(d\Wh{h_{i}}\Wh{\pi}(\Wh{h_{i}})) \left(
\frac{2\cosh \beta \sum_{i}^{c}\Wh{h}_{i}}{\prod_{i}^{c}2\cosh\beta \Wh{h}_{i}  }
\right)^{n}.\label{eq4:I3}
\end{eqnarray}
The functions $\pi(h_{i})$ and $\Wh{\pi}(\Wh{h}_{i})$ are determined by the 
saddle-point conditions.
The saddle point conditions for $\pi$ and $\Wh{\pi}$ yield
\begin{eqnarray}
&&\Wh{\pi}(\Wh{h})=\int \prod_{i=1}^{k-1}(dh_{i} \pi(h_{i}))
\left[
\delta \left(\Wh{h}-\frac{1}{\beta}\tanh^{-1}(\tanh \beta J \prod_{i}^{k-1}\tanh \beta h_{i} 
) 
\right)
\right]_{\V{J}} \label{eq4:pi},\\
&&\pi(h)\propto \int \prod_{i}^{c-1}(d \Wh{h_{i}} \Wh{\pi}(\Wh{h}_{i}))
\delta \left(h-\sum_{i}^{c-1}\Wh{h}_{i} \right)
\left(
\frac{2\cosh \beta h}{\prod_{i}^{c-1}2 \cosh \beta \Wh{h}_{i}}
\right)^n. \label{eq4:pihat} 
\end{eqnarray}
These relations are identical to those of the cavity field $h$ and bias
$\Wh{h}$ of BLs. Substituting solutions of eqs.\ (\ref{eq4:pi}) and
(\ref{eq4:pihat}) to eq.\ (\ref{eq4:betheg}), we obtain an explicit
expression of $\phi(n)$.

Next, we turn to the derivation of eq.\ (\ref{eq4:betheg}).

The first procedure is to 
calculate the number of possible configurations of a random
graph.
Let us define $a_{\mu}$ as a characteristic function which takes $1$ when a
coupling exists between a $k$-spin 
combination $\mu =\{l_{1},l_{2},\cdots,l_{k}\}$ and $0$ otherwise. 
The number of possible graphs is expressed by using $a_{\mu}$ as
\begin{equation}
G=\sum_{\{a_{\mu}\}}\prod_{l=1}^{N}\delta\left(\sum_{a_{\mu_{l}}}a_{\mu_{l}}-c\right),
\label{eq4:G}
\end{equation}
where $\mu_{l}=\{l,l_{2},\cdots,l_{k}\}$. 
This can be transformed as
\begin{eqnarray}
&&G=\frac{1}{(2\pi i)^N}\oint \prod_{l=1}^{N}\left(\frac{1}{z_{l}}\right)
\sum_{\{a_{\mu}\}}\prod_{l=1}^{N}z_{l}^{\sum_{\mu_{l}}a_{\mu_{l}}-c} \notag \\
&&=\frac{1}{(2\pi i)^N}\oint \prod_{l=1}^{N}\left(\frac{1}{z_{l}^{c+1}}\right)
\exp\left(\sum_{\mu}\log\left(1+z_{l_{1}}z_{l_{2}}
\cdots z_{l_{k}}\right) \right)
,
\end{eqnarray}
where we used an identity 
\begin{equation}
\frac{1}{2 \pi i}\oint dz\frac{1}{z}z^n=\delta_{n,0}, 
\end{equation}
for the first formula. We expand
$\log\left( 1+ z_{l_{1}}z_{l_{2}} \cdots z_{l_{k}} \right)$ 
and keep only the first-order term 
$z_{l_{1}}z_{l_{2}}\cdots z_{l_{k}}$ since 
higher-order terms correspond to
multiple interactions for a combination $\mu$. Hence, 
\begin{eqnarray}
\hspace{-2cm}G=\frac{1}{(2\pi i)^N}\oint \prod_{l=1}^{N}\left(\frac{1}{z_{l}^{c+1}}\right)
\exp\left(\sum_{\mu}z_{l_{1}}z_{l_{2}}
\cdots z_{l_{k}} \right).\label{eq4:Gexp}
\end{eqnarray}
We use an asymptotic formula 
$k!\sum_{\mu}z_{l_{1}}z_{l_{2}} \cdots z_{l_{k}}\approx (\sum_{l=1}^{N}z_{l})^k$ 
 for large $N$
to introduce a variable 
$q=(\sum_{l=1}^{N}z_{l})/N$. Expressing this constraint by a delta
function and using the Fourier expression, we get
\begin{eqnarray}
G\approx 
\frac{1}{(2\pi i)^N}
\int dq d\Wh{q}
\oint \prod_{l=1}^{N}\left(\frac{1}{z_{l}^{c+1}}\right)
\exp\left(\Wh{q}\left(\sum_{l}z_{l}-Nq\right)+ \frac{N^k}{k!}q^k \right).\label{eq4:G1}
\end{eqnarray}
The integration with respect to $z_{l}$ 
can be performed and eq.\ (\ref{eq4:G1}) 
reads
\begin{equation}
G\approx 
\frac{1}{(2\pi i)^N}
\int dq d\Wh{q}
\exp\left( N\log\frac{\Wh{q}^c}{c!}-N\Wh{q}q+ \frac{N^k}{k!}q^k \right).
\end{equation}
The saddle-point method yields the asymptotic behavior of $G$ as
\begin{eqnarray}
&&\Wh{q}=\frac{N^{k-1}}{(k-1)!}q^{k-1},\,\, q=\frac{c}{\Wh{q}},\\
&&\frac{1}{N}\log G=
\frac{c}{k}-c+c\log c+
\frac{c}{k}\log\frac{N^{k-1}}{c(k-1)!}-\log c!.\label{eq4:logG}
\end{eqnarray}

Secondly, we calculate $[Z^n]_{\V{J}}$. The expression using $a_{\mu}$ is
given by
\begin{equation}
[Z^n]_{\V{J}}=\frac{1}{G}
\sum_{\{a_{\mu}\}}\prod_{l=1}^{N}\delta\left(\sum_{a_{\mu_{l}}}a_{\mu_{l}}-c\right)\Tr{\{\V{S}\} } 
\left[
\exp\left(\beta \sum_{\mu}a_{\mu}J_{\mu}
\sum_{\alpha=1}^{n} 
S_{l_{1}}^{\alpha}
S_{l_{2}}^{\alpha} 
\cdots
S_{l_{k}}^{\alpha}\right)
\right]_{\V{J}}.
\end{equation}
Using the same transformation as eqs.\ (\ref{eq4:G})-(\ref{eq4:Gexp}),
we get 
\begin{equation}
[Z^n]_{\V{J}}=\frac{1}{G}\frac{1}{(2 \pi i)^N}\oint \prod_{l}^{N}\left(
dz_{l}\frac{1}{z_{l}^{c+1}}\right) \Tr{\{\V{S}\} }\exp\left(
\sum_{\mu}z_{l_{1}}z_{l_{2}}\cdots
z_{l_{k}}\left[e^{\beta J_{\mu}\sum_{\alpha} S_{l_{1}}^{\alpha}S_{l_{2}}^{\alpha} \cdots
S_{l_{k}}^{\alpha} }\right]_{\V{J}}
\right).
\end{equation}
We introduce the following transformation
\begin{equation}
\left[e^{\beta J_{\mu}\sum_{\alpha} S_{l_{1}}^{\alpha}S_{l_{2}}^{\alpha} \cdots
S_{l_{k}}^{\alpha} }\right]_{\V{J}}=
\Tr{\{\V{ \sigma }\}}
\left[e^{\beta J_{\mu}\sum_{\alpha} 
\sigma_{1}^{\alpha} \sigma_{2}^{\alpha} \cdots
\sigma_{k}^{\alpha} }
\right
]_{\V{J}}
\prod_{i=1}^{k}\delta_{(\V{\sigma}_{i},\V{S}_{l_{i}}) },
\end{equation}
where $\sigma=\pm 1$ denote the additional spin variable and $\delta(\V{\sigma}_{i},\V{S}_{l_{i}} )$ is the indicator function which takes $1$ when 
the spin variables $\V{\sigma}_{i}$ completely accord with $\V{S}_{l_{i}}$ and 
$0$ otherwise.
This formula leads to
\begin{eqnarray}
&&\hspace{-15mm}\sum_{\mu}z_{l_{1}}z_{l_{2}}\cdots
z_{l_{k}}\left[e^{\beta J_{\mu}\sum_{\alpha} S_{l_{1}}^{\alpha}S_{l_{2}}^{\alpha} \cdots
S_{l_{k}}^{\alpha} }\right]_{\V{J}}
=\Tr{\{\V{\sigma}\}}\sum_{\mu}
\prod_{i}^{k}\left(z_{l_{i}}
\delta_{(\V{\sigma}_{i},\V{S}_{l_{i}})}\right)
\left[e^{\beta J_{\mu}\sum_{\alpha} 
\sigma_{1}^{\alpha} \sigma_{2}^{\alpha} \cdots
\sigma_{k}^{\alpha} }
\right]_{\V{J}}
\\
&&\approx
\frac{N^k}{k!}
\Tr{\{\V{\sigma}\}}
\prod_{i}^{k}
\left(Q(\V{\sigma}_{i})\right)
\left[e^{\beta J_{\mu}\sum_{\alpha} 
\sigma_{1}^{\alpha} \sigma_{2}^{\alpha} \cdots
\sigma_{k}^{\alpha} }
\right
]_{\V{J}},
\end{eqnarray}
where 
\begin{equation}
Q(\V{\sigma}_{i})=\frac{1}{N}\sum_{l}z_{l}\delta_{(\V{\sigma}_{i},\V{S}_{l})}.\label{eq4:Qconstraint}
\end{equation}
For each configuration $\V{\sigma}$, we impose the constraint 
(\ref{eq4:Qconstraint}) by a delta function and introduce its
Fourier expression by employing the conjugate variable
$\Wh{Q}(\V{\sigma})$.  
The result is 
\begin{eqnarray}
&&\hspace{-20mm}[Z^n]_{\V{J}}=\frac{1}{G}\frac{1}{(2\pi i)^2}
\oint \prod_{l}^{N}\left(dz_{l}\frac{1}{z_{l}^{c+1}}\right)
\Tr{\{\V{S}\} } \int 
\prod_{\V{\sigma}}\left(
dQ(\V{\sigma})d\Wh{Q}(\V{\sigma})
\right)
\notag\\
&&\hspace{-10mm}
\times\exp\Bigg\{
\Tr{\V{\sigma}}\Wh{Q}(\V{\sigma})\left(
\sum_{l}z_{l}\delta_{(\V{\sigma},\V{S}_{l})}-NQ(\V{\sigma})
\right) \notag
\\
&&\hspace{5mm}
+\Tr{\{\V{\sigma}\}}
\frac{N^k}{k!}\prod_{i}^{k}(Q(\V{\sigma}_{i}))
\left[e^{\beta J_{\mu}\sum_{\alpha} 
\sigma_{1}^{\alpha} \sigma_{2}^{\alpha} \cdots
\sigma_{k}^{\alpha} }
\right
]_{\V{J}}
\Bigg\}
,
\end{eqnarray}
Performing spin trace with respect to $\V{S}$ 
and integration with respect to $z_{l}$, we
get
\begin{eqnarray}
&&\hspace{-15mm}[Z^n]_{\V{J}}=\frac{1}{G}
\int 
\prod_{\V{\sigma}}\left(
dQ(\V{\sigma})d\Wh{Q}(\V{\sigma})
\right)
\notag\\
&&\hspace{-20mm}
\times\exp \left\{
N\log
\frac{ 
\Tr{\V{\sigma}}\Wh{Q}(\V{\sigma}) }{c}
-
N\Tr{\V{\sigma}}\Wh{Q}(\V{\sigma})Q(\V{\sigma})
+
\Tr{\{\V{\sigma}\}}
\frac{N^{k}}{k!}\prod_{i}^{k}(Q(\V{\sigma}_{i}))
\left[e^{\beta J_{\mu}\sum_{\alpha} 
\sigma_{1}^{\alpha} \sigma_{2}^{\alpha} \cdots
\sigma_{k}^{\alpha} }
\right
]_{\V{J}}
\right\}
.\label{eq4:Z}
\end{eqnarray}
The saddle-point method is again useful to evaluate this
equation. Under the RS ansatz of the saddle-point, $Q$ and $\Wh{Q}$ are
generally written in the form
\begin{eqnarray}
&&Q(\V{\sigma})=q\int dh \pi (h)\prod_{\alpha}^{n}\left(
\frac{1+\tanh \beta h \sigma^{\alpha}}{2}
\right),\\
&&
\Wh{Q}(\V{\sigma})=\Wh{q}\int d\Wh{h} \Wh{\pi} (\Wh{h})\prod_{\alpha}^{n}\left(
\frac{1+\tanh \beta \Wh{h} \sigma^{\alpha}}{2}
\right).
\end{eqnarray}
Substituting these formulas to eq.\ (\ref{eq4:Z}), we can calculate
all the terms in the argument
of exponential. The result is 
\begin{equation}
\Tr{\{\V{\sigma}\} }
\frac{N^k}{k!}\prod_{i}^{k}(Q(\V{\sigma}_{i}))
\left[e^{\beta J_{\mu}\sum_{\alpha} 
\sigma_{1}^{\alpha} \sigma_{2}^{\alpha} \cdots
\sigma_{k}^{\alpha} }
\right
]_{\V{J}}=\frac{N^k}{k!}q^k I_{1}, \,\, 
\Tr{\{\V{\sigma}\} } \Wh{Q}(\V{S})Q(\V{S})=q \Wh{q}I_{2}, \,\,
\Tr{\{\V{\sigma}\} } \Wh{Q}(\V{S})=\Wh{q}^c I_{3},
\end{equation}
where $I_{1}$-$I_{3}$ are defined in eqs.\ (\ref{eq4:I1})-(\ref{eq4:I3}). 
Then, eq.\ (\ref{eq4:Z}) is rewritten as
\begin{equation}
\frac{1}{N}\log [Z^n]_{\V{J}}={\rm Extr}\left\{
\frac{N^{k-1}}{k!}q^k I_{1}-q\Wh{q}I_{2}+c\log \Wh{q}-\log c! +\log I_{3}
\right\}-\frac{1}{N}\log G.\label{eq4:RRGmoment}
\end{equation}
Taking the saddle-point conditions with respect to $q$ and $\Wh{q}$, 
we get 
\begin{eqnarray}
\Wh{q}qI_{2}=c=\frac{N^{k-1}}{(k-1)!}q^{k}I_{1},\,\,\, 
\Wh{q}=\frac{c}{qI_{2}}=\frac{c}{I_{2} }
\left(
\frac{I_{1}N^{k-1}}{c(k-1)!}
\right)^{1/k},
\end{eqnarray}
Substituting these conditions and eq.\ (\ref{eq4:logG}) into 
eq.\ (\ref{eq4:RRGmoment}),
we find that irrelevant constants are canceled out and 
obtain eq.\ (\ref{eq4:betheg}). 

The free energy is easily obtained from eq.\ (\ref{eq4:betheg})
\begin{equation}
\left. \Part{\phi^{\rm RRG}(n)}{n}{}\right|_{n=0}=-\beta f^{\rm RRG}=
\frac{c}{k}I_{1}^{'}-c I_{2}^{'}+I_{3}^{'},\label{eq4:RRGf}
\end{equation}
where 
\begin{eqnarray}
&&I_{1}^{'}=\int \prod_{i=1}^{k}(dh_{i}\pi(h_{i}))
\log 
\left(
2\cosh \beta J\frac{1+\tanh \beta J \prod_{i}^{k}\tanh \beta h_{i} }{2^{k+1}}
\right),
\\
&&I_{2}^{'}=\int d\Wh{h}dh \Wh{\pi}(\Wh{h})  \pi(h)
\log \left(
\frac{1+\tanh \beta \Wh{h}  \tanh \beta h}{4} \right),
\\
&&I_{3}^{'}=
\int \prod_{i=1}^{c}(d\Wh{h_{i}}\Wh{\pi}(\Wh{h_{i}}))
\log \left(
\frac{2\cosh \beta \sum_{i}^{c}\Wh{h}_{i}}{\prod_{i}^{c}2\cosh\beta \Wh{h}_{i}  }
\right).
\end{eqnarray}
These distributions $\pi(h)$ and $\Wh{\pi}(\Wh{h})$ satisfy 
eqs.\ (\ref{eq4:pi}) and (\ref{eq4:pihat}) in $n=0$. 
Note that the first term in eq.\ (\ref{eq4:RRGf}) corresponds to the 
bond contribution (\ref{eq4:fij}) and the second and third terms 
are the site contribution (\ref{eq4:fi}), 
which shows the correspondence between the free energy of an RRG 
$f^{\rm RRG}$ 
and 
that of internal part of a BL $f_{\rm I}$ in eq.\ (\ref{eq4:fin}).
 Although 
the second and third terms in eq.\ (\ref{eq4:RRGf}) take 
different mathematical expressions, 
the saddle-point conditions with respect to $\pi(h)$ and $\Wh{\pi}(\Wh{h})$
guarantee the accordance of these terms.

\section{Location of replica zeros of the width-$2$ ladder}\label{app:proof}
We prove that 
all RZs of a $2\times L$ ladder lie on the line 
${\rm Im}(y) =\pi/2$ for any $L$. 
We introduce the notation 
\begin{equation}
p_{l}(x)=p_{l;0}=\frac{n_{l}(x)}{d_{l}(x)},
\end{equation}
where $d_{l}$ and $n_{l}$ are polynomials of $x=e^y$ and 
$n_{l}(x)/d_{l}(x)$ is assumed to be irreducible.
The outline of the proof is as follows. First we present the general
solution of $p_{l}$ and show that
the denominator $d_{l}$ has $2F((l+1)/2)$ roots which are all
purely imaginary. 
The function $F(l)$ is the floor function, which is defined to return 
the maximum integer $i$ in the range $i \leq l$.     
Also, we show that 
the number of nontrivial solutions of $\Xi_{l}=0$ is 
equal to $2F((l+1)/2)$ and
$\Xi_{l}$ can be factorized as $C_{l}(x)d_{l}(x)$, where $C_{l}(x)$ is
a polynomial of $x$. From the correspondence of the numbers of the roots, 
we conclude that all the zeros of $\Xi_{l}$ are 
equivalent to the roots of $d_{l}(x)$ and $C_{l}(x)$ takes the form $ax^{b}$. 
  
The iteration for $p_{l}$ (\ref{eq4:T0piJ}) has 
a solvable form and its general solution is given by
\begin{equation}
p_{l}=\frac{2(4^l-h(x)^l)}{4^{l}(2+x^2-x\sqrt{x^2+8})-(2+x^2+x\sqrt{x^2+8})h(x)^l},\label{eq4:p_l}
\end{equation}
where
\begin{equation}
h(x)\equiv -4-x(x+\sqrt{x^2+8})=4\frac{x+\sqrt{x^2+8} }{x-\sqrt{x^2+8}}.\label{eq4:h(x)}
\end{equation}
The roots of the numerator in eq.\ (\ref{eq4:p_l}) can be easily calculated as 
\begin{equation}
x=\left\{
		    \begin{array}{cc}
		     \pm 2\sqrt{2}i  \hspace{2cm} (l=2m+1) \\
		 \hspace{-0.5cm}  0, \pm 2\sqrt{2}i \hspace{2cm} (l=2m)
		    \end{array}
			  \right.\label{eq4:nroots},
\end{equation}
where $i$ denotes the imaginary 
unit and $m$ is a natural number. Then, we concentrate on finding 
the roots of the denominator in eq.\ (\ref{eq4:p_l}) 
except for those of the numerator (\ref{eq4:nroots}). 
From numerical observations in sec.\ \ref{sec4:RZ^ladder}, 
we found that any of the roots $x^{*}$,
which satisfy
$\Xi_{l}(x^{*})=0$, are purely imaginary and bounded 
by $|x^{*}|\leq 2\sqrt{2}$. Hence, we assume these conditions  
and perform the variable transformation
$z=-xi$. Equating the denominator of eq.\ (\ref{eq4:p_l}) to $0$, we get 
\begin{equation}
\left(\frac{h(-iz)}{4}\right)^l=\left(\frac{\sqrt{8-z^2}+iz}{\sqrt{8-z^2}-iz} \right)^l
=\frac{2-z^2-i\sqrt{8-z^2}}{2-z^2+i\sqrt{8-z^2}}\label{eq4:droots}.
\end{equation}
We now enumerate the
number of solutions under conditions that $z$ is real and bounded as $-2\sqrt{2} \leq z \leq 2 \sqrt{2}$. Under these conditions, we can transform eq.\ (\ref{eq4:droots}) into a simple form by using the polar representation. The result is 
\begin{equation}
e^{i(2\theta_{1}-\pi)l}=e^{i2\theta_{2}},\label{eq4:droots2}
\end{equation} 
where  
$\sqrt{8-z^2}+iz=r_{1}e^{i\theta_{1}}\,\,(-\pi<\theta_{1}\leq\pi)$
 and 
$r_{2}e^{i\theta_{2}}=2-z^2-i\sqrt{8-z^2}\,\,(-\pi<\theta_{2}\leq\pi)$.
While $z$ varies from
$-2\sqrt{2}$ to $2\sqrt{2}$ continuously, the radius $r_{1}$ stays at
a constant $2 \sqrt{2}$ and the argument $\theta_{1}$ varies
from $-\pi/2$ to $\pi/2$ in the positive direction. In the same
situation, $\theta_{2}$ changes from $+\pi$ to $-\pi$ in the negative
direction. The radius $r_{2}$ is not constant, but is finite in this 
range.
The variables $\theta_{1}$ and $\theta_{2}$ are obviously 
continuous and monotonic functions of $z$.      
Therefore, the argument of the left-hand side of eq.\ (\ref{eq4:droots2})
starts from $\theta=0$ and rotates with angle $2l\pi$ 
in the positive direction and
the counterpart of the right-hand side varies from
the same point $\theta=0$ to $-4\pi$. 
This means that there are $l+1$ values of $z$ where the
factor $(2\theta_{1}(z)-\pi)l$ becomes equal to $2\theta_{2}(z)$ 
except for
trivial solutions $z=\pm 2\sqrt{2}$. When $l$ is even, 
these solutions contain a trivial solution $z=0$, which can also be 
confirmed from eq.\ (\ref{eq4:p_l}). 
Hence, the number of nontrivial roots
of $d_{l}$ becomes $l+1$ for odd $l$ and $l$ for even $l$, which is
equivalent to $2F((l+1)/2)$. 

As already noted,
 the number of nontrivial solutions of $\Xi_{l}=0$ is equal
to $2F((l+1)/2)$. This can be understood by considering 
that the number of terms of $[Z^n]_{\V{J}}$ is
 determined by the maximum number of defects $n_{d}$.
In the $2\times l$ ladder case, the value of $n_{d}$ is given by 
$F((l+1)/2)$ and the number of terms is $n_{d}+1$. The highest degree of
the relevant polynomials for RZs comes from the difference between the
highest and lowest ground-state energies and is  
given by $2n_{d}=2F((l+1)/2)$, which 
yields the number of nontrivial solutions of $\Xi_{l}=0$.

Finally, we prove that $\Xi_{l}$ takes the form $A_{l}x^{b_{l}} d_{l}(x)$ by
 induction. From eqs.\ (\ref{eq4:T0piJ}) and (\ref{eq4:Xi^ladder}) with the
 initial conditions $p_{0;0}=0,\,\,\Xi_{0}=x$, 
we derive 
\begin{equation}   
p_{1}=\frac{1}{x^2+1},\,\, \Xi_{1}= \frac{1}{2}x^2 (x^2+1),
\end{equation}
which satisfies the desired form. Assuming that 
the condition $\Xi_{l}=Ax^{b_{l}} d_{l}(x)$ 
is true for $l=k$, we substitute this expression into eq.\ (\ref{eq4:Xi^ladder})
to get 
\begin{eqnarray}
&&\Xi_{k+1}=Ax^{b_{k}}d_{k}x^3\left\{
\frac{n_{k}}{d_{k}}+\frac{1}{2}\left(1+\frac{1}{x^2}
\right) \left(1-\frac{n_{k}}{d_{k}}\right)
\right\}\cr
&&=\frac{1}{2}Ax^{b_{k}+1}\left\{
(x^2-1)n_{k}+(1+x^2)d_{k}\right\}.\label{eq4:Xi_{k+1}}
\end{eqnarray}
Equation (\ref{eq4:T0piJ}) can be written as  
\begin{equation}
p_{k+1}=\frac{d_{k}-n_{k}}{(x^2-1)n_{k}+(1+x^2)d_{k}}=\frac{n_{k+1}}{d_{k+1}},
\end{equation}
which gives
\begin{equation}
(x^2-1)n_{k}+(1+x^2)d_{k}=c_{k+1}(x)d_{k+1}(x),
\end{equation}
where $c_{k+1}$ is a polynomial and satisfies
 $c_{k+1}=(d_{k}-n_{k})/n_{k+1}$. Substituting this relation,
 we can rewrite eq.\ (\ref{eq4:Xi_{k+1}}) as
\begin{equation}
\Xi_{k+1}=\frac{1}{2}Ax^{b_{k}+1}c_{k+1}(x)d_{k+1}(x).
\end{equation}
As we have already shown, the number of nontrivial zeros of
$\Xi_{k+1}$ is equal to that of $d_{k+1}$. This means that $c_{k+1}$ cannot have
nontrivial roots and hence $c_{k+1}$ takes the form
$Ax^b$. This completes the proof by induction and demonstrates our proposition that
all RZs for the $2\times L$ ladder have a constant imaginary 
part $i\pi/2$.

\section{Rate function for a CT with $c=3$}\label{app:rate}
We here calculate the generating function
$\phi_{L}(y)$ for finite $L$. 
Consider an $L$-generation branch of a $c=3$ CT. 
An explicit form $\phi_{L}(y)$ is
easily derived from eq.\ (\ref{eq4:T0Zh1}) as 
\begin{equation}
\phi_{L}(x)=\frac{ 2^L }{ 2^{L+1}-1 } \phi_{0}
+\frac{ 2^{L+1}}{2^{L+1}-1}(1-2^{-L})\log x
+\frac{1}{4-2^{-L+1}}\sum_{i=0}^{L-1} \frac{\log f_{i}}{2^{i}},
\end{equation}
where $x=e^{y}$ and
\begin{equation}
\phi_{0}=\log \Xi_{0},\,\, f_{i}=f_{i}(x,p_{i;0})=1-\frac{1}{2}(1-x^{-2})(1-p_{i;0})^2,
\end{equation}
using the same notations as in section \ref{sec4:GF}.  
The rate function with finite generations $L$ is given by
\begin{eqnarray}
R_{L}(x)=\frac{2^L}{2^{L+1}-1}\left(\phi_{0}-x\log x \frac{d \phi_{0}}{d x}\right) \cr
+\frac{1}{4-2^{-L+1}} \sum_{i=0}^{L-1}\frac{1}{2^{i}f_{i}} \left(
f_{i}\log f_{i}-C_{i}(x)x \log x  
\right), \label{eq4:sigmaL}
\end{eqnarray}
where the factor $C_{i}(x)$ is given by
\begin{equation}
C_{i}(x)=
\Part{f_{i}}{p_{i;0}}{} \frac{d p_{i;0}}{d x}+\Part{f_{i}}{x}{}
=(1-x^{-2})(1-p_{i;0}) \frac{d p_{i;0}}{d x} -x^3(1-p_{i;0})^2
.
\end{equation}

Let us denote $R_{\infty}(x)=\lim_{L \to \infty} R_{L}(x)$. 
Because the inequality $R_{\infty}\leq 0$ always holds, the 1RSB
transition does not occur as long as the condition
$R_{\infty}(x)=R(x)$ is satisfied. 

In the range $y \geq 0 \Leftrightarrow  1 \leq x$, 
the factor $f_{i}$ 
is bounded as $1/2\leq f_{i} \leq 1$. This guarantees the uniform convergence of $\phi_{L}(x)$. 
The boundedness of $(d p_{i;0}/d x)$ can also be shown 
with some calculations. 
These conditions guarantee that 
$R_{L}(x)$ converges to a function $R_{\infty}$ uniformly. 
Hence, from elementary calculus, 
the equality $R(x)=R_{\infty}(x)$ holds, which implies the
absence of 1RSB. 
The same conclusion is more easily
derived for a BL 
because $f_{i}$ does not depend on $i$.

\section{AT condition for the $(k,c)=(2,3)$ Bethe lattice}\label{app:AT^BL}
We derive the AT condition for a BL. Especially, 
we explain the case $(k, c) = (2, 3)$ in
detail. The extension to general $(k, c)$ 
is somewhat involved but straightforward.
To evaluate $P_{(G\rightarrow 0)}$, we construct the transition
matrix of our random-walk problem.  
For a given $(\Wh{h}_{g},\Wh{h}_{g+1})$, 
the posterior distribution of $r_g$ is given
as
\begin{equation}
p(r_{g}|\Wh{h}_{g})=p(r_{g},\Wh{h}_{g})/p(\Wh{h}_{g})
\propto e^{ y(|r_g+\Wh{h}_{g}|-|r_{g}|-|\Wh{h}_{g}|)}  p(r_{g}),
\label{eq4:p(r|h)}
\end{equation}
where $p(r_{g})$ is the prior distribution of $r_{g}$. This enables us to
derive the concrete expression of $p(r_{g}|\Wh{h}_{g})$, summarized
in Table \ref{tab:pb}.
\begin{table}[hbt]
    \begin{center}
\begin{tabular}{c|c|c|c}
\noalign{\hrule height 0.8pt}
$ r_{g}$ $\backslash$ $\Wh{h}_{g}$& 1  & 0 & $-1$  \\
\hline
1 & 
 $\displaystyle \frac{1-p_{b}}{(1+p_{b})+(1-p_{b})e^{-2y}}$
 & $\displaystyle \frac{1-p_{b}}{2}$&
 $\displaystyle \frac{(1-p_{b})e^{-2y}}{(1+p_{b})+(1-p_{b})e^{-2y}}$ \\
\hline
0 & $\displaystyle \frac{2p_{b}}{(1+p_{b})+(1-p_{b})e^{-2y}}$ &
 $\displaystyle p_{b}$
 &$\displaystyle \frac{2p_{b}}{(1+p_{b})+(1-p_{b})e^{-2y}}$ \\
\hline
$-1$ & 
 $\displaystyle \frac{(1-p_{b})e^{-2y}}{(1+p_{b})+(1-p_{b})e^{-2y}}$
& $\displaystyle \frac{1-p_{b}}{2}$&
 $\displaystyle \frac{1-p_{b}}{(1+p_{b})+(1-p_{b})e^{-2y}}$ \\
\noalign{\hrule height 0.8pt}
\end{tabular}
\end{center}
\caption{Values of $p(r_{g}|\Wh{h}_{g})$ for $(k,c)=(2,3)$. The symbol $p_{b}$ is
 the probability that the cavity bias takes the value $0$. } \label{tab:pb}
\end{table}
We can distinguish three states of the walker at the $g$-step as follows:
\begin{description}
\item[$\ket{1}$:] The walker has already vanished. 
\item[$\ket{2}$:] The walker survives and $|\Wh{h}_{g}|=1$. 
\item[$\ket{3}$:] The walker survives and $|\Wh{h}_{g}|=0$.
\end{description}
Hence, using the relation (\ref{eq4:transition}), 
the transition matrix $T$ can be written as
\begin{equation}
T=
 \left(      
   \begin{array}{ccc}
    1 & p_{1,1} & \frac{1}{2}p_{1,0} \times 2 \\
    0 & p_{0,1} & \frac{1}{2}p_{1,0} \times 2 \\
    0 & p_{-1,1} & p_{0,0}
	 \end{array}
 \right),
\end{equation}  
where $p_{r_{g},\Wh{h}_{g}}$ represents $p(r_{g}|\Wh{h}_{g})$ and  
the condition $p_{r_{g},\Wh{h}_{g}}=p_{-r_{g},-\Wh{h}_{g}}$ applies. 
When $|\Wh{h}_{g}+r_{g}|=1 $ and  $|\Wh{h}_{g}|=0$, 
the states $\ket{1}$ and $\ket{2}$ occur 
with equal probability $1/2$, 
while $\ket{2}$ is always 
chosen when $|\Wh{h}_{g}+r_{g}|=1 $ and $|\Wh{h}_{g}|=1$.
This is due to the correlation between 
$\Wh{h}_{g}$ and $\delta \Wh{h}_{g}$.
To see this, let us consider the
evolutional equations for biases and perturbations. 
Their explicit forms are 
\begin{eqnarray}
&&\Wh{h}_{g-1}={\rm sgn}\left (J_g (\Wh{h}_g+r_g) \right ),\\ 
&&\delta \Wh{h}_{g-1}=\Part{ \Wh{h}_{g-1} }{\Wh{h}_{g}}{} \delta \Wh{h}_{g}.
\end{eqnarray} 
From eqs.\ (\ref{BPzerotemp}) and 
(\ref{eq4:transition}), 
we can obtain 
all possible values of $\Wh{h}_{g-1}$ and $\partial \Wh{h}_{g-1}/\partial \Wh{h}_{g}$, and summarize the result under the condition $J_{g}=1$
in Tables \ref{tab:bias} and \ref{tab:dbias}. 
A notation $q_{g}(\pm|u)$ represents the conditional probability for $\sgn{\delta \Wh{h}_{g}}=\pm $ under the condition $\Wh{h}_{g}=u$. 
\begin{table}[hbt]
\hspace{-10mm}
\begin{minipage}[t]{0.48\hsize}
\begin{center}
\hspace{12mm}
\begin{tabular}{c|c|c|c}
\noalign{\hrule height 0.8pt}
\vspace{0.2mm}
$ r_{g}$ $\backslash$ $\Wh{h}_{g}$& 1  & 0 & $-1$  \\
\hline
\multirow{2}{*}{1}& \multirow{2}{*}{1} & \multirow{2}{*}{1}&\multirow{2}{*}{0}\\           &                    & &   \\
\hline
\multirow{2}{*}{0}& \multirow{2}{*}{1} & \multirow{2}{*}{0}&\multirow{2}{*}{-1}\\           &                    & &   \\
\hline
\multirow{2}{*}{-1}& \multirow{2}{*}{0} & \multirow{2}{*}{-1}&\multirow{2}{*}{-1}\\           &                    & &   \\
\noalign{\hrule height 0.8pt}
\end{tabular}
\hspace{20mm}
\caption{Values of $\Wh{h}_{g-1}$ with $J_{g}=1$.} \label{tab:bias}
\end{center}
\end{minipage}
\hspace{2mm}
 \begin{minipage}[t]{0.48\vsize}
\vspace{-21mm} 
\begin{center}
\hspace{-12mm}
\begin{tabular}{c|c|c|c}
\noalign{\hrule height 0.8pt}
\vspace{0.2mm}
$ r_{g}$ $\backslash$ $\Wh{h}_{g}$& 1  & 0 & $-1$  \\
\hline

\multirow{2}{*}{1}& \multirow{2}{*}{0} & 0 ($q_{g}(+|0)$)& \multirow{2}{*}{1}\\
                  &                    & 1 ($q_{g}(-|0)$)&   \\
\hline

\multirow{2}{*}{0}& 0 ($q_{g}(+|1)$) &\multirow{2}{*}{1}& 1 ($q_{g}(+|-1)$)\\
                  & 1 ($q_{g}(-|1)$) &                  & 0 ($q_{g}(-|-1)$) \\

\hline
\multirow{2}{*}{-1}&\multirow{2}{*}{1} & 1 ($q_{g}(+|0)$)& \multirow{2}{*}{0}\\
                   &                   & 0 ($q_{g}(-|0)$)&   \\
\noalign{\hrule height 0.8pt}
\end{tabular}
\hspace{20mm}\caption{Values and probabilities ($q_{g}(\pm|u)$) of $\partial \Wh{h}_{g-1}/\partial \Wh{h}_{g}$ with $J_{g}=1$.} \label{tab:dbias}
\end{center}
\end{minipage}
\end{table}
Comparing the Tables \ref{tab:bias} and \ref{tab:dbias}, 
we can find that 
when $\Wh{h}_{g-1}=1$, $\delta \Wh{h}_{g-1}=
(\partial \Wh{h}_{g-1}/\partial \Wh{h}_{g}) \delta h_{g}$ becomes never positive. 
Similarly, 
when $\Wh{h}_{g-1}=-1$ holds, 
$\delta \Wh{h}_{g-1}$ is positive. 
It is easy to see that these facts hold for $J_{g}=-1$.
Hence, an inequality 
$\Wh{h}_{g-1}\cdot \delta h_{g-1}\leq 0$ is 
necessarily required, 
which explains 
the transition rules between the states $\ket{1}$ and $\ket{2}$.

The matrix $T$ has three eigenvalues: 
$\lambda_{1}=1,\,\lambda_{2},$ and $\lambda_{3}$.  
The eigenvector of the 
largest eigenvalue $\lambda_{1}=1$ corresponds to the state $\ket{1}$ or
the vanishing state. 
Hence, the surviving probability $P_{(G\rightarrow 0)}$ is given by 
$1-\braket{1}{G}$, where $\ket{G}$ is the state of the walker at the $G$ step.
For large $G$, the relevant state is of the second-largest eigenvalue
$\lambda_{2}$, and we get 
\begin{equation}
P_{(G \rightarrow 0)}\approx \lambda_{2}^{G}.
\end{equation}
Using the stationary solution (\ref{eq4:pb2c3}), we obtain
$P_{(G\rightarrow 0)}$ as a function of $x=e^{y}$. 
The AT condition becomes 
\begin{equation}
\chi_{SG}\propto \sum_{G}(k-1)^G (c-1)^G P_{(G\rightarrow 0)} \rightarrow \infty
\Leftrightarrow 
(k-1)(c-1)\lambda_{2}>1.
\end{equation}
This condition is easily examined numerically and we can verify that 
the AT instability occurs at $y_{AT}\approx 0.54397$ for $(k,c)=(2,3)$. 

Before closing this section, we give some brief comments on general $(k, c)$.
For general $c$ under the case $k = 2$, the evolution equation of the cavity bias becomes
\begin{eqnarray}
\Wh{h}_{g-1}=\frac{1}{\beta}\tanh^{-1}
\left(
\tanh(\beta J_{g})
\tanh
\left\{
\beta
\left(
\Wh{h}_{g}+\sum_{j=1}^{c-1}r_{g,j}
\right)
\right\}
\right)
\nonumber
\\
\to
{\rm sgn}\left(
J_{g}\left(
\Wh{h}_{g}+\sum_{j=1}^{c-1}r_{g,j}
\right)
\right)
\hspace{3mm}
(\beta\to \infty)
\end{eqnarray}
This equation implies that the table \ref{tab:dbias} is modified and becomes a new table between
$\Wh{h}_{g}$ and $r_{g}=\sum_{j=1}^{c-1}r_{g,j}$. 
Other discussions are almost the same as the case $c=3$. 
Using the notation $p(r_{g}|\Wh{h}_{g})=p_{r_g, \Wh{h}_g}$
and the symmetry $p_{r_{g},\Wh{h}_g} = p_{-r_g,-\Wh{h}_g}$
as the $c = 3$ case, the
transition matrix $T$ can be written as
\begin{equation}
T=
 \left(      
   \begin{array}{ccc}
    1 & \sum_{r=1}^{c-1}p_{r,1}+\sum_{r=2}^{c-1}p_{-r,1} 
		& \sum_{r=2}^{c-1}(p_{r,0}+p_{-r,0})+\frac{1}{2}(p_{1,0}+p_{-1,0})  \\
    0 & p_{0,1} & \frac{1}{2}(p_{1,0}+p_{-1,0}) \\
    0 & p_{-1,1} & p_{0,0}
	 \end{array}
 \right).
\end{equation}  
Note that the term $p_{-2,1}$ in the second column is included in the transition rate from $\ket{2}$ to $\ket{1}$. This is because the inequality 
$\Wh{h}_{g-1}\cdot \delta h_{g-1}\leq 0$ holds in the general $c$ cases.
The second largest eigenvalue $\lambda_{2}$ of $T$ determines the AT condition by $(c-1)\lambda_{2}=1$. 

For the $k=3$ case, the update rule of the cavity bias is 
\begin{eqnarray}
\Wh{h}_{g-1}=\frac{1}{\beta}\tanh^{-1}
\left(
\tanh(\beta J_{g})
\tanh(\beta f_{g})
\tanh
\left\{
\beta
\left(
\Wh{h}_{g}+\sum_{j=1}^{c-1}r_{g,j}
\right)
\right\}
\right)
\nonumber
\\
\to
{\rm sgn}\left(
J_{g}f_{g}\left(
\Wh{h}_{g}+\sum_{j=1}^{c-1}r_{g,j}
\right)
\right)
\hspace{3mm}
(\beta\to \infty),\label{eq4:k3gc}
\end{eqnarray}
where $f_{g}$ denotes the cavity field coming from the other branch of the tree and $r_{g}$ again represents $\sum_{j=1}^{c-1}r_{g,j}$. Since the cavity field $f_{g}$ is uncorrelated with the biases $(\Wh{h}_{g},\{r_{g,j} \})$, it is sufficient to consider the correlation among the biases $(\Wh{h}_{g},\{r_{g,j} \})$ as eq.\ (\ref{eq4:p(r|h)}). This fact leads to a similar list to table \ref{tab:dbias}. Difference between the $k = 2$ and $3$ cases appears
in the transition matrix. According to eq.\ (\ref{eq4:k3gc}), when the cavity field $f_{g}$ equals to $0$, $\partial \Wh{h}_{g-1}/\partial \Wh{h}_{g}$ becomes 
$0$  and the walker vanishes independently of the values of $(\Wh{h}_{g},\{r_{g,j} \})$. This yields the transition matrix as 
\begin{equation}
T=
 \left(      
   \begin{array}{ccc}
    1 & 1-(1-p_{f})(p_{0,1}+p_{-1,1})	& 1-(1-p_{f})
\left(
\frac{1}{2}(p_{1,0}+p_{-1,0})+p_{0,0}
\right)
\\
    0 & (1-p_{f})p_{0,1} & (1-p_{f})\frac{1}{2}(p_{1,0}+p_{-1,0}) \\
    0 & (1-p_{f})p_{-1,1} & (1-p_{f})p_{0,0}
	 \end{array}
 \right),
\end{equation}  
where $p_{f}$ denotes the probability that the cavity field $f_{g}$ becomes $0$\footnote{For the $(k, c) = (3, 3)$ case, the probability $p_{f}$ is already given as the convergent solution of eq.\ (\ref{eq4:33CFD}),
or the solid line in fig.\ \ref{fig:apb3c3}.}.
The second largest
eigenvalue of this matrix $\lambda_{2}$ again gives the AT condition 
by $(3-1)(c-1)\lambda_{2}=1$.

\chapter{Calculations for chapter 5}
\section{Divergence point of the spin-glass susceptibility at zero temperature }\label{app:AT^BL2}
Our starting point is the self-consistent equation 
satisfied by the 
convergent CBD $\Wh{P}(\Wh{h})$ in the absence of the 
external field
\begin{equation}
\Wh{P}(\Wh{h}_{i})=\int 
\prod_{j=1}^{c-1}
\left(
d\Wh{h}_{j}\Wh{P}(\Wh{h}_{j})
\right)
\left[
\delta\left(
\Wh{h}-\frac{1}{\beta}\tanh^{-1}\left\{
\tanh \beta J \tanh 
\left(\beta 
\sum_{i=1}^{c-1}\Wh{h}_{i} 
\right)
\right\}
\right)
\right]_{\V{J}},\label{eq:cavity-bias distribution}
\end{equation}
Analytical assessment of the solution of this equation 
is possible only at zero temperature as in appendix \ref{app:AT^BL}, 
which enables us to treat the problem in an analytical manner. 

The evolution equation of the cavity bias is given by 
\begin{eqnarray}
&&\Wh{h}_{g-1}=\frac{1}{\beta} 
\tanh^{-1}\left (\tanh(\beta J_g) \tanh(\beta (\Wh{h}_g+r_g))\right )\cr 
&&\to  \left \{
\begin{array}{ll}
{\rm sgn}\left (J_g (\Wh{h}_g+r_g) \right ) &(\,\, |\Wh{h}_g+r_g| \ge 1 \,\,) \cr
J_g (\Wh{h}_g+r_g)  &(\,\, \mbox{otherwise} \,\,) 
\end{array} \right .
(\,\,\beta\to \infty\,\,), 
\label{eq:BPzerotemp}
\end{eqnarray}
where $J_g$ denotes the coupling between sites $g-1$ and $g$. 
This relation implies 
the cavity fields and biases become integers, 
because we are treating the $\pm J=1$ model. 
Hence, the general form of the CBD can be written as
\begin{equation}
\Wh{P}(\Wh{h})=\sum_{k=-1}^{1}p_{k}\delta(\Wh{h}-k).\label{eq:Pofh}
\end{equation}
For the BL with $c=3$, 
eq.\ (\ref{eq:cavity-bias distribution}) is rewritten as
\begin{eqnarray}
&&p_{1}=p(p_{1}^2+2p_{0}p_{1})+(1-p)(p_{-1}^2+2p_{0}p_{-1}),\\
&&p_{0}=p_{0}^2+2p_{1}p_{-1},\\
&&p_{-1}=p(p_{-1}^2+2p_{0}p_{-1})+(1-p)(p_{1}^2+2p_{0}p_{1}).
\end{eqnarray}
These equations have a ferromagnetic solution  
for $ 7/8\leq p\leq 1$, the explicit form of which is given by
\begin{eqnarray}
&&p_{1}=\frac{4p-3}{2(2p-1)}+\frac{1}{2}\sqrt{\frac{32p^2-52p+21}{(2p-1)^2} },\label{eq:p1}\\&&p_{0}=\frac{1}{2}(2p_{1}+1)-\frac{1}{2}\sqrt{12p_{1}^2-4p_{1}+1},\label{eq:p0}\\
&&p_{-1}=1-p_{1}-p_{0},\label{eq:pm1}
\end{eqnarray}
and have another solution $p_{-1}=p_{1}=p_{0}=1/3$ for $p<7/8$, 
which can be regarded as the RS spin-glass solution. 

Equations (\ref{eq:BPzerotemp}) and (\ref{eq:Pofh}) means
\begin{eqnarray}
\left |
\frac{\partial \Wh{h}_{g-1}}{
\partial \Wh{h}_g}
\right |=
\left \{
\begin{array}{ll}
0 & (\,\, |\Wh{h}_{g}+r_g|> 1 \,\,)\\
0 \mbox{ or $1$} & (\,\, |\Wh{h}_{g}+r_g|= 1\,\,) \\
1 & (\,\, {\rm otherwise}\,\,)
\end{array}
\right.,  \label{eq:transition}
\end{eqnarray}
where the value of $0$ or $1$ for the case of $|\Wh{h}_{g}+r_{g}|=1$ is determined depending on the value of $\Wh{h}_{g}$. 
Equation (\ref{eq:transition}) indicates that 
the assessment of eq.\ (\ref{chainrule})
is analogous to an analysis of a random-walk 
which is bounded by absorbing walls. 
We denote by $P_{(G \to 0)}$ the probability that 
$\left |\partial \Wh{h}_{g-1}/\partial \Wh{h}_g\right |$ 
never vanishes during the walk from $G$ to $0$ and 
the value of 
$\prod_{g=1}^G
|\partial \Wh{h}_{g-1}/\partial \Wh{h}_g|$ is kept to unity. 
This indicates that 
the critical condition at zero temperature becomes 
\begin{eqnarray}
\log (c-1) +\lim_{G \to \infty} \frac{1}{G} 
\log P_{(G \rightarrow 0)} =0.
\label{AT}
\end{eqnarray}

To evaluate the surviving probability $P_{(G \rightarrow 0)}$, 
let us formulate this random-walk problem. Since we are now considering 
the case that the ferromagnetic bias exists, which is different from 
appendix \ref{app:AT^BL},
we should in principle distinguish the following seven states of the walker at the $g$-step:
\begin{description}
\item[$\ket{1}$:] The walker has already vanished. 
\item[$\ket{2\pm}$:] The walker survives under the conditions $\Wh{h}_{g}=1$ and $ \sgn{ \delta \Wh{h}_{g} }=\pm $. 
\item[$\ket{3\pm}$:] The walker survives under the conditions $\Wh{h}_{g}=-1$ and $\sgn{\delta \Wh{h}_{g}}=\pm $.
\item[$\ket{4\pm}$:] The walker survives under the conditions $\Wh{h}_{g}=0$ and $ \sgn{\delta \Wh{h}_{g}}=\pm $.
\end{description} 
These seven states, however, can be reduced to four states 
by considering the relation 
$\Wh{h}_{g-1}\cdot \delta h_{g-1}\leq 0$ which can be 
derived by the same discussions as in appendix \ref{app:AT^BL}.
As a result, the state $\ket{2\pm}$ and $\ket{3\pm}$ are reduced to
$\ket{2}\equiv \ket{2-}$ and $\ket{3}\equiv\ket{3+}$, respectively.
On the other hand, when $\Wh{h}_{g-1}=0$ 
the sign of $\delta \Wh{h}_{g}$ does not change 
with probability $p$ or change with probability $1-p$,
which leads to the following equations
\begin{eqnarray}
q_{g-1}(+|0)=pq_{g}(+|0)+(1-p)q_{g}(-|0),\\  
q_{g-1}(-|0)=pq_{g}(-|0)+(1-p)q_{g}(+|0),
\end{eqnarray}
where $q_{g}(\pm|u)$ represents the conditional probability for $\sgn{\delta \Wh{h}_{g}}=\pm $ under the condition $\Wh{h}_{g}=u$, which is the same notation 
as appendix \ref{app:AT^BL}.
Under the assumption that 
the initial perturbation is completely random, 
we find that $q_{g}(\pm|0)$ equals to $1/2$ for 
$\forall{g}$. 
Because of this symmetry, 
we need not to distinguish the states $\ket{4\pm}$ and 
hereafter write as $\ket{4}$.
Considering the above discussions and 
using the relation (\ref{eq:transition}), we can write 
the transition matrix $T$ of the walker as
\begin{equation}
T=
 \left(      
   \begin{array}{cccc}
    1 & p_{1} & p_{-1} & \frac{1}{2}\left(p_{1}+p_{-1}\right) \\
    0 & pp_{0} & (1-p)p_{0}& \frac{1}{2}\left(pp_{1}+(1-p)p_{-1}\right) \\
    0 & (1-p)p_{0} & p p_{0}&\frac{1}{2}\left(pp_{-1}+(1-p)p_{1}\right) \\
    0 &p_{-1} & p_{1} & p_{0}
	 \end{array}
 \right),\label{eq:T}
\end{equation}  
where $T_{ij}$ denotes the transition probability from $\ket{j}$ to $\ket{i}$.
When $|r_{g}|=1$ and $\Wh{h}_{g}=0$, the probability flow from 
$\ket{4}$ to $\ket{1}$ exists with the weight $1/2$ 
while there is no such flow when $r_{g}=0$ and 
$|\Wh{h}_{g}|=1$. This is the consequence from the above considerations.    
This matrix has four eigenvalues $\lambda_{1}=1,\lambda_{2},\lambda_{3}$ and $\lambda_{4}$. The eigenvector of the largest eigenvalue $\lambda_{1}=1$ corresponds to the state $\ket{1}$ or the vanishing state. Hence, the surviving probability $P_{(G\to0)}$ is given by $1-\braket{1}{g=0}$, where $\ket{g=0}$ is the state of the walker at the $g=0$ step. For large $G$, the relevant state is of the second-largest eigenvalue $\lambda_{2}$ and $P(G\to0)\propto \lambda_{2}^{G}$, 
which leads to the critical condition
\begin{equation}
(c-1)\lambda_{2}=1.
\end{equation}
For $c=3$, this condition is easily examined by using eqs.\ (\ref{eq:p1})-(\ref{eq:pm1}) and (\ref{eq:T}). 
The resultant transition point is $p=0.92067$, 
which well agrees with the results given in the main text.


\end{document}